\documentclass[12pt]{article}
 \usepackage{array}
\usepackage{enumerate}
\usepackage{amsmath}
\usepackage{booktabs}
\usepackage{hyperref}
\usepackage{makecell}
\usepackage{siunitx}
\renewcommand{\arraystretch}{1.2}

\usepackage{listings}
\usepackage{color}
\usepackage{float} 

\usepackage{amssymb,amsmath}
\usepackage{listings} 
\usepackage{url,graphicx,subfig}  
\usepackage{mathrsfs} 
\usepackage{bm} 
\usepackage[margin=1.5cm]{geometry}
\usepackage{pdflscape} 
\usepackage{booktabs}
\usepackage{adjustbox} 
\usepackage{tikz}
\usepackage{tikz}
\usetikzlibrary{decorations.pathmorphing} 
\usetikzlibrary{decorations.markings}     
\usetikzlibrary{decorations.text}         
\usetikzlibrary{decorations.pathreplacing} 
\usepackage{booktabs}

\newtheorem{Remark}{Remark}

\newcommand{\bxi}{\mbox{\boldmath$\xi$}}
\newcommand{\bpi}{\mbox{\boldmath$\pi$}}

\newcommand{\balpha}{\mbox{\boldmath$\alpha$}}

\title{Quasi-birth-and-death processes evolving within trees: Applications to comparative phylogenetics}

\date{\normalsize November 5, 2025}

\author{
Habtu Kiros Nigus\thanks{University of Tasmania, Australia, email: habtukiros.nigus@utas.edu.au}
\and
	Barbara R. Holland\thanks{Australian Research Council Centre of Excellence for Plant Success, University of Tasmania, Australia, email: barbara.holland@utas.edu.au}
\and
	Ma{\l}gorzata M. O'Reilly\thanks{University of Tasmania, Australia, email: malgorzata.oreilly@utas.edu.au}	
}

\begin{document}

%
%

\maketitle

\begin{abstract}
We consider a quasi-birth-and-process (QBD) that duplicates itself at some fixed times within a tree that contains information about duplication times and potentially partially observed states.  We analyse a continuous trait by discretising it to obtain the QBD level variable. Then, the phase variable is used to model the dynamics of the underlying environment.

Here, we extend the framework of Soewongsono et al.~\cite{soewongsono2025matrix} to enable a more general analysis. We develop an efficient recursive algorithm for computing the likelihood of an observed tree under this model and construct several numerical examples to illustrate its application potential. Through our synthetic data examples, we show a range of potential behaviours that could be modelled with this approach. Further, we apply the framework to two empirical examples from comparative phylogenetics (the evolution of range area and body size traits across a phylogeny of 49 mammals) to gain different insights into the evolution of these continuous traits. In this setting duplication of the QBD represents speciation and continuous trait evolution is modelled in a discretised state space. In our empirical examples, we explore the impact of different parameter choices on the corresponding likelihood of observing a given phylogenetic tree and the observed levels at its tips.

\end{abstract}




\section{Introduction}\label{Sec:Intro}

We consider a stochastic process in which a quasi-birth-and-death process (QBD) duplicates itself at some time, and from then on the two copies evolve independently. The information about the evolution of the resulting family of the QBDs is then recorded in the form of a tree, in which each node corresponds to a duplication, and a branch starts at the time of a duplication event. Suppose that the data in the form of a tree topology with some observable data at the tips is given, which contains the branch lengths. We are interested in finding suitable QBDs that maximise the likelihood of observing the tree and states at the tips of the tree. For an overview of the key results for QBDs in the theory of matrix-analytic methods the reader is referred to Latouche and Ramaswami~\cite{latouche1999introduction}, Neuts~\cite{Neuts_1981}, Ramaswami~\cite{rama96new}, He~\cite{he2014fundamentals}, Joyner and Fralix~\cite{2016JF}, Bean, Pollett and Taylor~\cite{QSLDQBDNEW}, Bright and Taylor~\cite{BrighTaylorNEW}, and Phung-Duc et al.~\cite{phung2010simpleNEW}.

Such a process of duplicating QBDs was previously described by Soewongsono et al. in~\cite{soewongsono2025matrix}, who used it to study gene duplication and loss within a species tree. In their paper the authors used a specific QBD with a special structure that matched the biological situation of gene duplication, but here we wish to establish more general results that could be applied to a range of problems in queuing theory and comparative phylogenetics. So in this paper, we develop methodology that could be used to model a wide variety of behaviors with QBD parameters fit via numerical optimization.

One area where such a duplicating QBD process could be applied is comparative phylogenetics and this is what motivated the development of this paper. Comparative phylogenetics studies the evolution of traits along an evolutionary tree.
Understanding the evolution of traits in phylogenetic trees is crucial for several reasons. It is important for conservation biology, as it helps in designing effective strategies. 
It helps predict how species might respond to different stresses, such as climate change, habitat fragmentation, or disease spread. 
Influential models for single continuous traits include the Brownian motion model introduced in this context by Felsenstein \cite{felsenstein1985phylogenies} and the Ornstein Uhlenbeck (OU) model in Hansen~\cite{hansen1997stabilizing} and later in Butler and King~\cite{butler2004phylogenetic}.
Brownian motion models a trait evolving neutrally, i.e. without directional selection. 
The OU models are popular for modeling a trait under stabilizing selection, i.e. a trait for which there is some optimal value such that fitness declines as you get further away from the optimum. OU processes have both a stochastic drift component and a deterministic mean-reverting component which moves the trait value towards the optimum at a rate proportional to how far away from the optimum it currently is.
More recent papers have investigated models where the trait can evolve according to different regimes in different clades on the evolutionary tree.
For example the SURFACE method of Ingram and Mahler~\cite{ingram2013surface} considers different OU models operating in different parts of the evolutionary tree. Beaulieu et al~\cite{beaulieu2012modeling} expanded the OU model to allow for different drift rates, selection strengths, and trait optima in different parts of the tree. However, these models have faced criticism for being over-parameterized and statistically unidentifiable as noted by Ho and An{\'e} in~\cite{ho2014intrinsic}.
QBD models seem to be a promising area to explore further in this context as they provide a flexible structure with phases representing environmental regimes (which could be hidden) and levels corresponding to a trait of interest, thereby allowing modeling of trait evolution in which transition rates may depend on level, phase, or both.

The key contributions of our paper are:
\begin{itemize}
\item An approach to modelling the evolution of traits through a duplicating QBD process.
\item A novel efficient algorithm for recursive computation of the likelihood of observing a tree, which is more convenient than the algorithm in Soewongsono et al.~\cite{soewongsono2025matrix}. 
\item A range of QBD models for the evolution of traits within species and their analysis.
\item Application examples to synthetic data that illustrate the modelling potential to a wide range of behaviours and the methodology to draw useful insights from probabilistic metrics, such as drift.
\item Applications to two empirical examples from comparative phylogenetics.
\end{itemize}

This paper is organised as follows. Section~\ref{sec:OurModel} introduces the QBD model and details the discretisation of a continuous trait evolving along a single  branch. Section~\ref{sec:Theory} describes the algorithms we developed to compute the tree likelihood. In Section~\ref{sec:QBDmodels}, we describe a suite of QBDs and discuss the numerical optimisation strategy. In Sections~\ref{sec:NumExSynthetic} and~\ref{sec:NumExData} we present applications of the QBD models to the analysis of synthetic and empirical datasets, respectively, illustrating model performance and biological insights obtained. Finally, in Section~\ref{Sec:Conclusions} we discuss potential directions for future research.

\section{Trait evolving along a single branch: QBD model}\label{sec:OurModel}

To model the evolution of a continuous trait of species evolving along a single branch within a phylogenetic tree, as described by Soewongsono et al. in~\cite{soewongsono2025matrix}, we apply the following approach.

First, assuming that the values of the continuous trait are bounded and have a value between $a$ and $b$, we discretise the interval $[a,b]$ into levels labelled $n=0,\ldots,N$, so that level $n=0,\ldots,N-1$ corresponds to the interval $[a+\frac{n(b-a)}{N+1},a+\frac{(n+1)(b-a)}{N+1})$, and level $N$ corresponds to the interval $[a+\frac{N(b-a)}{N+1},b]$.

Next, we construct a quasi-birth-and-death process (QBD) $\{(X(t),\varphi(t)):t\geq 0\}$~\cite{he2014fundamentals,latouche1999introduction} such that the level variable $X(t)$ represents the level of the trait at time $t$, while the phase variable $\varphi(t)$ models the underlying factors that control the behavior of the process such as the direction and strength of selection on the trait.

The QBD is a Markov chain process  with state space 
\begin{eqnarray*}
\{
(n,i):n=0,1,2\ldots,N;i=1,\ldots,K
\},
\end{eqnarray*}
some initial distribution $\balpha=[\alpha_{(n,i)}]$, $\alpha_{(n,i)}=\mathbb{P}(X(0)=n,\varphi(0)=i)$, and generator 
\begin{eqnarray*}\scriptsize
	{\bf Q}
	= 
	[{\bf Q}^{[n,n']}]
	=
	\begin{bmatrix}
		{\bf Q}^{[0,0]} & {\bf Q}^{[0,1]} & {\bf 0} & \cdots & \cdots & \cdots & \cdots & {\bf 0}\\
		{\bf Q}^{[1,0]} & {\bf Q}^{[1,1]} & {\bf Q}^{[1,2]} & {\bf 0} & \cdots & \cdots & \cdots & {\bf 0}\\
		{\bf 0} & {\bf Q}^{[2,1]} & {\bf Q}^{[2,2]} & {\bf Q}^{[2,3]} & \cdots &  \cdots & \cdots & {\bf 0}\\
		\vdots & \vdots & \vdots & \vdots & \cdots & \cdots & \cdots & \vdots\\
		{\bf 0} & {\bf 0} & {\bf 0} & {\bf 0} & \cdots & {\bf Q}^{[N-1,N-2]} & {\bf Q}^{[N-1,N-1]} & {\bf Q}^{[N-1,N]}\\
		{\bf 0} & {\bf 0} & {\bf 0} & {\bf 0} & \cdots & {\bf 0} & {\bf Q}^{[N,N-1]} & {\bf Q}^{[N,N]}
	\end{bmatrix},
\end{eqnarray*}
where matrices ${\bf Q}^{[n,n^{'}]}=[q_{(n,i)(n',j)}]_{i,j=1,2,...,K}$ record transition rates $(n,i)\to(n',j)$.

We assume that the state space of the QBD is irreducible and so, since the state space of the QBD is finite, there exists the stationary distribution $\bpi=[\pi(n,\varphi)]$, $\pi(n,\varphi)=\lim_{t\to\infty}\mathbb{P}(X(t)=n,\varphi(t)=\varphi)$.

Below, we state the key expressions useful for the evaluation of the distribution at time $t$ summarised in Aksamit, O’Reilly, Palmowski~\cite{Aksamit2024}. 
Suppose that a QBD starts from level $n_0$ in some phase~$i_0=1,\ldots,K$ according to the initial distribution $\balpha=[\alpha_j]$, $\alpha_j=\mathbb{P}(\varphi(0)=j)$.
Then the distribution at time $t$ vector ${\bf f}(t)=[{\bf f}_n(t)]_{n=0,1,\ldots,N}$ records probabilities
$[{\bf f}_n(t)]_j = \mathbb{P}_{\balpha}(X_t=n,\varphi(t)=j)
$ that the process is in state $(n,j)$ at time $t$. To evaluate ${\bf f}_n(t)$, we may invert the Laplace Transforms $[\widetilde{\bf f}_n(s)]_j= \int_{0}^{\infty} e^{-st} \mathbb{P}_{{\balpha}}(X_t=n,\varphi(t)=j)dt$ using standard numerical inversion algorithms (see e.g. Den~Iseger~\cite{DenIseger_2006} or Horvath et al.~\cite{horvath2020numerical}), by applying
\begin{eqnarray*}
\widetilde{\bf f}_{n_0}(s)&=&
-{\balpha}
\left(
({\bf Q}^{[n_0,n_0]}-s{\bf I})
+
\widehat {\bf R}^{(n_0-1)}(s){\bf Q}^{[n_0-1,n_0]}
+
\widetilde{\bf R}^{(n_0+1)}(s){\bf Q}^{[n_0+1,n_0]}
\right)^{-1}
\\
\widetilde{\bf f}_{n}(s)&=&
-{\balpha}
{\bf H}^{n_0,n}(s)
\left(
({\bf Q}^{[n,n]}-s{\bf I})
+
\widehat {\bf R}^{(n-1)}(s){\bf Q}^{[n-1,n]}
+
\widetilde{\bf R}^{(n+1)}(s){\bf Q}^{[n+1,n]}
\right)^{-1},\quad n>n_0
,
\\
\widetilde{\bf f}_{n}(s)&=&
-{\balpha}
{\bf G}^{n_0,n}(s)
\left(
({\bf Q}^{[n,n]}-s{\bf I})
+
\widehat {\bf R}^{(n-1)}(s){\bf Q}^{[n-1,n]}
+
\widetilde{\bf R}^{(n+1)}(s){\bf Q}^{[n+1,n]}
\right)^{-1},\quad n<n_0
\end{eqnarray*}
where
\begin{eqnarray*}
	\widetilde{\bf R}^{(n)}(s) &=& -{\bf Q}^{[n-1,n]}({\bf Q}^{[n,n]}-s{\bf I}+\widetilde{\bf R}^{(n+1)}(s){\bf Q}^{[n+1,n]})^{-1},\quad \widetilde{\bf R}^{(N+1)}(s)={\bf 0},\\
 \widehat {\bf R}^{(n)}(s) &=& -{\bf Q}^{[n+1,n]}(\widehat{\bf R}^{(n-1)}(s){\bf Q}^{[n-1,n]}+{\bf Q}^{[n,n]}-s{\bf I})^{-1},\quad \widehat {\bf R}^{(-1)}(s)={\bf 0},
\end{eqnarray*}
and, for any $k$ such that $n+1\leq n+k\leq N$,
\begin{eqnarray*}
{\bf H}^{n,n+k}(s)&=&
{\bf H}^{n,n+1}(s)
{\bf H}^{n+1,n+2}(s)
\times
\cdots
\times
{\bf H}^{n+k-1,n+k}(s)
,\\
{\bf H}^{n,n+1}(s)&=&
-({\bf Q}^{[n,n]}-s{\bf I}
+{\bf Q}^{[n,n-1]}{\bf H}^{n-1,n}(s)
)^{-1}
{\bf Q}^{[n,n+1]}
,\quad {\bf H}^{-1,0}(s)={\bf 0},
\end{eqnarray*}
and, for any $k$ such that  $0\leq n-k\leq n-1$,
\begin{eqnarray*}
{\bf G}^{n,n-k}(s)&=&
{\bf G}^{n,n-1}(s)
{\bf G}^{n-1,n-2}(s)
\times
\cdots
\times
{\bf G}^{n-k+1,n-k}(s),
\\
{\bf G}^{n,n-1}(s)&=&
-({\bf Q}^{[n,n]}-s{\bf I}
+{\bf Q}^{[n,n+1]}{\bf G}^{n+1,n}(s)
)^{-1}
{\bf Q}^{[n,n-1]},
\quad {\bf G}^{N+1,N}(s)={\bf 0}.
\end{eqnarray*}

\section{Likelihood of observing certain traits at the tips of a phylogenetic tree: Algorithmic approach}\label{sec:Theory}

We develop a methodology for the computation of the likelihood of the phylogenetic tree, given the tree topology with branch lengths and trait values observed at the tips. In the initial Algorithm 1, built on the ideas developed by Soewongsono et al. in~\cite{soewongsono2025matrix}, we assume that distance from the root of the tree to each tip is the same for all tips. This implies the tree is clock-like. In the phylogenetics literature this is referred to as the ultrametric condition. In the improved Algorithm 2, we remove this assumption, and also develop a more general idea, leading to a more convenient computation.

\subsection{Algorithm 1}\label{Subsec:Alg1}

First, consider a simple motivating example: a phylogenetic tree with two branches, each of length $t$, that started with a branching event at time $t_B<t$. To model this, assume a QBD with initial distribution row vector ${\bf f}(t_B)$. Suppose that we observe trait levels $n_L$ and $n_R$ at the left and right tip of the tree, respectively. Then the corresponding likelihood is,
\begin{eqnarray}\label{eq:like_simple}
{\bf f}(t_B)
\times 
\Big(
 \left(
 {\bf P}_{n_L}(t-t_B)
  \times 
{\bf 1}
\right)
	\odot
 \left(
{\bf P}_{n_R}(t-t_B)
\times 
{\bf 1}
\right)
\Big)
,
\end{eqnarray}
where ${\bf 1}$ is a column vector of ones and ${\bf P}(t)$ is the conditional distribution matrix at time $t$,
\begin{eqnarray}
{\bf P}(t)
&=&
\left[ 
P_{(n,i),(n',j)}(t)
\right]_{n,n'=0,\ldots,N;,i,j=1,\ldots,K} 
\end{eqnarray}
such that the $(n,i),(n',j)$ entry of ${\bf P}(t)$ is given by
\begin{eqnarray}
P_{(n,i),(n',j)}(t)
&=&
\mathbb{P}
(X(t)=n',\varphi(t)=j
\ |\ 
X(0)=n,\varphi(0)=i
)
,
\end{eqnarray}
and for any $m$ (e.g. $m=n_L$ or $m=n_R$),
\begin{eqnarray}
{\bf P}_m(t)
&=&
\left[ 
P_{(n,i),(n',j)}(t)
\right]_{n'=m,n=0,\ldots,N;,i,j=1,\ldots,K} 
\end{eqnarray}
is a matrix collecting the conditional probabilities corresponding to observing trait level $m$ at time $t$. 


Note that $ \left(
 {\bf P}_{n_L}(t-t_B)
  \times 
{\bf 1}
\right)
	\odot
 \left(
{\bf P}_{n_R}(t-t_B)
\times 
{\bf 1}
\right)
$ is a column vector recording the conditional likelihood of observing the traits $n_L$ and $n_R$ at the same time $t$, and so~\eqref{eq:like_simple} is a scalar.

\begin{figure}[!htbp]
\centering
			\begin{tikzpicture}[>=stealth,redarr/.style={->},scale=1.5]

    \draw (3.5,9.5) node[anchor=north, below=-0.17cm] {{\color{black} $T^*$ }};
    \draw [solid] (3.5,8.9) -- (3.5,9.1);


			\draw [dashed] (-1.0,9) -- (6,9);
			\draw (-1.5,9) node[anchor=north, below=-0.3cm] {{\color{black} $t_1$ }};

            \draw (1.5,9.25) node[anchor=north, below=-0.3cm] {{\color{black} ${T^*}^{(0,1)}(t_1,t)$ }};
            \draw (5.5,9.25) node[anchor=north, below=-0.3cm] {{\color{black} ${T^*}^{(1,1)}(t_1,t)$ }};


             \draw (-0.4,7) node[anchor=north, below=-0.3cm] {{\color{black} ${T^*}^{(0,2)}(t_2,t)$ }};
             \draw (1.6,7) node[anchor=north, below=-0.3cm] {{\color{black} ${T^*}^{(1,2)}(t_2,t)$ }};
             \draw (4.6,7) node[anchor=north, below=-0.3cm] {{\color{black} ${T^*}^{(2,2)}(t_2,t)$ }};

			\draw [dashed] (-1.0,8) -- (6,8);
			\draw (-1.5,8) node[anchor=north, below=-0.3cm] {{\color{black} $t_2$ }};

			\draw [dashed] (-1.0,5.5) -- (6,5.5);
			\draw (-1.5,5.5) node[anchor=north, below=-0.3cm] {{\color{black} $t$ }};

			\draw [green,very thick] (1.5,9) -- (5.5,9);


			\draw [green,very thick] (5.5,9) -- (5.5,5.5);

             \filldraw  (5.4,9) rectangle (5.6,8);
             \filldraw [fill=none] (5.4,8) rectangle (5.6,5.5);

			\draw [green,very thick] (1.5,9) -- (1.5,8);

            \filldraw (1.4,9) rectangle (1.6,8);

			\draw [green,very thick] (0.5,8) -- (2.5,8);

			\draw [green,very thick] (2.5,8) -- (2.5,5.5);

             \filldraw [fill=none] (2.4,8) rectangle (2.6,5.5);
			
			\draw [green,very thick] (0.5,8) -- (0.5,5.5);

             \filldraw [fill=none] (0.4,8) rectangle (0.6,5.5);
						
			\end{tikzpicture}
			\caption{Species tree $T^*$ (figure adapted from Soewongsono et al.~\cite{soewongsono2025matrix} to the model considered here). Branching event at time $t_1$ results in two subtrees, ${T^*}^{(0,1)}(t_1,t)$ and ${T^*}^{(1,1)}(t_1,t)$. Next, another branching event occurs at time $t_2$, which results in two subtrees of ${T^*}^{(0,1)}(t_1,t)$, being ${T^*}^{(0,2)}(t_2,t)$ and ${T^*}^{(1,2)}(t_1,t)$, respectively. The corresponding part of the ${T^*}^{(1,1)}(t_1,t)$ that starts at time $t_2$ is denoted ${T^*}^{(2,2)}(t_2,t)$.}
			\label{fig:RecNotation}
	\end{figure}
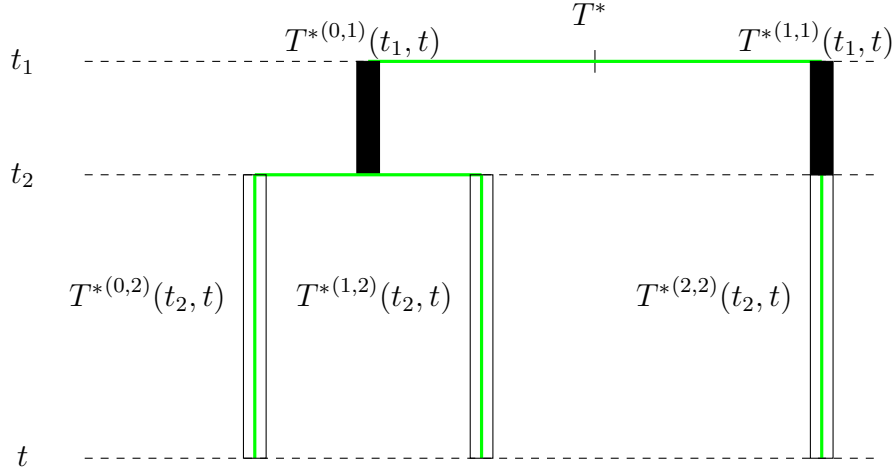

We generalise the above idea to the recursive formula for the likelihood of the phylogenetic tree as follows. Denote by $f(T^*; (t_1,\ldots,t_B);t)$ the (scalar) likelihood of the tree $T^*$ with $B+1$ different species and branching events at times $t_1,\ldots,t_B$ respectively, that occurred before the current time $t$. Further, denote by ${\bf f}({T^*}^{(k,b)}(t_b,t))$ the conditional (row vector) likelihood of the $(k,b)$-th subtree that starts at time $t_b$, $b=0,\ldots,B$, and ends at time $t$, counting $k=0,1,\ldots ,b$ from the left to the right on the tree, see Figure~\ref{fig:RecNotation}.

We then have the following recursive formula, 
\begin{eqnarray}
		f(T^*;
		(t_1,\ldots,t_B);t)
	&=&
  		{\bf f}(t_1)
	\times
	\left({\bf f}({T^*}^{(0,1)}(t_1,t)) \odot {\bf f}({T^*}^{(1,1)}(t_1,t))\right),
	\label{rec_all_Paper2}
\end{eqnarray}
where, for $b=1,\ldots,B-1$, assuming that the speciation at time $t_b$ occurs on the $\ell_b$-th subtree ${T^*}^{(\ell_b,b-1)}$ for some $\ell_b=0,\ldots,b-1$,
\begin{eqnarray}
	{\bf f}({T^*}^{(k,b)}(t_b,t))&=&
	\left\{
	\begin{array}{ll}
 {\bf P}(t_{b+1}-t_b)\times 
	{\bf f}({T^*}^{(k,b+1)}(t_{b+1},t))
	& \quad k< \ell_b,\bigskip
	\\
	\left(
 {\bf P}(t_{b+1}-t_b)\times
	{\bf f}({T^*}^{(k,b+1)}(t_{b+1},t))
	\right)
	&\\
	\odot
	\left(
  {\bf P}(t_{b+1}-t_b)\times
	{\bf f}({T^*}^{(k +1,b+1)}(t_{b+1},t))
	\right)
	& \quad k= \ell_b,\bigskip
	\\
  {\bf P}(t_{b+1}-t_b)\times
	{\bf f}({T^*}^{(k+1,b+1)}(t_{b+1},t))
	& \quad k> \ell_b,
	\end{array}	
	\right.
	\label{rec_n_Paper2}
	\end{eqnarray}
 and
 \begin{eqnarray}
	{\bf f}({T^*}^{(k,B)}(t_B,t))&=&
	\left\{
	\begin{array}{ll}
 {\bf P}_{m(k,B)}(t-t_B)\times {\bf 1}
	& \quad k< \ell_B,\bigskip
	\\
 \left(
 {\bf P}_{m(k,B)}(t-t_B)
  \times 
{\bf 1}
\right)
	\odot
 \left(
{\bf P}_{m(k+1,B)}(t-t_B)
\times 
{\bf 1}
\right)
	& \quad k= \ell_B,\bigskip
	\\
 {\bf P}_{m(k+1,B)}(t-t_B)\times {\bf 1}
	& \quad k> \ell_B,
	\end{array}	
	\right.
	\label{rec_N_Paper2}
	\end{eqnarray}
where $m(k,B)$ is the level (trait) observed at the $k$-th leftmost tip of the phylogenetic tree with $B$ branching events at times $t_1,\ldots,t_B$ and a root at time $t_0$.

\subsection{Algorithm 2}\label{Subsec:Alg2}

The required input to Algorithm 1 is the information about the speciation times $t_1,\ldots,t_B$, the ordering of the subtrees $({T^*}^{(k,b)}(t_b,t)$ for $k=0,\ldots,b$, and the assumption is that all external branches terminate at the same time~$t$. Since trees in which external branches may terminate at different times often occur, for example, in datasets where we have information on trait values for fossils, we may remove this assumption and rewrite~\eqref{rec_N_Paper2} as
\begin{eqnarray}
	{\bf f}({T^*}^{(k,B)}(t_N,t))&=&
	\left\{
	\begin{array}{ll}
 {\bf P}_{m(k,B)}(b_k)\times {\bf 1}
	& \quad k< \ell_N,\bigskip
	\\
 \left(
 {\bf P}_{m(k,B)}(b_k)
  \times 
{\bf 1}
\right)
	\odot
 \left(
{\bf P}_{m(k+1,B)}(b_{k+1})
\times 
{\bf 1}
\right)
	& \quad k= \ell_B,\bigskip
	\\
 {\bf P}_{m(k+1,N)}(b_k)\times {\bf 1}
	& \quad k> \ell_B,
	\end{array}	
	\right.
	\label{rec_N_Paper2mod}
	\end{eqnarray}
where we additionally assume that $b_k$ is the full length of the external branch $k$, rather than the length $(t-t_B)$ of its part from the most recent speciation event on the tree at time $t_B$, unlike in Algorithm 1.

Furthermore, it is more convenient to represent the input in a format in which we record the parent nodes, the children nodes, the connections between the various nodes, and the distances between them. For example, the topology of the tree in Figure~\ref{fig:DataExample1} corresponding to data discussed later in Section~\ref{sec:NumExSynthetic}, can be represented in a matrix form as
\begin{equation}
\begin{array}{ccc}
\text{Parent node}&\text{Child node}&\text{Distance}\\\hline\hline
5& 1& 1\\\hline
5& 2& 1\\\hline
6& 5& 1\\\hline
6& 3& 2\\\hline
7& 6& 1\\\hline
7& 4& 3
\end{array}\label{eq:inout}
\end{equation}
where parent node $5$ points at child $1$, with the distance between them being $1$, and so on.

Therefore, in our improved Algorithm 2, using the input matrix such as in~\eqref{eq:inout}, we follow these steps. 
\begin{enumerate}[(1)]
    \item First, for each tip of the phylogenetic tree, evaluate the column vector recording the conditional likelihood of observing the trait level at that tip, given the distribution of the phases at the start of the corresponding branch.
    \item Next, for each pair of the tips with the same parent, take Hadamard product of their column vectors, resulting in one column vector for each such pair.
    \item Further, left-multiply each column vector by a conditional probability matrix corresponding to the branching event directly above, resulting in a column vector again.
    \item Repeat (2) until left with two column vectors. Take their Hadamard product and left-multiply by the initial distribution (row vector) of phases at the parent node.
\end{enumerate}

So, rather than perform recursive computations for each subinterval $(t_{b-1},t_b)$ as in~\eqref{rec_n_Paper2}, in this improved Algorithm 2, we perform computations for each branch considering its full length, similarly to the approach applied to a different Markov model by Felsenstein in~\cite{felsenstein1981evolutionary}. 

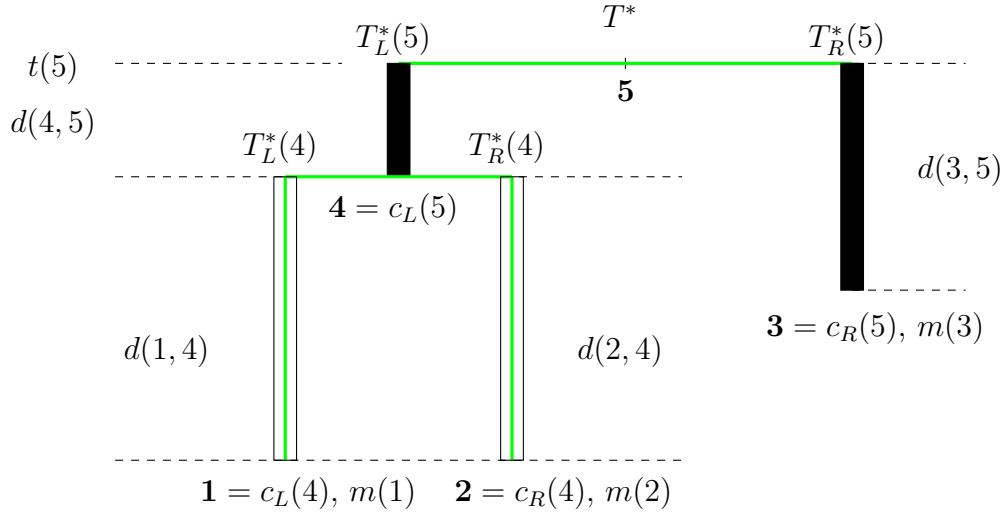
\begin{figure}[H]
\centering
			\begin{tikzpicture}[>=stealth,redarr/.style={->},scale=1.5]

    \draw (3.5,9.5) node[anchor=north, below=-0.17cm] {{\color{black} $T^*$ }};


			\draw [dashed] (-1.0,9) -- (1,9);
			\draw (-1.5,9) node[anchor=north, below=-0.3cm] {{\color{black} $t(5)$ }};

            \draw (1.5,9.25) node[anchor=north, below=-0.3cm] {{\color{black} $T^*_L(5)$ }};
            \draw (5.5,9.25) node[anchor=north, below=-0.3cm] {{\color{black} $T^*_R(5)$ }};

       \draw (-1.5,8.5) node[anchor=north, below=-0.3cm] {{\color{black} $d(4,5)$ }};
      \draw (0.5,8.3) node[anchor=north, below=-0.3cm] {{\color{black} $T^*_L(4)$ }};
      \draw (2.5,8.3) node[anchor=north, below=-0.3cm] {{\color{black} $T^*_R(4)$ }};
         \draw (6.5,8.1) node[anchor=north, below=-0.3cm] {{\color{black} $d(3,5)$ }};

            \draw (-0.5,6.5) node[anchor=north, below=-0.3cm] {{\color{black} $d(1,4)$ }};
            \draw (3.5,6.5) node[anchor=north, below=-0.3cm] {{\color{black} $d(2,4)$ }};
    \draw [dashed] (-1.0,5.5) -- (4,5.5);
    \draw [dashed] (2.0,8) -- (4,8);

 \draw (0.75,5.25) node[anchor=north, below=-0.3cm] {{\color{black} ${\bf 1}=c_L(4)$, $m(1)$ }};
\draw (3,5.25) node[anchor=north, below=-0.3cm] {{\color{black} ${\bf 2}=c_R(4)$, $m(2)$ }};
\draw (5.75,6.7) node[anchor=north, below=-0.3cm] {{\color{black} ${\bf 3}=c_R(5)$, $m(3)$ }};
\draw (1.5,7.75) node[anchor=north, below=-0.3cm] {{\color{black} ${\bf 4}=c_L(5)$ }};
\draw (3.55,8.75) node[anchor=north, below=-0.3cm] {{\color{black} ${\bf 5}$ }};

\draw (3.5,8.95) -- (3.5,9.05);

			\draw [dashed] (-1.0,8) -- (1,8);

			\draw [dashed] (5.5,7) -- (6.5,7);
            \draw [dashed] (5.5,9) -- (6.5,9);

			\draw [green,very thick] (1.5,9) -- (5.5,9);



             \filldraw  (5.4,9) rectangle (5.6,7);

			\draw [green,very thick] (1.5,9) -- (1.5,8);

            \filldraw (1.4,9) rectangle (1.6,8);

			\draw [green,very thick] (0.5,8) -- (2.5,8);

			\draw [green,very thick] (2.5,8) -- (2.5,5.5);

             \filldraw [fill=none] (2.4,8) rectangle (2.6,5.5);
			
			\draw [green,very thick] (0.5,8) -- (0.5,5.5);

             \filldraw [fill=none] (0.4,8) rectangle (0.6,5.5);

			\end{tikzpicture}
			\caption{Species tree $T^*$ (notation in Algorithm 2): The tree starts at time $t(5)$ with parent node $5$ who has (left and right) children nodes $4=c_L(5)$ and $3=c_R(5)$. Nodes $1$, $2$, and $3$ are the tips of the tree, with $1=c_L(4)$ and $2=c_R(4)$ being the (left and right) children nodes of their parent node $4$. Trait levels $m(1)$, $m(2)$ and $m(3)$ are observed at tips $1$, $2$, and $3$, respectively. Tip $3$ is some fossil record.}
			\label{fig:RecNotation2}
	\end{figure}

Now, denote by $t(k)$ the time corresponding to node $k$ at which a speciation occurs, resulting in two subtrees, the left subtree $T^*_L(k)$ and the right subtree $T^*_R(k)$, with corresponding likelihoods $ {\bf f}(T^*_L(k)) $ and $ {\bf f}(T^*_R(k)) $ respectively. Also, denote by $c_L(k)$ and $c_R(k)$, the two children of node $k$, being the nodes below $k$ on the left and the right subtree, respectively. Further, let $d(c,p)$ be the distance between child node $c$ and its parent $p$ (the length of the branch that starts at parent node $p$ and terminates at child node $c$), see Figure~\ref{fig:RecNotation2}.

Then, the recursive formula for Algorithm 2 is written as,
\begin{eqnarray}
		f(T^*;p)
	&=&
  		{\bf f}(t(p))
	\times
	\left(
    {\bf f}(T^*_L(p)) 
    \odot 
   {\bf f}(T^*_R(p) 
    \right),
	\label{rec_all_Paper2A2}
\end{eqnarray}
where $p$ is the parent node at the root of the tree corresponding to branching event at time $t(p)$, and for any subtree $T^*_{\bullet}(k)$, $\bullet=L,R$, corresponding to an internal node $k$,
\begin{eqnarray}
	{\bf f}(T^*_{\bullet}(k))&=&
	\left(
 {\bf P}(d(c_{\bullet}(k),k))\times
	{\bf f}(T^*_L(c_{\bullet}(k))
	\right)
	\odot
	\left(
  {\bf P}(d(c_{\bullet}(k),k))\times
	{\bf f}(T^*_R(c_{\bullet}(k))
	\right)
\label{rec_n_Paper2A2}
\end{eqnarray}
whenever $c_{\bullet}(k)$ is also an internal node, and
\begin{eqnarray}
	{\bf f}(T^*_{\bullet}(k))&=&
 {\bf P}_{m(c_{\bullet}(k))}(d(c_{\bullet}(k),k))
    \times 
{\bf 1}
\label{rec_n_Paper2A2b}
\end{eqnarray}
whenever $c_{\bullet}(k)$ is one of the tips of the tree and $m(c_{\bullet}(k))$ is the trait level observed at that tip.

In the practical application of the recursive formula for Algorithm 2, we evaluate the quantities in the direction from the tips to the root of the tree. For example, to evaluate the likelihood of the tree depicted in Figure~\ref{fig:RecNotation2}, we first evaluate the likelihoods of the two (left and right) subtrees starting from the internal node $4$,
$${\bf f}(T^*_L(4))=
{\bf P}_{m(1)}(d(1,4))\times{\bf 1}
,\quad 
{\bf f}(T^*_R(4))=
{\bf P}_{m(2)}(d(2,4))\times{\bf 1}
$$
then the two (left and right) subtrees starting from the internal node $5$,
$${\bf f}(T^*_L(5))=
{\bf P}(d(4,5))\times
\left(
{\bf f}(T^*_L(4))
\odot 
{\bf f}(T^*_R(4))
\right)
,\quad
{\bf f}(T^*_R(5))=
{\bf P}_{m(3)}(d(3,5))\times{\bf 1},
$$
and finally,
$${\bf f}(T^*;5)=	{\bf f}(t(5))
	\times
	\left(
    {\bf f}(T^*_L(5))
	\odot
{\bf f}(T^*_R(5))
\right).$$

\begin{Remark}
We note that our algorithm can be readily modified to include information about the internal nodes by replacing block matrices ${\bf P}_{m(c_{\bullet}(k))}(d(c_{\bullet}(k),k))$ with a suitable range of columns.
\end{Remark}


\section{Numerical Examples: QBD models}\label{sec:QBDmodels}

Below, we present several examples to illustrate the application of our methodology for computing the likelihood of observing the phylogenetic species tree given its topology and trait values observed at its tips.
In the examples of the QBD models considered in this section, we assume that the block matrices ${\bf Q}^{[n,n+1]}$, ${\bf Q}^{[n,n-1]}$, and ${\bf Q}^{[n,n]}$ are the same for all $n=1,\ldots.N-1$, and that ${\bf Q}^{[0,0]}={\bf Q}^{[n,n]}+{\bf Q}^{[n,n-1]}$ and ${\bf Q}^{[N,N]}={\bf Q}^{[n,n]}+{\bf Q}^{[n,n+1]}$, with $N=100$.

Let $\bxi$ be the stationary distribution of the Markov chain with generator ${\bf A}={\bf Q}^{[n,n+1]}+{\bf Q}^{[n,n-1]}+{\bf Q}^{[n,n]}$. Denote by 
\begin{equation}
\gamma=\bxi{\bf Q}^{[n,n+1]}{\bf 1}-\bxi{\bf Q}^{[n,n-1]}{\bf 1} 
\label{eq:Drift}
\end{equation}
the drift measure of the QBD, which can be positive, negative, or zero. Positive drift $\gamma>0$ indicates the tendency of the trait to increase, while negative drift $\gamma>0$ indicates the tendency of the trait to decrease, in the long-run. Zero drift $\gamma=0$ indicates no tendency to increase or decrease in the long run. 

In all examples, to fit model parameters, we use the following three methods: manual (trial and error), Nelder-Mead, and  Broyden–Fletcher–Goldfarb–Shanno (BFGS) algorithm, see Section~\ref{Sec:Optimisation}.

\subsection{Description of QBD Models (Table~\ref{tab:QBD_overview})}


\begin{table}
    \centering
    \small
    \begin{adjustbox}{max width=\linewidth}
    \begin{tabular}{ccccc}
        \toprule
        Model & Number of Phases & ${\bf Q}^{[n,n+1]}$ & ${\bf Q}^{[n,n-1]}$ & ${\bf Q}^{[n,n]}$  \\ 
        \midrule
        \hypertarget{QBD0}{QBD0} & 2 & 
        $\begin{bmatrix} \lambda/2 & \lambda/2 \

\\[1mm] 0 & 0 \end{bmatrix}$ & 
        $\begin{bmatrix} 0 & 0 \

\\[1mm] \mu/2 & \mu/2 \end{bmatrix}$ & 
        $\begin{bmatrix} -\lambda & 0 \

\\[1mm] 0 & -\mu \end{bmatrix}$ \\
        \\
        \hypertarget{QBD1}{QBD1} & 3 &
        $\begin{bmatrix} 0 & \lambda_1 & \lambda_2 \\ 0 & 0 & 0 \\ \lambda_3 & 0 & 0 \end{bmatrix}$ &
        $\begin{bmatrix} 0 & 0 & 0 \\ \mu/3 & \mu/3 & \mu/3 \\ 0 & 0 & 0 \end{bmatrix}$ &
        $\begin{bmatrix} -(\lambda_1+\lambda_2) & 0 & 0 \\ 0 & -\mu & 0 \\ 0 & 0 & -\lambda_3 \end{bmatrix}$         \\
        \\
        \hypertarget{QBD2}{QBD2} & 4 & 
        $\begin{bmatrix}
        0 & \lambda_1 & \lambda_2 & \lambda_3\\
        0 & 0 & 0 & 0 \\
        \lambda_4 & \lambda_5 & 0 & 0 \\
        0 & 0 & 0 & 0
        \end{bmatrix}$ &
        $\begin{bmatrix}
        0 & 0 & 0 & 0 \\
        \mu_1 & 0 & \mu_2 & \mu_3 \\
        0 & 0 & 0 & 0 \\
        \mu_4 & \mu_5 & 0 & 0
        \end{bmatrix}$ &
        $\begin{bmatrix}
        -(\lambda_1+\lambda_2+\lambda_3) & 0 & 0 & 0 \\
        0 & -(\mu_1+\mu_2+\mu_3) & 0 & 0 \\
        0 & 0 & -(\lambda_4+\lambda_5) & 0 \\
        0 & 0 & 0 & -(\mu_4+\mu_5)
        \end{bmatrix}$ \\
        \\
        \hypertarget{QBD3}{QBD3} & 5 &
        $\begin{bmatrix}
        r_1 & r_1 & r_1 & r_1 & r_1\\
        r_2 & r_2 & r_2 & r_2 & r_2\\
        0 & 0 & 0 & 0 & 0\\
        0 & 0 & 0 & 0 & 0\\
        0 & 0 & 0 & 0 & 0
        \end{bmatrix}$ &
        $\begin{bmatrix}
        0 & 0 & 0 & 0 & 0\\
        0 & 0 & 0 & 0 & 0\\
        r_3 & r_3 & r_3 & r_3 & r_3\\
        r_4 & r_4 & r_4 & r_4 & r_4\\
        0 & 0 & 0 & 0 & 0
        \end{bmatrix}$ &
        $\begin{bmatrix}
        -(5r_1) & 0 & 0 & 0 & 0\\
        0 & -(5r_2) & 0 & 0 & 0\\
        0 & 0 & -(5r_3) & 0 & 0\\
        0 & 0 & 0 & -(5r_4) & 0\\
        r_5 & r_6 & r_7 & r_8 & -(r_5+r_6+r_7+r_8)
        \end{bmatrix}$ \\
        \\
        \hypertarget{QBD4}{QBD4} & 5 &
        $\begin{bmatrix}
        \alpha r_1 & (1-\alpha)r_1 & 0 & 0 & 0\\
        (1-\alpha)r_2/2 & \alpha r_2 & (1-\alpha)r_2/2 & 0 & 0\\
        0 & 0 & 0 & 0 & 0\\
        0 & 0 & 0 & 0 & 0\\
        0 & 0 & 0 & 0 & 0
        \end{bmatrix}$ &
        $\begin{bmatrix}
        0 & 0 & 0 & 0 & 0\\
        0 & 0 & 0 & 0 & 0\\
        0 & 0 & 0 & 0 & 0\\
        0 & 0 & (1-\alpha)r_4/2 & \alpha r_4 & (1-\alpha)r_4/2\\
        0 & 0 & 0 & (1-\alpha)r_5 & \alpha r_5
        \end{bmatrix}$ &
        $\begin{bmatrix}
        -r_1 & 0 & 0 & 0 & 0\\
        0 & -r_2 & 0 & 0 & 0\\
        0 & (1-\beta)r_3 & -r_3 & \beta r_3 & 0\\
        0 & 0 & 0 & -r_4 & 0\\
        0 & 0 & 0 & 0 & -r_5
        \end{bmatrix}$ 
        \\
        \bottomrule
    \end{tabular}
    \end{adjustbox}
    \caption{Summary of QBD models and their associated block matrices. Each model defines transition dynamics across discrete levels and phases. The matrix ${\bf Q}^{[n,n+1]}$ governs transitions to the next higher level, while ${\bf Q}^{[n,n-1]}$ governs transitions to the next lower level. Rows correspond to phases at the current level, and columns to phases at the destination level.}
    \label{tab:QBD_overview}
\end{table}

A simple QBD with two phases (\protect\hyperlink{QBD0}{QBD0}), as shown in Table~\ref{tab:QBD_overview}, models random transitions between two states. 
The level increases at rate $\lambda$ when the process is in phase~1 and decreases at rate $\mu$ when in phase~2.
When $\lambda = \mu$, this model exhibits zero drift and is intended to be a discrete analogue of Brownian motion.

\protect\hyperlink{QBD1}{QBD1} has three phases with both upward and downward transitions. 
At level $n$, the process in phase~1 may increase to level $n+1$, transitioning to phase~2 or phase~3 at rates $\lambda_1$ and $\lambda_2$, respectively.
Conversely, a process in phase~3 at level $n$ can increase to level $n+1$ in phase~1 at rate $\lambda_3$. Downward transitions occur from phase~2, where the process moves to level $n-1$ in any of the three phases with equal probability, each at rate $\mu/3$. 

\protect\hyperlink{QBD2}{QBD2} extends the structure of \protect\hyperlink{QBD1}{QBD1} by introducing a fourth phase. At level~$n$, a process in phase~1 can transition to level~$n+1$, entering phase~2, phase~3, or phase~4 at rates $\lambda_1$, $\lambda_2$, and $\lambda_3$, respectively.
Similarly, a process in phase~3 at level~$n$ can transition upward to phase~1 or phase~2 at rates $\lambda_4$ and $\lambda_5$.
Downward transitions occur from phase~2, where the process moves to level~$n-1$, entering phase~1, phase~3, or phase~4 at rates $\mu_1$, $\mu_2$, and $\mu_3$, respectively.

\protect\hyperlink{QBD3}{QBD3} has five phases with two allowing only upward transitions and two allowing only downward transitions. 
At level $n$, a process in phase~1 or phase~2 can increase to level $n+1$, transitioning to any phase at rates $r_1$ and $r_2$, respectively. Downward transitions occur from phase~3 or phase~4 to level $n-1$, again allowing transitions to any phase at rates $r_3$ and $r_4$. 
The diagonal block ${\bf Q}^{[n,n]}$ includes rates $r_5$ through $r_8$, which govern within-level persistence in each phase. 

\protect\hyperlink{QBD4}{QBD4} has five phases with two allowing only upward transitions and two allowing only downward transitions. We introduce a parameter $\alpha$ that governs the tendency to stay in the same phase upon changing level.
At level $n$, a process in phase~1 can increase to level $n+1$, remaining in the same phase with rate $\alpha r_1$ or moving to phase~2 with rate $(1-\alpha) r_1$. 
Similarly, a process in phase~2 at level $n$ can increase to level $n+1$, remaining in phase~2 with rate $\alpha r_2$, or moving to phase~1 or phase~3 with rate $(1-\alpha) r_2/2$ for each transition.  
Downward transitions occur from phase~4, where the process can move from level $n$ to level $n-1$ in phase~3, phase~5, or remain in phase~4 with rates $(1-\alpha) r_4/2$, $(1-\alpha) r_4/2$, and $\alpha r_4$, respectively. 
Similarly, a process in phase~5 at level $n$ can decrease to level $n-1$, remaining in phase~5 with rate $\alpha r_5$ or moving to phase~4 with rate $(1-\alpha) r_5$.
Also, a process in phase~3 at level $n$ can stay in the same level $n$ phase~3 with a rate $r_3$, remaining in the same level but moving into phase~2 or phase~4 with rate $(1-\beta) r_3$ or moving to phase~4 with rate $\beta r_3$. Values of $\beta < 0.5$ give a process with negative drift and values $\beta > 0.5$ give positive drift.
 For all examples in this paper we take $\alpha = 0.9$, so that changes in level usually do not result in a change in phase. 

Across all datasets and QBD models, we assume a uniform distribution of the ancestral (parent) trait value.

\subsection{Search for parameters to maximise the likelihood}\label{Sec:Optimisation}

To numerically optimise the parameters of the QBDs we used both the Nelder-Mead~\cite{nelder1965simplex} and BFGS~\cite{fletcher2000practical} methods. These were chosen as they do not require a gradient function (Nelder-Mead is derivative free and BFGS is a quasi-Newton method that uses gradient information to approximate the Hessian). Our parameters are all nonnegative (they are rates), but to avoid doing constrained optimization, we used the common technique of optimizing the log-transformed parameters.

As numerical optimisation can be prone to getting stuck in local optima, we applied a multi-start optimisation approach in which a different set of initial parameters is used at the start of each iteration of these algorithms.
Further, to avoid underflow (i.e. numbers less than machine precision which would affect the correct computation of the likelihood), in place of~\eqref{rec_N_Paper2mod}, we apply our factor-scaled likelihood evaluations with
\begin{eqnarray}
	{\bf f}_M({T^*}^{(k,B)}(t_N,t))&=&
	\left\{
	\begin{array}{ll}
 {\bf P}_{m(k,B)}(b_k)\times {\bf 1}
	& \quad k< \ell_N,\bigskip
	\\
    M\times
 \left(
 {\bf P}_{m(k,B)}(b_k)
  \times 
{\bf 1}
\right)
	\odot
 \left(
{\bf P}_{m(k+1,B)}(b_{k+1})
\times 
{\bf 1}
\right)
	& \quad k= \ell_B,\bigskip
	\\
 {\bf P}_{m(k+1,N)}(b_k)\times {\bf 1}
	& \quad k> \ell_B,
	\end{array}	
	\right.
	\label{rec_N_Paper2modFactor}
\end{eqnarray}
for some large $M$, and then use
\begin{eqnarray}
	{\bf f}({T^*}^{(k,B)}(t_1,t))&=&
    {\bf f}_M({T^*}^{(k,B)}(t_1,t))
    / M^B,
\end{eqnarray}
or equivalently,
\begin{eqnarray}
log
\big(
	{\bf f}({T^*}^{(k,B)}(t_1,t)
    \big)&=&
    log
    \big(
    {\bf f}_M({T^*}^{(k,B)}(t_1,t))
    \big)
    -
    B\times log(M).
\end{eqnarray}

\section{Numerical Examples: Synthetic Data}\label{sec:NumExSynthetic}

We constructed two small synthetic datasets, each consisting of four tips and three internal nodes (Figures~\ref{fig:DataExample1} and~\ref{fig:DataExample2}), to explore alternative patterns of trait evolution.
In the first dataset, two closely related tips sharing a recent common ancestor exhibit low trait values, whereas the two tips branching closer to the root display higher values.
In the second dataset, this pattern is reversed, with more basally branching taxa exhibiting lower trait values.

\subsection{Synthetic Dataset~1}\label{sec:SynthData}

\begin{figure}[H]
	\centering
	\begin{tikzpicture}[xscale=2.1, yscale=2, >=stealth, redarr/.style={->}]
		\draw [dashed, gray] (1,9.5) -- (9,9.5);
		\draw [dashed, gray] (1,8.5) -- (9,8.5);
		\draw [dashed, gray] (1,7.5) -- (9,7.5);
		\draw [dashed, gray] (1,6.46) -- (9,6.46);

		\draw [dashed,black, ultra thick] (5.25,10) -- (5.25,9.5);
        
		\draw [green, ultra thick] (5.25,9.5) -- (4,9.5);
        
		\draw [green, ultra thick] (4,9.5) -- (4,8.5);
        	\node at (4.3,9) {\LARGE \textcolor{purple}{1}};	
		\draw [blue, ultra thick] (3.1,8.5) -- (4,8.5);

		\draw [blue, ultra thick] (3.1,8.5) -- (3.1,7.5);
		\node at (3.4,8) {\LARGE \textcolor{purple}{1}};

		\draw [purple, ultra thick] (2.3,7.5) -- (3,7.5);
		\node at (2.6,7) {\LARGE \textcolor{purple}{1}};
		\node[draw=purple, fill=purple!20, circle, inner sep=3pt] at (2.3,6.3) {\LARGE 1};
\node at (2.6,6.3) {\LARGE \textcolor{cyan}{3}};
		\draw [purple, ultra thick] (2.3,7.5) -- (2.3,6.5);

		\draw [purple, ultra thick] (2.7,7.5) -- (3.9,7.5);
		\draw [purple, ultra thick] (3.9,7.5) -- (3.9,6.5);
		\node at (4.2,7) {\LARGE \textcolor{purple}{1}};
		\node[draw=purple, fill=purple!20, circle, inner sep=3pt] at (3.9,6.3) {\LARGE 2};
\node at (4.22,6.3) {\LARGE \textcolor{cyan}{7}};
		\draw [blue, ultra thick] (4,8.5) -- (5.1,8.5);
		\draw [blue, ultra thick] (5.1,8.5) -- (5.1,6.5);
		\node at (5.4,8) {\LARGE \textcolor{purple}{2}};
		\node[draw=blue, fill=blue!20, circle, inner sep=3pt] at (5.1,6.3) {\LARGE 3};
\node at (5.46,6.3) {\LARGE \textcolor{cyan}{12}};
		\draw [green, ultra thick] (5.26,9.5) -- (6.41,9.5);
		\draw [green, ultra thick] (6.41,9.5) -- (6.41,6.5);
		\node at (6.71,9) {\LARGE \textcolor{purple}{3}};
		\node[draw=green, fill=green!20, circle, inner sep=3pt] at (6.4,6.3) {\LARGE 4};
        \node at (6.76,6.3) {\LARGE \textcolor{cyan}{16}};

\node[draw=purple, fill=purple!20, circle, inner sep=3pt] at (3.1,7.3) {\LARGE \textcolor{blue}{5}};

		\node[draw=purple, fill=purple!20, circle, inner sep=3pt] at (4,8.3) {\LARGE \textcolor{blue}{6}};

        \node[draw=purple, fill=purple!20, circle, inner sep=3pt] at (5.25,9.3) {\LARGE \textcolor{blue}{7}};
        
	\end{tikzpicture}
	\caption{Phylogenetic tree with nodes $1,\ldots,7$, including tips $1,\ldots,4$, internal nodes $5,6$ and parent node $7$.
Branch lengths are indicated along each edge. Observed trait values are indicated to the right of each tip.}
	\label{fig:DataExample1}
\end{figure}
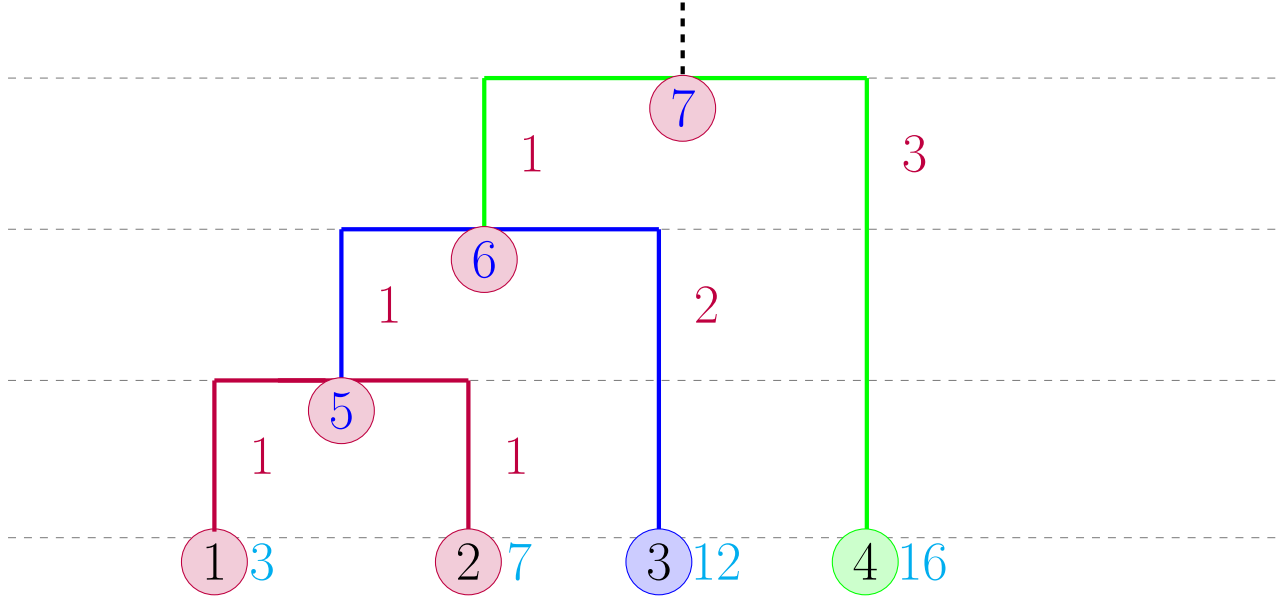

In our examples in this section, we assume a tree topology with $4$ tips and $3$ internal nodes as depicted in Figure~\ref{fig:DataExample1}.

\subsubsection{\protect\hyperlink{QBD3}{QBD3} -- Preliminaries: The effect of the mean drift $\gamma$}\label{sec:Model5phases}

First, to study the effect of the mean drift $\gamma$, we consider \protect\hyperlink{QBD3}{QBD3} from Table~\ref{tab:QBD_overview} with 
\begin{eqnarray}
{\bf Q}^{[n,n+1]} =
\begin{bmatrix}
r_1 & r_1 & r_1 & r_1 & r_1\\
r_2 & r_2 & r_2 & r_2 & r_2\\
0 & 0 & 0 & 0 & 0\\
0 & 0 & 0 & 0 & 0\\
0 & 0 & 0 & 0 & 0
\end{bmatrix}
,\quad
{\bf Q}^{[n,n-1]} = 
\begin{bmatrix}
0 & 0 & 0 & 0 & 0\\
0 & 0 & 0 & 0 & 0\\
r_3 & r_3 & r_3 & r_3 & r_3\\
r_4 & r_4 & r_4 & r_4 & r_4\\
0 & 0 & 0 & 0 & 0
\end{bmatrix}
,
\nonumber\\
{\bf Q}^{[n,n]} = 
\begin{bmatrix}
 -5r_1 & 0 & 0  & 0 & 0\\
 0 & -5r_2 & 0  & 0 & 0\\
 0 & 0 & -5r_3  & 0 & 0\\
 0 & 0 & 0  & -5r_4 & 0\\
r_5 & r_6 & r_7 & r_8 & -(r_5+r_6+r_7+r_8)
\end{bmatrix}
,
\label{eq:prelimData1}
\end{eqnarray}
and the range of parameters in Table~\ref{tab:Synexamplet1}. 
A more detailed set of results is presented in Figures~\ref{StatLevelsModel5phases}–\ref{StatTraitsModel5phasesOverallParrentTraitsOnly} in Appendix~\ref{sec:OutputQBD5phases}.

\begin{table}[H]
    \centering
    \begin{tabular}{ccrccc}
       \toprule
        ${\bf r}$\# & ${\bf r}=[r_1\  \ldots \ r_8]$ & drift $\gamma$ &Likelihood &LogLikelihood &AICc\\ 
        \midrule
        $1$    & $[2\ 1\ 4\ 3\ 2\ 1\ 6\ 5]$  & $-1.0169$& $6.9049\times 10^{-22}$ &$-48.7246$&84.6492\\ 
               $2$   & $[10\     3\     5\     5\     4\     6\     1\     1]$   &  $2.2472$ &$2.7730\times 10^{-15}$& $-33.5188$&54.2376
\\ 
        $3$   & $[12\     7\     8\    8\    4\     6 \    1\     1]$   & $3.2909$ & $ 1.1775\times 10^{-12}$&$ -27.4676$ &42.1352\\ 
         $4$  & $[10\     3\     8\     8\    9\     10\     7\    7]$ &  $0.7455$& $4.1574\times 10^{-13}$&$-28.5087$& 44.2174 \\
        $5$  & $[9\    6\     7\     7\    5\     7\     3\    3]$  &  $1.6928$  &$7.6748\times 10^{-13}$&$-27.8957$&42.9914
        \\
        6    & ${\bf r}$\#1$\ \times10$  & $-10.1695$&$3.0103\times 10^{-10}$ &$-21.9238$&31.0476\\
         7    & ${\bf r}$\#2$\ \times10$  & $22.4719$&$8.2780\times 10^{-15}$ &$-32.4252$&52.0504\\
         8    & ${\bf r}$\#3$\ \times10$  & $32.9089$&$7.8435\times 10^{-19}$ &$-41.6894$&70.5788\\
         9    & ${\bf r}$\#4$\ \times10$  & $7.4553$& $4.4464\times 10^{-9}$ & $ -19.2312$&25.6624\\
         10    & ${\bf r}$\#5$\ \times10$  & $16.9279$&$4.9853\times 10^{-11}$ & $-23.7219$&34.6438\\
        
        \bottomrule
    \end{tabular}
    \caption{The likelihood of the Synthetic Dataset~1 (Figure~\ref{fig:DataExample1}) for the different parameters of \protect\hyperlink{QBD3}{QBD3}, assuming uniform initial distribution of levels and phases at the root of the tree, and $[a,b]=[3,20]$.}
\label{tab:Synexamplet1}
\end{table}

We first conducted a preliminary analysis using manually selected rate vectors, ${\bf r}$~$\#1–\#10$, as provided in Table~\ref{tab:Synexamplet1}, to illustrate the range of behaviours and insights offered by QBD models exhibiting zero drift, positive drift, or negative drift.
The corresponding drift and likelihood results are reported in the respective columns of Table~\ref{tab:Synexamplet1} and are further illustrated in Figures~\ref{StatLevelsModel5phases}–\ref{StatTraitsModel5phasesOverallParrentTraitsOnly} in Appendix~\ref{sec:OutputQBD5phasessynJUST_QBD} and \ref{sec:OutputQBD5phasessyn1}.
Table~\ref{tab:Synexamplet1} shows that the best parameter set ${\bf r}$~$\#9$ has positive drift. One reason why a model with positive drift might be preferred, even though the mean of the trait values observed at the tips is 9.5 which is less than the halfway point between 3 and 20, is that the two longer branches end in higher trait values and must be reached independently, whereas the two smaller trait values can be explained by a shared path from the root (node 7) to their common ancestor (node 5).

In the QBD model with ${\bf r}$ vector $\#1$ in Table~\ref{tab:Synexamplet1}, which represents a negative mean drift, the stationary distribution of levels (Figure~\ref{StatLevelsModel5phases}) peaks at 0. 
Similarly, the stationary distribution of traits (Figure~\ref{StatTraitsModel5phases}) becomes concentrated near the trait’s minimum.
This pattern reflects the tendency of negative drift to reduce levels (trait values) over time.
Furthermore, in a QBD with a negative mean drift, the likelihood of observing tips $i=1,\ldots,4$ given initial trait~$x$ reaches its maximum to the
right of the observed tip trait values~(Figure~\ref{StatTraitsModel5phasesOverallcherries}). 
This is because the initial trait values would decrease due to negative drift. 

In contrast, for QBD models exhibiting positive mean drift (corresponding to ${\bf r}$ vectors $\#2$–$\#4$), increasing the drift strength~($\gamma$) progressively shifts the stationary distribution of levels (Figure~\ref{StatLevelsModel5phases}) toward the upper boundary.
Similarly, the stationary distribution of trait values (Figure~\ref{StatTraitsModel5phases}) becomes increasingly concentrated near the maximum level, indicating a long-term tendency for the trait to increase.
This is expected, as positive drift drives levels upward, causing trait values to approach the upper boundary over time.
Furthermore, in QBD models with a positive mean drift, the likelihood of observing tips $i=1,\ldots,4$ given initial trait~$x$ reaches its 
maximum to the
left of the observed tip trait values~(Figure~\ref{StatTraitsModel5phasesOverallcherries}). 
This is expected, as positive drift causes the trait values to increase over time from their initial states.

For the QBD model with negative mean drift, the likelihood of observing the phylogenetic tree given an initial trait value $x$ and phase~$\varphi$~ (Figure~\ref{StatTraitsModel5phasesOverallParrentTrait}) is higher for phases $3$ and $4$, which correspond to transitions toward lower levels.
Also, the initial trait values $x$ for QBD models with negative drift are higher than those for QBDs with positive mean drift (Figure~\ref{StatTraitsModel5phasesOverallcherries}). This is because, under negative drift, higher initial trait values tend to decrease over time, resulting in lower trait values at the tips.
In contrast, for QBD models with a positive mean drift, the likelihood of observing the phylogenetic tree given initial trait~$x$ and phase~$\varphi$ (Figure~\ref{StatTraitsModel5phasesOverallParrentTrait}) is highest for upward-moving phases $1$ and $2$.
For instance, in the QBD model with vector $r\#1$ (negative mean drift), the maximum likelihood occurs in phase~$\varphi=3$ at trait value $x=13$, illustrating a scenario where traits originate at higher values and gradually drift downward.
Conversely, in the QBD model with ${\bf r}$ vector $\#4$~(positive mean drift), the peak likelihood is found in phase~$\varphi=1$ at trait value $x=9$, illustrating a scenario where traits originate at lower values and gradually drift upward. 
Moreover, as the strength of the positive drift increases traits are likely to start from lowest trait value at the top~(Figure~\ref{StatTraitsModel5phasesOverallParrentTraitsOnly}).

\subsubsection{\protect\hyperlink{QBD0}{QBD0}: with $2$ phases}\label{Mod0:2phasessyndata2}

Recall the simple QBD with 2 phases in Table~\ref{tab:QBD_overview} such that
\begin{align}
{\bf Q}^{[n,n+1]} &=
\begin{bmatrix}
\lambda/2 & \lambda/2 \\
0 & 0
\end{bmatrix}, \quad
{\bf Q}^{[n,n-1]} =
\begin{bmatrix}
0 & 0 \\
\mu/2 & \mu/2 
\end{bmatrix}, \quad
{\bf Q}^{[n,n]} =
\begin{bmatrix}
 -\lambda & 0 \\
 0 & -\mu
\end{bmatrix}
\label{eq:Mod0:2phasessyndata2}
\end{align}
where $\lambda$ is the rate of moving up in phase~$1$, and $\mu$ is the rate of moving down in phase~$2$.

We applied {\protect\hyperlink{QBD0}{QBD0}} to the analysis of Synthetic Dataset 1 shown in Figure~\ref{fig:DataExample1}. We set $\lambda = \mu$ to enforce neutral trait evolution without directional bias. The Nelder–Mead optimization produced parameter estimates that maximized the log-likelihood and minimized the $AICc$ (Table~\ref{tab:exampleModel0Syntheticdata2}).

\begin{table}[H]
    \centering
    \begin{tabular}{cccccc}
        \toprule
        Method & $\lambda = \mu$ & Drift $\gamma$ & Likelihood & LogLikelihood & AICc \\ 
        \midrule
        Manual       & $10$           & $0$ & $4.8771 \times 10^{-22}$ & $-49.0723$ & $114.1446$ \\ 
        Nelder-Mead  & $576.3080$     & $0$ & $1.2336 \times 10^{-8}$  & $-18.2108$ & $52.4216$  \\ 
        BFGS         & $53925584.9996$& $0$ & $9.6098 \times 10^{-9}$  & $-18.4605$ & $52.9210$  \\ 
        \bottomrule
    \end{tabular}
    \caption{The likelihood of the Synthetic Dataset~1 (Figure~\ref{fig:DataExample1}) for the different parameters of \protect\hyperlink{QBD0}{QBD0}, assuming uniform initial distribution of levels and phases at the root of the tree, and $[a,b] = [3, 20]$.}
    \label{tab:exampleModel0Syntheticdata2}
\end{table}

\subsubsection{\protect\hyperlink{QBD1}{QBD1}: with $3$ phases}\label{Mod1:3phases}

Next, we consider the QBD with $3$ phases in Table~\ref{tab:QBD_overview} such that
\begin{align}
{\bf Q}^{[n,n+1]} =
\begin{bmatrix}
0 & \lambda_1 & \lambda_2 \\
0 & 0 & 0 \\
\lambda_3 & 0 & 0
\end{bmatrix}
,\quad 
{\bf Q}^{[n,n-1]} =
\begin{bmatrix}
0 & 0 & 0 \\
\mu/3 & \mu/3 & \mu/3 \\
0 & 0 & 0
\end{bmatrix}
,\quad
{\bf Q}^{[n,n]} =
\begin{bmatrix}
 -(\lambda_1 +\lambda_2) & 0 & 0 \\
0 & -\mu & 0 \\
 0 & 0 & -\lambda_3
\end{bmatrix}
\label{eq:Mod1:3phasessyndata2}
\end{align}
and apply it to the analysis of dataset shown in Figure~\ref{fig:DataExample1}.

 Both the Nelder–Mead and BFGS optimisation methods converged on solutions that exhibit positive drift (Table~\ref{tab:exampleQBD13phasessyntheticdata2}). 
 The BFGS algorithm yielded optimal maximum-likelihood estimates with positive drift ($\gamma=13.3295$), achieving the highest log-likelihood and the minimum AICc scores~(Table~\ref{tab:exampleQBD13phasessyntheticdata2}).

\begin{table}[H]
    \centering
    \small
    \begin{tabular}{ccccccc}
        \toprule
        Method & $[\lambda_1,\lambda_2,\lambda_3]$ and $\mu$ & Drift $\gamma$ & Likelihood & LogLikelihood& AICc  \\
        \midrule
        Manual       & $[10,\ 0.1,\ 0.1]$ and $10$ & $0.0037$   & $1.2049 \times 10^{-47}$ & $-108.0351$ &184.0702\\
        Nelder-Mead  & $[282.5302,\ 418439145.5456,\ 10]$ and $10$ & $20.0000$  & $3.3185 \times 10^{-21}$ & $-47.1548$&62.3096 \\
        BFGS         & $[1.6023,\ 14.5773,\ 89.2153]$ and $2.3682$ & $13.3295$  & $3.2841 \times 10^{-10}$ & $-21.8368$&11.6736 \\
        \bottomrule
    \end{tabular}
    \caption{The likelihood of the Synthetic Dataset~1 (Figure~\ref{fig:DataExample1}) for the different parameters of the \protect\hyperlink{QBD1}{QBD1}, assuming uniform initial distribution of levels and phases at the root of the tree, and $[a,b] = [3, 20]$.}
    \label{tab:exampleQBD13phasessyntheticdata2}
\end{table}

\subsubsection{\protect\hyperlink{QBD2}{QBD2}: with $4$ phases} \label{Mod2:4phases}

We then applied the QBD model with $4$ phases in Table~\ref{tab:QBD_overview} such that
\begin{align}
{\bf Q}^{[n,n+1]} &=
\begin{bmatrix}
0 & \lambda_1 & \lambda_2 & \lambda_3\\
0 & 0 & 0 & 0 \\
\lambda_4 & \lambda_5 & 0 & 0 \\
0 & 0 & 0 & 0
\end{bmatrix}, \quad
{\bf Q}^{[n,n-1]} =
\begin{bmatrix}
0 & 0 & 0 & 0 \\
\mu_1 & 0 & \mu_2 & \mu_3 \\
0 & 0 & 0 & 0 \\
\mu_4 & \mu_5 & 0 & 0
\end{bmatrix}, \notag \\
{\bf Q}^{[n,n]} &=
\begin{bmatrix}
 -(\lambda_1+\lambda_2+\lambda_3) & 0 & 0 & 0 \\
 0 & -(\mu_1+\mu_2+\mu_3) & 0 & 0 \\
 0 & 0 & -(\lambda_4+\lambda_5) & 0 \\
 0 & 0 & 0 & -(\mu_4+\mu_5)
\end{bmatrix}.
\label{eq:Mod2:4phasessyndata2}
\end{align}
to the analysis of dataset shown in Figure~\ref{fig:DataExample1}. 

The Nelder–Mead optimisation method converged on parameter estimates with a negative drift, while the BFGS algorithm produced estimates with very small positive drift. 
None of the parameter estimates found for this model performed as well in terms of AICc as \protect\hyperlink{QBD1}{QBD1} suggesting that numerical optimization is struggling to fit the 10 parameter model.

\begin{table}[H]
\centering
\renewcommand{\arraystretch}{1.15}
\setlength{\tabcolsep}{4pt}

\begin{adjustbox}{max width=\textwidth}
\begin{tabular}{
    c
    >{\raggedright\arraybackslash}m{7.6cm}
    >{\raggedright\arraybackslash}m{7.6cm}
    >{\centering\arraybackslash}m{2.6cm}
    >{\centering\arraybackslash}m{3.1cm}
    >{\centering\arraybackslash}m{2cm}
    >{\centering\arraybackslash}m{1.5cm}
}
\toprule
Method & $[\lambda_1,\ldots,\lambda_5]$ & $[\mu_1,\ldots,\mu_5]$ & Drift $\gamma$ & Likelihood & Log-Likelihood & AICc \\ 
\midrule
Manual &
$[10,\ 0.1,\ 0.1,\ 1,\ 1]$ &
$[10,\ 0.1,\ 0.1,\ 1,\ 1]$ &
0 &
\num{1.4784e-101} &
$-232.1701$ &
$452.9116$ \\ 

Nelder–Mead &
\makecell[l]{$[42.6160,\ 0.0003,\ 0.0000,\ 0.0000,\ 1.0098]$}&
\makecell[l]{$[1.5126,\ 0.0000, \ 310726,\ 0.0001,\ 1.0248]$} &
$-2.0497$ &
\num{4.6238e-42} &
$-95.1166$ &
$178.8046$ \\ 

BFGS &
$[2.1530,\ 4.9817,\ 6.0689,\ 2.2324,\ 1.5064]$ &
$[0.0990,\ 0.0833,\ 0.0593,\ 0.0451,\ 0.0763]$ &
\num{0.0159} &
\num{1.4344e-68} &
$-156.2150$ &
$301.0014$ \\  
\bottomrule
\end{tabular}
\end{adjustbox}

\caption{The likelihood of the Synthetic Dataset~1 (Figure~\ref{fig:DataExample1}) for the differnet parameters of \protect\hyperlink{QBD2}{QBD2}, assuming uniform initial distribution of levels and phases at the root of the tree, and $[a,b] = [3,20]$.}
\label{tab:exampleModel2syn1data2}
\end{table}

\subsubsection{Comparison of QBD Model Fits for Synthetic Dataset~1~(Figure~\ref{fig:DataExample1})}

Of the three models we evaluated, the {\protect\hyperlink{QBD}{QBD1}} model, fitted using BFGS optimisation, yielded maximum-likelihood estimates with a positive drift and provided the best fit to Synthetic Dataset~1 (minimum $AICc = 11.6736$; Table~\ref{tab:exampleQBD13phasessyntheticdata2}).
This result indicates that the observed trait values at the tips are best explained by a process that tends to increase toward the upper boundary of the trait range ($b = 20$).

\subsection{Synthetic Dataset~2}\label{sec:SynthData3}

\begin{figure}[H]
	\centering
	\begin{tikzpicture}[xscale=2.1, yscale=2, >=stealth, redarr/.style={->}]
		\draw [dashed, gray] (1,9.5) -- (9,9.5);
		\draw [dashed, gray] (1,8.5) -- (9,8.5);
		\draw [dashed, gray] (1,7.5) -- (9,7.5);
		\draw [dashed, gray] (1,6.46) -- (9,6.46);

		\draw [dashed,black, ultra thick] (5.25,10) -- (5.25,9.5);
        
		\draw [green, ultra thick] (5.25,9.5) -- (4,9.5);
        
		\draw [green, ultra thick] (4,9.5) -- (4,8.5);
        	\node at (4.3,9) {\LARGE \textcolor{purple}{1}};	
		\draw [blue, ultra thick] (3.1,8.5) -- (4,8.5);

		\draw [blue, ultra thick] (3.1,8.5) -- (3.1,7.5);
		\node at (3.4,8) {\LARGE \textcolor{purple}{1}};

		\draw [purple, ultra thick] (2.3,7.5) -- (3,7.5);
		\node at (2.6,7) {\LARGE \textcolor{purple}{1}};
		\node[draw=purple, fill=purple!20, circle, inner sep=3pt] at (2.3,6.3) {\LARGE 1};
\node at (2.7,6.3) {\LARGE \textcolor{cyan}{16}};
		\draw [purple, ultra thick] (2.3,7.5) -- (2.3,6.5);

		\draw [purple, ultra thick] (2.7,7.5) -- (3.9,7.5);
		\draw [purple, ultra thick] (3.9,7.5) -- (3.9,6.5);
		\node at (4.2,7) {\LARGE \textcolor{purple}{1}};
		\node[draw=purple, fill=purple!20, circle, inner sep=3pt] at (3.9,6.3) {\LARGE 2};
\node at (4.3,6.3) {\LARGE \textcolor{cyan}{12}};
		\draw [blue, ultra thick] (4,8.5) -- (5.1,8.5);
		\draw [blue, ultra thick] (5.1,8.5) -- (5.1,6.5);
		\node at (5.4,8) {\LARGE \textcolor{purple}{2}};
		\node[draw=blue, fill=blue!20, circle, inner sep=3pt] at (5.1,6.3) {\LARGE 3};
\node at (5.46,6.3) {\LARGE \textcolor{cyan}{7}};
		\draw [green, ultra thick] (5.26,9.5) -- (6.41,9.5);
		\draw [green, ultra thick] (6.41,9.5) -- (6.41,6.5);
		\node at (6.71,9) {\LARGE \textcolor{purple}{3}};
		\node[draw=green, fill=green!20, circle, inner sep=3pt] at (6.4,6.3) {\LARGE 4};
        \node at (6.76,6.3) {\LARGE \textcolor{cyan}{3}};

\node[draw=purple, fill=purple!20, circle, inner sep=3pt] at (3.1,7.3) {\LARGE \textcolor{blue}{5}};

		\node[draw=purple, fill=purple!20, circle, inner sep=3pt] at (4,8.3) {\LARGE \textcolor{blue}{6}};

        \node[draw=purple, fill=purple!20, circle, inner sep=3pt] at (5.25,9.3) {\LARGE \textcolor{blue}{7}};
        
	\end{tikzpicture}
	\caption{Phylogenetic tree with nodes $1,\ldots,7$, including tips $1,\ldots,4$, internal nodes $5,6$ and parent node $7$.
       Branch lengths are indicated along each edge. Observed trait values are indicated to the right of each tip.}
	\label{fig:DataExample2}
\end{figure}
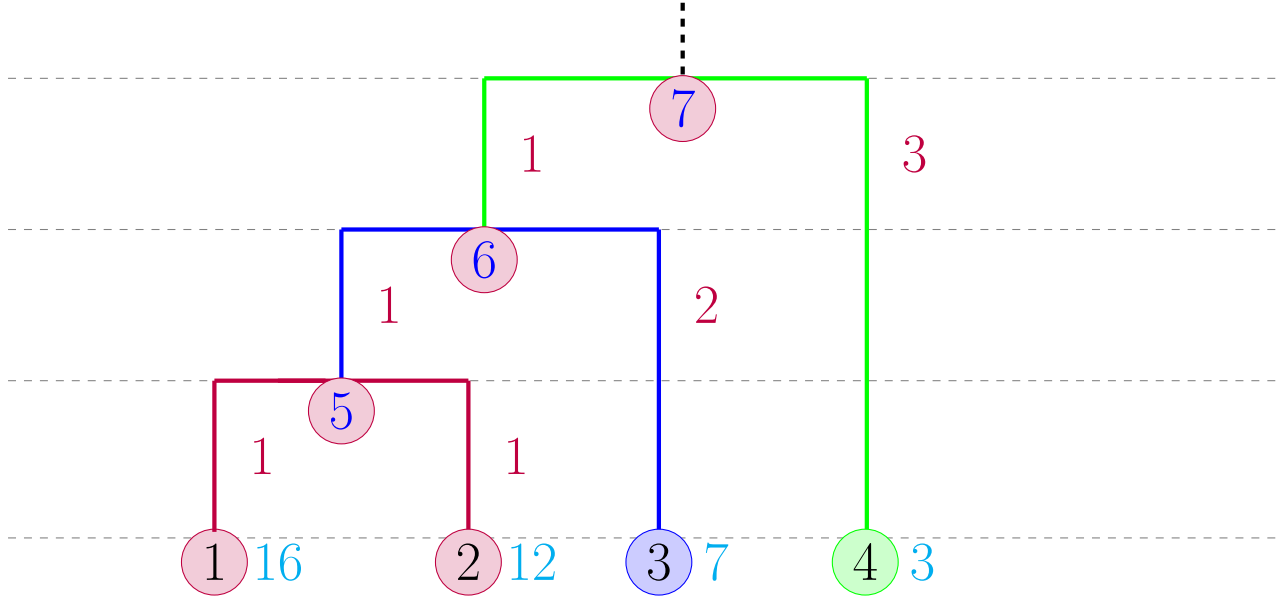

\subsubsection{\protect\hyperlink{QBD3}{QBD3} -- Preliminaries: The effect of the mean drift $\gamma$}\label{sec:Model5phasesData2}

First, to study the effect of the mean drift $\gamma$, we consider a QBD with the block matrices in~\eqref{eq:prelimData1} for a selected range of parameters, the output is presented in Table~\ref{tab:Synexamplet2}.

\begin{table}[H]
    \centering
    \begin{tabular}{cccccc}
        \toprule
        ${\bf r}$\# & ${\bf r} = [r_1\ \ldots\ r_8]$ & Drift $\gamma$ & Likelihood & LogLikelihood &AICc\\ 
        \midrule
        $1$  & [$2$ $1$ $4$ $3$ $2$ $1$ $6$ $5$]       & $-1.0169$  & \bf $6.1049 \times 10^{-23}$ & $-51.1504$&89.5008 \\ 
        $2$  & [$10$ $3$ $5$ $5$ $4$ $6$ $1$ $1$]      & $2.2472$   & $1.5023 \times 10^{-17}$     & $-38.7370$ &64.6740\\ 
        $3$  & [$12$ $7$ $8$ $8$ $4$ $6$ $1$ $1$]      & $3.2909$   & $7.8723 \times 10^{-15}$     & $-32.4754$ &52.1508\\ 
        $4$  & [$10$ $3$ $8$ $8$ $9$ $10$ $7$ $7$]     & $0.7455$   & $5.4629 \times 10^{-14}$     & $-30.5382$ &48.2764\\ 
        $5$  & [$9$ $6$ $7$ $7$ $5$ $7$ $3$ $3$]       & $1.6928$   & \bf $5.6480 \times 10^{-14}$ & \bf $-30.5049$&48.2098 \\ 
        $6$  & ${\bf r}$\#1$\ \times 10$              & $-10.1695$ & \bf $2.0926 \times 10^{-8}$  &  $-17.6823$& 22.5646\\ 
        $7$  & ${\bf r}$\#2$\ \times 10$              & $22.4719$  & $6.6542 \times 10^{-17}$     & $-37.2487$ &61.6974\\ 
        $8$  & ${\bf r}$\#3$\ \times 10$              & $32.9089$  & \bf $1.0621 \times 10^{-21}$ & \bf $-48.2940$ &83.7880\\ 
        $9$  & ${\bf r}$\#4$\ \times 10$              & $7.4553$   & $7.6177 \times 10^{-10}$     & $-20.9954$ &29.1908\\ 
        $10$ & ${\bf r}$\#5$\ \times 10$              & $16.9279$  & $2.5789 \times 10^{-12}$     & $-26.6837$&40.5674 \\ 
        \bottomrule
    \end{tabular}
 \caption{The likelihood of the Synthetic Dataset~2 (Figure~\ref{fig:DataExample2}) for the different parameters of \protect\hyperlink{QBD3}{QBD3}, assuming uniform initial distribution of levels and phases at the root of the tree, and $[a,b]=[3,20]$.}
\label{tab:Synexamplet2}
\end{table}

To investigate the extent to which the structure of the tree affects the preference for different levels of drift we used the same set of rate parameters explored in Table 2 but applied to Synthetic Dataset 2 where the order of the traits is reversed compared to Synthetic Dataset 1. 
The corresponding drift and the likelihood results for these rates for the \protect\hyperlink{QBD4}{QBD4} model are provided in Table~\ref{tab:Synexamplet2} and are further illustrated in Appendix~\ref{outputSyntheticDataExample2}
In this case we found that the best performing parameter set, ${\bf r}\#6$, has negative drift. This difference to what we saw for Synthetic Dataset 1 is interesting, we note that in synthetic dataset 2 the two lowest trait values have to be reached independently, whereas the two larger trait values can be reached by a shared path. Combined with the results from synthetic dataset 1, this indicates that the structure of the tree, as well as the observed trait values, influences the preferred level of drift.

Consistent with Synthetic Dataset~1 (Figure~\ref{fig:DataExample1}), in Synthetic Dataset~2 (Figure~\ref{fig:DataExample2}), the likelihood of observing tips $i=1,\ldots,4$ given the initial trait~$x$ peaks to the right of the observed traits under negative drift and to the left under positive drift.

For Synthetic Dataset~2 (Figure~\ref{fig:DataExample2}), the likelihood of observing the tree under a QBD model with negative drift was highest when starting in phase~$\varphi=4$, with levels ranging from 40 to 100~(Figure~\ref{StatTraitsModel5phasesOverallParrentLevelData2}).
In contrast, in a QBD model with small positive drift, the most likely starting phase was  phase~$\varphi=1$, with initial levels ranging from 10 to 50~(Figure~\ref{StatTraitsModel5phasesOverallParrentLevelData2}). 
Phase~$\varphi=3$ corresponds to transitions toward lower levels, whereas phase~$\varphi=1$ corresponds to transitions toward upper levels. 
Furthermore, initial trait values $x$ for QBD models with a negative drift were higher than those for QBD models with a positive mean drift (Figure~\ref{StatTraitsModel5phasesOverallParrentTraitsOnlyData2}).

\subsubsection{\protect\hyperlink{QBD0}{QBD0}: with $2$ phases}\label{Mod0:2phases}

 We then applied a simple two-phase QBD model, summarised in Table~\ref{tab:QBD_overview} and defined by the generator in Equation~\eqref{eq:Mod0:2phasessyndata2}, to Synthetic Dataset~2 (Figure~\ref{fig:DataExample2}). 
The best set of parameters was found using Nelder–Mead optimisation (Table~\ref{tab:exampleModel0Synthetic}).

\begin{table}[H]
    \centering
    \begin{tabular}{ccccccc}
        \toprule
        Method &  $\lambda = \mu$ & Drift $\gamma$ & Likelihood & LogLikelihood & AICc \\ 
        \midrule
        Manual        & $10$        & $0$ & $3.5576 \times 10^{-22}$ & $-49.3878$ & 114.7756 \\ 
        Nelder-Mead  & $1945.2130$ & $0$ & $9.6234 \times 10^{-9}$  & $-18.4591$ & 52.9182 \\
        BFGS          & $300$       & $0$ & $6.6114 \times 10^{-9}$  & $-18.8345$ & 53.6690 \\   
        \bottomrule
    \end{tabular}
    \caption{The likelihood of the Synthetic Dataset~2 (Figure~\ref{fig:DataExample2}) for the different parameters of \protect\hyperlink{QBD0}{QBD0}, assuming uniform initial distribution of levels and phases at the root of the tree, and $[a,b] = [3,20]$. }
    \label{tab:exampleModel0Synthetic}
\end{table}

\subsubsection{\protect\hyperlink{QBD}{QBD1}: with $3$ phases}\label{Mod1:3phases}
 
We also applied the {\protect\hyperlink{QBD1}{QBD1}} model, summarised in Table~\ref{tab:QBD_overview} and defined by the generator in equation ~\eqref{eq:Mod1:3phasessyndata2}, to Synthetic Dataset~2 (Figure~\ref{fig:DataExample2}).
The best-fitting parameter estimates were obtained using BFGS optimisation; the resulting parameter estimates yielded zero drift~(Table~\ref{tab:exampleModel1Synthetic}). Based on our arguments above, here we expected the model to prefer negative drift values. However, numerical exploration (a million random choices of the parameters) suggests that the structure of this model makes it impossible to achieve negative drift. This is presumably because the level can only go downwards when in phase 2, and when it does it will change to phase 1 or 3 with 2/3 probability and from these phases it can only go up. Whereas when it goes up a level it is possible to remain in phase 1 or 3 and then go up again.

\begin{table}[H]
    \centering
    \scriptsize
    \begin{tabular}{ccccccc}
        \toprule
         Method & $[\lambda_1,\lambda_2,\lambda_3]$ and $\mu$ & Drift $\gamma$ & Likelihood & LogLikelihood& AICc  \\
        \midrule
            Manual       & $[10,\ 0.1,\ 0.1]$ and $10$ & $0.0037$   & $8.9698 \times 10^{-69}$ & $-156.6845$ &281.3690\\
  Nelder-Mead&$[137641922.0985,\ 0.0479,\ 1158.7438]$ and  $21155242.3867$&$1.6127 \times 10^{-06}$&$1.0008 \times 10^{-08}$&-18.4199&4.8398\\
  BFGS& $[366976194.7948,\ 0.0362,\  910.2207]$ and $50293301.1721$ &$3.5913\times 10^{-7}$&$1.0012 \times 10^{-08}$&$-18.4195$&4.8390\\
  
    \bottomrule
    \end{tabular}
    \caption{The likelihood of the Synthetic Dataset~2 (Figure~\ref{fig:DataExample2}) for the different parameters of \protect\hyperlink{QBD1}{QBD1}, assuming uniform initial distribution of levels and phases at the root of the tree, and $[a,b]=[3,20]$.}
    \label{tab:exampleModel1Synthetic}
\end{table}

\subsubsection{\protect\hyperlink{QBD2}{QBD2}: with $4$ phases} \label{Mod2:4phases}
 
We also applied \protect\hyperlink{QBD2}{QBD2} model with $4$ phases, summarised in Table~\ref{tab:QBD_overview} and defined in equation~\ref{eq:Mod2:4phasessyndata2}, to the dataset shown in Figure~\ref{fig:DataExample2}.
Optimisation using the BFGS algorithm yielded parameter estimates with a small positive drift whereas a parameter estimation using Nelder-Mead yielded a parameter estimates with small negative drift.
However, similar to Synthetic Dataset~1 (Figure~\ref{fig:DataExample1}), numerical optimisation appears to struggle to fit the 10-parameter model as evidenced by the large AICc values compared to other models~(Table \ref{tab:exampleModel2syn2}).

\begin{table}[H]
    \centering
    \scriptsize
    \begin{adjustbox}{max width=\linewidth}
    \begin{tabular}{lcccccc}
        \toprule
        Method & $[\lambda_1,\ldots,\lambda_5]$ & $[\mu_1,\ldots,\mu_5]$ & Drift $\gamma$ & Likelihood & Log-Likelihood &AICc\\ 
        \midrule
        Manual & 
        $[10,\ 0.1,\ 0.1,\ 1,\ 1]$ & 
        $[10,\ 0.1,\ 0.1,\ 1,\ 1]$ & 
        $\approx 0$ & 
        $6.0005 \times 10^{-102}$ & 
        $-233.0718$&$491.9331$ \\ 
        
        Nelder–Mead & 
        $[0.2134,\ 0.8380,\ 4.4182,\ 0.3612,\ 4.9838]$ & 
        $[0.0027,\ 0.0074,\ 0.0043,\ 0.0036,\ 0.0072]$ & 
        $-0.0048$ & 
        $1.7562 \times 10^{-103}$ & 
        $-236.6031$&$498.9957$ \\
        BFGS & 
        $[2.1530,\ 4.9817,\ 6.0689,\ 2.2324,\ 1.5064]$ & 
        $[0.0990,\ 0.0833,\ 0.0593,\ 0.0451,\ 0.0763]$ & 
        $0.0159$ & 
        $1.2312\times 10^{-46}$ & 
        $-105.7109$&$237.2113$ \\
        \bottomrule
    \end{tabular}
    \end{adjustbox}
    \caption{ The likelihood of the Synthetic Dataset~2 (Figure~\ref{fig:DataExample2}) for the different parameters of \protect\hyperlink{QBD2}{QBD2}, assuming uniform initial distribution of levels and phases at the root of the tree, and $[a,b]=[3,20]$.}    \label{tab:exampleModel2syn2}
\end{table}

\subsubsection{Comparison of QBD Model Fits for Synthetic Dataset~2~(Figure~\ref{fig:DataExample2})}

The zero-drift {\protect\hyperlink{QBD}{QBD1}} model provided the best fit to Synthetic Dataset~2 (Figure~\ref{fig:DataExample2}), yielding the minimum $AICc = 4.839$. However, as numerical methods did not appear to do a good job of exploring models that allow negative drift we don't read too much into this.

\section{Numerical Examples: Empirical Data}\label{sec:NumExData}

\begin{figure}[H]
\centering     \includegraphics[width=\textwidth]{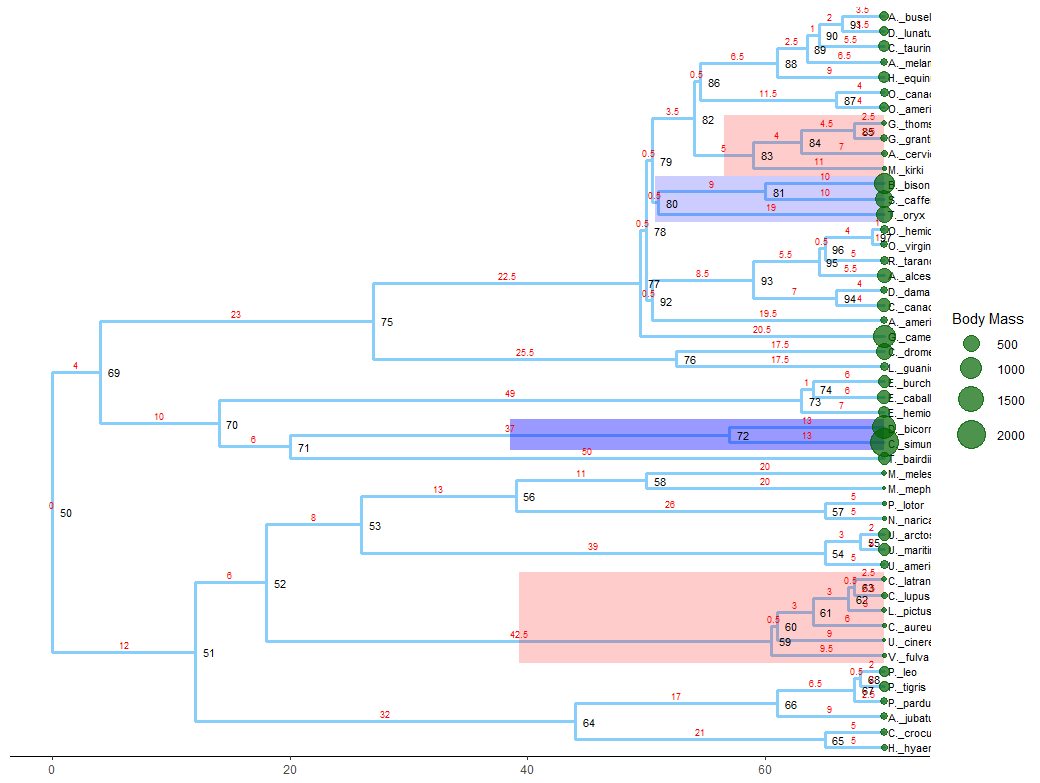}
        \caption{
        A phylogenetic tree of 49 mammal species, with branch lengths shown on the edges in red, nodes in black, and body mass values (kg) displayed at the tips using circles of varying size. 
        Trait values range from $a = 2.5$ to $b = 2000$. Circle sizes are proportional to body mass, with smaller circles representing smaller species and larger circles representing larger species.}
        \label{DataExample2}
 \end{figure}

In this section, we analyse a phylogenetic tree representing the evolutionary relationships among 49 mammal species, as depicted in Figures~\ref{DataExample2} and \ref{DataExample3}. 
Trait measurements, home-range area (km²) and body mass (kg), are observed at the terminal tips. 
The datasets originate from Garland et al.~\cite{Garland1992} and are available through the phytools R package~\cite{revell2012phytools}. For the analysis we applied a selection of different parameter values for \protect\hyperlink{QBD3}{QBD3}  ($5$ phases) to explore different values of mean drift, we also compare to \protect\hyperlink{QBD0}{QBD0} (2 phases, no drift) and a selection of parameters for \protect\hyperlink{QBD4}{QBD4} ($5$ phases).


\subsection{Empirical Dataset 1: Body mass ($kg$)}


\subsubsection{Histogram of observed body mass}

\begin{figure}[H]
\centering     \includegraphics{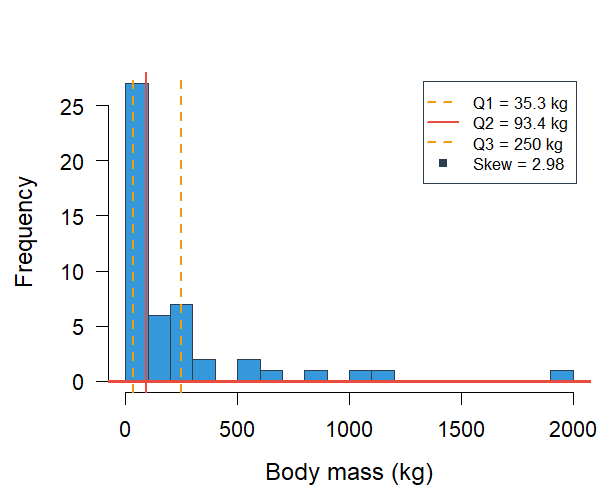}

        \caption{Histogram of body mass ($kg$) showing positive skew with most values concentrated at the lower ranges. }
        \label{DataExample1}
 \end{figure}

The histogram in Figure~\ref{DataExample1} reveals a right-skewed distribution of body mass, showing that the majority of species are smaller-bodied.
Observed body mass values ranges from $2.5\mathrm{kg}$ to $2000\mathrm{kg}$, with approximately $75\%$ of species falling below $250\mathrm{kg}$~(Figure~\ref{DataExample1}). 
We further aim to use the QBD model to underscore whether the observed trait values at the tips were driven by negative drift or zero drift, implying they were already small at their ancestral states, or whether they were initially very small and increased due to positive drift.

\subsubsection{\protect\hyperlink{QBD3}{QBD3} with $5$ phases}

We performed the analysis for the ${\bf r}$ vectors $\#1-\#5$ as shown in Table~\ref{tab:exampleDATA2} (${\bf r}$ values and drift are analogous to Table~\ref{tab:Synexamplet1}). The output of this analysis is also presented in  Figures~\ref{StatTraitsModel5phasesOverallcherriesEmpirical}-\ref{StatTraitsModel5phasesOverallParrentTraitsOnlyEmpirical} in Appendix~\ref{outputEmpiricalDataExample2}.

\begin{table}[H]
    \centering
    \begin{tabular}{ccccc}
        \toprule
       ${\bf r}$\# & ${\bf r}=[r_1\  \ldots \ r_8]$ & drift $\gamma$ &Likelihood &LogLikelihood\\ 
        \midrule
        1    & [2 1 4 3 2 1 6 5]  & $-1.0169$&
        $1.3951\times10^{-70}$&$  -160.8480$\\ 
               2   & [10     3     5     5     4     6     1     1]   &  $2.2472$ &$2.8247\times10^{-180}$  &$-413.4269$
\\ 
        3   & [12     7     8    8    4     6     1     1 ]  & $3.2909$ &$3.0596\times10^{-227}$  &$-521.5685$ \\ 
         4  & [10     3     8     8    9     10     7    7] &  $0.7455$&$2.0960\times10^{-94}$   &$-215.7030$\\
        5  & [9    6     7     7    5     7     3    3]  &  $1.6928$  &$7.2798\times10^{-128}$   &$-292.7458$
        \\
        
        \bottomrule
    \end{tabular}
    \caption{The likelihood of the Empirical Dataset~1 (Figure~\ref{DataExample2}) for the different parameters of \protect\hyperlink{QBD3}{QBD3}, assuming uniform initial distribution of levels and phases at the root, and $a=2.5$, $b=2000$.}
    \label{tab:exampleDATA2}
\end{table}

\subsubsection{\protect\hyperlink{QBD0}{QBD0}: with $2$ phases}

\begin{table}[H]
    \centering
    \begin{tabular}{cccc}
        \toprule
         $\lambda=\mu$ & drift $\gamma$ &Likelihood &LogLikelihood \\ 
        \midrule
         $10$  & 0&$7.8129\times 10^{-75}$ &$-170.6381$\\ 
        
        \bottomrule
    \end{tabular}
 \caption{The likelihood of the Empirical Dataset~1 (Figure~\ref{DataExample2}) for the \protect\hyperlink{QBD0}{QBD0} model, assuming uniform initial distribution of levels and phases at the root of the tree, and $a=2.5$, $b=2000$.}
    \label{tab:exampleModel0EmpiricalData2}
\end{table}

\subsubsection{\protect\hyperlink{QBD4}{QBD4} with $5$ phases}

We then applied the {\protect\hyperlink{QBD4}{QBD4}} model with $5$ phases, summarised in Table~\ref{tab:QBD_overview}, with $\alpha=0.9$, for the various values of parameters $\beta$, with
\begin{align}
{\bf Q}^{[n,n+1]} &=
\begin{bmatrix}
\alpha r_1 & (1-\alpha)r_1 & 0 & 0 & 0 \\
\frac{(1-\alpha)r_2}{2} & \alpha r_2 & \frac{(1-\alpha)r_2}{2} & 0 & 0 \\
0 & 0 & 0 & 0 & 0 \\
0 & 0 & 0 & 0 & 0 \\
0 & 0 & 0 & 0 & 0
\end{bmatrix} , \notag \\
{\bf Q}^{[n,n-1]} &=
\begin{bmatrix}
0 & 0 & 0 & 0 & 0 \\
0 & 0 & 0 & 0 & 0 \\
0 & 0 & 0 & 0 & 0 \\
0 & 0 & \frac{(1-\alpha)r_4}{2} & \alpha r_4 & \frac{(1-\alpha)r_4}{2} \\
0 & 0 & 0 & (1-\alpha)r_5 & \alpha r_5
\end{bmatrix} , \notag \\
{\bf Q}^{[n,n]} &=
\begin{bmatrix}
 -r_1 & 0 & 0 & 0 & 0 \\
 0 & -r_2 & 0 & 0 & 0 \\
 0 & (1-\beta)r_3 & -r_3 & \beta r_3 & 0 \\
 0 & 0 & 0 & -r_4 & 0 \\
 0 & 0 & 0 & 0 & -r_5
\end{bmatrix}
.
\label {eq:QBD:5phasesbiological}
\end{align}
The output is presented in Table~\ref{tab:exampleModel4emp1}.

\begin{table}[H]
    \centering
    
    \begin{tabular}{cccccc}
        \toprule
        $\alpha$ & $\beta$ & $[r_1 \ldots r_5]$ & drift $\gamma$ & Likelihood & LogLikelihood \\ 
        \midrule

         $0.9$ & $0.70$ & 
        $[2,\ 1,\ \frac{1}{2},\ 1,\ 2]$ 
        & $-0.4444$ & $1.3296 \times 10^{-66}$ & $-151.6857$ \\

               $0.9$ & $0.50$ & 
        $[1,\ 1,\ 1,\ 1,\ 1]$ 
        & $0$ & $1.2656 \times 10^{-65}$ & $-149.4325$ \\

               $0.9$ & $0.70$ & 
        $[1,\ 1,\ 1,\ 1,\ 1]$ 
        & $-0.3871$ & $ 1.7394 \times 10^{-65}$ & $-149.1145$ \\

               $0.9$ & $0.30$ & 
        $[1,\ 1,\ 1,\ 1,\ 1]$ 
        & $0.3871$ & $1.2126 \times 10^{- 65}$ & $-151.7779$ \\

              \bottomrule
    \end{tabular}
    \caption{The likelihood of the Empirical Dataset~1 (Figure~\ref{DataExample2}) for the \protect\hyperlink{QBD4}{QBD4} model, assuming uniform initial distribution of levels and phases at the root of the tree, and $a=2.5$, $b=2000$.}
    \label{tab:exampleModel4emp1}
\end{table}


       

\subsection{Empirical Dataset 2: Home range area ($km^{2}$)}
 \begin{figure}[H]
\centering        \includegraphics[width=\textwidth]{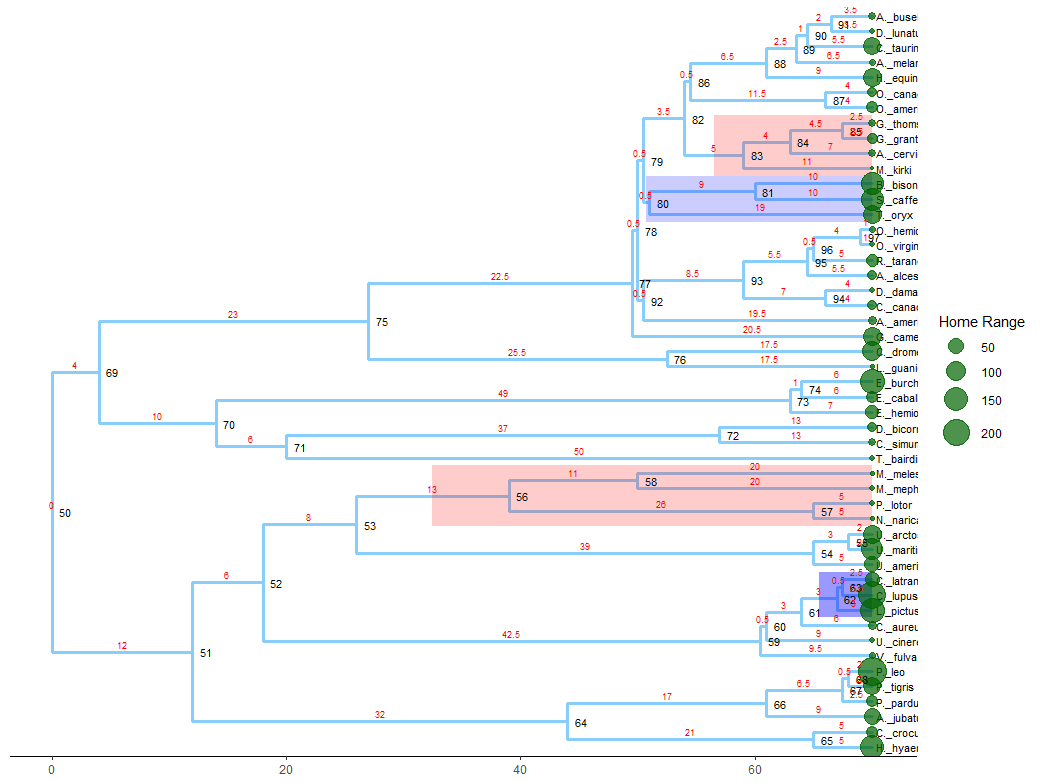}
        \caption{
     Phylogenetic tree of 49 mammal species showing home range area values ($\mathrm{km}^{2}$) at the tips. 
     Trait values range from $a = 0.043$ to $b = 236$. 
     Circle sizes are proportional to home range area, with smaller circles representing species occupying smaller territories and larger circles representing those with more extensive ranges.}
   
        \label{DataExample3}
 \end{figure}

\subsubsection{Histogram of observed home range area }

\begin{figure}[H]
\centering     \includegraphics{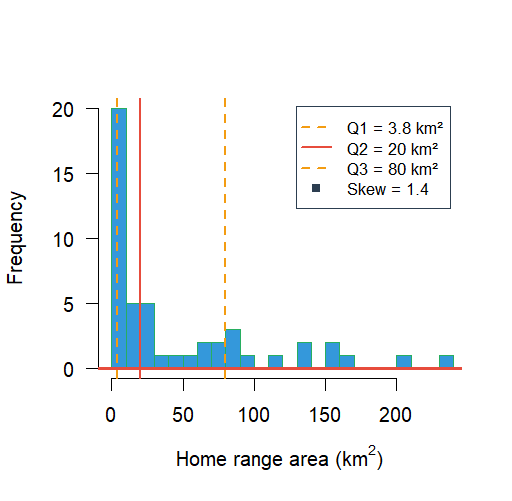}

        \caption{Histogram of home range area ($km^{2}$) showing positive skew with most values concentrated at the lower ranges. }
        \label{Histogram_Homerangearea}
 \end{figure}

Similar to the histogram of body mass ($\mathrm{kg}$), the histogram of home range area ($\mathrm{km}^{2}$) in Figure~\ref{Histogram_Homerangearea} reveals a right-skewed distribution, indicating majority of the species occupy relatively small ranges. 
Observed home range areas span from $0.043\,\mathrm{km}^{2}$ to $236\,\mathrm{km}^{2}$, with approximately $75\%$ of species fall below $80\,\mathrm{km}^{2}$. 

\subsubsection{\protect\hyperlink{QBD3}{QBD3} with $5$ phases}

We performed the analysis for the ${\bf r}$ vectors $\#1-\#5$ as shown in Table~\ref{tab:exampleDATA2B} (${\bf r}$ values and drift are analogous to Table~\ref{tab:Synexamplet1}). The output of this analysis is also presented in  Figures~\ref{StatTraitsModel5phasesOverallcherriesEmpiricalData3}-\ref{StatTraitsModel5phasesOverallParrentTraitsOnlyEmpiricalData3} in Appendix~\ref{outputEmpiricalDataExample3}.

\begin{table}[H]
    \centering
    \begin{tabular}{ccccc}
        \toprule
       ${\bf r}$\# & ${\bf r}=[r_1\  \ldots \ r_8]$ & drift $\gamma$ &Likelihood &LogLikelihood\\ 
        \midrule
        1    & [2 1 4 3 2 1 6 5]  & $-1.0169$&
        $4.6623\times 10^{-126}$&$-288.5862$\\ 
               2   & [10     3     5     5     4     6     1     1]   &  $2.2472$ &$2.1913\times 10^{-199}$  &$-457.4299$
\\ 
        3   & [12     7     8    8    4     6     1     1 ]  & $3.2909$ &$2.3053\times 10^{-236}$  & $-542.5748$ \\ 
         4  & [10     3     8     8    9     10     7    7] &  $0.7455$&$1.6930\times 10^{-114}$    &$ -261.9682$\\
        5  & [9    6     7     7    5     7     3    3]  &  $1.6928$  &$9.8633\times 10^{-143}$   &$-326.9809$
        \\
        
        \bottomrule
    \end{tabular}
    \caption{The likelihood of the Empirical Dataset~2 (Figure~\ref{DataExample2}) for the different parameters of \protect\hyperlink{QBD3}{QBD3}, assuming uniform initial distribution of levels and phases at the root of the tree, and $a=0.043$, $b=236$.}
   \label{tab:exampleDATA2B}
\end{table}

\subsubsection{\protect\hyperlink{QBD0}{QBD0}: with $2$ phases}

\begin{table}[H]
    \centering
    \begin{tabular}{ccccc}
        \toprule
        $\lambda=\mu$ & drift $\gamma$ &Likelihood &LogLikelihood \\ 
        \midrule
         $10$  & 0&$2.6742\times 10^{-125}$ &$-286.8395$\\ 
        
        \bottomrule
    \end{tabular}
 \caption{The likelihood of the Empirical Dataset~2 (Figure~\ref{DataExample3}) under the \protect\hyperlink{QBD0}{QBD0} model, assuming uniform initial distribution of levels and phases at the root of the tree, and $a=0.043$, $b=236$.}
    \label{tab:exampleModel0Empirical}
\end{table}

\subsubsection{\protect\hyperlink{QBD4}{QBD4} with $5$ phases}

We then aplied the {\protect\hyperlink{QBD4}{QBD4}} model with $5$ phases, summarised in Table~\ref{tab:QBD_overview} and generator in~\eqref{eq:QBD:5phasesbiological}, with $\alpha=0.9$, for the various values of parameters $\beta$. The output is presented in Table~\ref{tab:exampleModel4emp1EMpData2}.

\begin{table}[H]
    \centering
        \begin{tabular}{cccccc}
        \toprule
        $\alpha$ & $\beta$ & $[r_1 \ldots r_5]$ & drift $\gamma$ & Likelihood & LogLikelihood \\ 
        \midrule

        $0.9$ & $0.60$ & $[4,\ 2,\ \frac{1}{2},\ 1,\ 2]$ & $-0.2727$ & $2.3775 \times 10^{-135}$ & $-309.9829$ \\

               $0.9$ & $0.50$ & 
        $[1,\ 1,\ 1,\ 1,\ 1]$ 
        & $0$ & $ 1.3012 \times 10^{-226}$ & $-520.1209$ \\

               $0.9$ & $0.70$ & 
        $[1,\ 1,\ 1,\ 1,\ 1]$ 
        & $-0.3871$ & $ 1.5683 \times 10^{-226}$ & $-519.9343$ \\

               $0.9$ & $0.30$ & 
        $[1,\ 1,\ 1,\ 1,\ 1]$ 
        & $0.3871$ & $2.3095 \times 10^{-228}$ & $-524.1524$ \\

              \bottomrule
    \end{tabular}
    \caption{The likelihood of the Empirical Dataset~2 (Figure~\ref{DataExample3}) for the \protect\hyperlink{QBD4}{QBD4} model, assuming uniform initial distribution of levels and phases at the root of the tree, and $a=2.5$, $b=2000$.}
    \label{tab:exampleModel4emp1EMpData2}
\end{table}

\subsection{Comments on the Empirical Datasets~1 and~2}

For the two empirical datasets we tested a range of different parameter sets for models QBD3, QBD0 and QBD4. For QBD3 and QBD4 we picked some parameters that give negative drift and some that give positive drift. Recall that compared to QBD3, the $\alpha=0.9$ parameter in QBD4 enforces that the phases change infrequently compared to the level. 

For the body mass dataset with QBD3 the rank of the log likelihoods corresponded to the drift, with the parameters producing negative drift (${\bf r}$ $\#1$) preferred. For the body mass dataset with QBD4 the best parameter set also had negative drift but we note that the parameter set with zero drift and the parameter set with positive drift had a better likelihood than any version of QBD3. QBD0 with no drift did better than the positive drift variants of QBD3 but worse than all variants of QBD4.

For the home range data the rank of the log likelihoods corresponded with increasing absolute value of the drift with the best parameter set having the smallest drift. Most of the QBD3 parameter sets scored better than most of the QBD4 parameter sets. QBD0 performed better than all but one of the QBD3 parameter sets. The best fitting parameter set overall was a QBD3 option with positive drift.

\section{Conclusions}\label{Sec:Conclusions}

We considered a quasi-birth-and-death (QBD) process that undergoes binary branching at certain times to model the evolution of a continuous trait through discretisation.
We developed an efficient recursive algorithm for computing the likelihood of observing a phylogenetic tree under a duplicating quasi-birth-and-death (QBD) process.
We examined a range of QBD models to capture the evolution of trait dynamics, with parameters estimated via a multi-start optimisation strategy and factor-scaled likelihood evaluations to ensure numerical stability and mitigate underflow.
We applied various QBD models to synthetic datasets to illustrate the framework's general-purpose capability to illustrate a range of behaviours.
We also applied biologically motivated QBD structures to two empirical datasets to assess the evidence for positive versus negative drift in these traits. 
We compared the performance of different QBD structures for each dataset using AICc.
Overall, this methodology establishes a foundation for future extensions to other branching stochastic systems and broader applications in queueing and matrix-analytic contexts.
Future work may focus on the applications of stochastic fluid models to directly model the evolution of continuous traits without discretisation. 
While all the examples we explored in this paper used level-independent QBDs, we could also explore level-dependent QBDs to model stabilizing selection for comparative phylogenetic data.
For some comparative studies it is of interest to study how different traits coordinate with each other, in this context it would be very interesting to apply recent results by Aksamit et al.~\cite{Aksamit2025} who show that a random walk on a quadrant can be analysed as a one-dimensional QBD given an appropriate mapping of the state space.

\section{Statements and declarations}\label{Sec:Declarations}

\noindent{\bf Data availability}\\

The authors declare that the data supporting the findings of this study are available within the paper.\\

\noindent{\bf Authorship contribution statement}\\

This paper contributes to a chapter in the PhD project by Habtu Kiros Nigus~\cite{NigusPhD}. The following are the contributions of the authors, Habtu Kiros Nigus (HKN), Barbara R. Holland (BRH), Ma\l gorzata M. O'Reilly (MMO).
\begin{itemize}
\item Sections~\ref{sec:OurModel},~\ref{sec:NumExSynthetic} and~\ref{sec:NumExData}, Problem formulation and methodology development: HKN, BRH, and MMO;

\item Sections~\ref{sec:Theory}, Derivation of the theoretical expressions and algorithms: MMO;

\item Sections~\ref{sec:NumExSynthetic} and~\ref{sec:NumExData}, Coding: HKN, MMO;

\item Sections~\ref{sec:NumExSynthetic} and~\ref{sec:NumExData}, Data analysis: HKN;

\item Sections~\ref{sec:QBDmodels},~\ref{sec:NumExSynthetic} and~\ref{sec:NumExData}, Numerical analysis: HKN;

\item Conceptualisation, Biological background: HKN, BRH;

\item Conceptualisation, Mathematical background: HKN, BRH, MMO;

\item Write-up and edits: HKN, BRH,  MMO.

\end{itemize}

\bibliographystyle{abbrv}
\bibliography{arxiv_refs}

\newpage
\appendix


\section{The effect of the mean drift}
\label{sec:OutputQBD5phases}

To illustrate the effect of the mean drift on the likelihood of observing a given tree, we consider the \protect\hyperlink{QBD3}{QBD3} model in Section~\ref{sec:QBDmodels} with five phases, and evaluate the likelihood of the two synthetic datasets for a range of the QBD parameters. First, the stationary distribution of the QBD for the ${\bf r}$ vectors $\#1-\#4$ in Table~\ref{tab:Synexamplet1} (analogous to Table~\ref{tab:Synexamplet2}), is presented in Appendix~\ref{sec:OutputQBD5phasessynJUST_QBD}. Next, the output for Synthetic Dataset~1~(Figure~\ref{fig:DataExample1}) and Synthetic Dataset~2~(Figure~\ref{fig:DataExample2}) is presented in Section~\ref{sec:OutputQBD5phasessyn1} and Section~\ref{outputSyntheticDataExample2}, respectively.

We present a series of graphs illustrating the potential behaviours of the \protect\hyperlink{QBD3}{QBD3} model, emphasising the effect of mean drift and highlighting key insights obtained from Synthetic Datasets 1 and 2.
For this preliminary exploration, we selected parameter vectors designed to induce both negative and positive drift and fitted QBD models with these rates to each dataset.

Figures~\ref{StatLevelsModel5phases} to~\ref{StatTraitsModel5phases} in Appendix~\ref{sec:OutputQBD5phasessynJUST_QBD} depict the stationary distributions of the levels in the \protect\hyperlink{QBD3}{QBD3} model, showing the long-term behaviour of the process, whether it tends to converge towards the upper or lower boundary of the levels under positive or negative drift. 
We also include the stationary distributions of the phases, the states (joint level–phase combinations), and the traits (back-transformed from discretised levels) to illustrate how drift influences overall model dynamics.
The stationary distributions over states $(n, \varphi)$ indicate the long-run probability of occupying each level–phase combination, revealing both the tendency towards lower or higher levels and the corresponding environmental phase most often associated with them.

Figures~\ref{StatTraitsModel5phasesOverallcherries}--\ref{StatTraitsModel5phasesOverallParrentTraitsOnly} in Appendix~\ref{sec:OutputQBD5phasessyn1} illustrate the behaviours described below for Synthetic Dataset~1 (Figure \ref{fig:DataExample1}) when analysed under the \protect\hyperlink{QBD3}{QBD3} model.  Similarly, Figures~\ref{StatTraitsModel5phasesOverallcherriesData2}--\ref{StatTraitsModel5phasesOverallParrentTraitsOnlyData2} in Appendix~\ref{outputSyntheticDataExample2} show the corresponding behaviours for Synthetic Dataset~2 (Figure \ref{fig:DataExample2}) under the same \protect\hyperlink{QBD3}{QBD3} model.

Figure~\ref{StatTraitsModel5phasesOverallcherries} (Appendix~\ref{sec:OutputQBD5phasessyn1}) and Figure~\ref{StatTraitsModel5phasesOverallcherriesData2} (Appendix~\ref{outputSyntheticDataExample2}) depicts the likelihood of observing each tip ($i = 1, \ldots, 4$) given the initial trait value~$x$ at the start of the corresponding branch, illustrating the direction in which the likelihood peaks relative to the observed tip values, that is, whether it shifts to the right or left depending on the QBD process drift condition. 

Figures~\ref{StatTraitsModel5phasesOverallParrentLevel}--\ref{StatTraitsModel5phasesOverallParrentTrait} (Appendix~\ref{sec:OutputQBD5phasessyn1}) and Figures~\ref{StatTraitsModel5phasesOverallParrentLevelData2}--~\ref{StatTraitsModel5phasesOverallParrentTraitData2} (Appendix~\ref{outputSyntheticDataExample2}) illustrate the likelihood of observing the entire phylogenetic tree, conditional on the level~$n$ (trait~$x$) and phase~$\varphi$, highlighting the most probable level–phase (trait-phase) combinations, revealing both the likely initial levels (traits) and their associated environmental phases.

Finally, in Figure~\ref{StatTraitsModel5phasesOverallParrentTraitsOnly} (Appendix~\ref{sec:OutputQBD5phasessyn1}) and Figure~\ref{StatTraitsModel5phasesOverallParrentTraitsOnlyData2} (Appendix~\ref{outputSyntheticDataExample2}) we present the likelihood of observing the phylogenetic tree given the starting trait~$x$, under both positive and negative drift conditions, illustrating whether the process is more likely to have originated from higher or lower trait values depending on the direction of the drift.

\newpage
\subsection{The effect of the mean drift: \protect\hyperlink{QBD3}{QBD3} model
}
\label{sec:OutputQBD5phasessynJUST_QBD}

\begin{figure}[h]
\centering
\includegraphics[scale=0.4]{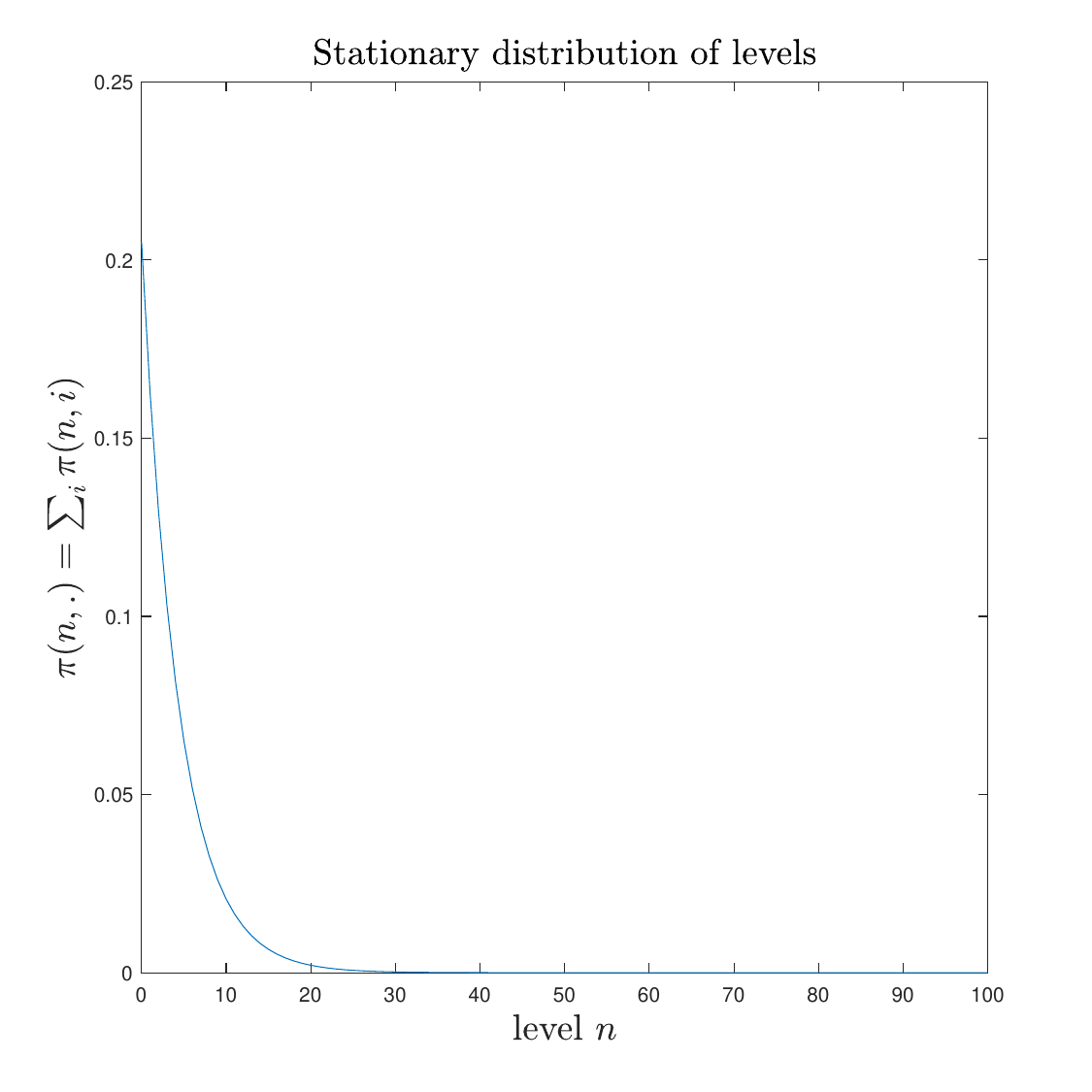}
\
\includegraphics[scale=0.4]{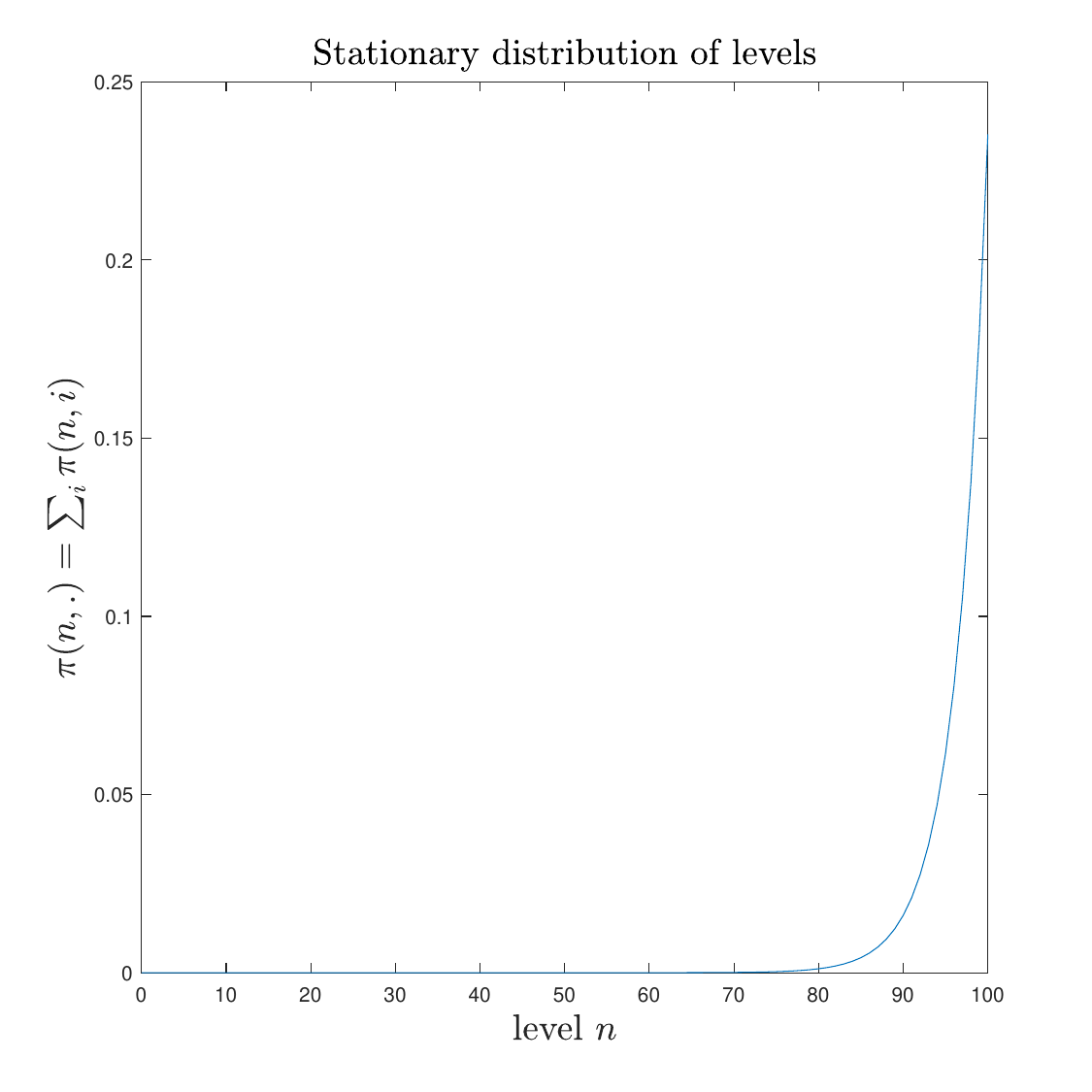}
\\
\includegraphics[scale=0.4]{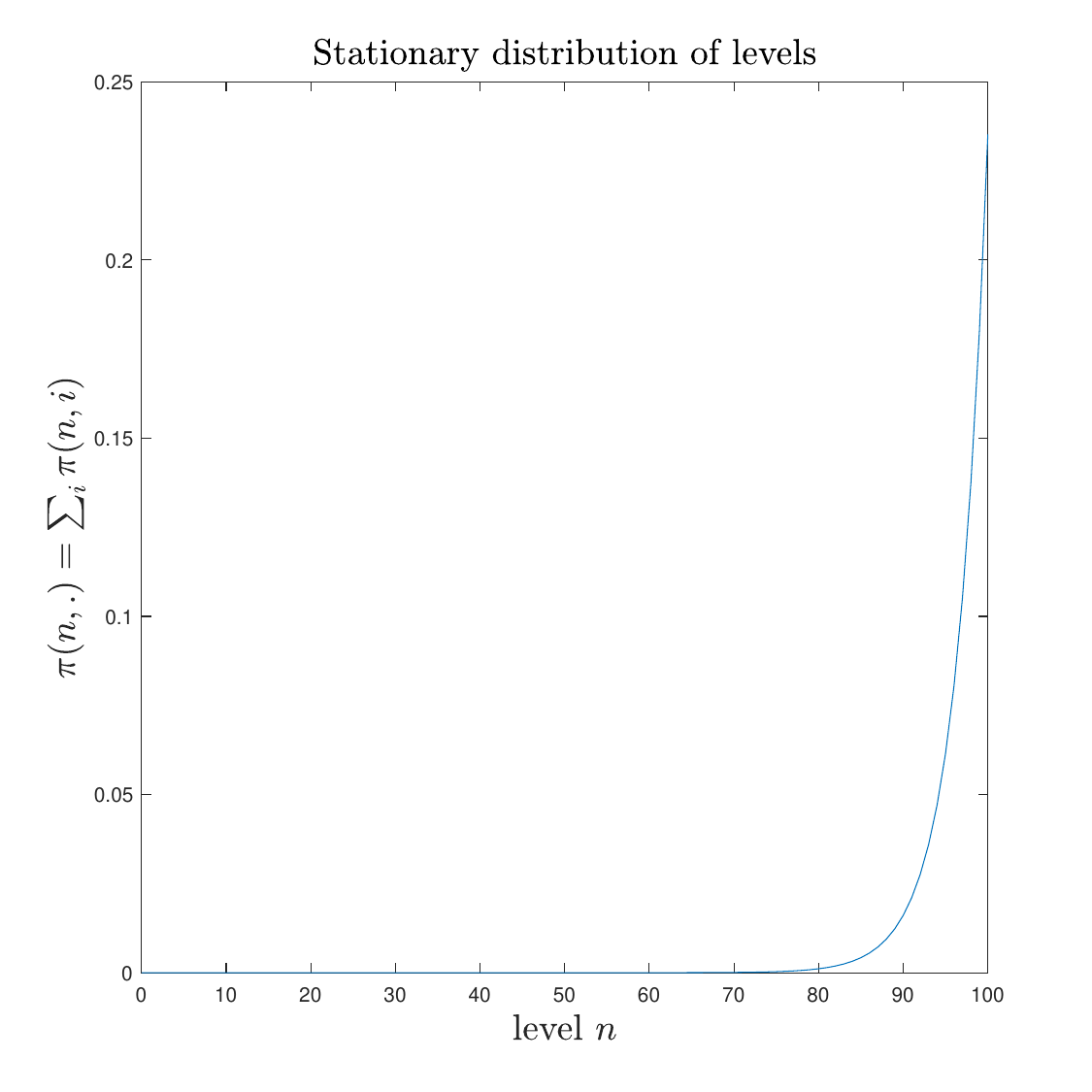}
\
\includegraphics[scale=0.4]{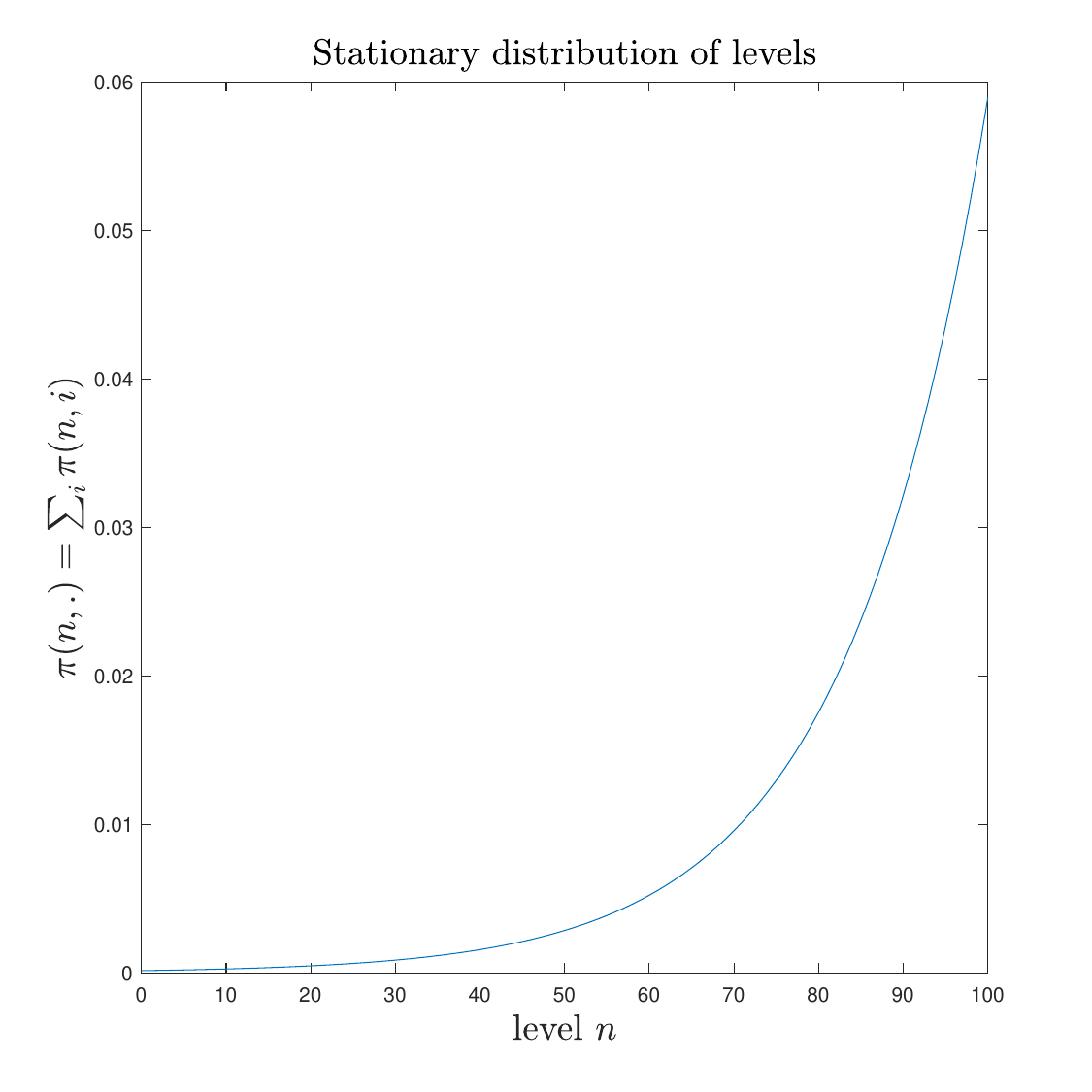}
\caption{From top left to the bottom right: Stationary distribution of the levels in the \protect\hyperlink{QBD3}{QBD3} model in Section~\ref{sec:QBDmodels} for the ${\bf r}$ vectors $\#1-\#4$ in Table~\ref{tab:Synexamplet1}.} 
\label{StatLevelsModel5phases}
\end{figure}

\begin{figure}
\centering
\includegraphics[scale=0.4]{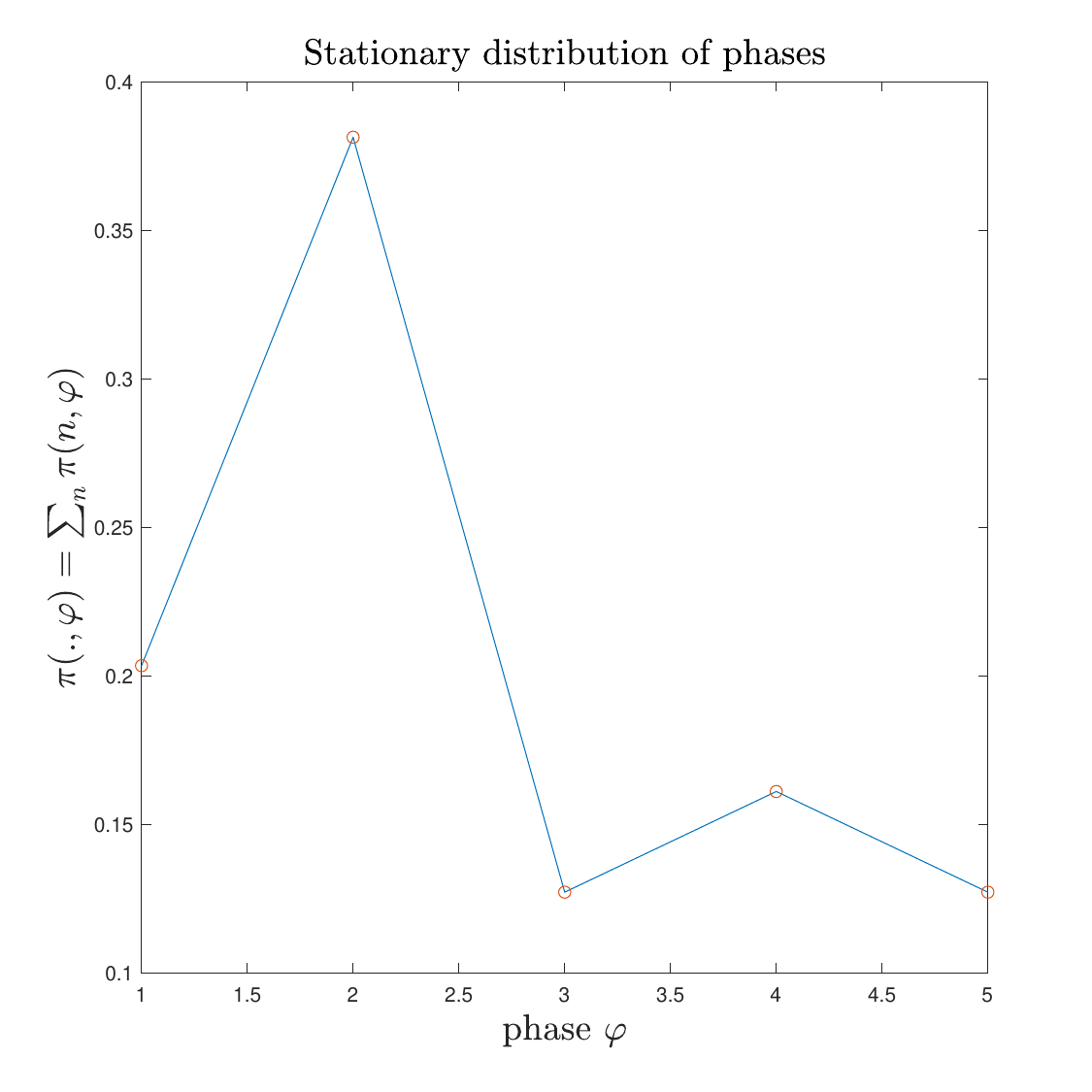}
\
\includegraphics[scale=0.4]{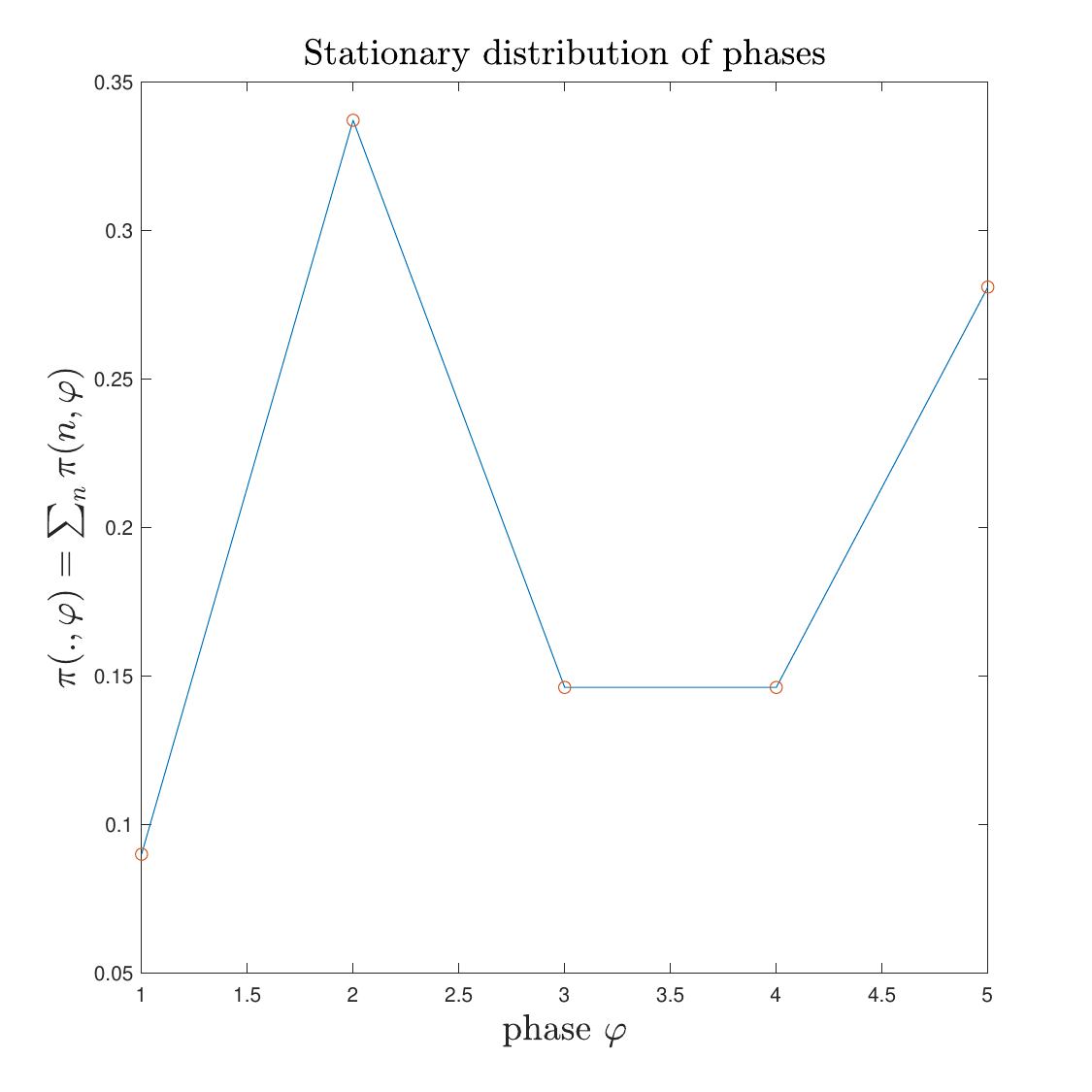}
\\
\includegraphics[scale=0.4]{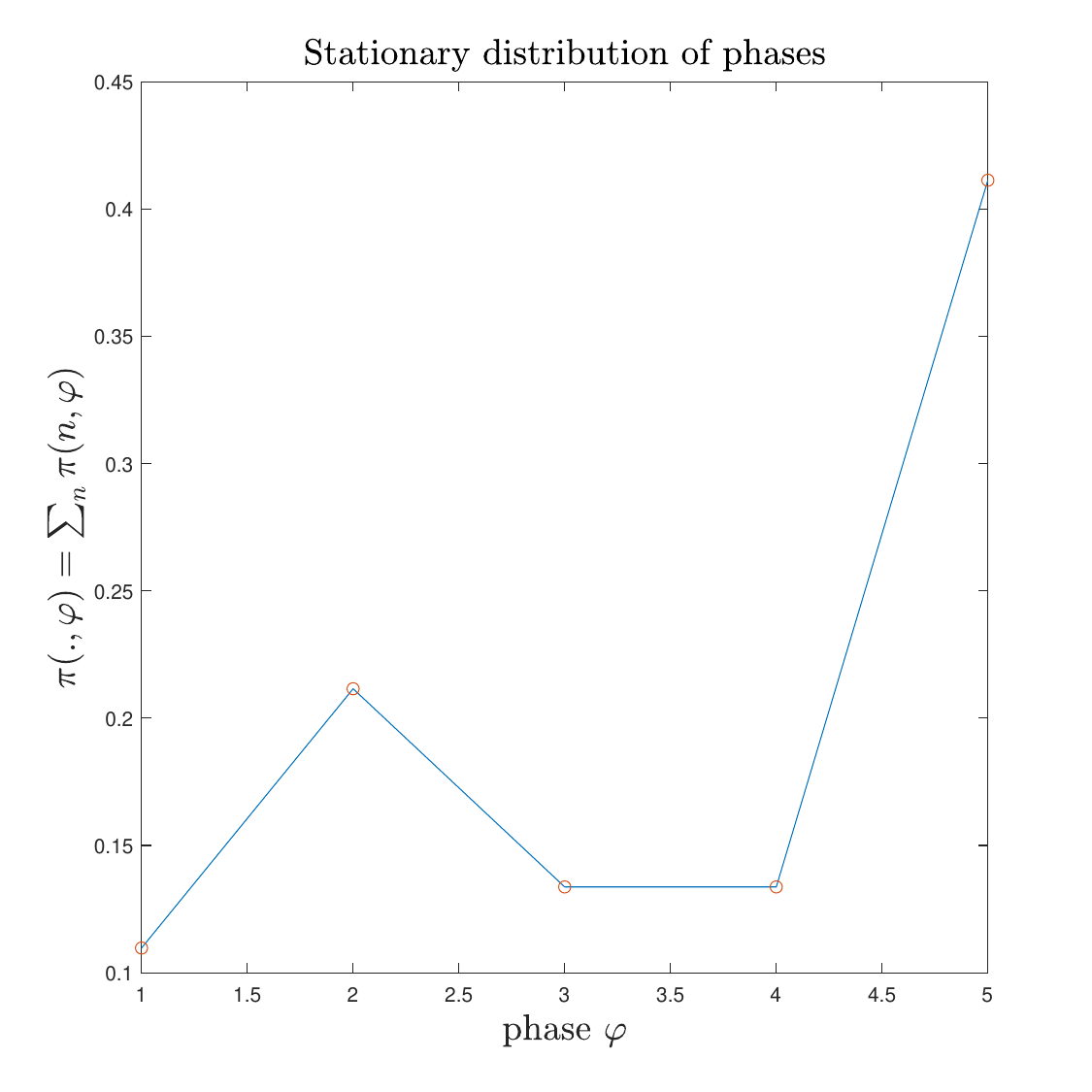}
\
\includegraphics[scale=0.4]{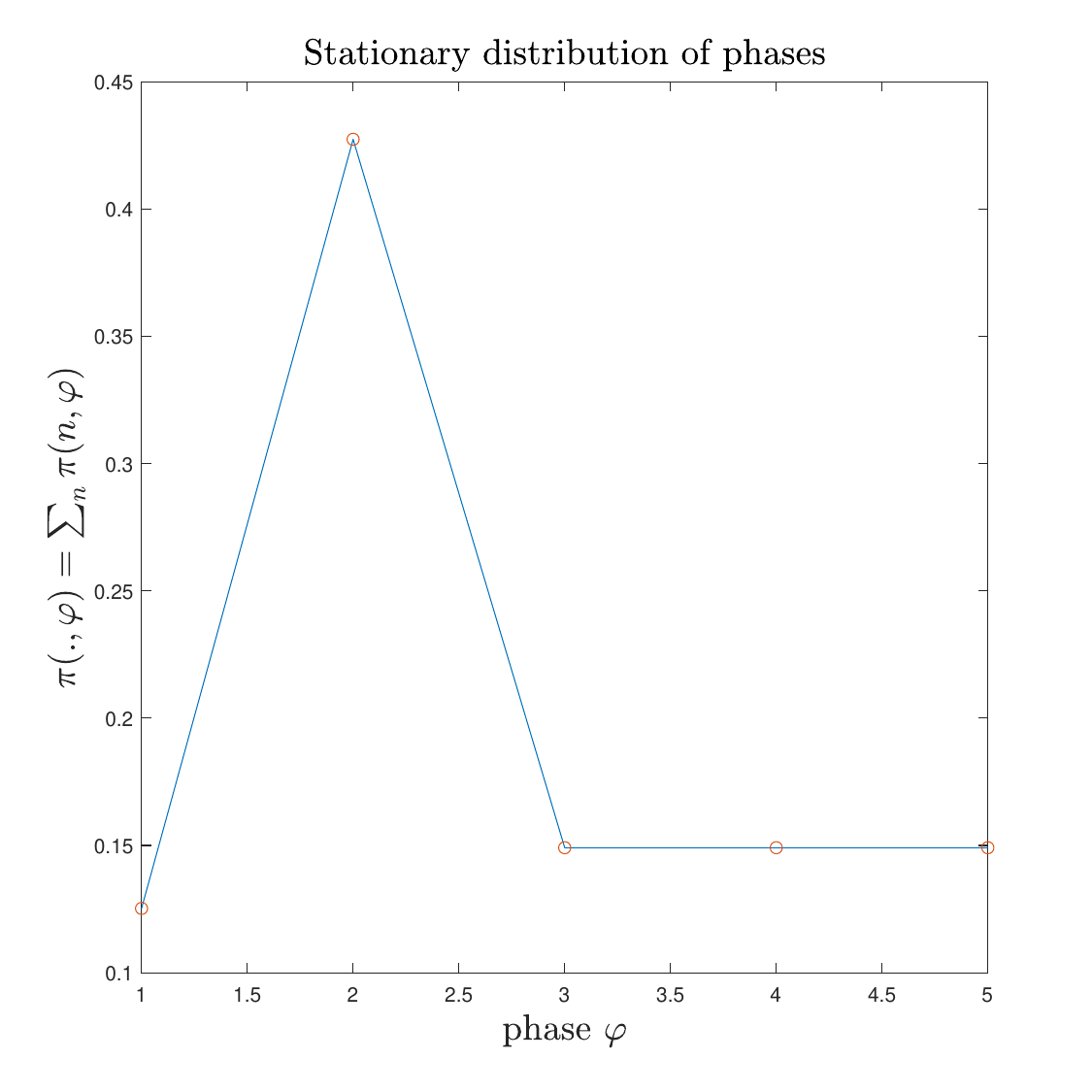}
\caption{From top left to the bottom right: Stationary distribution of the phases $\varphi=1,\ldots,5$ in the \protect\hyperlink{QBD3}{QBD3} model in Section~\ref{sec:QBDmodels} for the ${\bf r}$ vectors $\#1-\#4$ in Table~\ref{tab:Synexamplet1}.} 
\label{StatPhasesModel5phases}
\end{figure}

\begin{figure}
\centering
\includegraphics[scale=0.4]{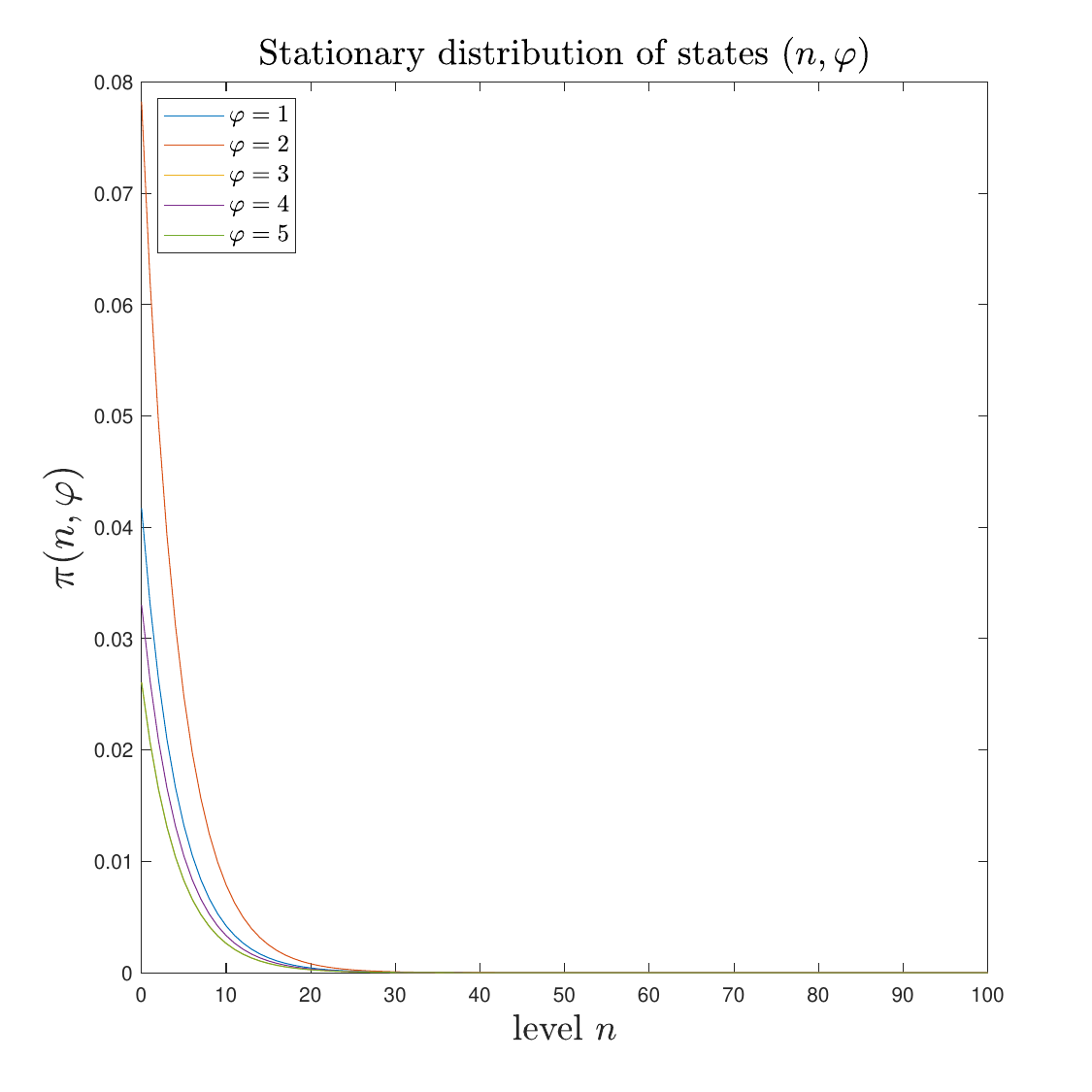}
\
\includegraphics[scale=0.4]{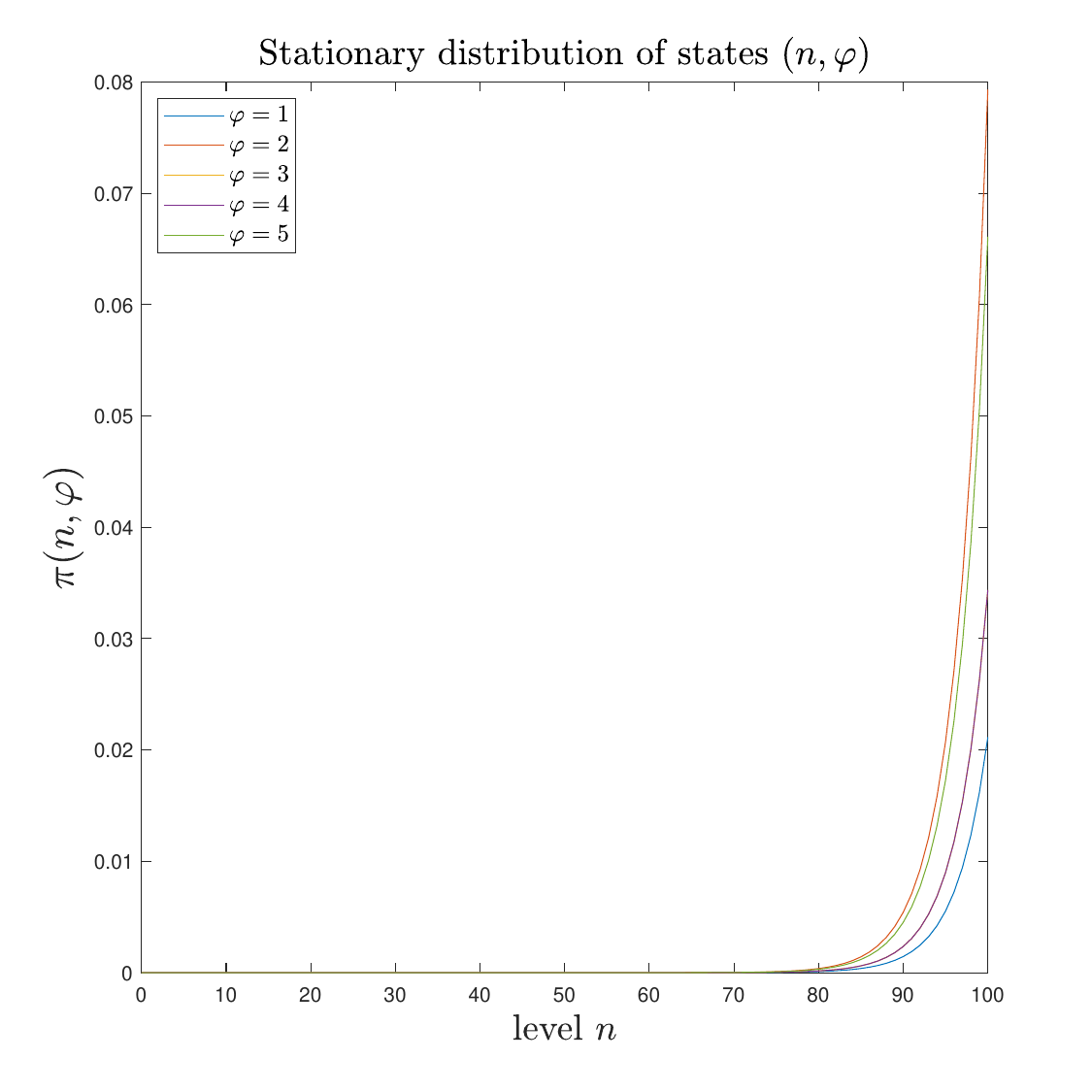}
\\
\includegraphics[scale=0.4]{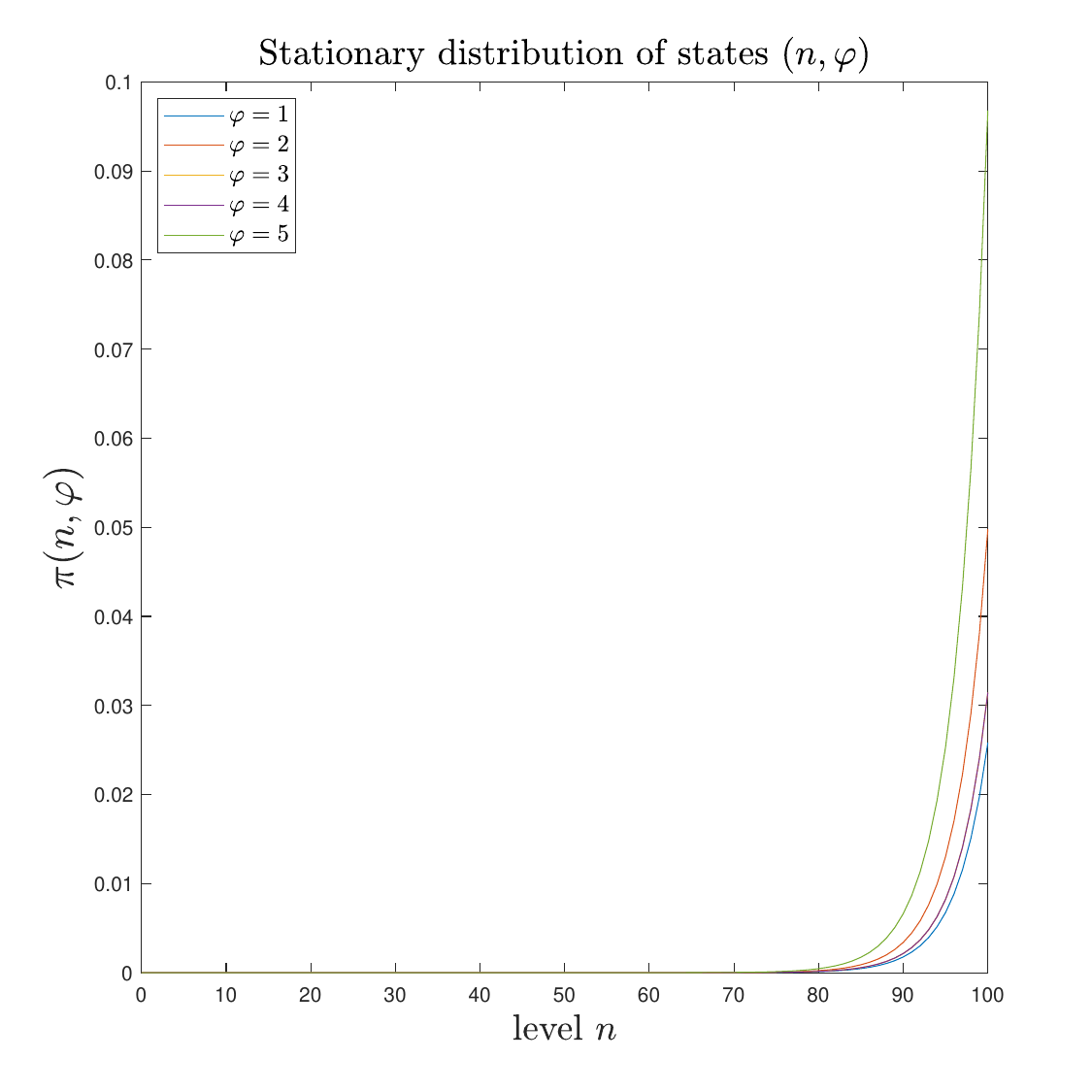}
\
\includegraphics[scale=0.4]{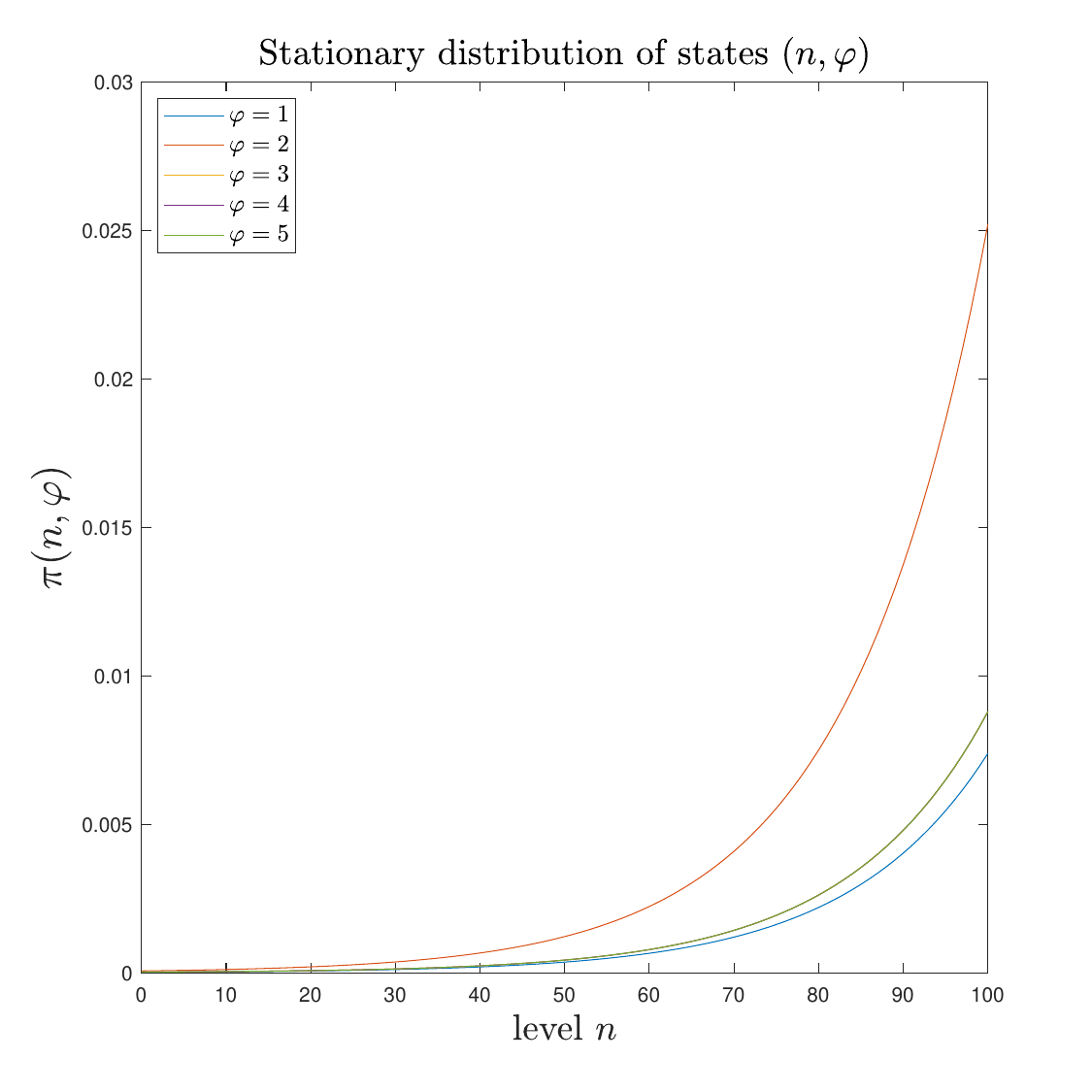}
\caption{From top left to the bottom right: Stationary distribution of the states in the \protect\hyperlink{QBD3}{QBD3} model in Section~\ref{sec:QBDmodels} for the ${\bf r}$ vectors $\#1-\#4$ in Table~\ref{tab:Synexamplet1}.} 
\label{StatStatesModel5phases}
\end{figure}

\begin{figure}
\centering
\includegraphics[scale=0.4]{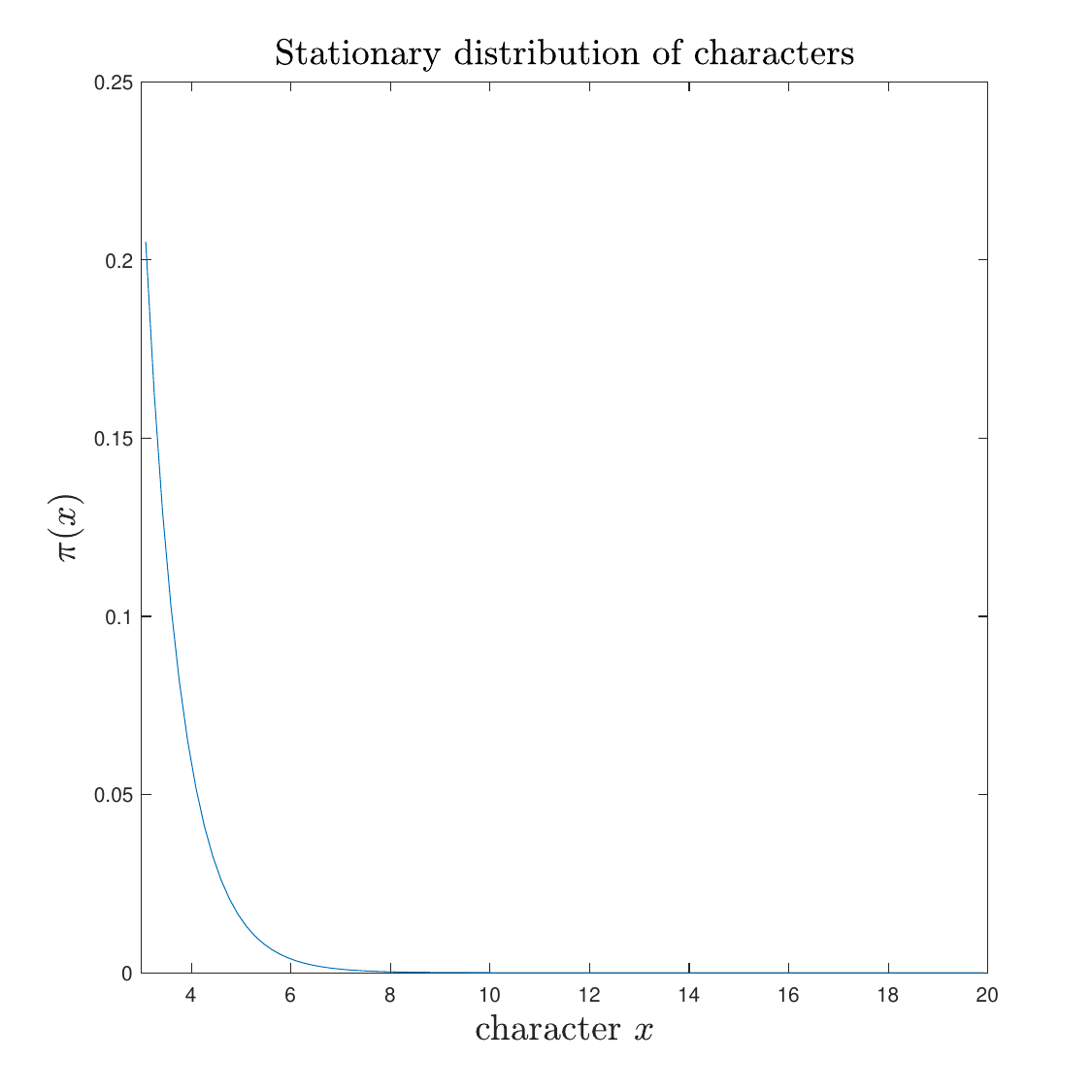}
\
\includegraphics[scale=0.4]{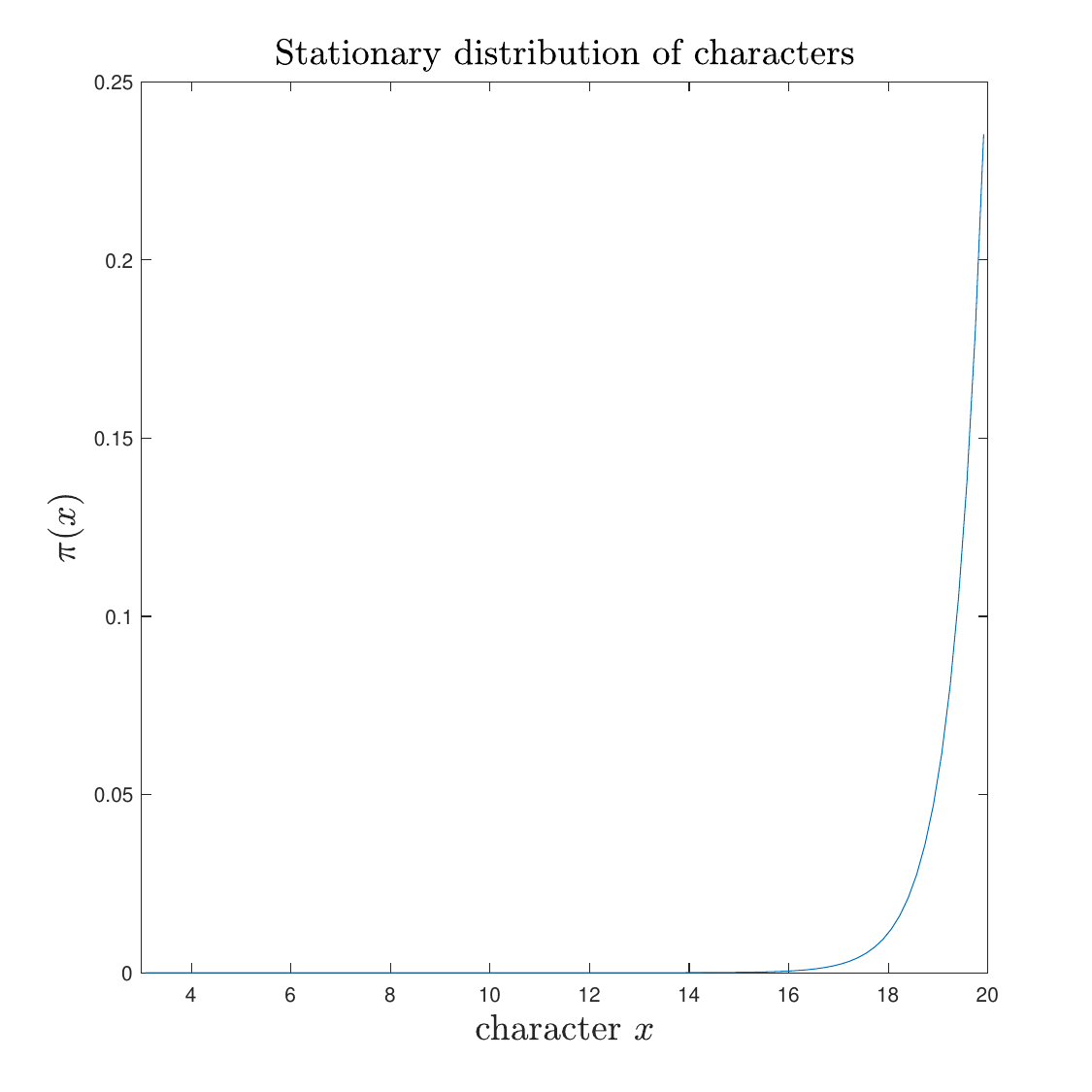}
\\
\includegraphics[scale=0.4]{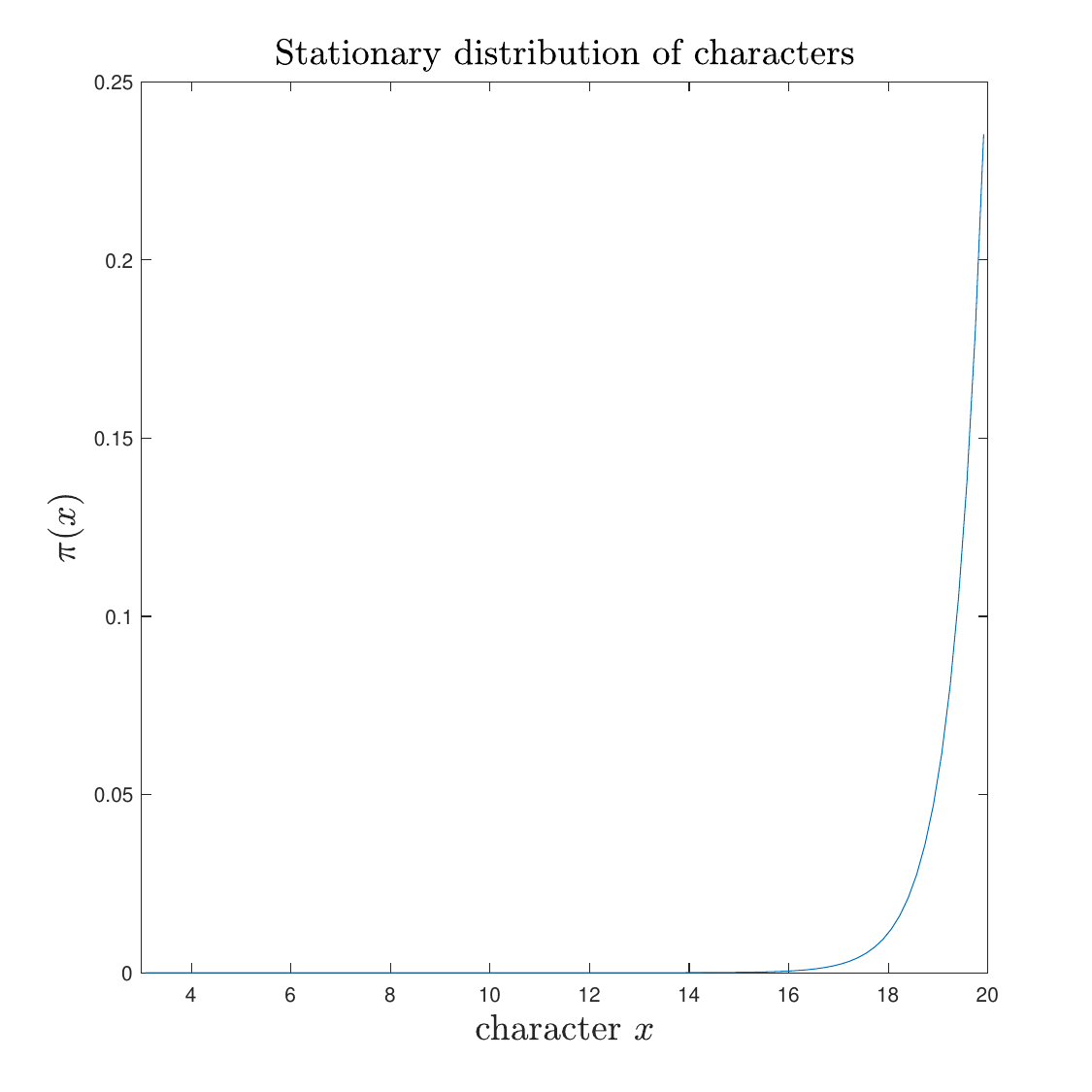}
\
\includegraphics[scale=0.4]{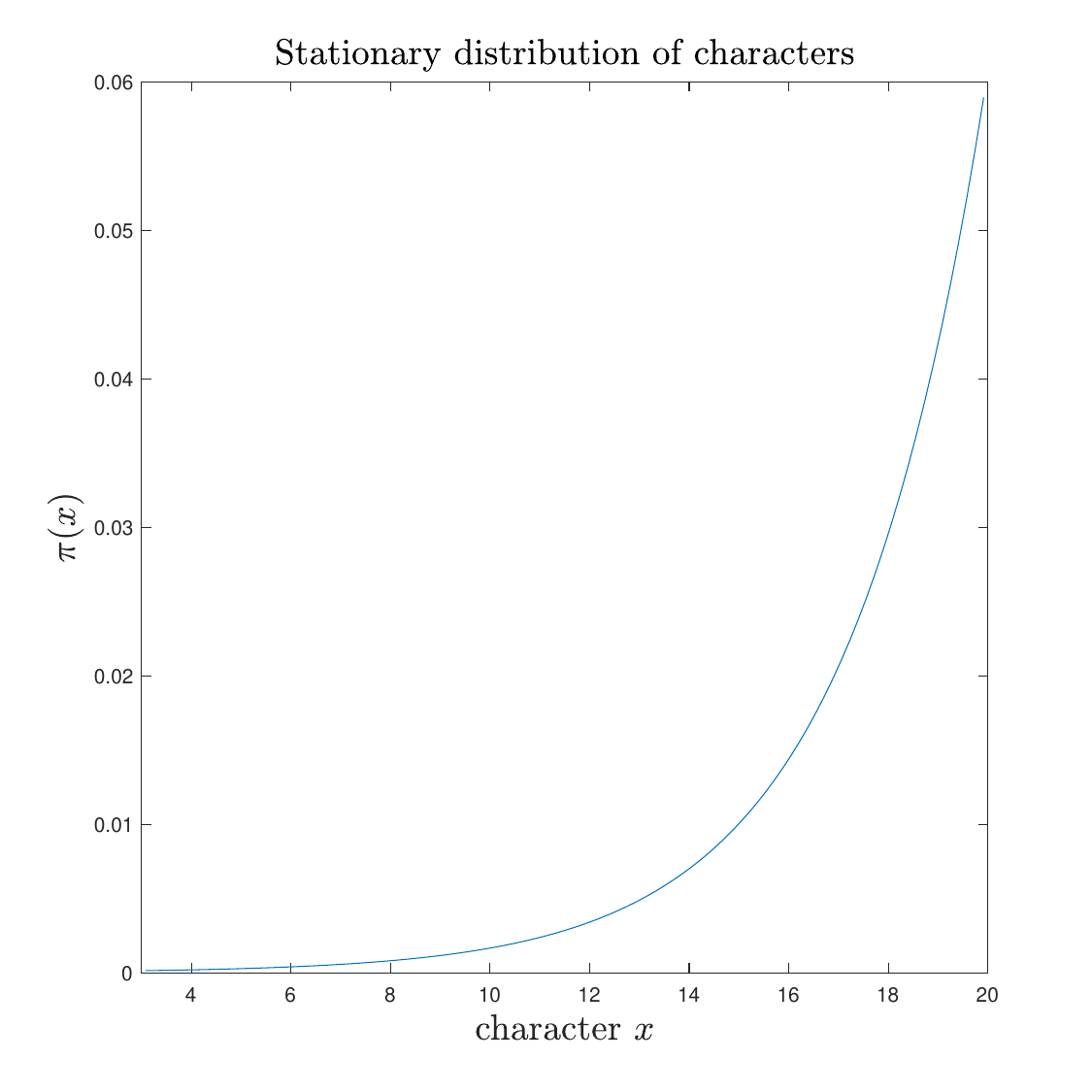}
\caption{From top left to the bottom right: Stationary distribution of the traits in the \protect\hyperlink{QBD3}{QBD3} model in Section~\ref{sec:QBDmodels} for the ${\bf r}$ vectors $\#1-\#4$ in Table~\ref{tab:Synexamplet1}.} 
\label{StatTraitsModel5phases}
\end{figure}

\newpage
\subsection{The effect of the mean drift: Synthetic Dataset~1 (Figure~\ref{fig:DataExample1})}
\label{sec:OutputQBD5phasessyn1}

\begin{figure}[h]
\centering
\includegraphics[scale=0.4]{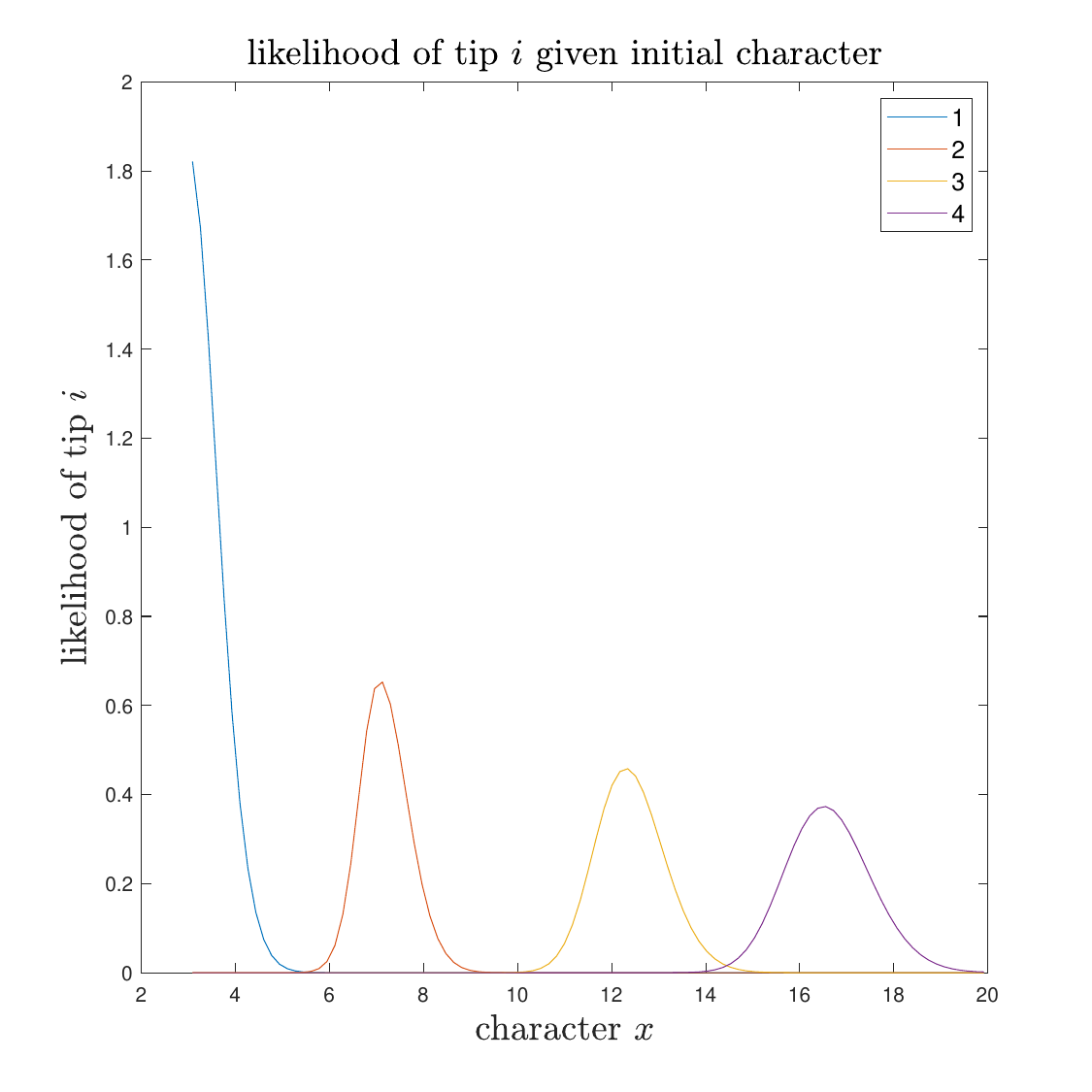}
\
\includegraphics[scale=0.4]{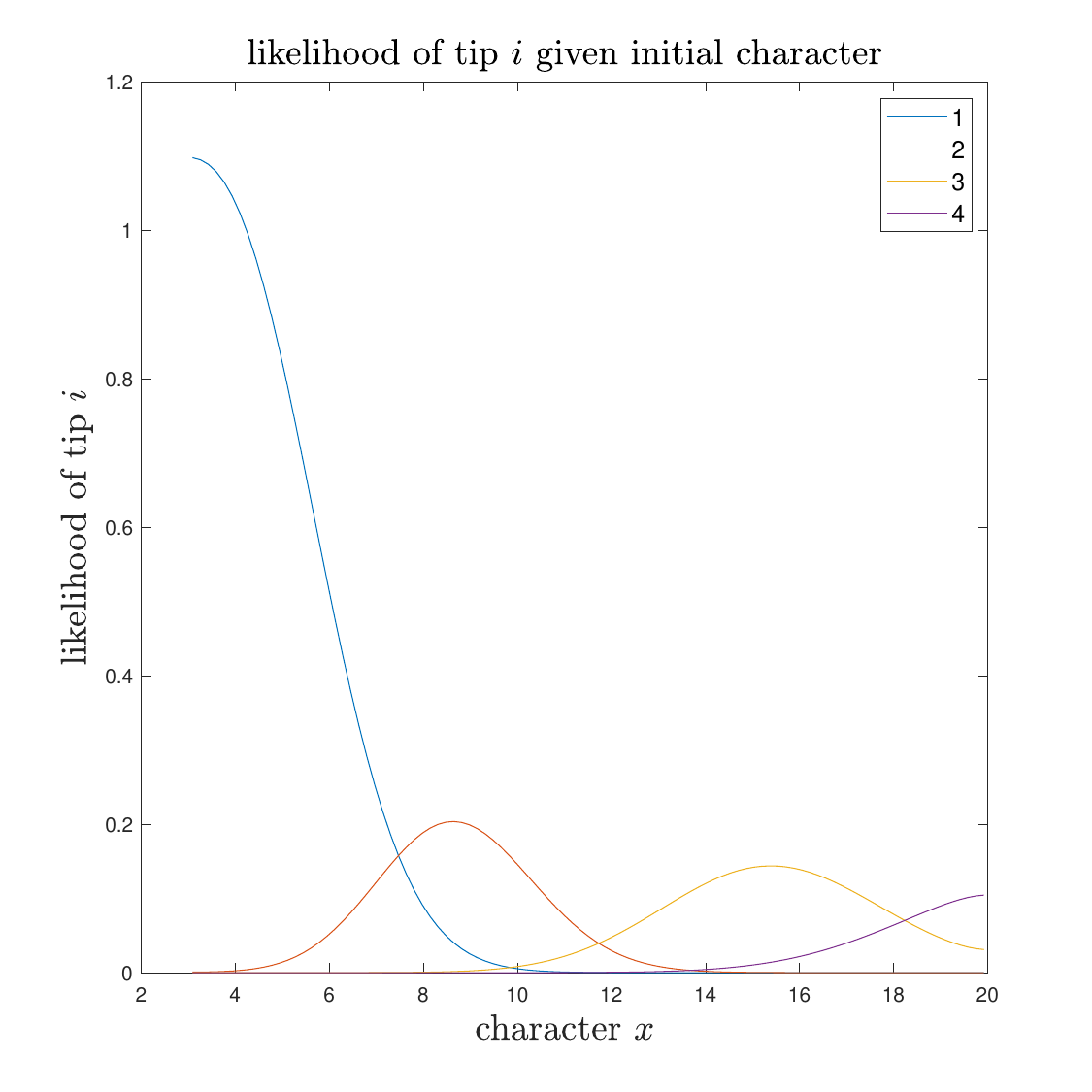}
\\
\includegraphics[scale=0.4]{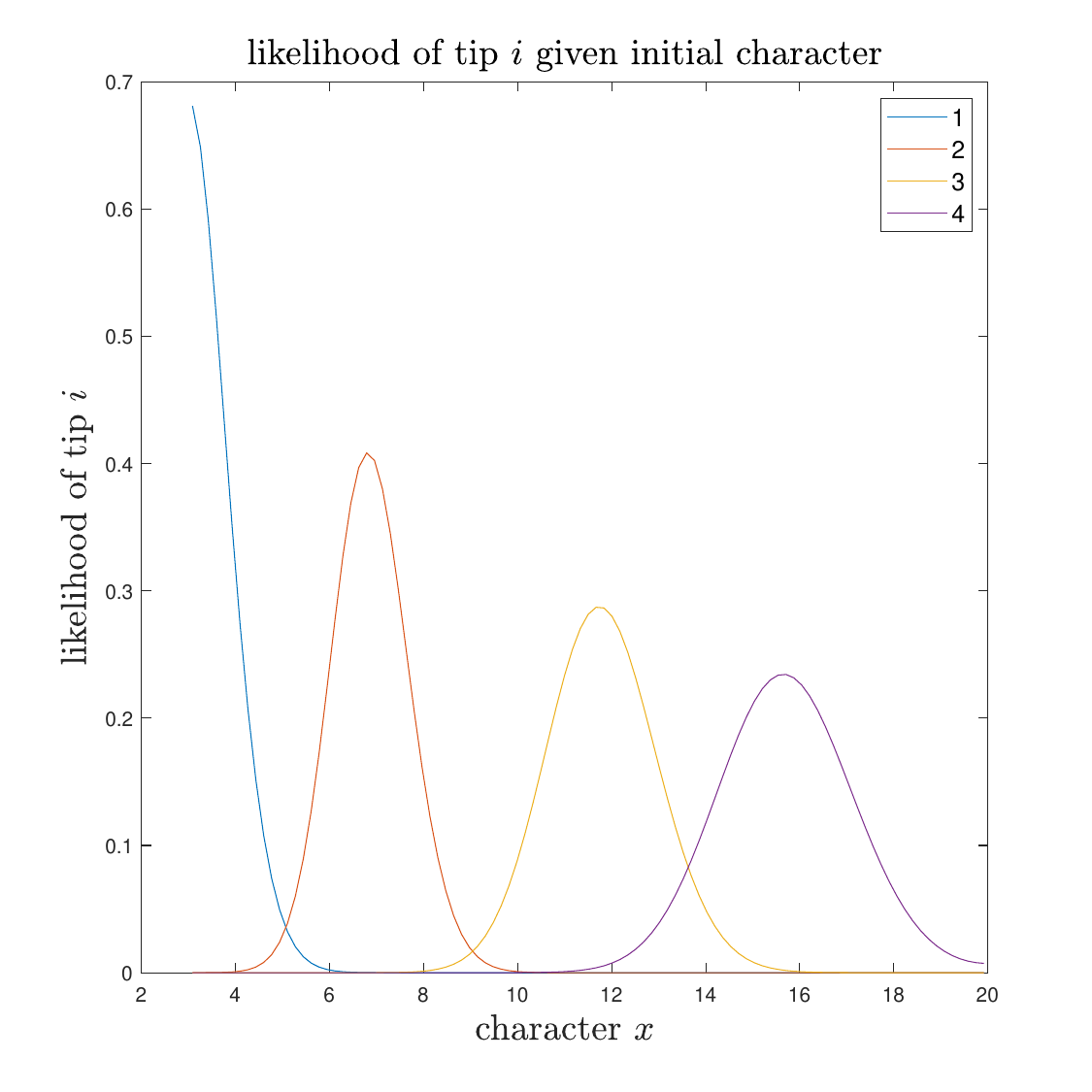}
\
\includegraphics[scale=0.4]{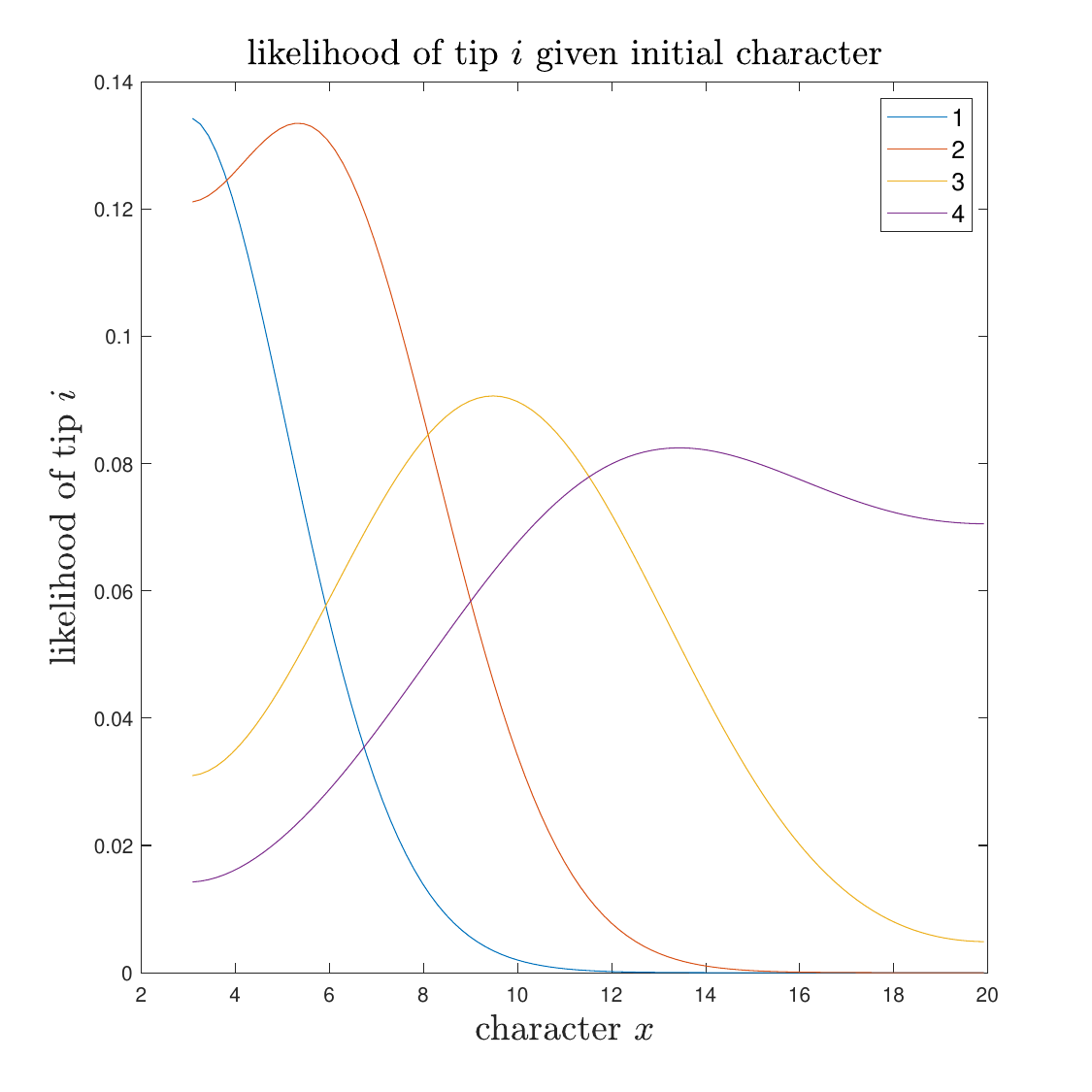}
\caption{From top left to the bottom right: The likelihood of observing tip $i=1,\ldots,4$ given trait observed at the start of the branch corresponding to tip $i$, in the \protect\hyperlink{QBD3}{QBD3} model in Section~\ref{sec:QBDmodels} and Synthetic Dataset~1 (Figure~\ref{fig:DataExample1}), for the ${\bf r}$ vectors $\#1,\#6, \#4$, and $\#9$ in Table~\ref{tab:Synexamplet1}.} 
\label{StatTraitsModel5phasesOverallcherries}
\end{figure}

\begin{figure}
\centering
\includegraphics[scale=0.4]{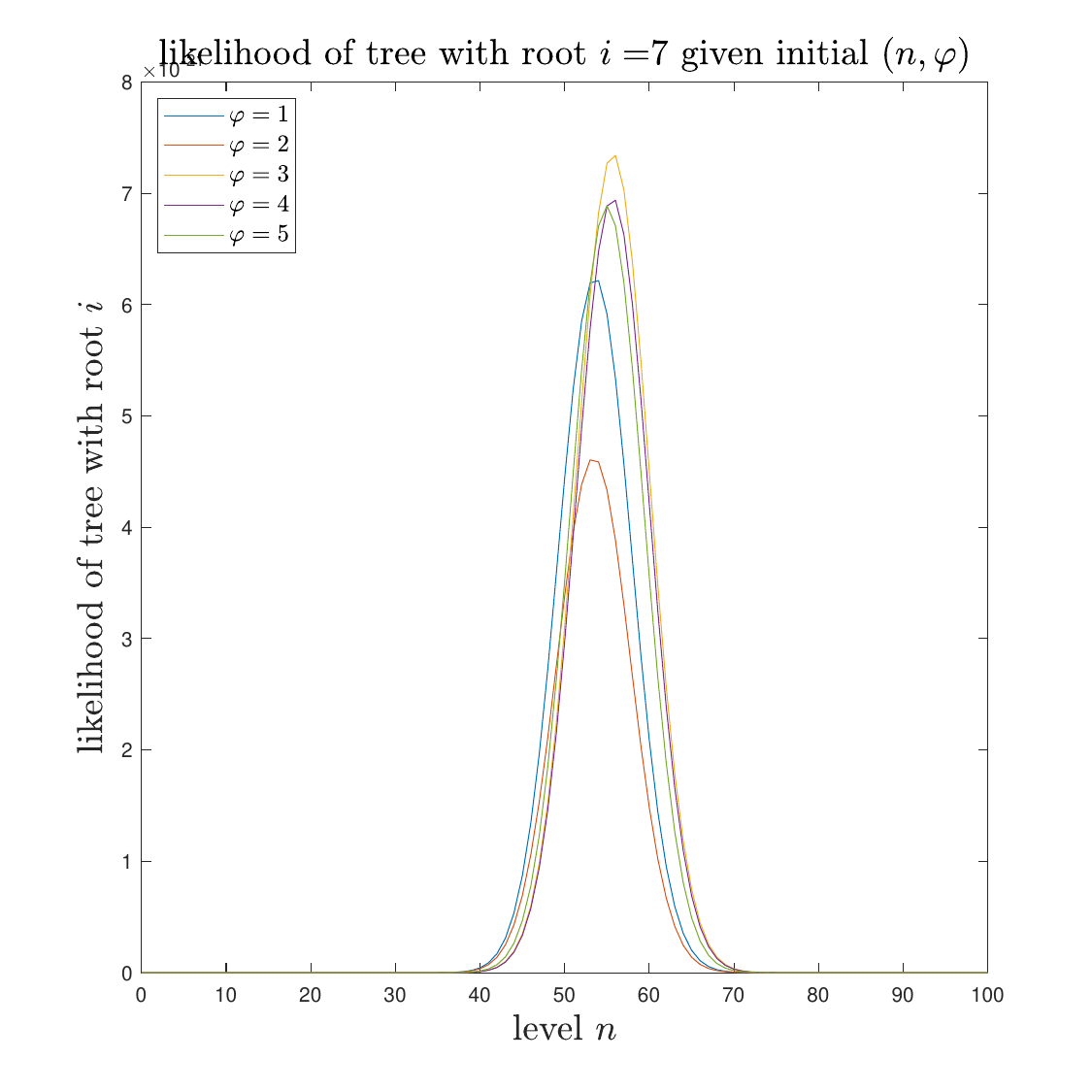}
\
\includegraphics[scale=0.4]{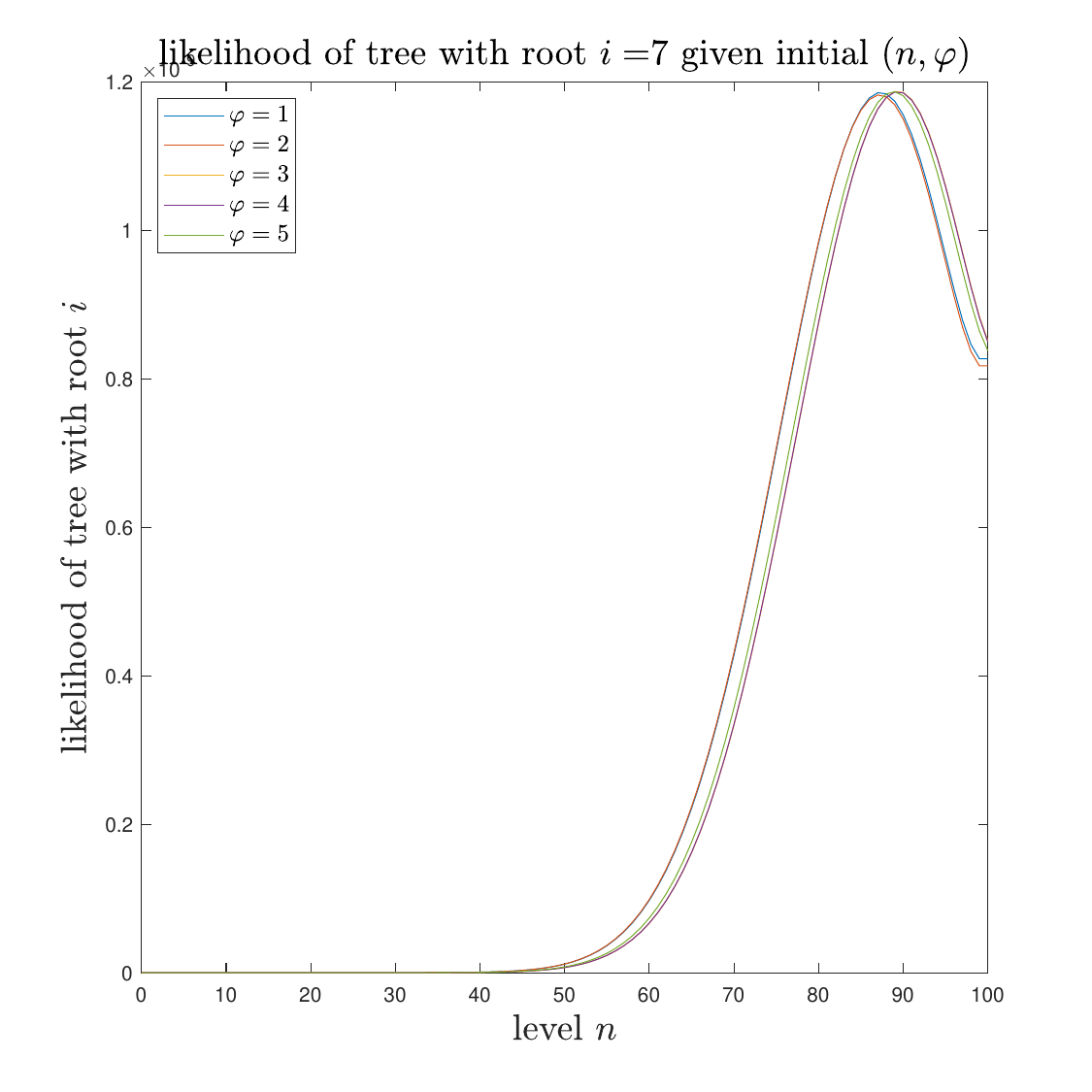}
\\
\includegraphics[scale=0.4]{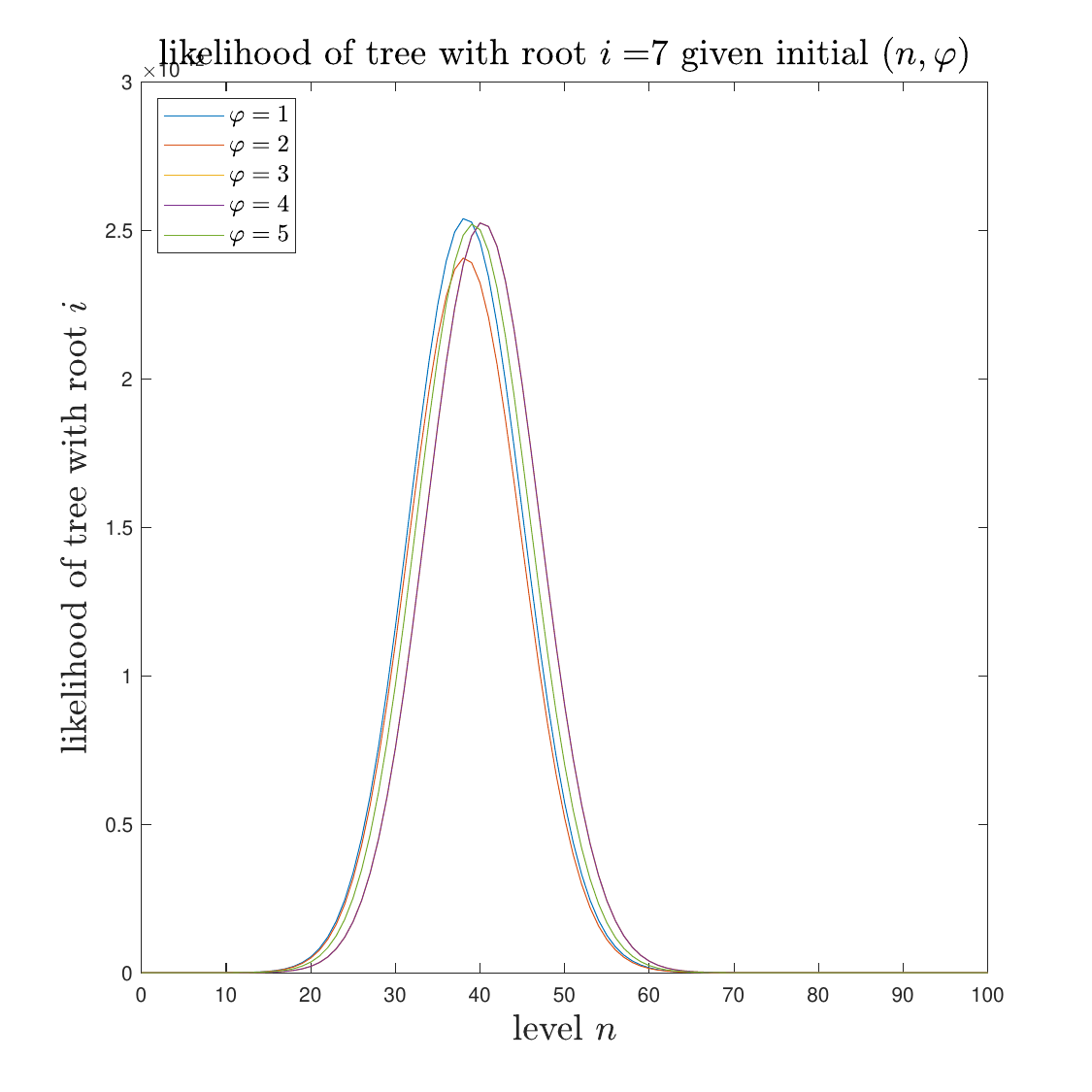}
\
\includegraphics[scale=0.4]{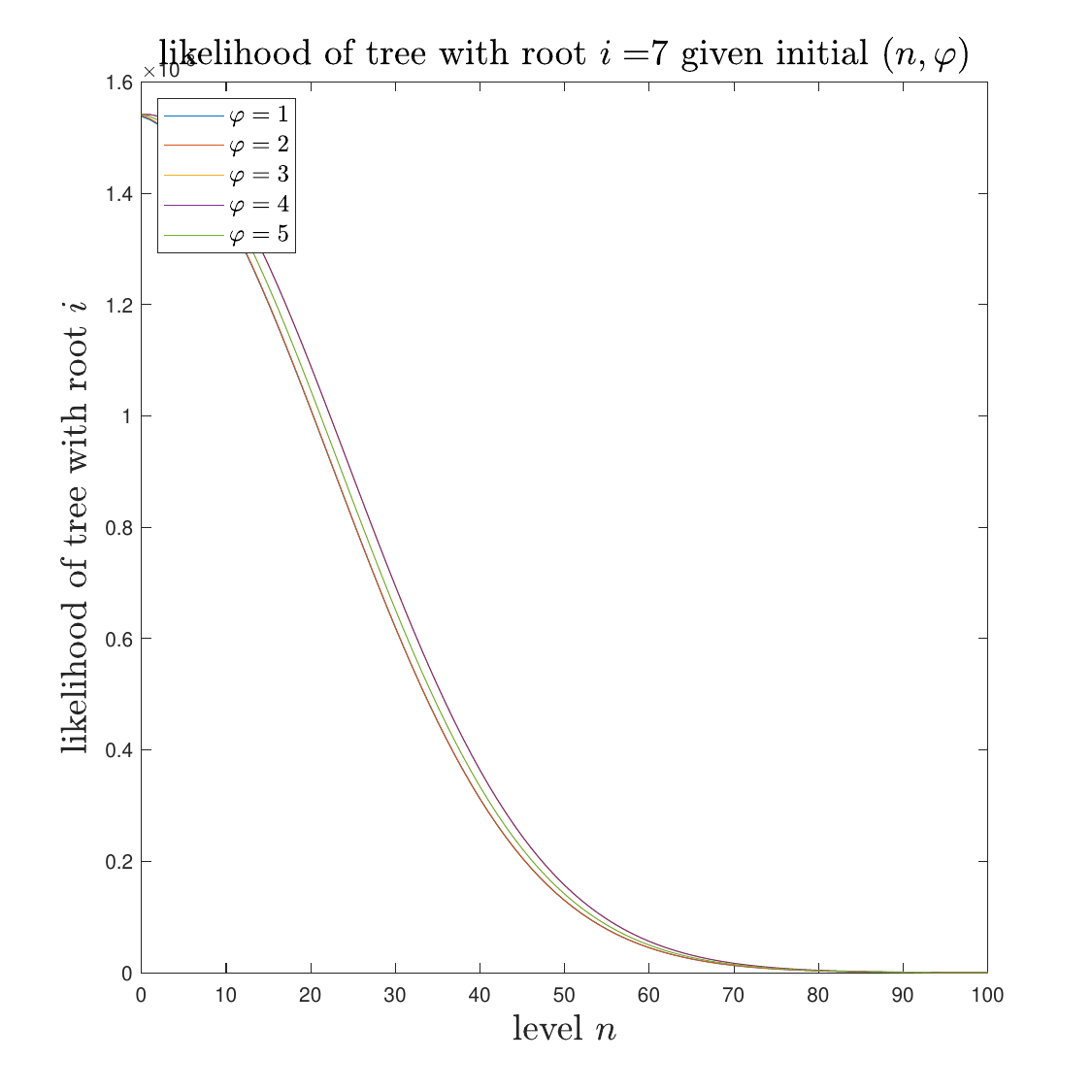}
\caption{From top left to the bottom right: The likelihood of observing the phylogenetic tree that started with parent $i=7$ given level $n$ and phase~$\varphi$ observed at the start of the tree, in the \protect\hyperlink{QBD3}{QBD3} model in Section~\ref{sec:QBDmodels} and Synthetic Dataset~1 (Figure~\ref{fig:DataExample1}), for the ${\bf r}$ vectors $\#1,\#6, \#4$, and $\#9$ in Table~\ref{tab:Synexamplet1}.} 
\label{StatTraitsModel5phasesOverallParrentLevel}
\end{figure}

\begin{figure}
\centering
\includegraphics[scale=0.4]{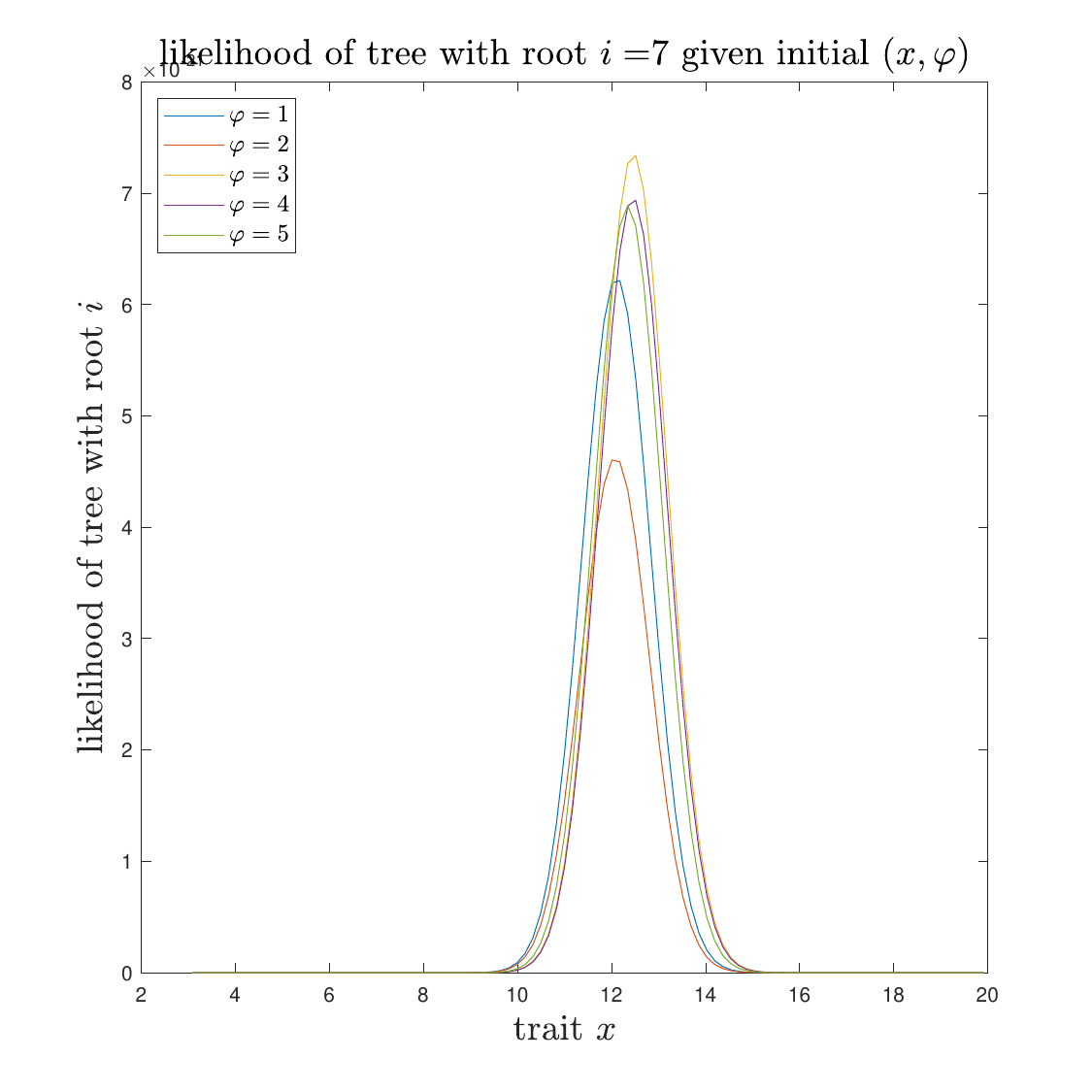}
\
\includegraphics[scale=0.4]{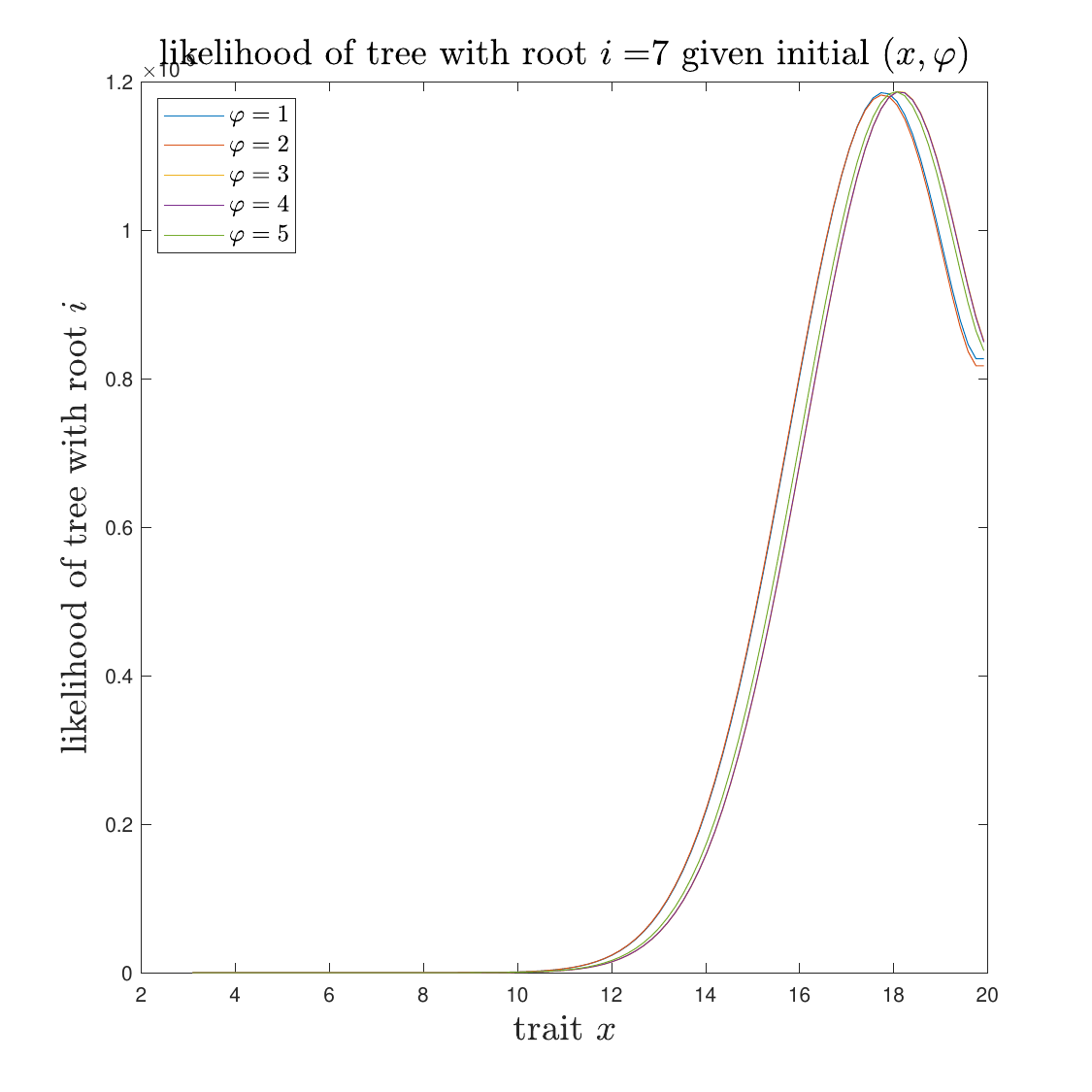}
\\
\includegraphics[scale=0.4]{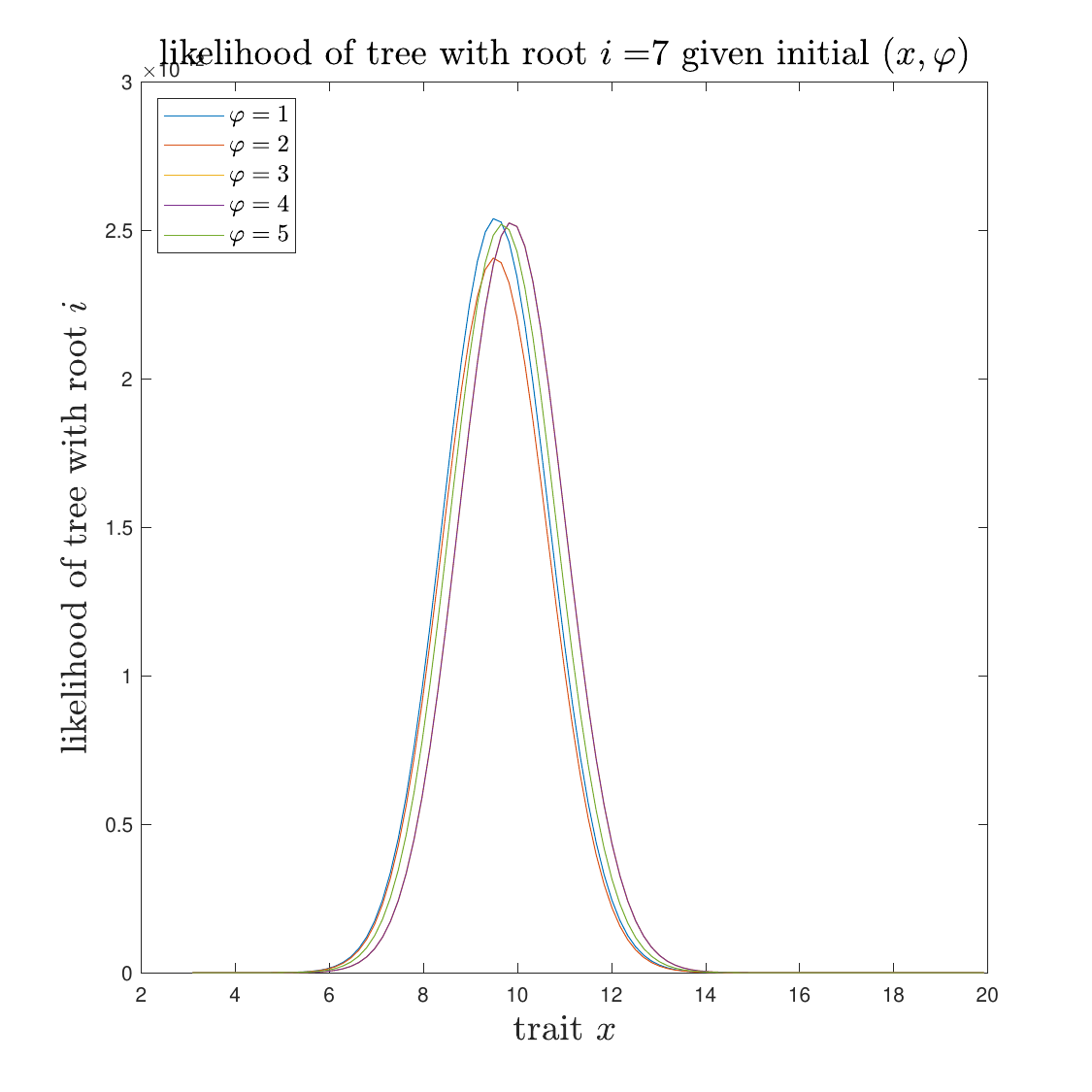}
\
\includegraphics[scale=0.4]{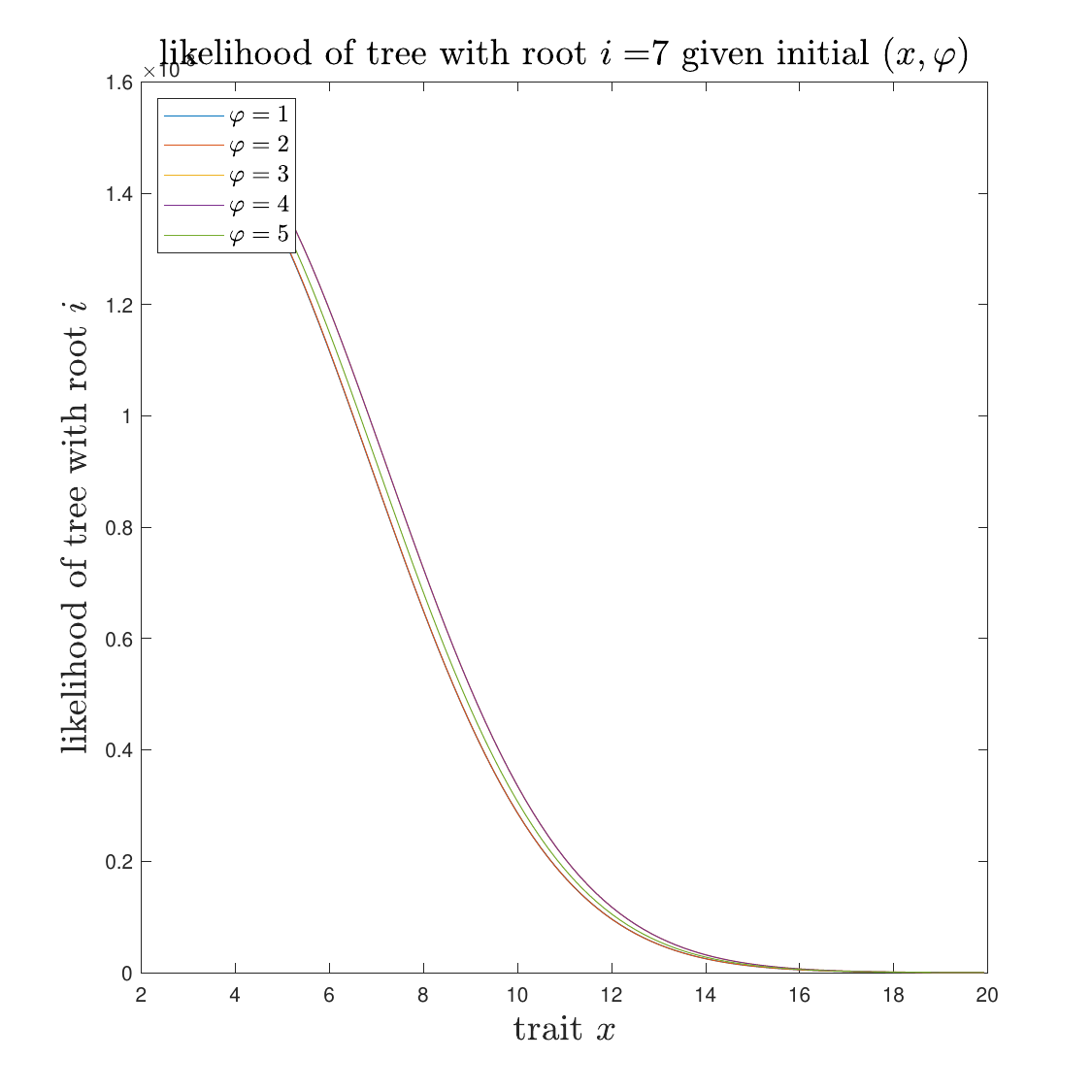}
\caption{From top left to the bottom right: The likelihood of observing the phylogenetic tree that started with parent $i=7$ given trait~$x$ and phase~$\varphi$ observed at the start of the tree, in the \protect\hyperlink{QBD3}{QBD3} model in Section~\ref{sec:QBDmodels} and Synthetic Dataset~1 (Figure~\ref{fig:DataExample1}), for the ${\bf r}$ vectors $\#1,\#6, \#4$, and $\#9$ in Table~\ref{tab:Synexamplet1}.} 
\label{StatTraitsModel5phasesOverallParrentTrait}
\end{figure}

\begin{figure}
\centering
\includegraphics[scale=0.4]{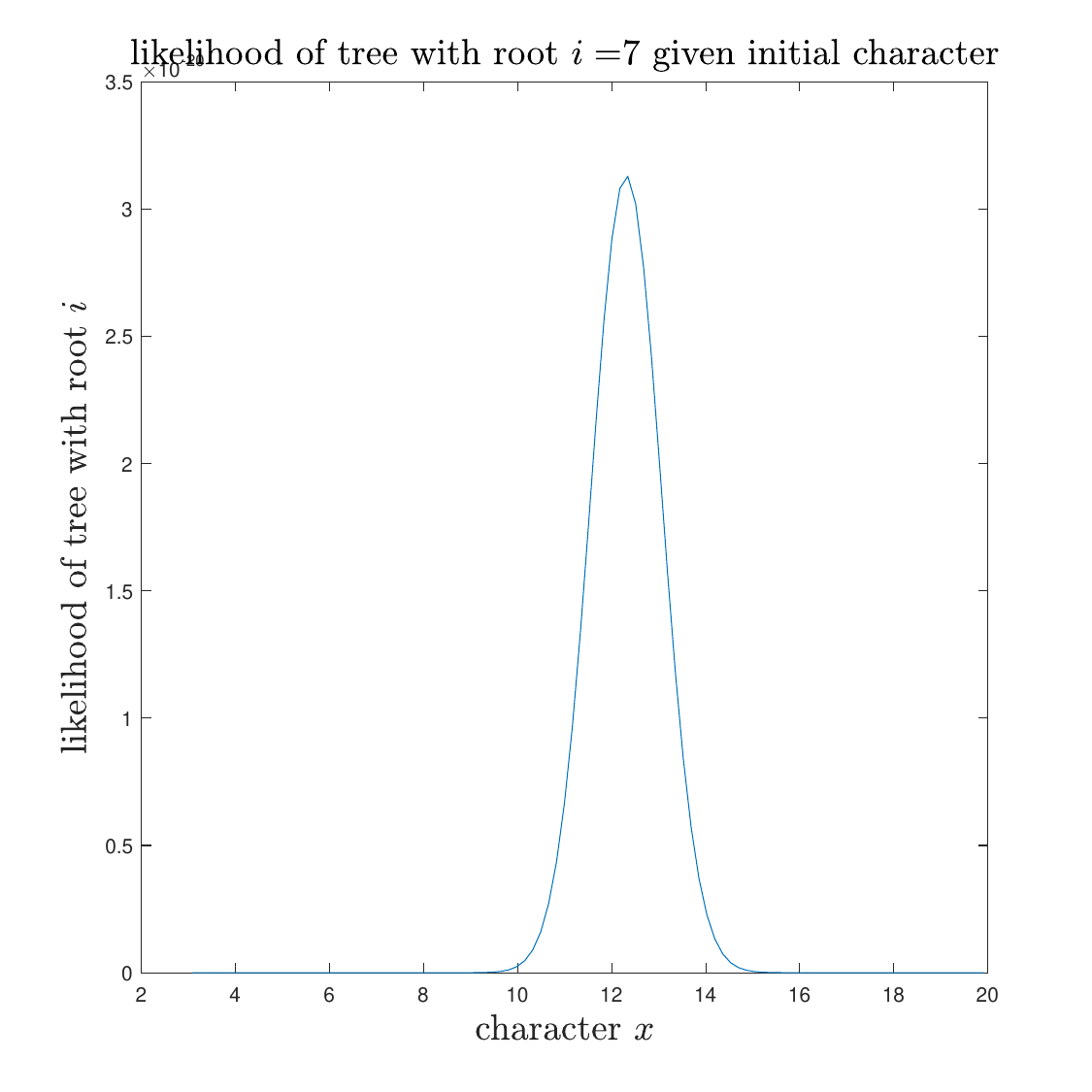}
\
\includegraphics[scale=0.4]{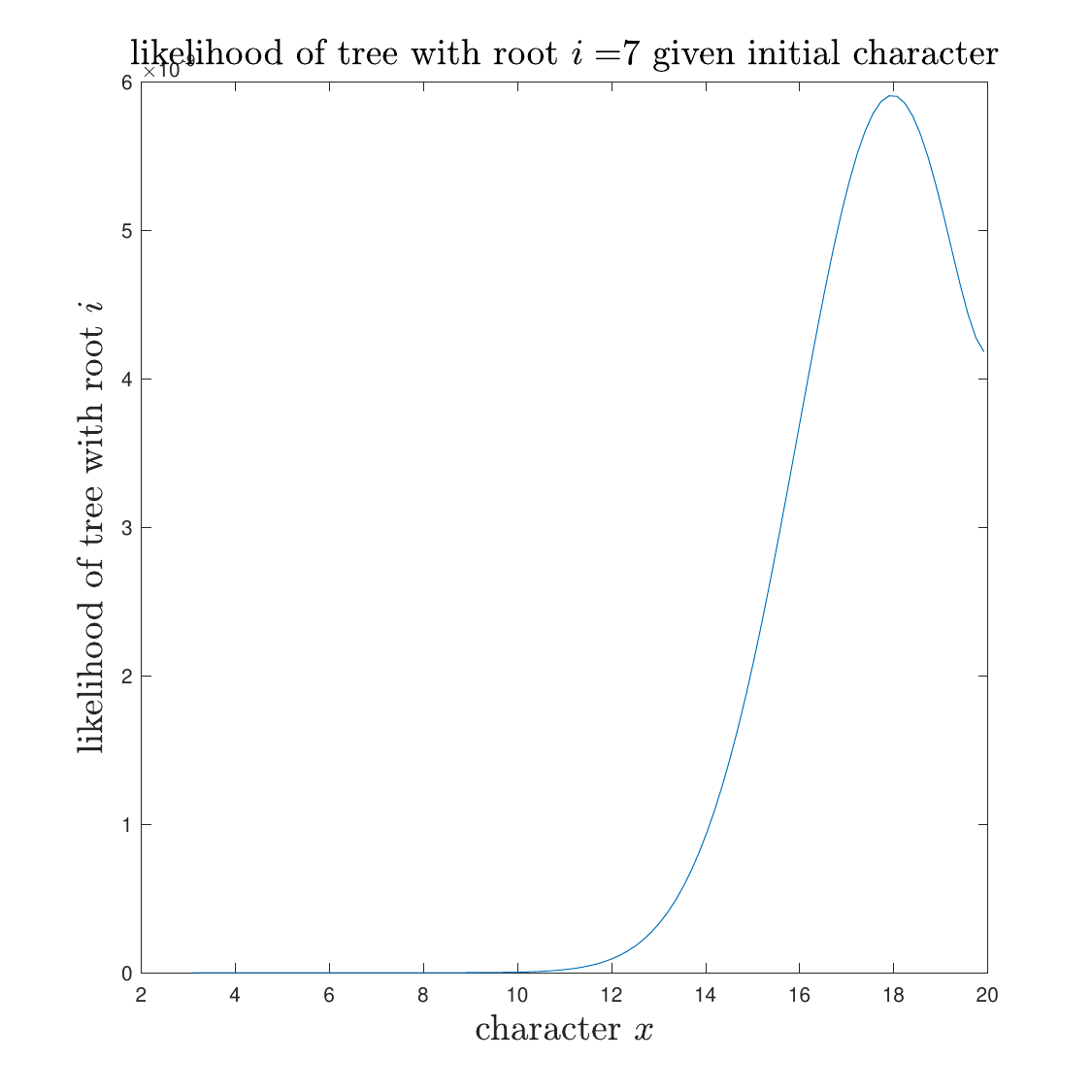}
\\
\includegraphics[scale=0.4]{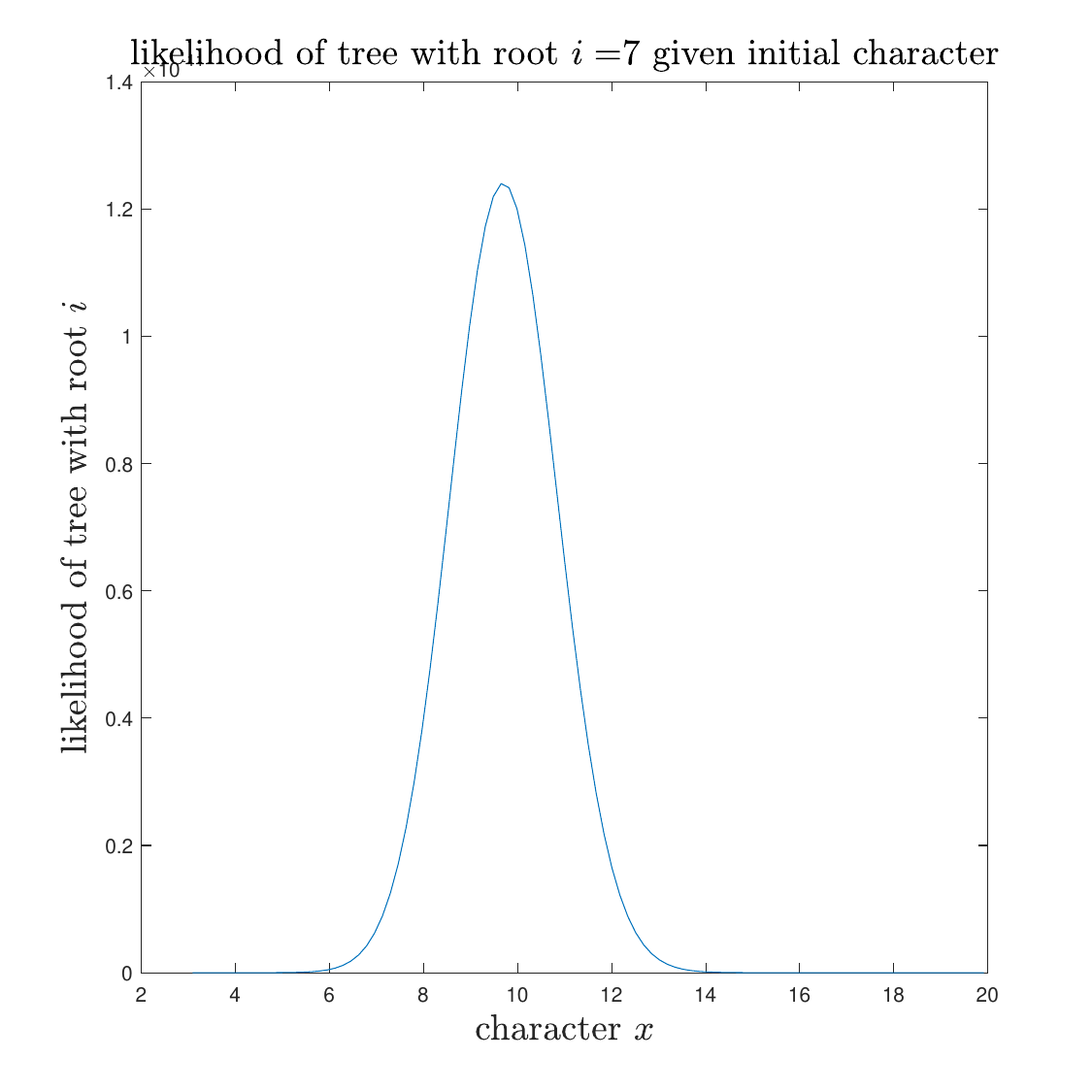}
\
\includegraphics[scale=0.4]{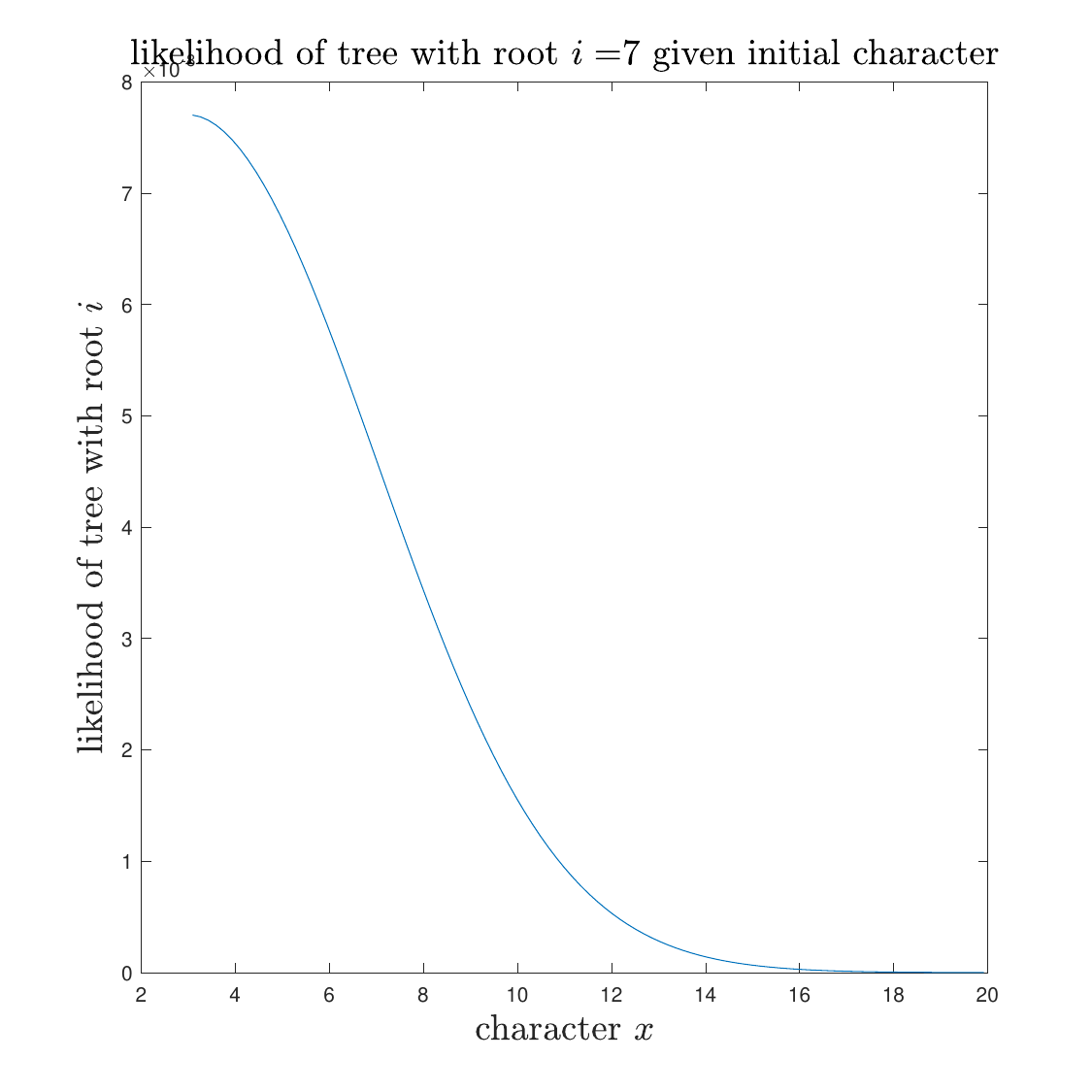}
\caption{From top left to the bottom right: The likelihood of observing the phylogenetic tree that started with parent $i=7$ given trait~$x$ observed at the start of the tree, in the QBD model in the \protect\hyperlink{QBD3}{QBD3} model in Section~\ref{sec:QBDmodels} and Synthetic Dataset~1 (Figure~\ref{fig:DataExample1}), for the ${\bf r}$ vectors $\#1,\#6, \#4$, and $\#9$ in Table~\ref{tab:Synexamplet1}.} 
\label{StatTraitsModel5phasesOverallParrentTraitsOnly}
\end{figure}

\newpage
\subsection{
The effect of the mean drift: Synthetic Dataset~2~(Figure~\ref{fig:DataExample2})} 
\label{outputSyntheticDataExample2}

\begin{figure}[h]
\centering
\includegraphics[scale=0.4]{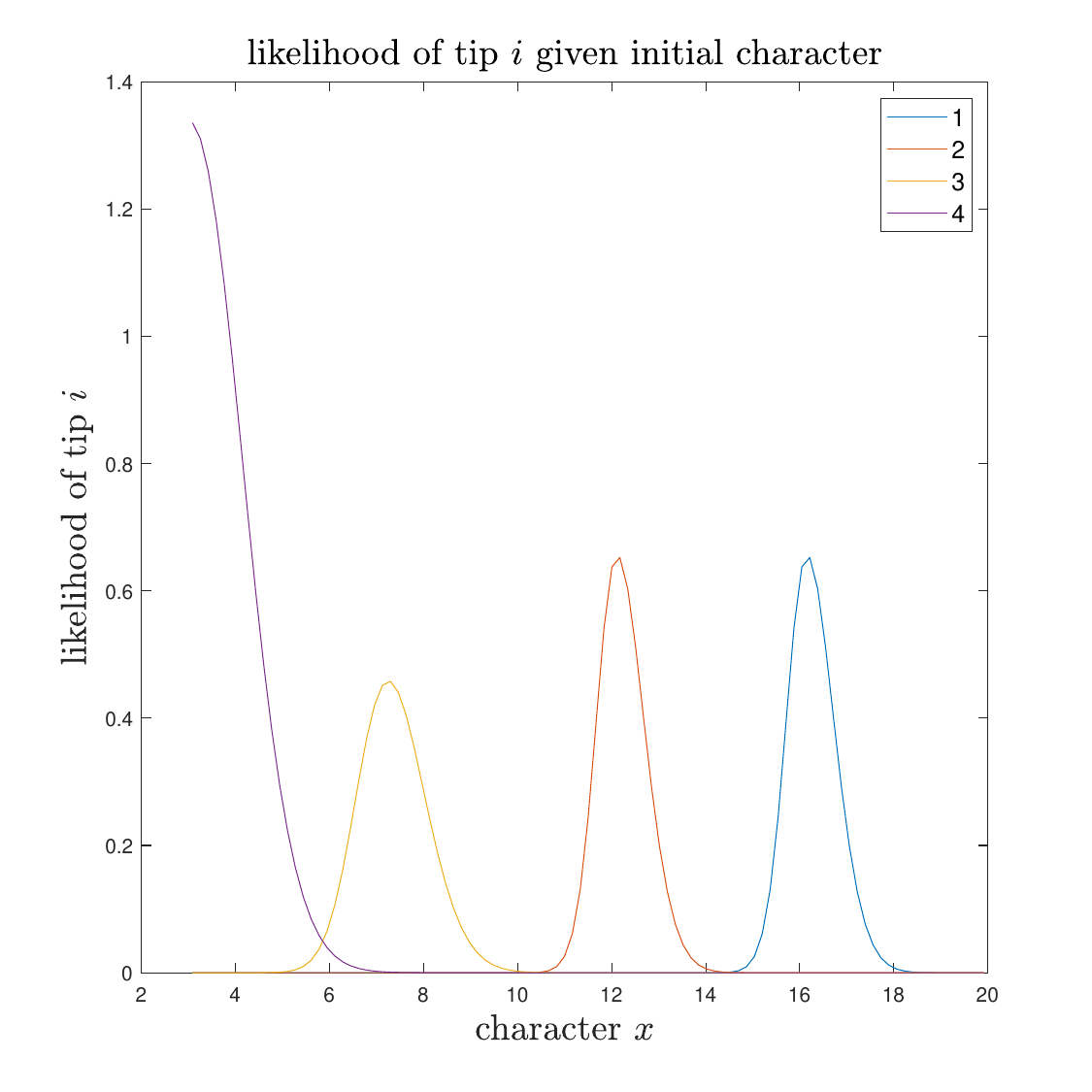}
\
\includegraphics[scale=0.4]{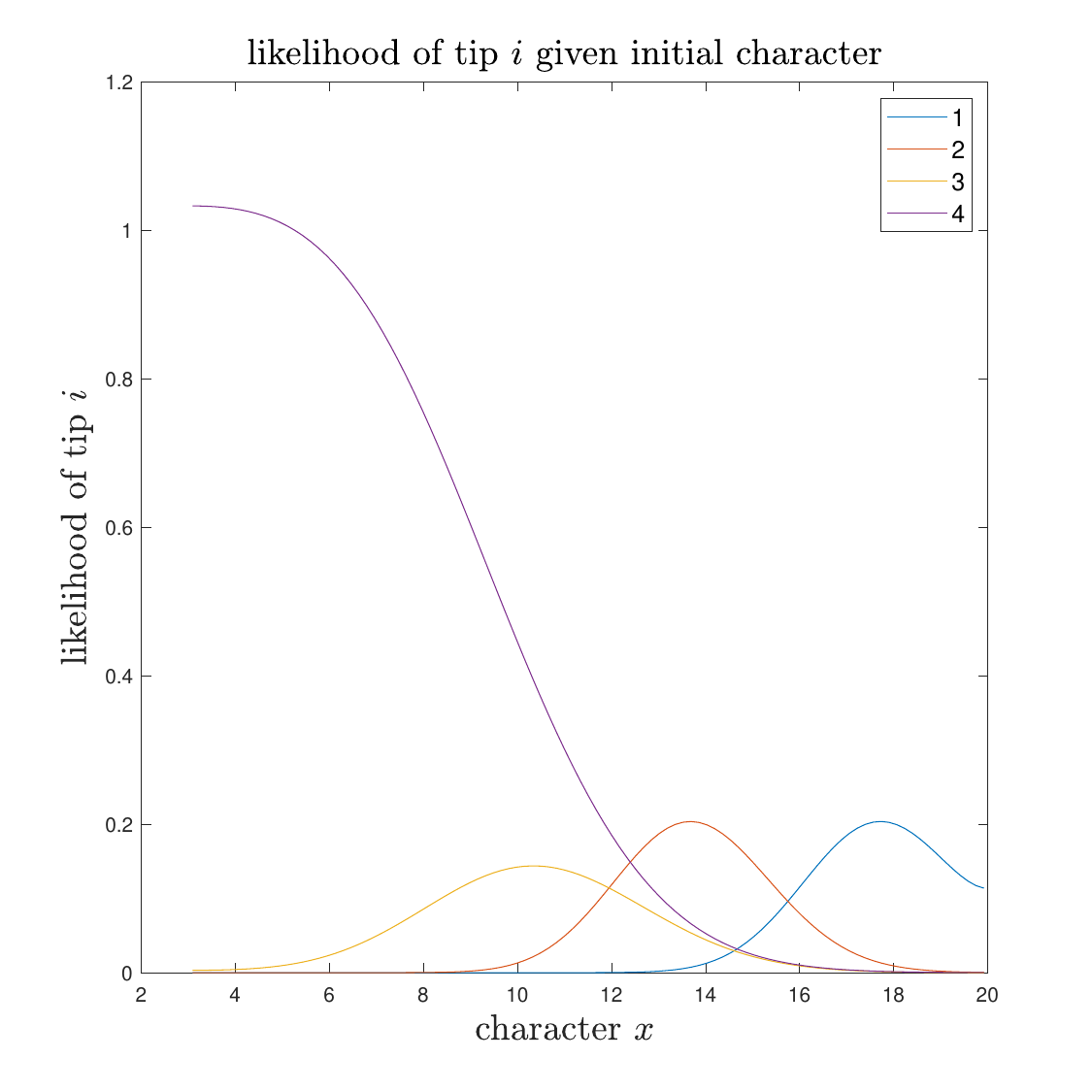}
\\
\includegraphics[scale=0.4]{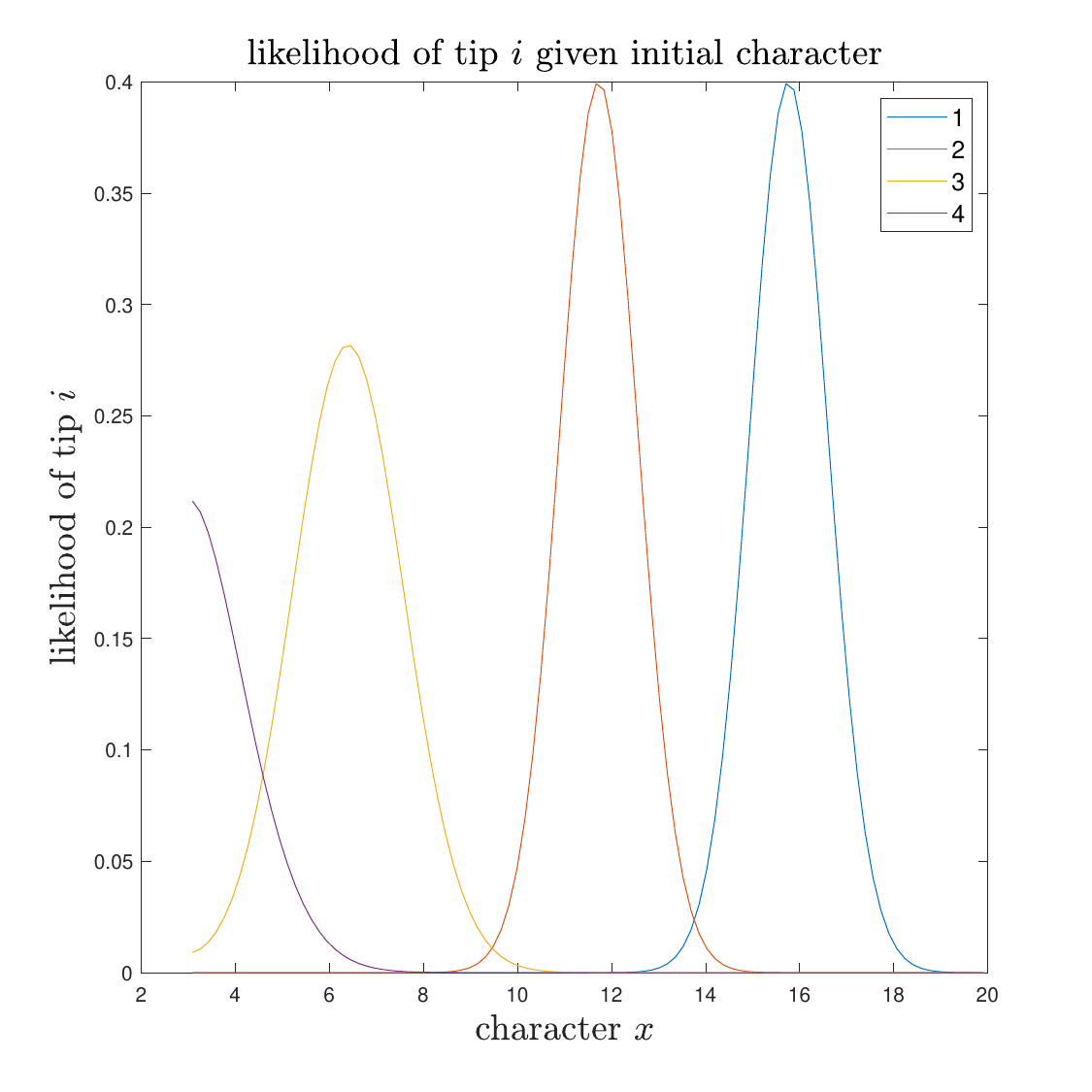}
\
\includegraphics[scale=0.4]{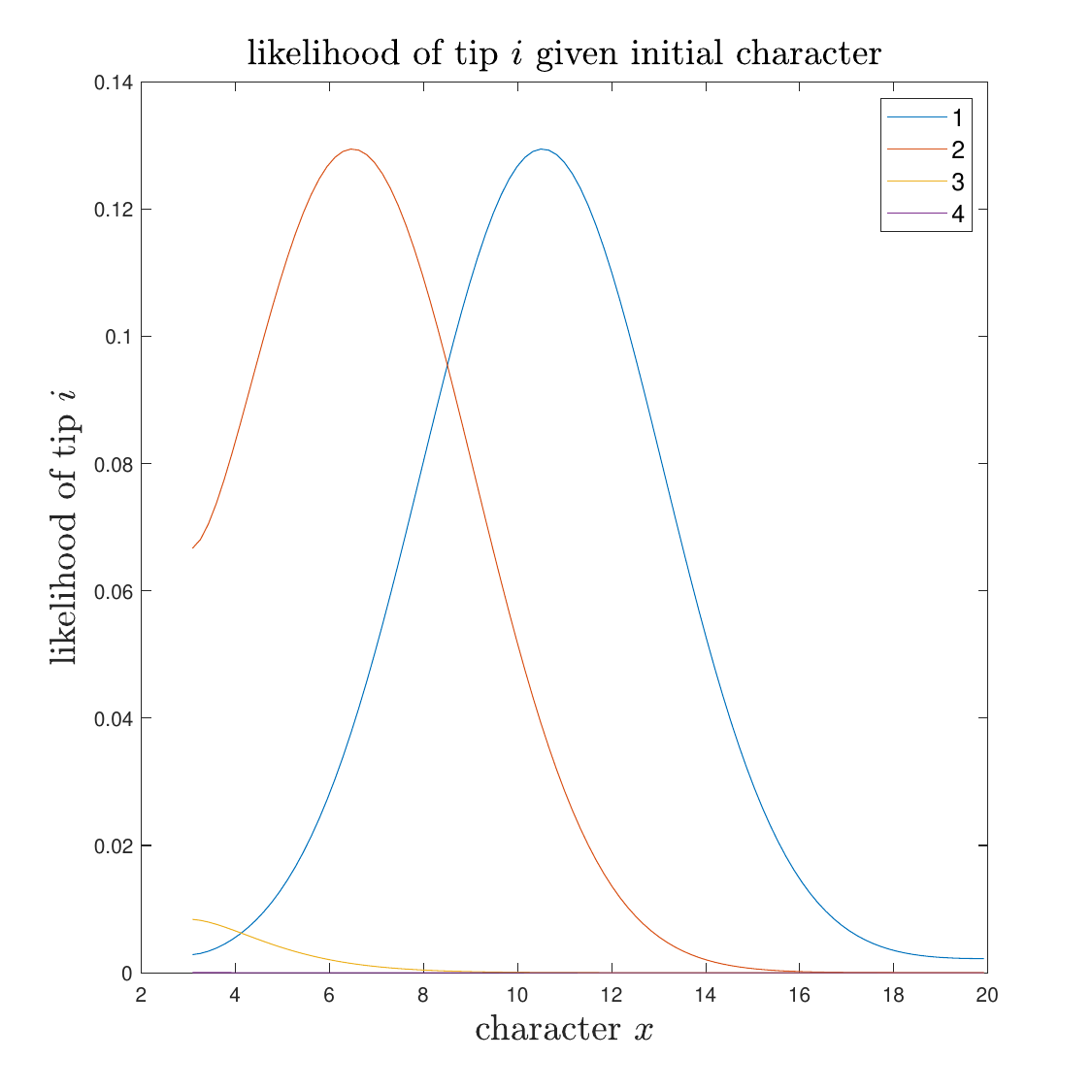}
\caption{From top left to the bottom right: The likelihood of observing tip $i=1,\ldots,4$ given trait observed at the start of the branch corresponding to tip $i$, in the \protect\hyperlink{QBD3}{QBD3} model in Section~\ref{sec:QBDmodels} and Synthetic Dataset~2 (Figure~\ref{fig:DataExample2}), for the ${\bf r}$ vectors $\#1,\#6, \#5$, and $\#8$ in Table~\ref{tab:Synexamplet2}.} 
\label{StatTraitsModel5phasesOverallcherriesData2}
\end{figure}

\begin{figure}
\centering
\includegraphics[scale=0.4]{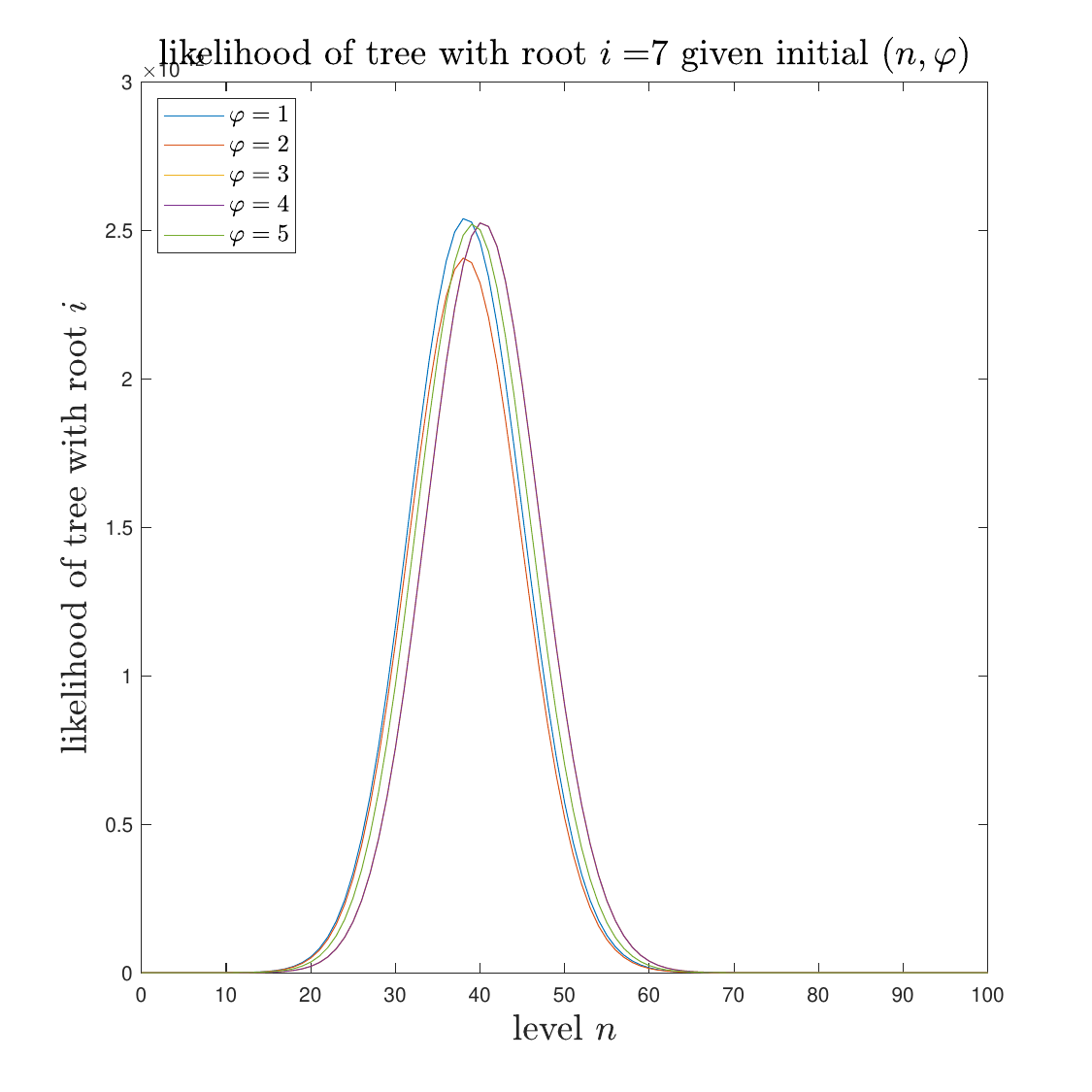}
\
\includegraphics[scale=0.4]{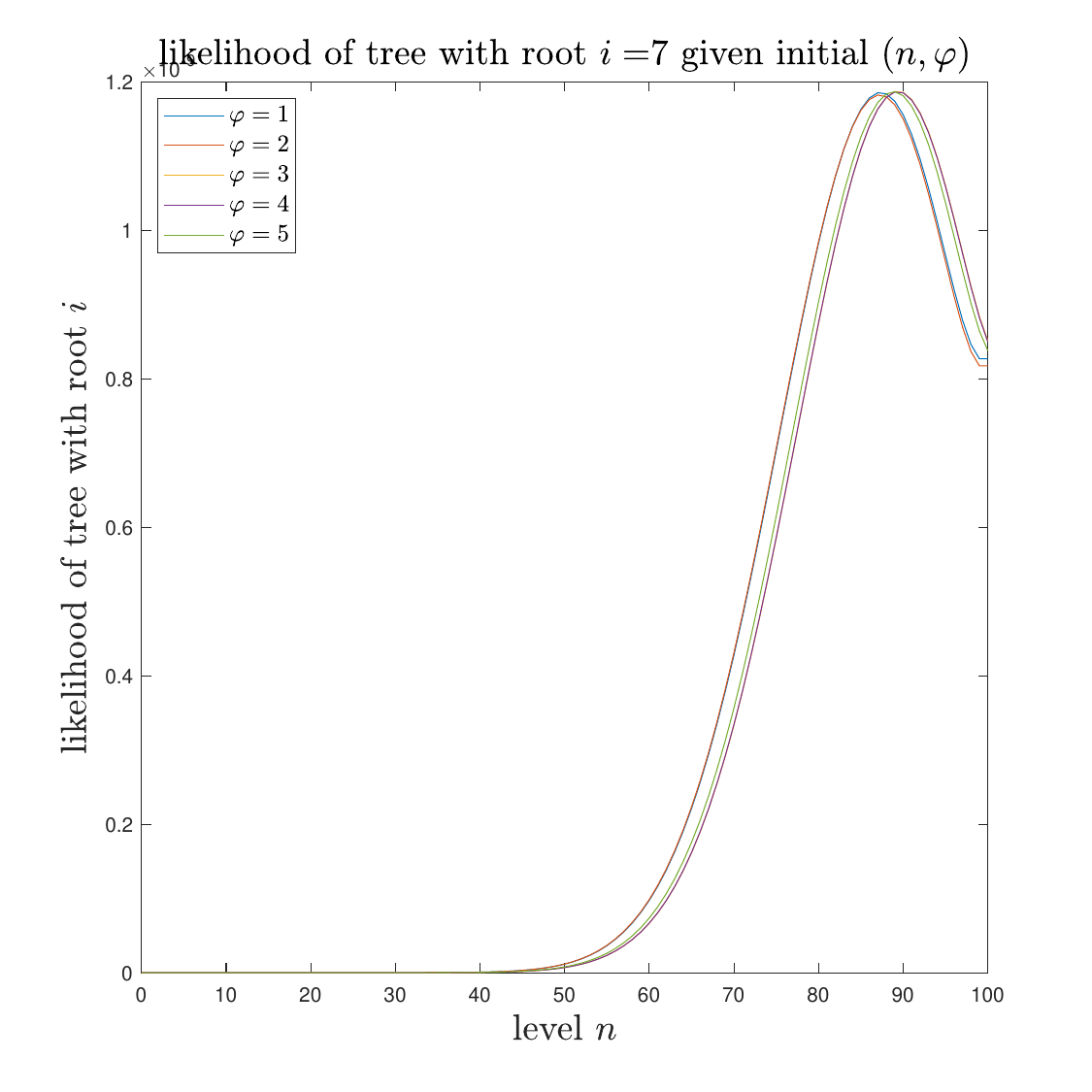}
\\
\includegraphics[scale=0.4]{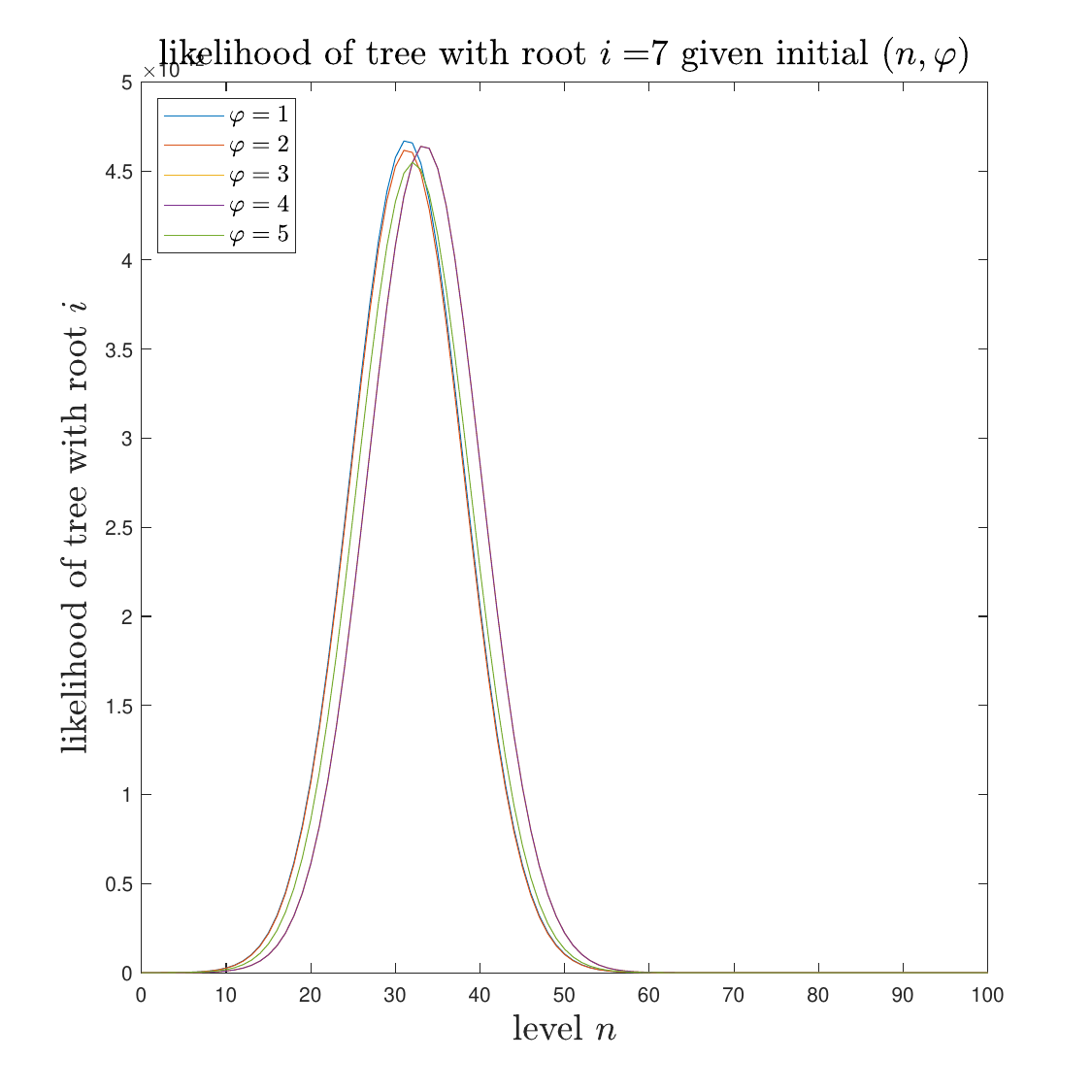}
\
\includegraphics[scale=0.4]{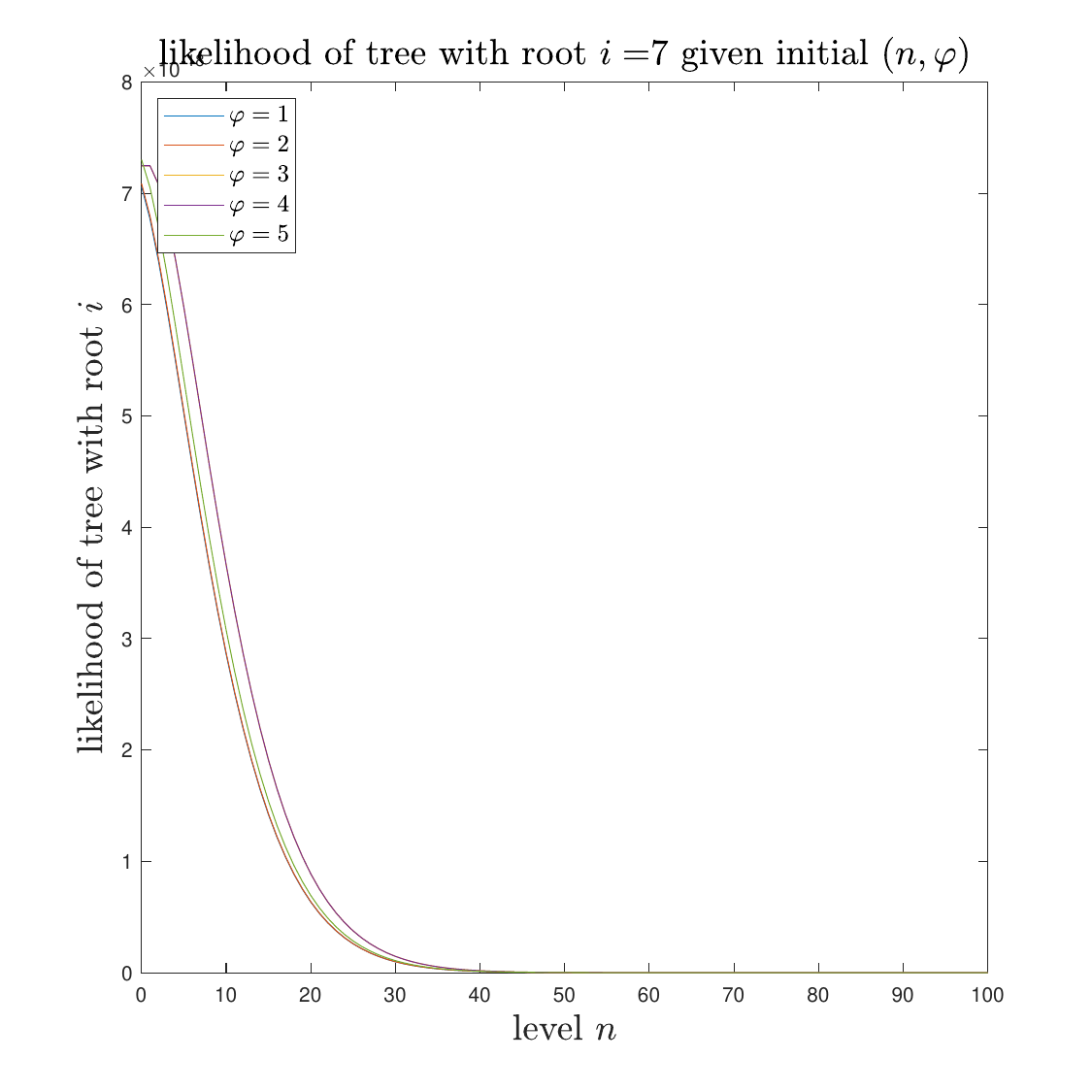}
\caption{From top left to the bottom right: The likelihood of observing the phylogenetic tree that started with parent $i=7$ given level $n$ and phase~$\varphi$ observed at the start of the tree, in the \protect\hyperlink{QBD3}{QBD3} model in Section~\ref{sec:QBDmodels} and Synthetic Dataset~2 (Figure~\ref{fig:DataExample2}), for the ${\bf r}$ vectors $\#1,\#6, \#5$, and $\#8$ in Table~\ref{tab:Synexamplet2}.} 
\label{StatTraitsModel5phasesOverallParrentLevelData2}
\end{figure}

\begin{figure}[H]
\centering
\includegraphics[scale=0.4]{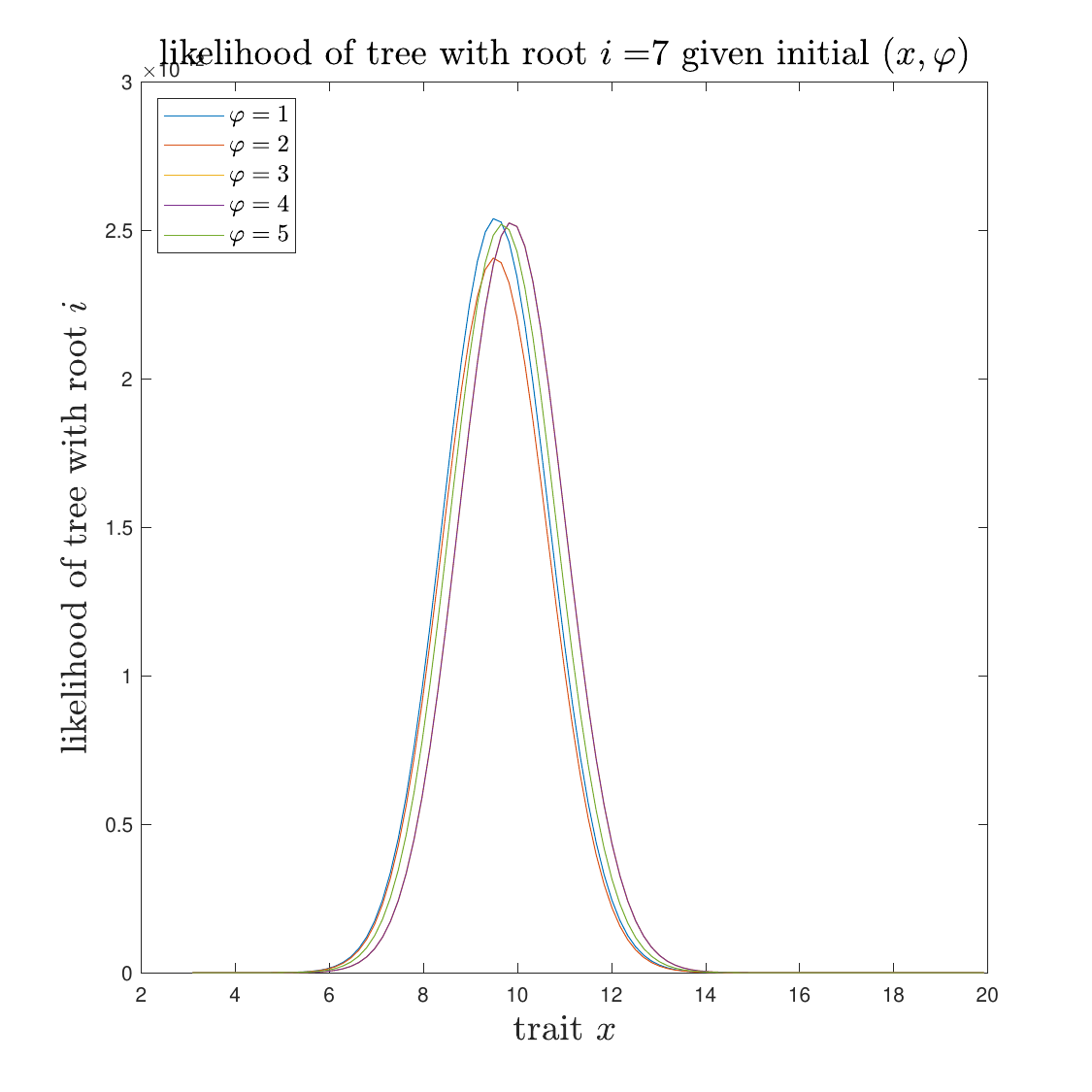}
\
\includegraphics[scale=0.4]{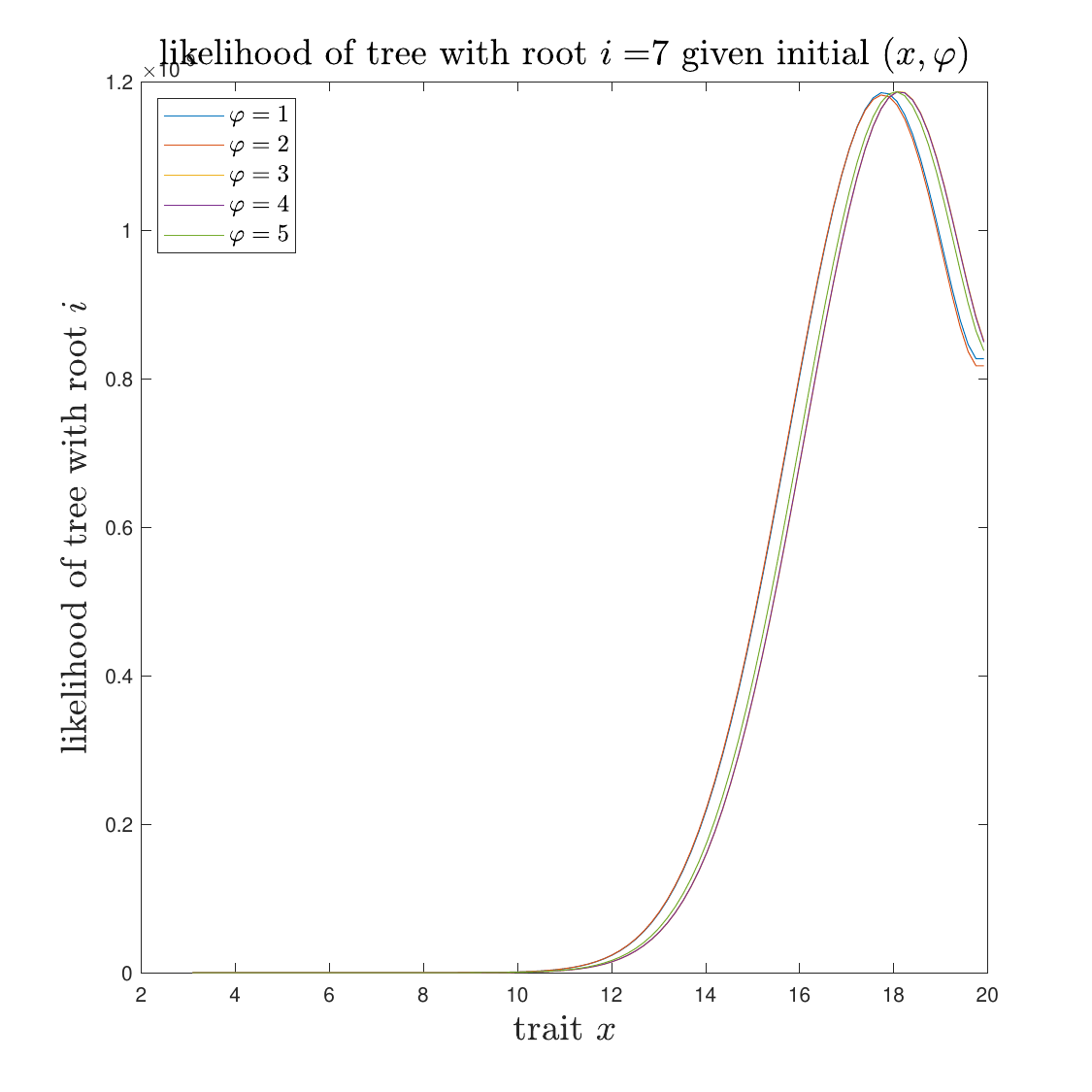}
\\
\includegraphics[scale=0.4]{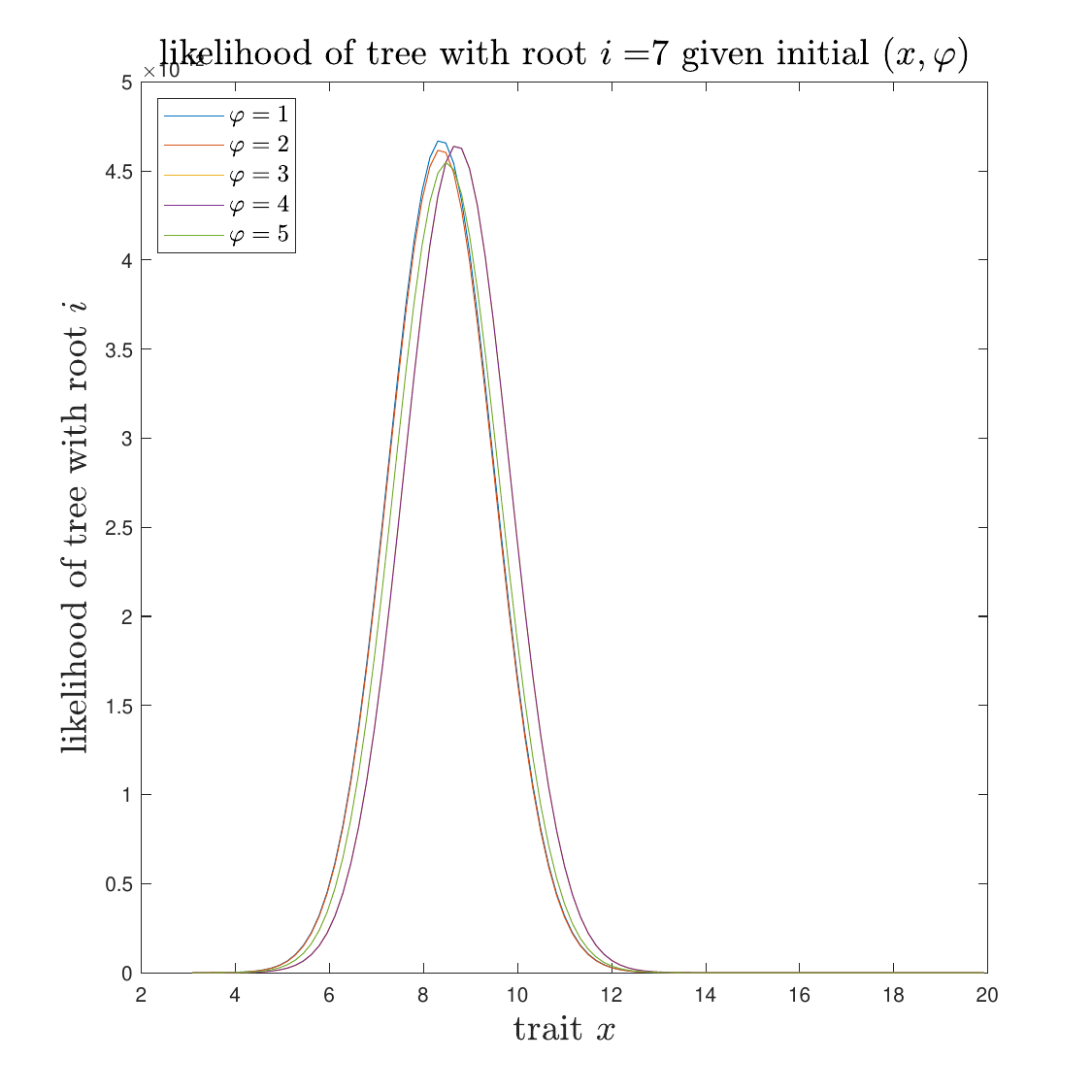}
\   
\includegraphics[scale=0.4]{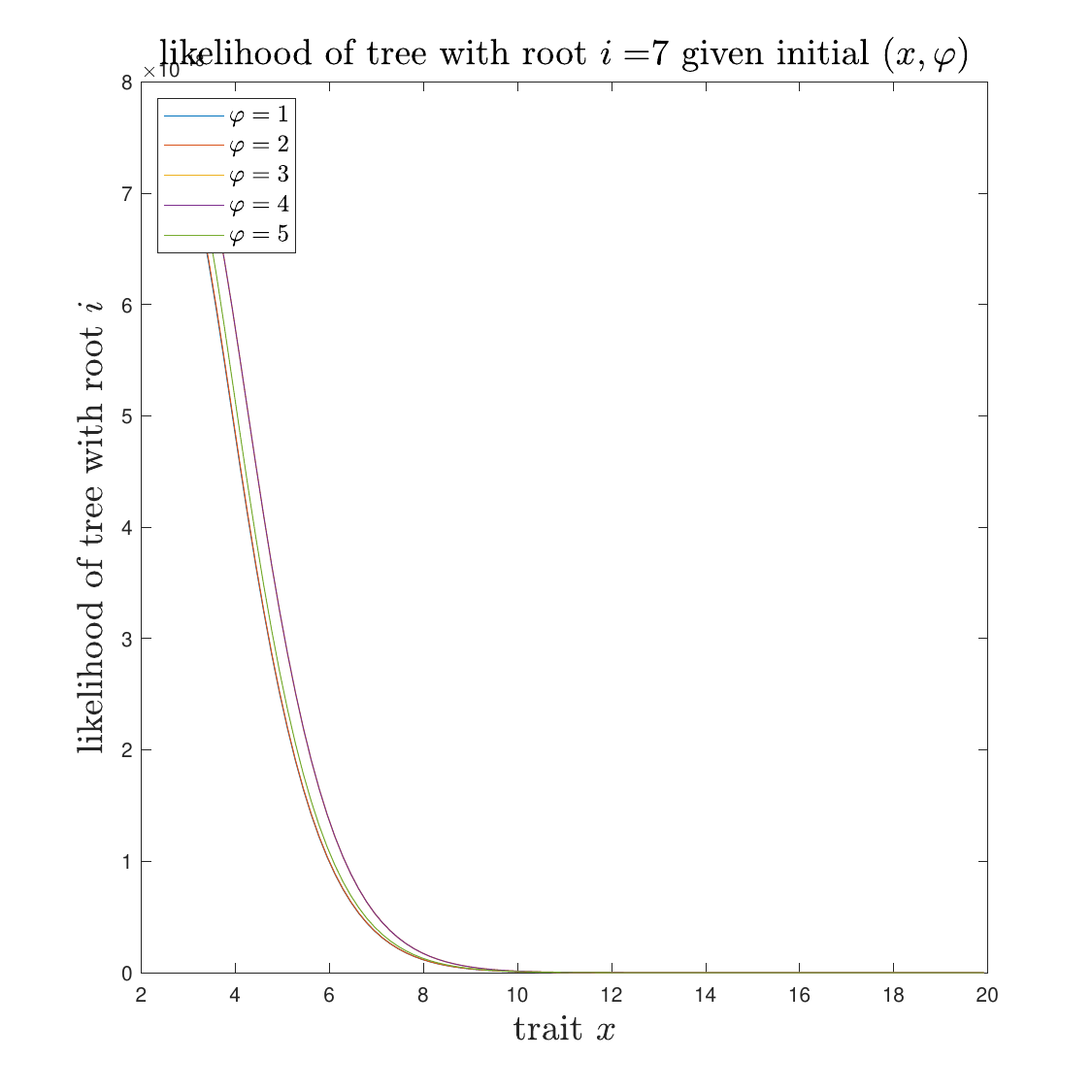}
\caption{From top left to the bottom right: The likelihood of observing the phylogenetic tree that started with parent $i=7$ given trait~$x$ and phase~$\varphi$ observed at the start of the tree, in the \protect\hyperlink{QBD3}{QBD3} model in Section~\ref{sec:QBDmodels}  and Synthetic Dataset~2 (Figure~\ref{fig:DataExample2}), for the ${\bf r}$ vectors $\#1,\#6, \#5$, and $\#8$ in Table~\ref{tab:Synexamplet2}.} 
\label{StatTraitsModel5phasesOverallParrentTraitData2}
\end{figure}

\begin{figure}
\centering
\includegraphics[scale=0.4]{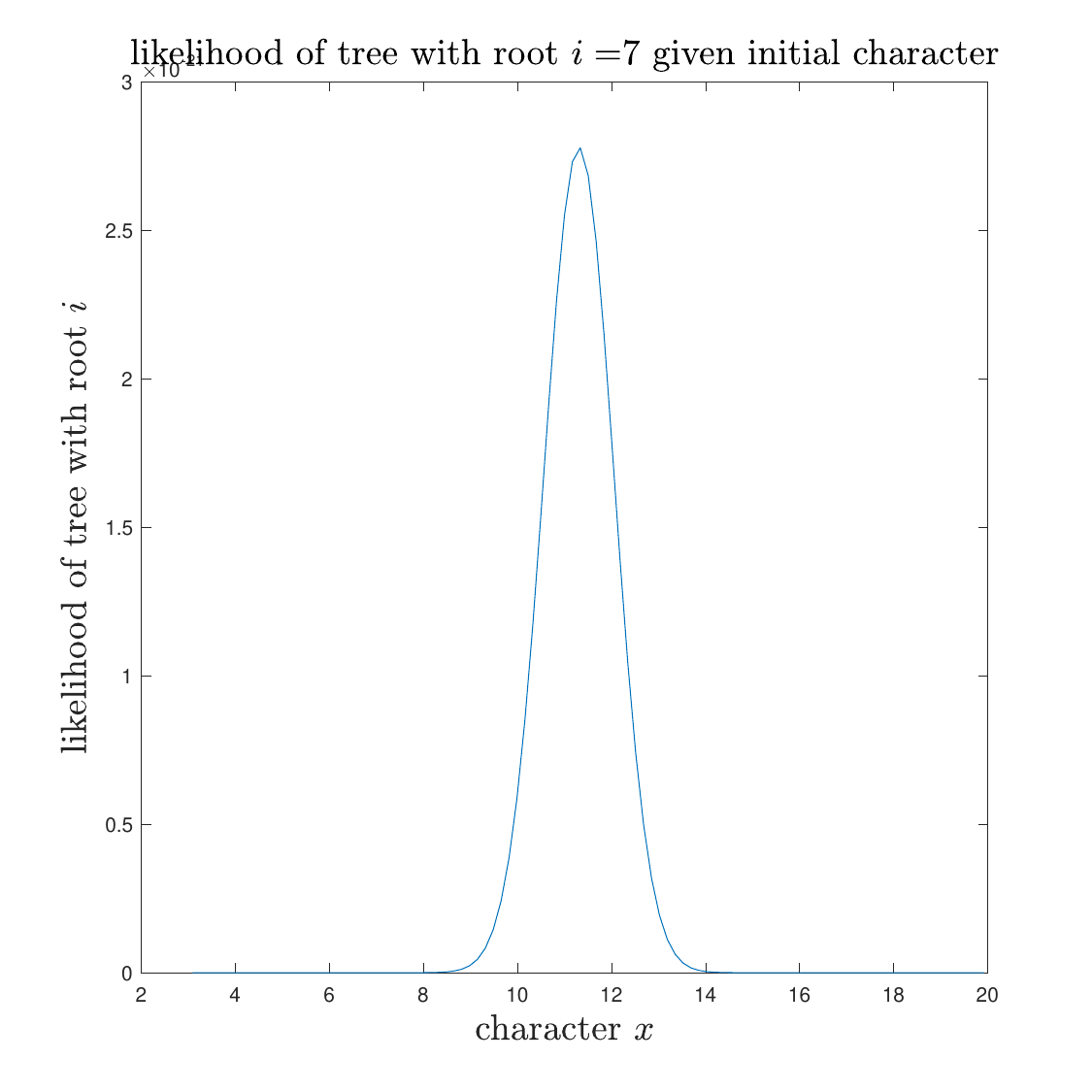}
\
\includegraphics[scale=0.4]{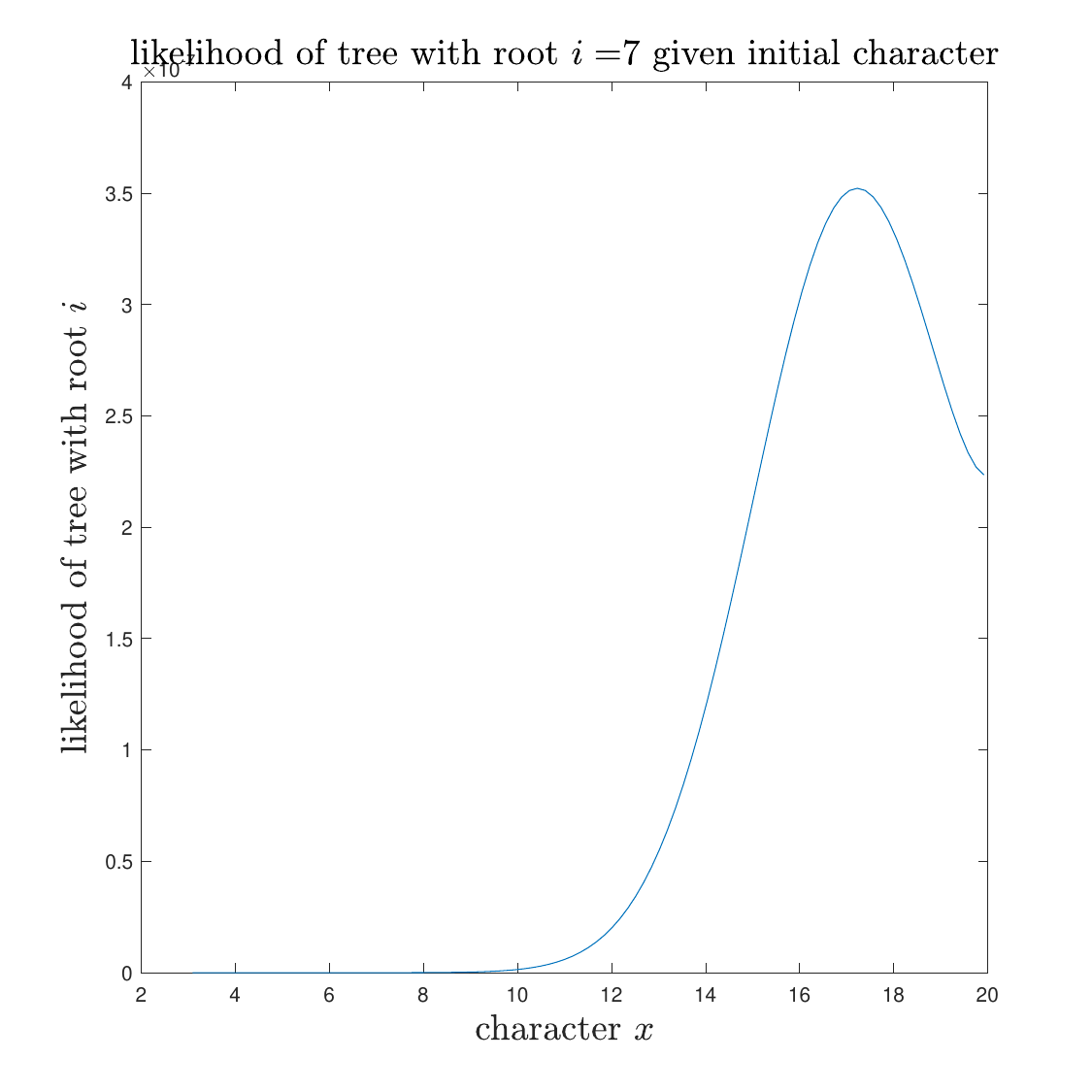}
\\
\includegraphics[scale=0.4]{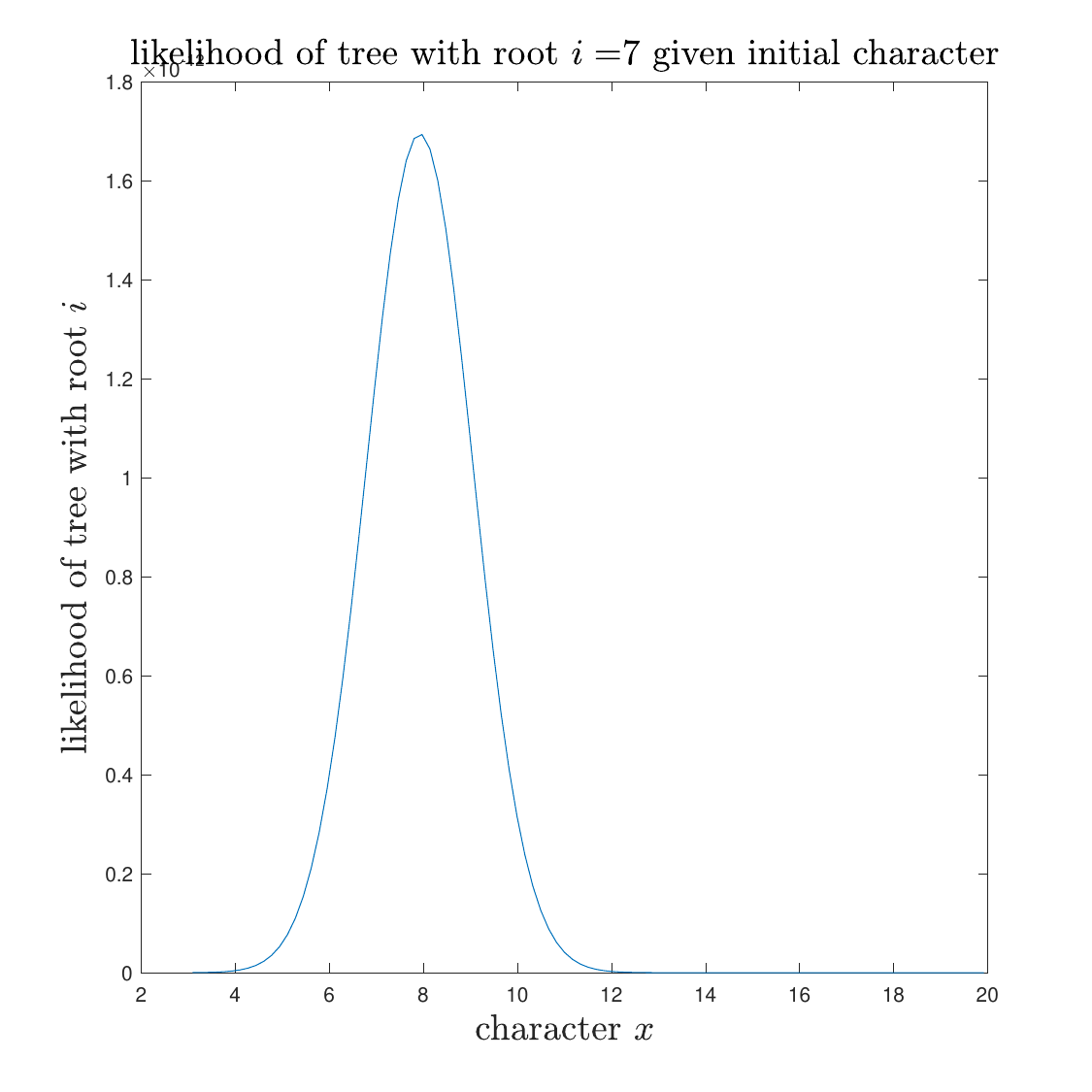}
\
\includegraphics[scale=0.4]{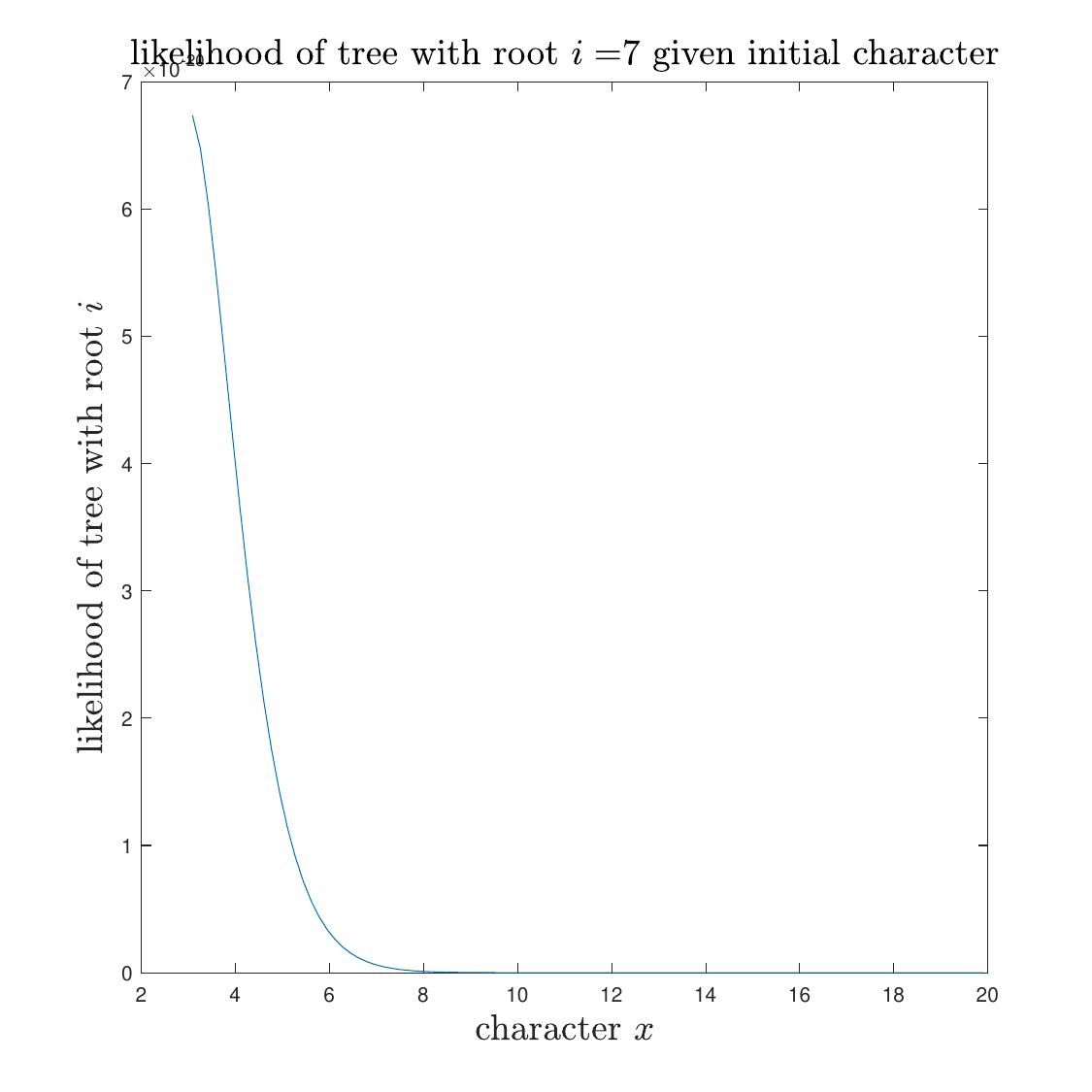}
\caption{From top left to the bottom right: The likelihood of observing the phylogenetic tree that started with parent $i=7$ given trait~$x$ observed at the start of the tree, in the \protect\hyperlink{QBD3}{QBD3} model in Section~\ref{sec:QBDmodels} and Synthetic Dataset~2 (Figure~\ref{fig:DataExample2}), for the ${\bf r}$ vectors $\#1,\#6, \#5$, and $\#8$ in Table~\ref{tab:Synexamplet2}.} 
\label{StatTraitsModel5phasesOverallParrentTraitsOnlyData2}
\end{figure}

\newpage
\section{Output: Empirical Dataset~1 (Figure~\ref{DataExample2})}
\label{outputEmpiricalDataExample2}

Here, we consider the \protect\hyperlink{QBD3}{QBD3} model in Section~\ref{sec:QBDmodels} with five phases, and evaluate the likelihood of the Empirical Dataset~1 for a range of the QBD parameters.

We can see in Figure 21 that the 4 parameter sets with positive drift (${\bf r}$ vectors $\#2-\#5$ in Table~\ref{tab:exampleDATA2}) give broadly similar sets of likelihood curves. Curves vary in their width depending on the height of the branch leading to tip $i$. It also seems that parameter sets with drift closer to zero produce broader curves whereas the two parameter sets with the largest (positive) drift have narrower curves centered around lower initial trait values.

In Figures 22, 23 and 24 we can again see that the behaviour seems to be dominated by the drift with the panels corresponding to ${\bf r}$ vectors $\#2-\#5$ all giving higher likelihood to root states where the level is near the lower boundary whereas panel one prefers root states with levels near the upper boundary.

\begin{figure}
\centering
\includegraphics[scale=0.4]{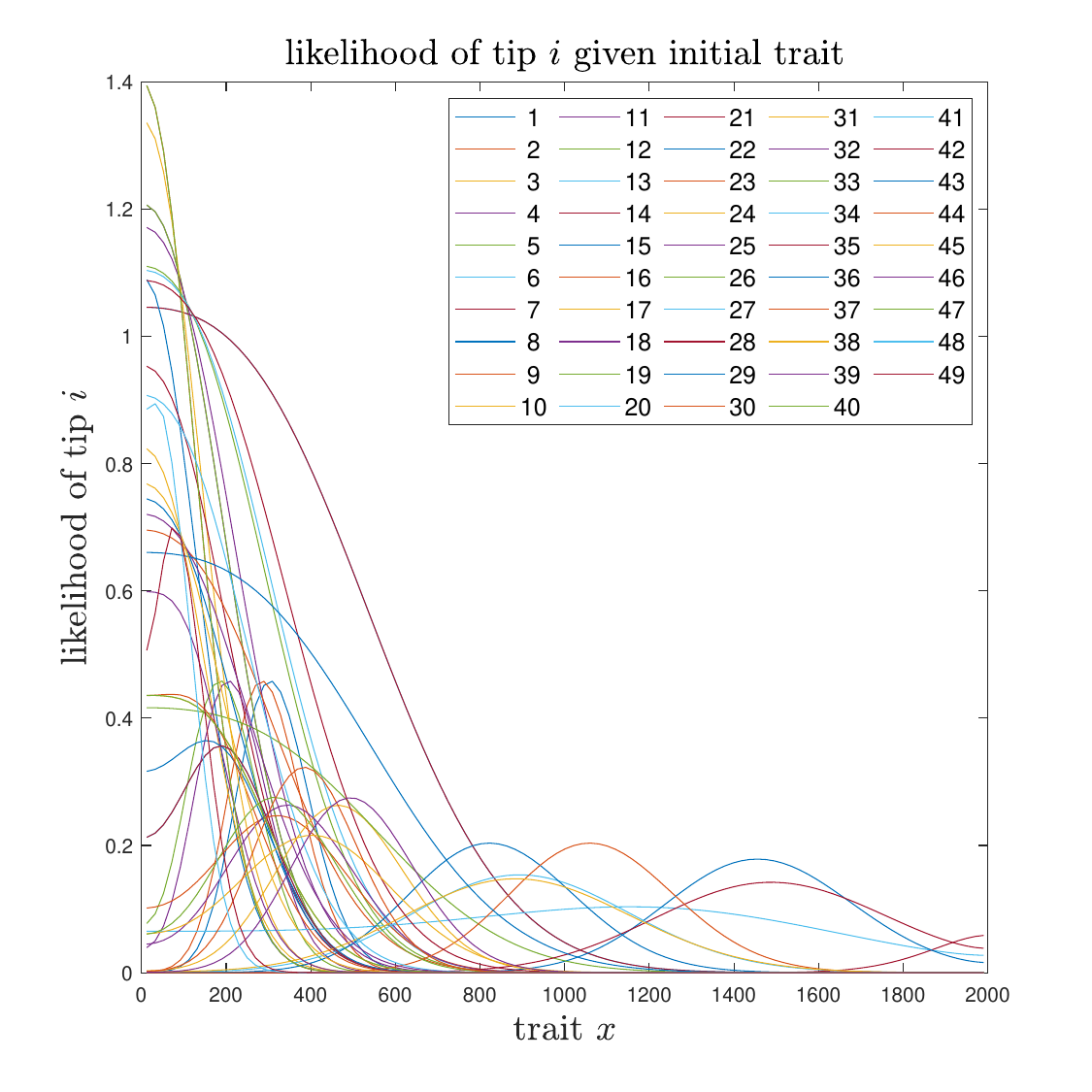}
\
\includegraphics[scale=0.4]{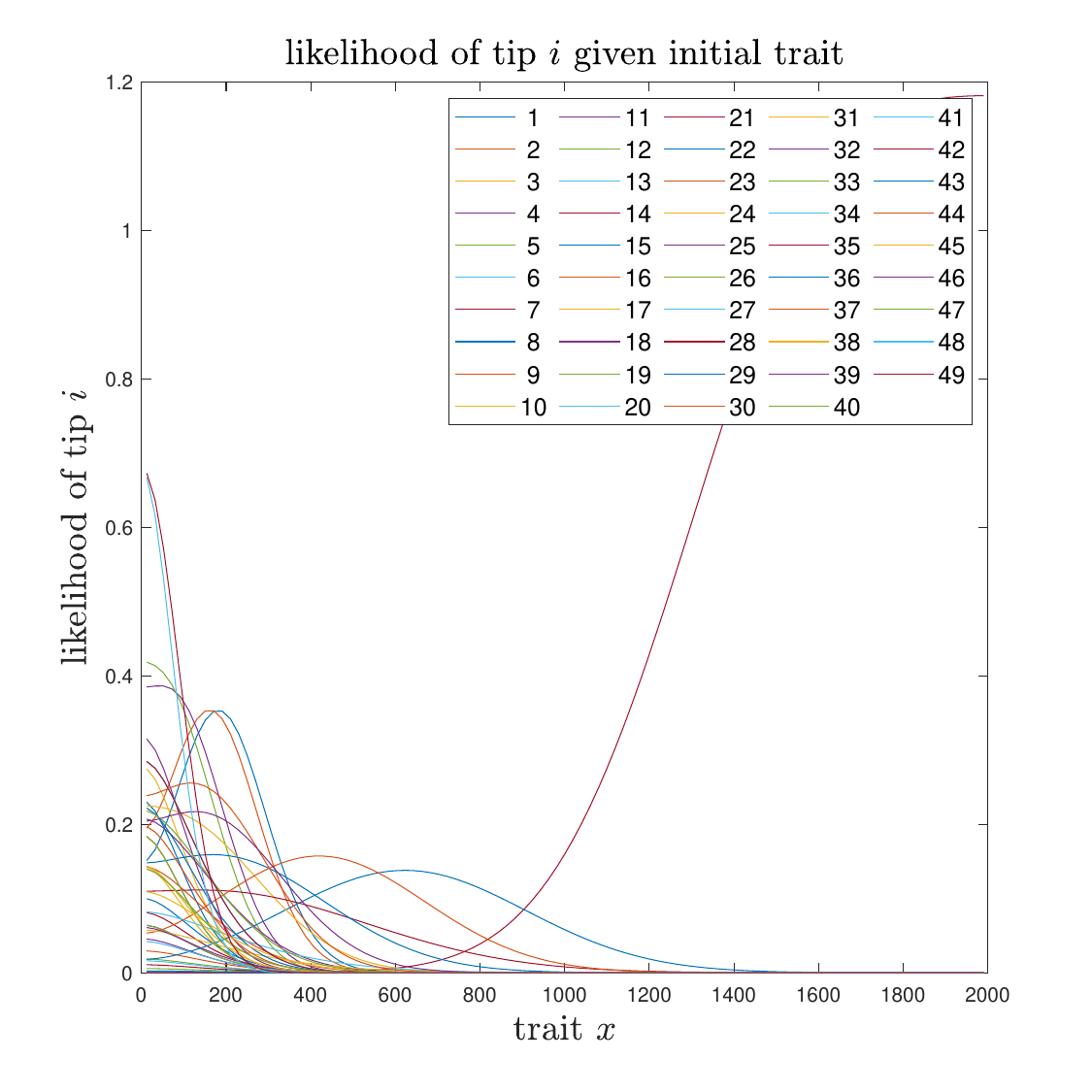}
\\
\includegraphics[scale=0.4]{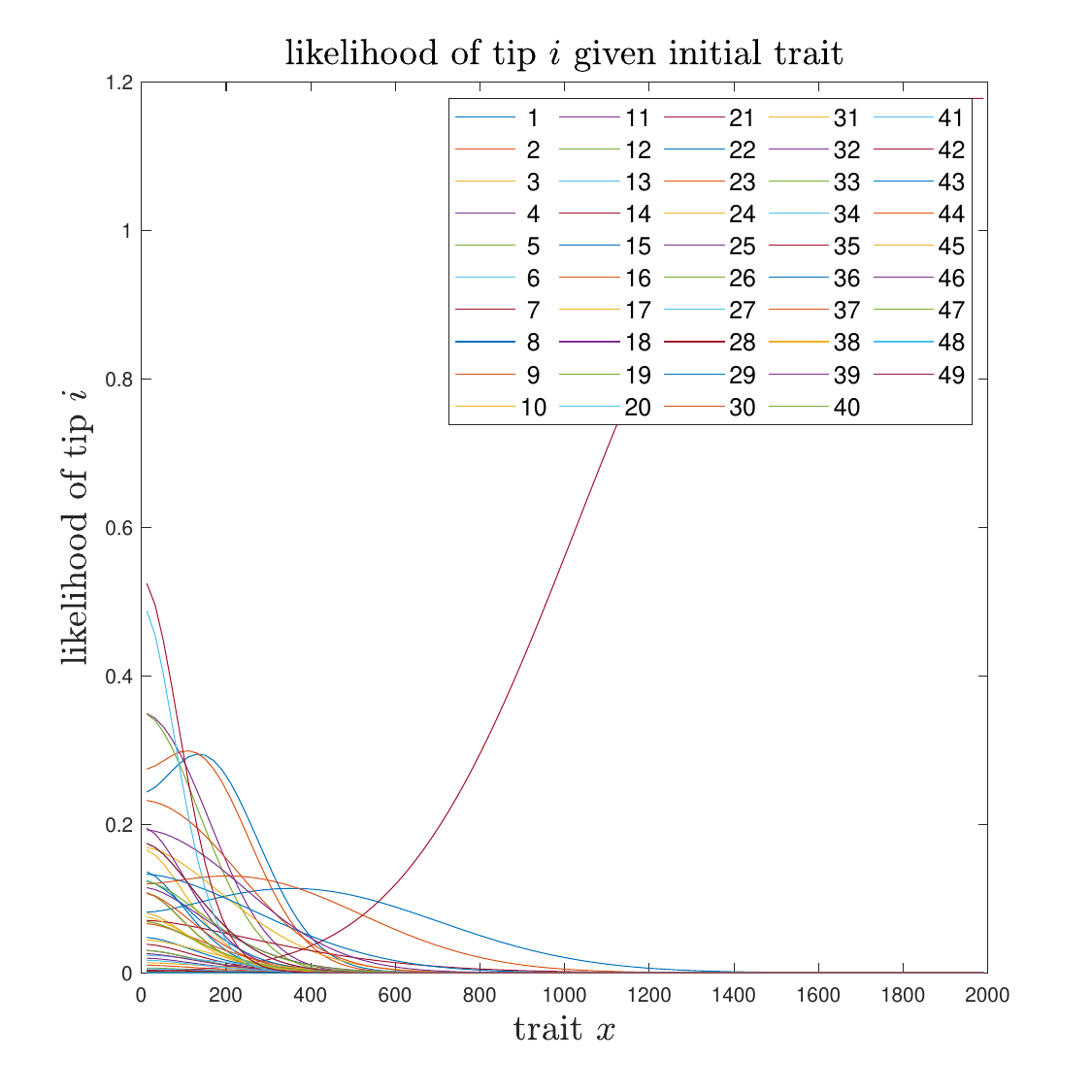}
\
\includegraphics[scale=0.4]{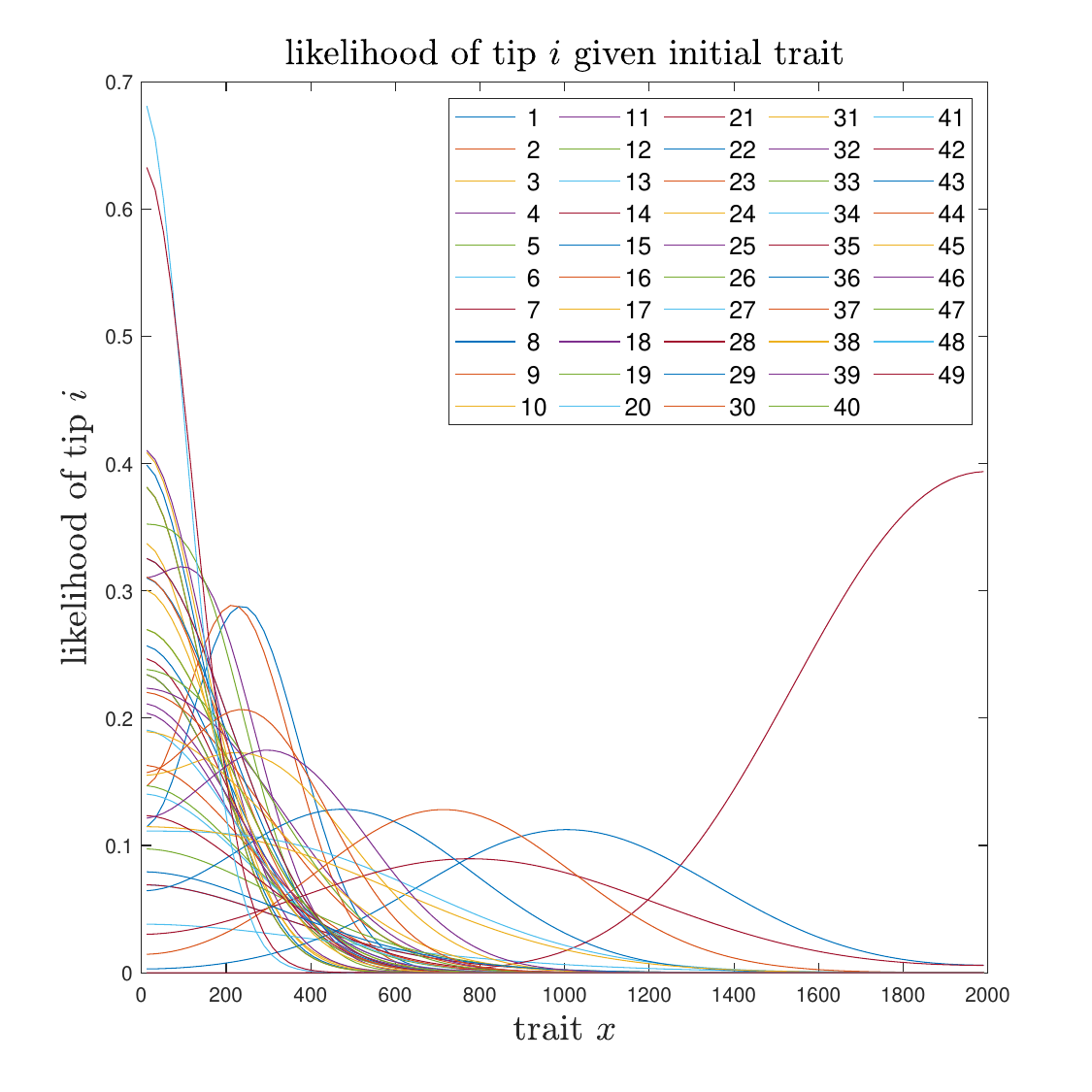}
\\
\includegraphics[scale=0.4]{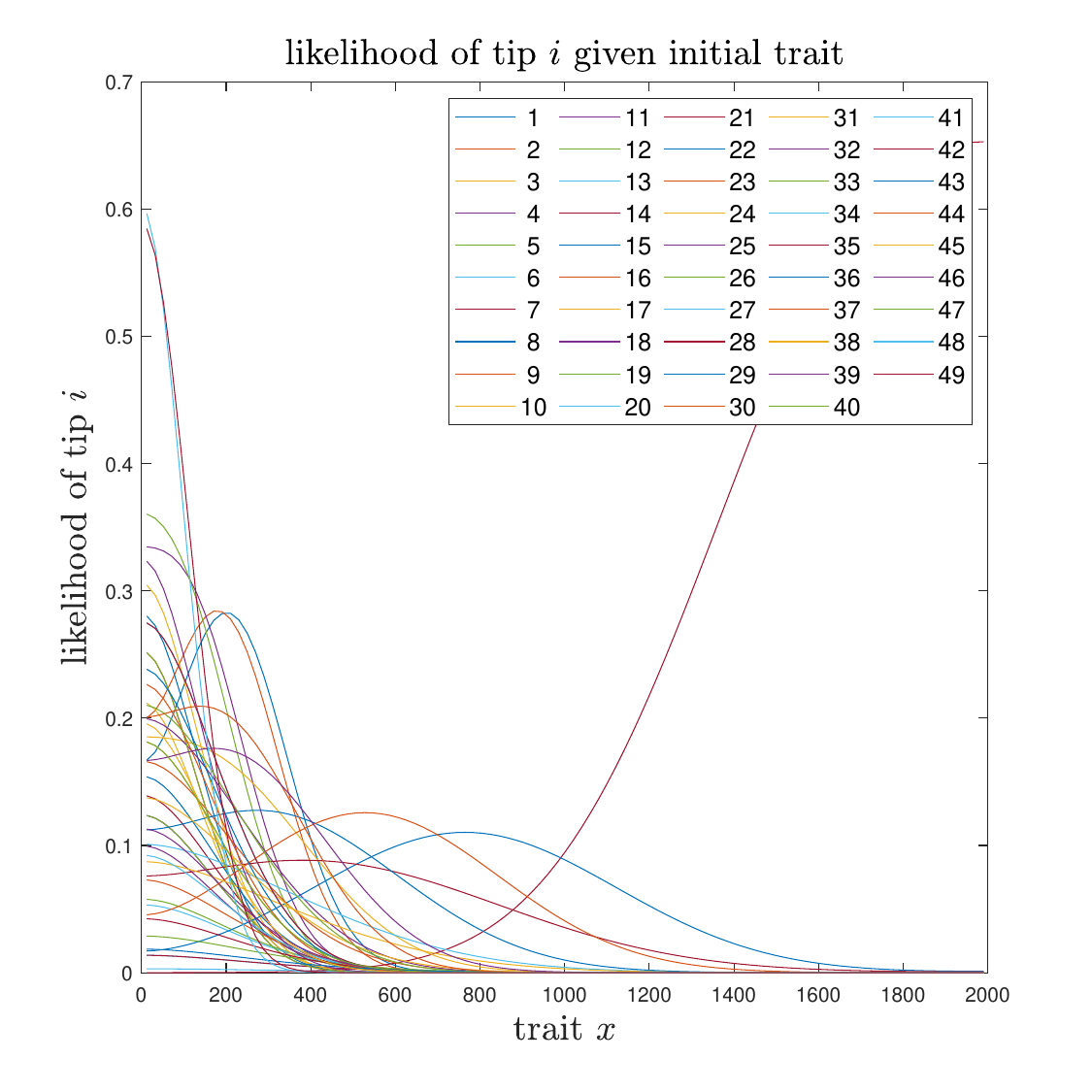}
\caption{From top left to the bottom right: The likelihood of observing tip $i$ given trait observed at the start of the branch corresponding to tip $i$, in the \protect\hyperlink{QBD3}{QBD3} model in Section~\ref{sec:QBDmodels} and Empirical Dataset~1 (Figure~\ref{DataExample2}), for the ${\bf r}$ vectors $\#1-\#5$ in Table~\ref{tab:exampleDATA2}.} 
\label{StatTraitsModel5phasesOverallcherriesEmpirical}
\end{figure}

\begin{figure}
\centering
\includegraphics[scale=0.4]{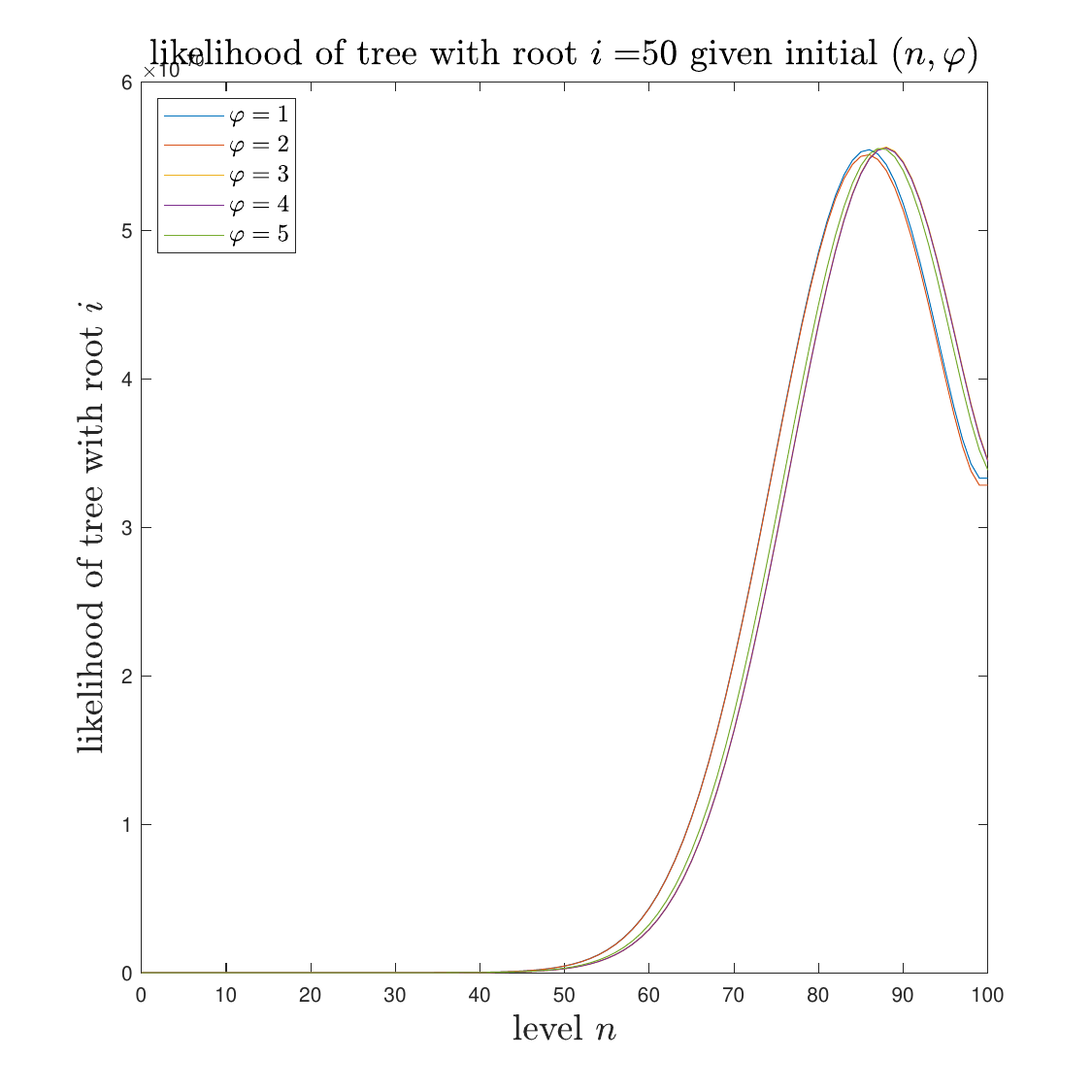}
\
\includegraphics[scale=0.4]{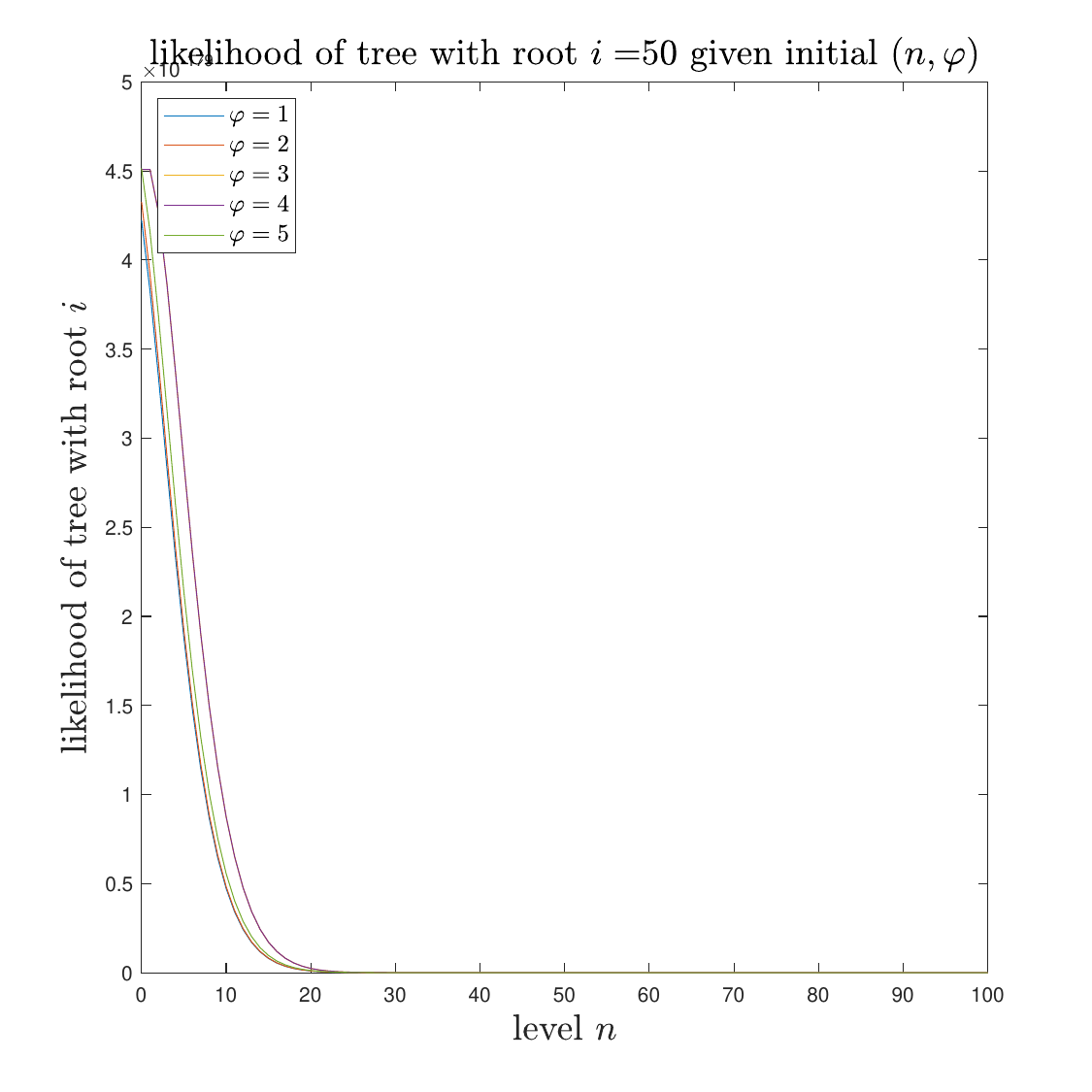}
\\
\includegraphics[scale=0.4]{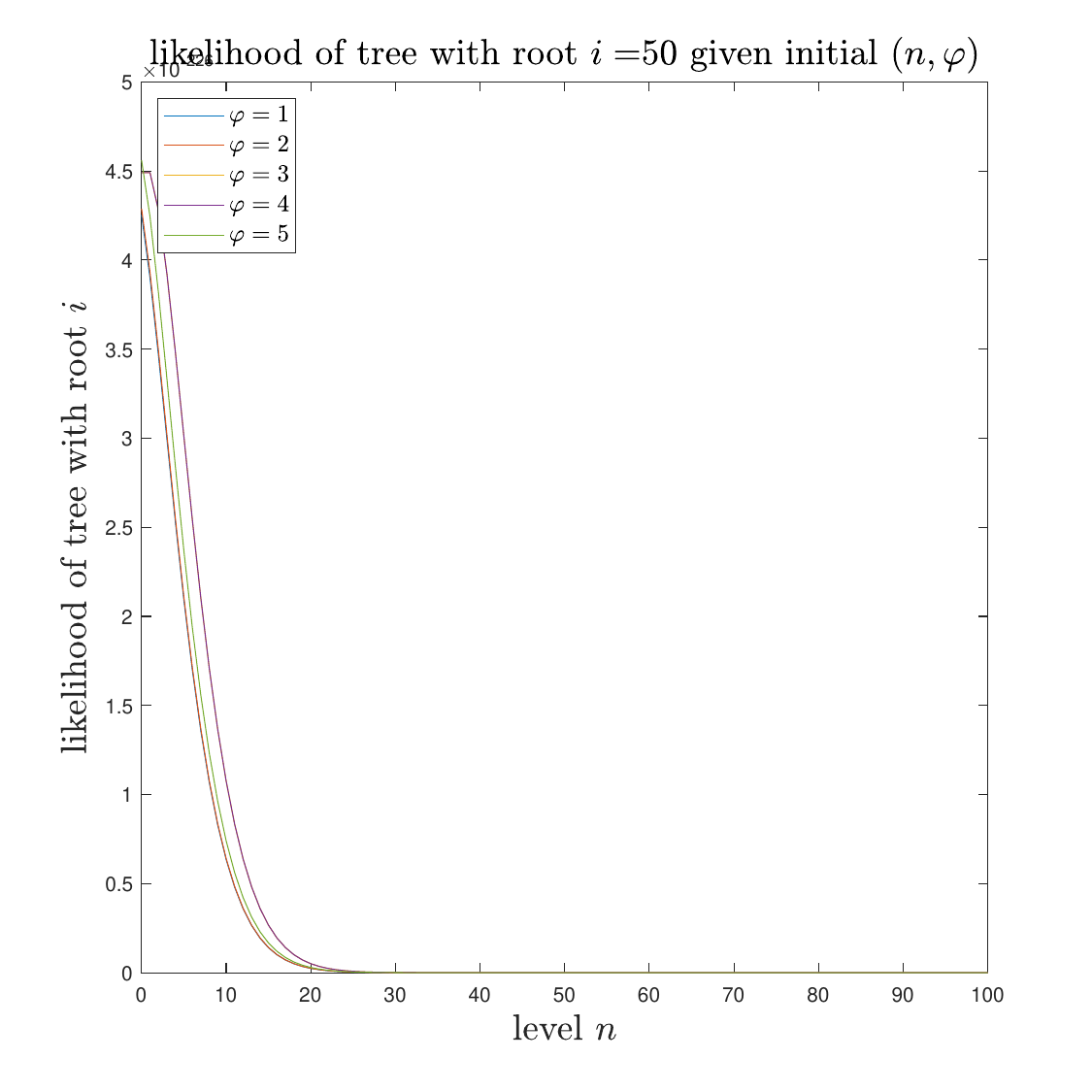}
\
\includegraphics[scale=0.4]{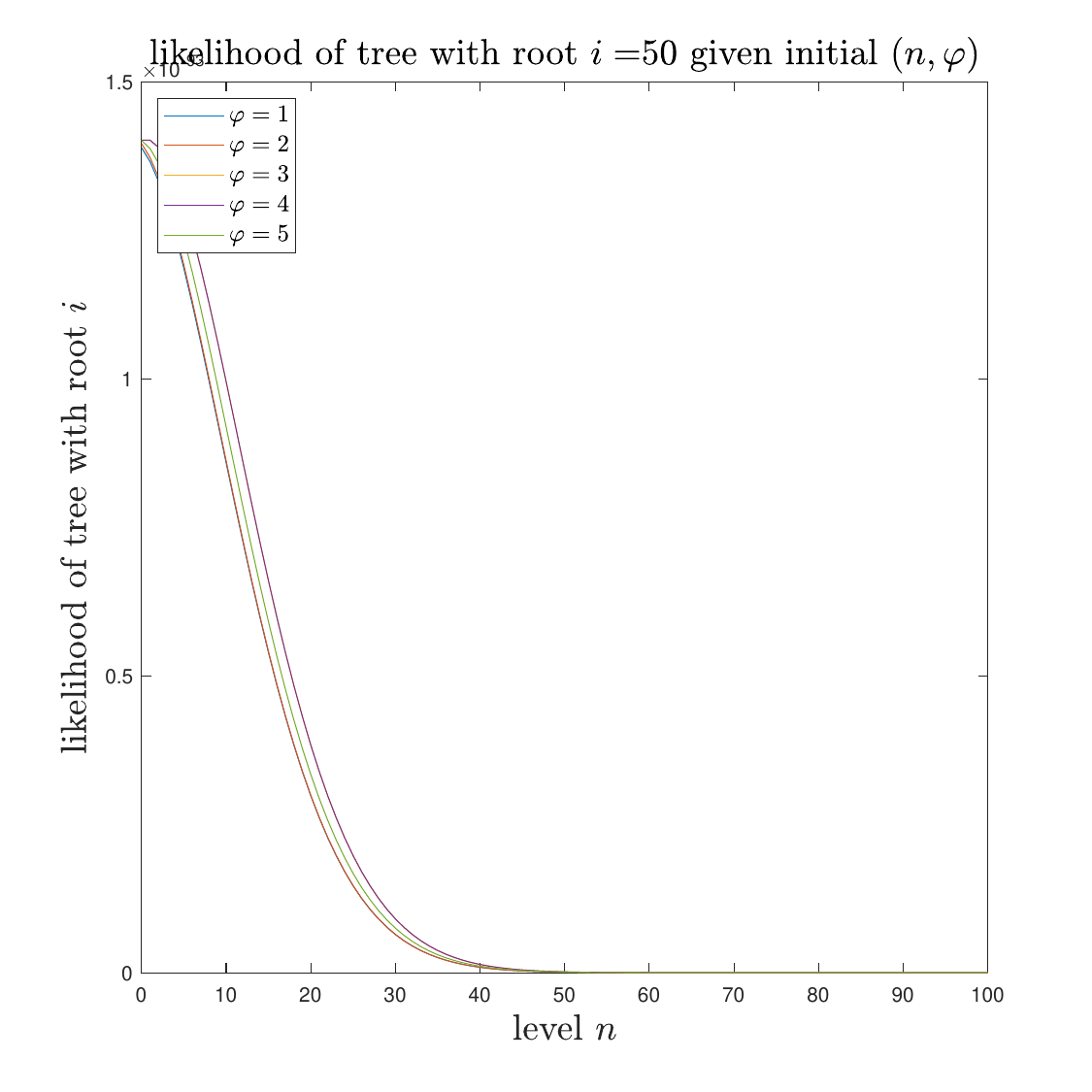}
\\
\includegraphics[scale=0.4]{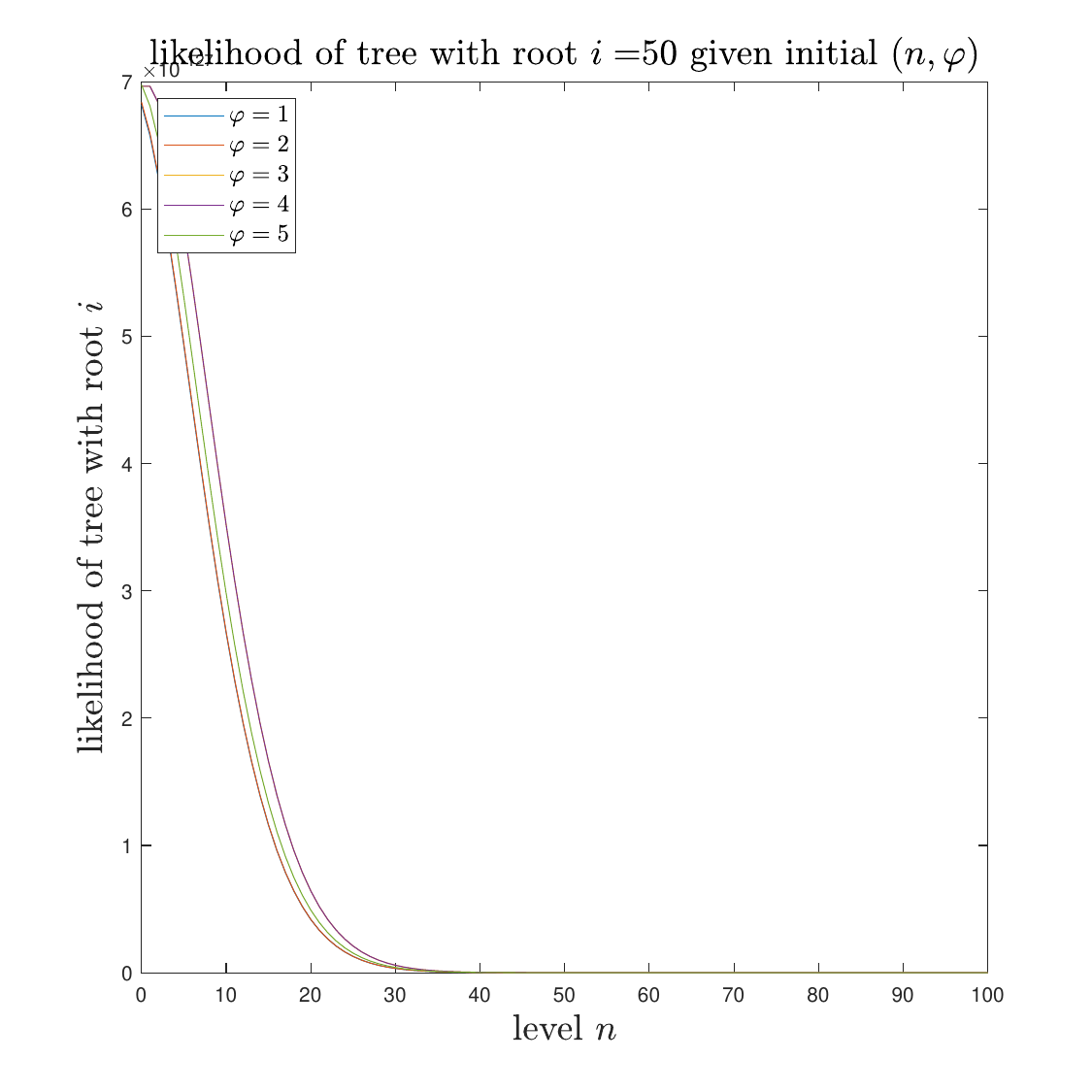}
\caption{From top left to the bottom right: The likelihood of observing the phylogenetic tree that started with parent $i$ given level $n$ and phase~$\varphi$ observed at the start of the tree, in the \protect\hyperlink{QBD3}{QBD3} model in Section~\ref{sec:QBDmodels} and Empirical Dataset~1 (Figure~\ref{DataExample2}), for the ${\bf r}$ vectors $\#1-\#5$ in Table~\ref{tab:exampleDATA2}.}
\label{StatTraitsModel5phasesOverallParrentEmpirical}
\end{figure}

\begin{figure}
\centering
\includegraphics[scale=0.4]{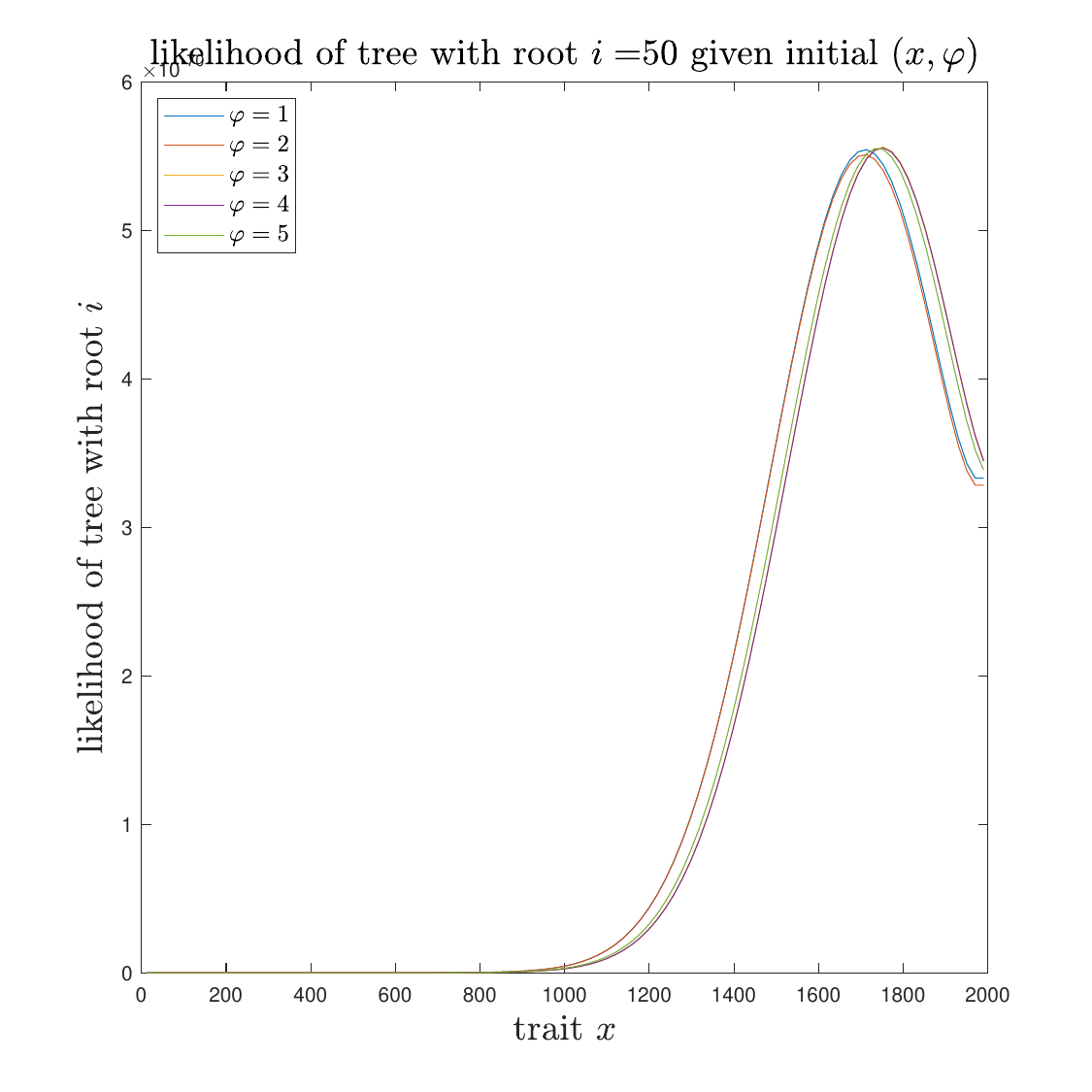}
\
\includegraphics[scale=0.4]{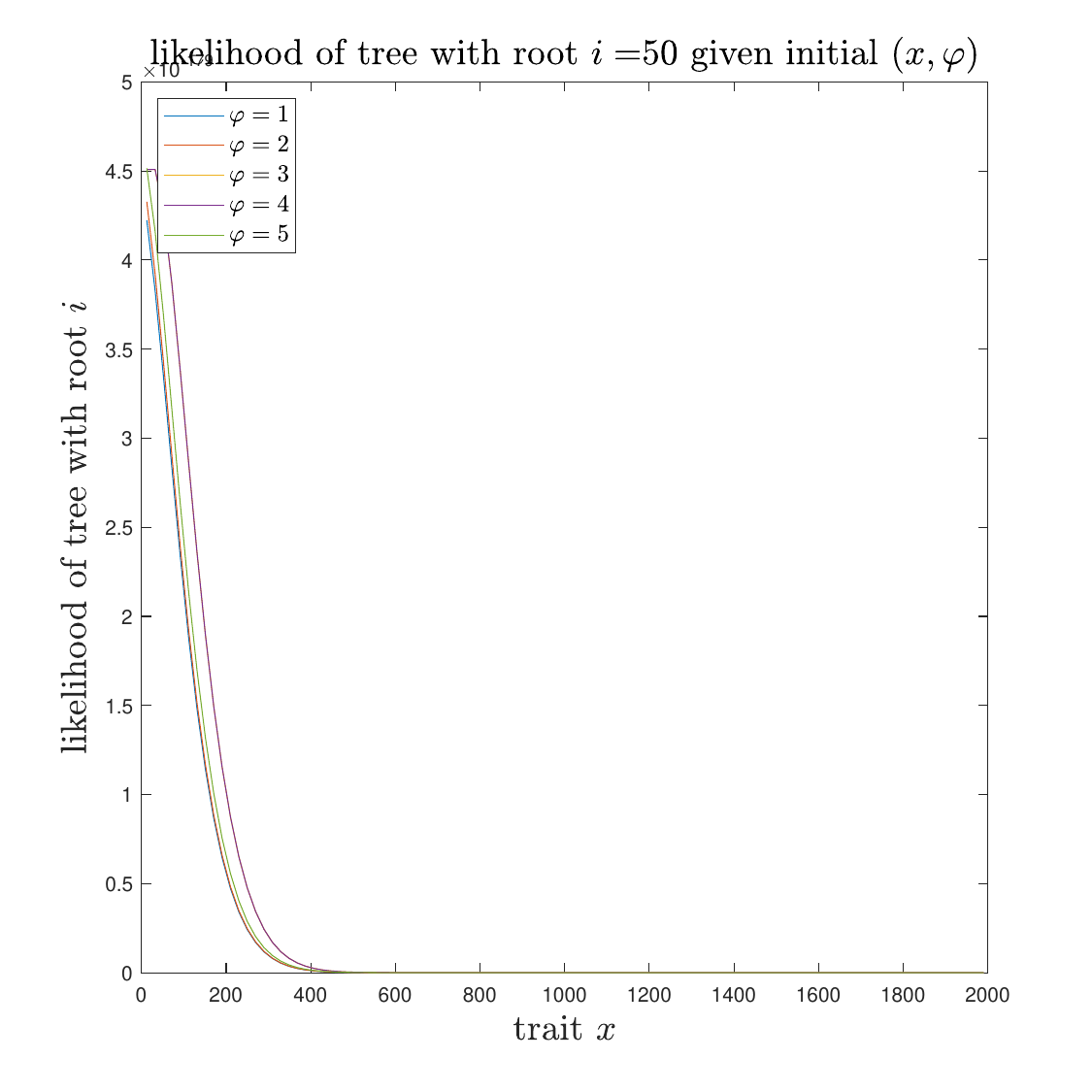}
\\
\includegraphics[scale=0.4]{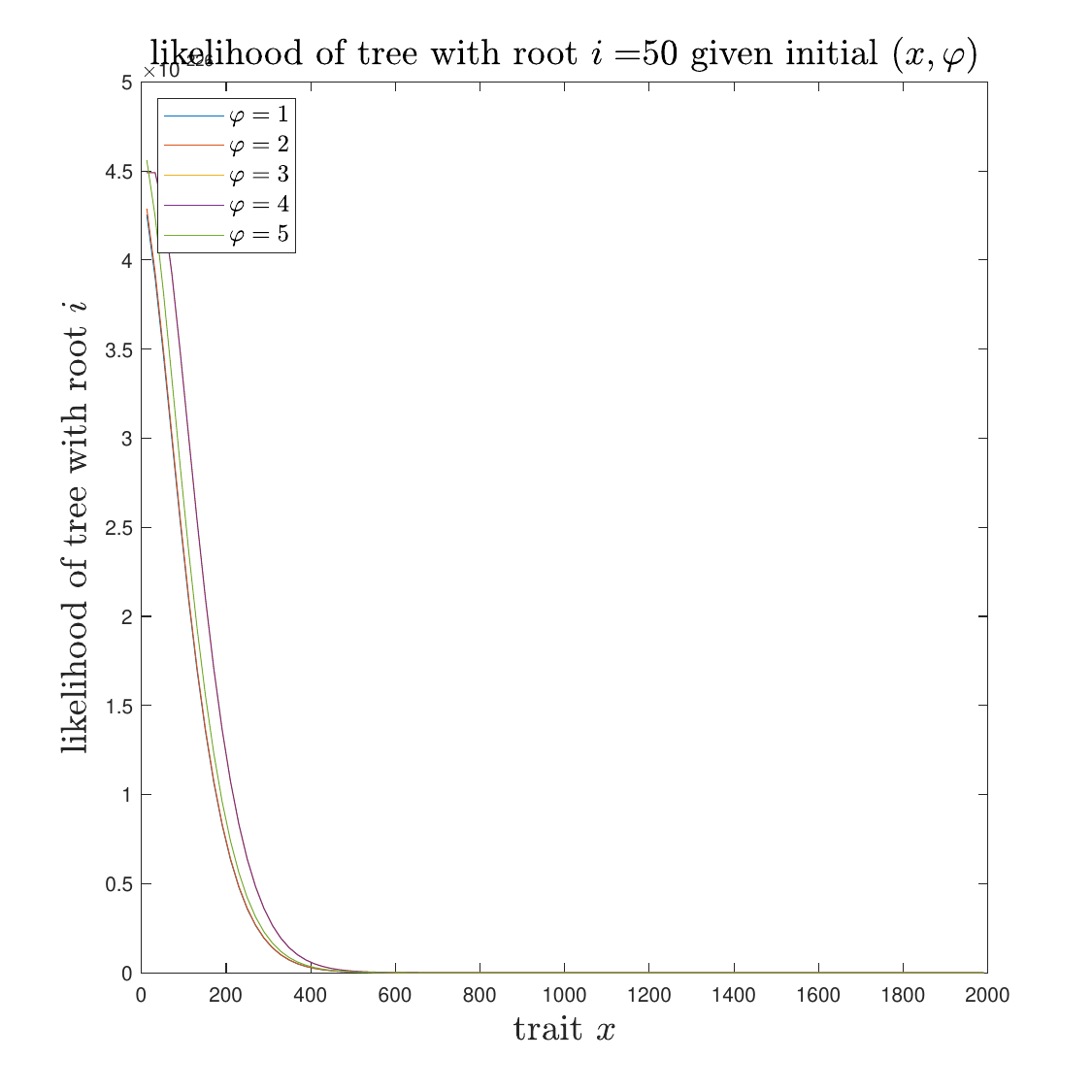}
\
\includegraphics[scale=0.4]{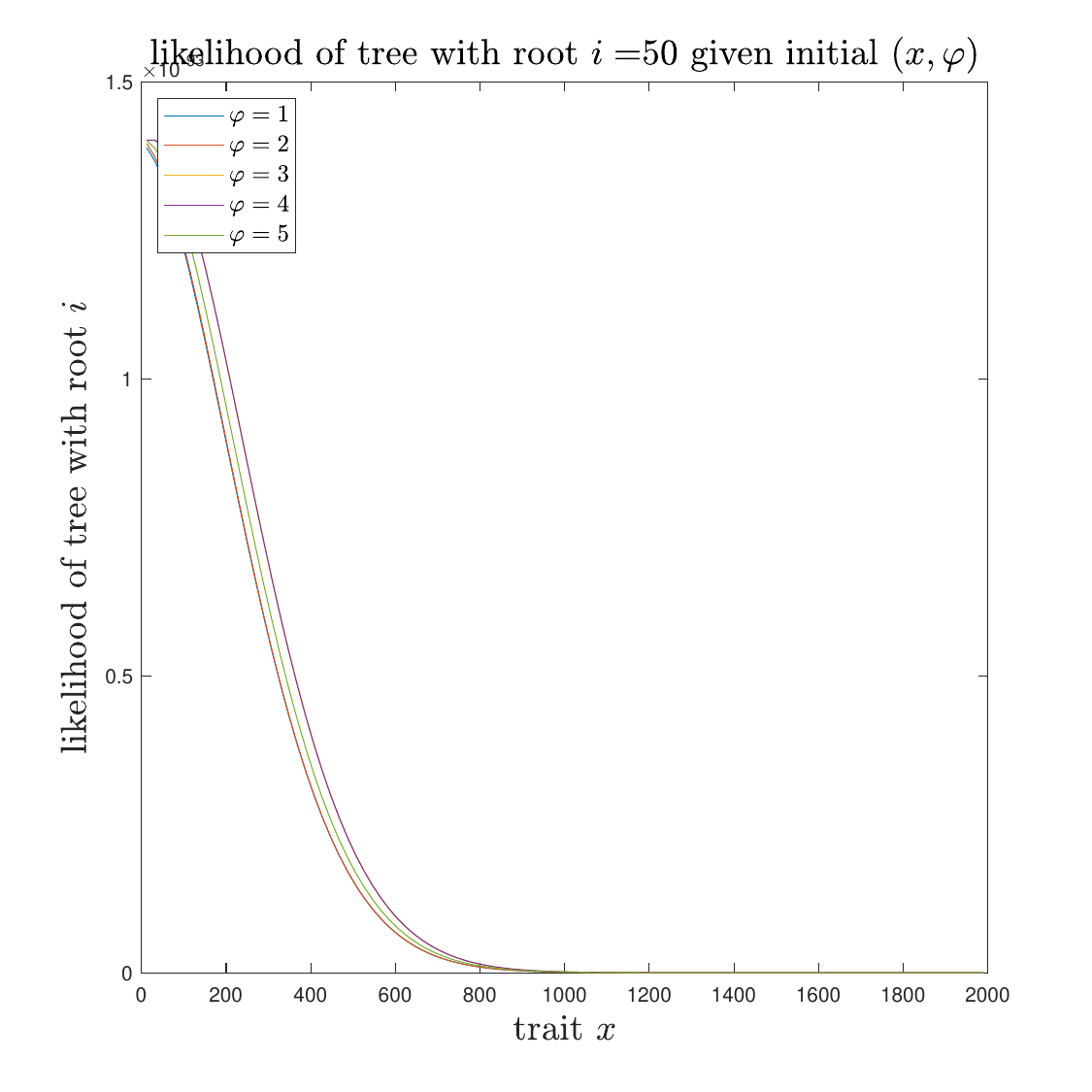}
\\
\includegraphics[scale=0.4]{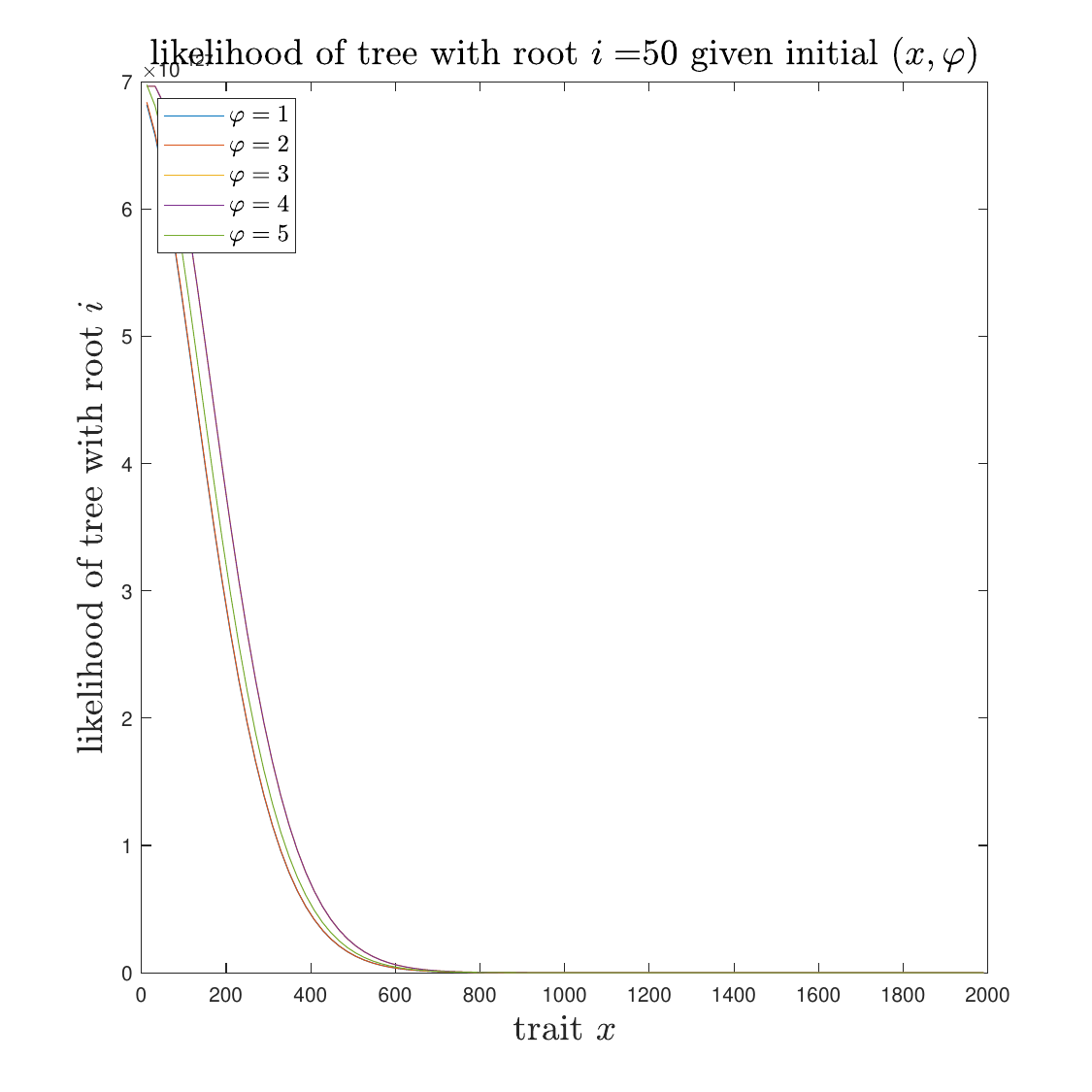}
\caption{From top left to the bottom right: The likelihood of observing the phylogenetic tree that started with parent $i$ given trait~$x$ and phase~$\varphi$ observed at the start of the tree, in the \protect\hyperlink{QBD3}{QBD3} model in Section~\ref{sec:QBDmodels} and Empirical Dataset~1 (Figure~\ref{DataExample2}), for the ${\bf r}$ vectors $\#1-\#5$ in Table~\ref{tab:exampleDATA2}.} 
\label{StatTraitsModel5phasesOverallParrentEmpirical}
\end{figure}

\begin{figure}
\centering
\includegraphics[scale=0.4]{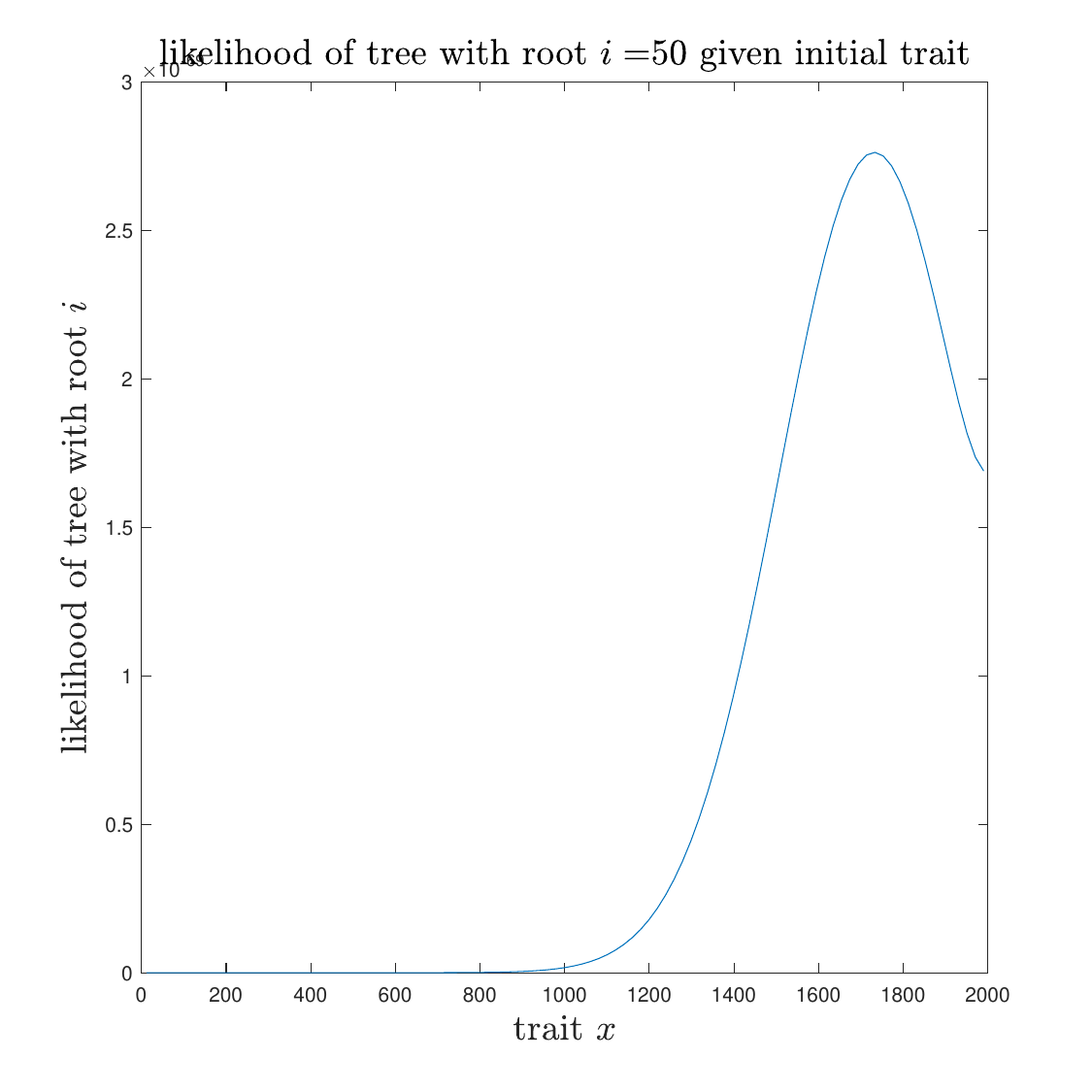}
\
\includegraphics[scale=0.4]{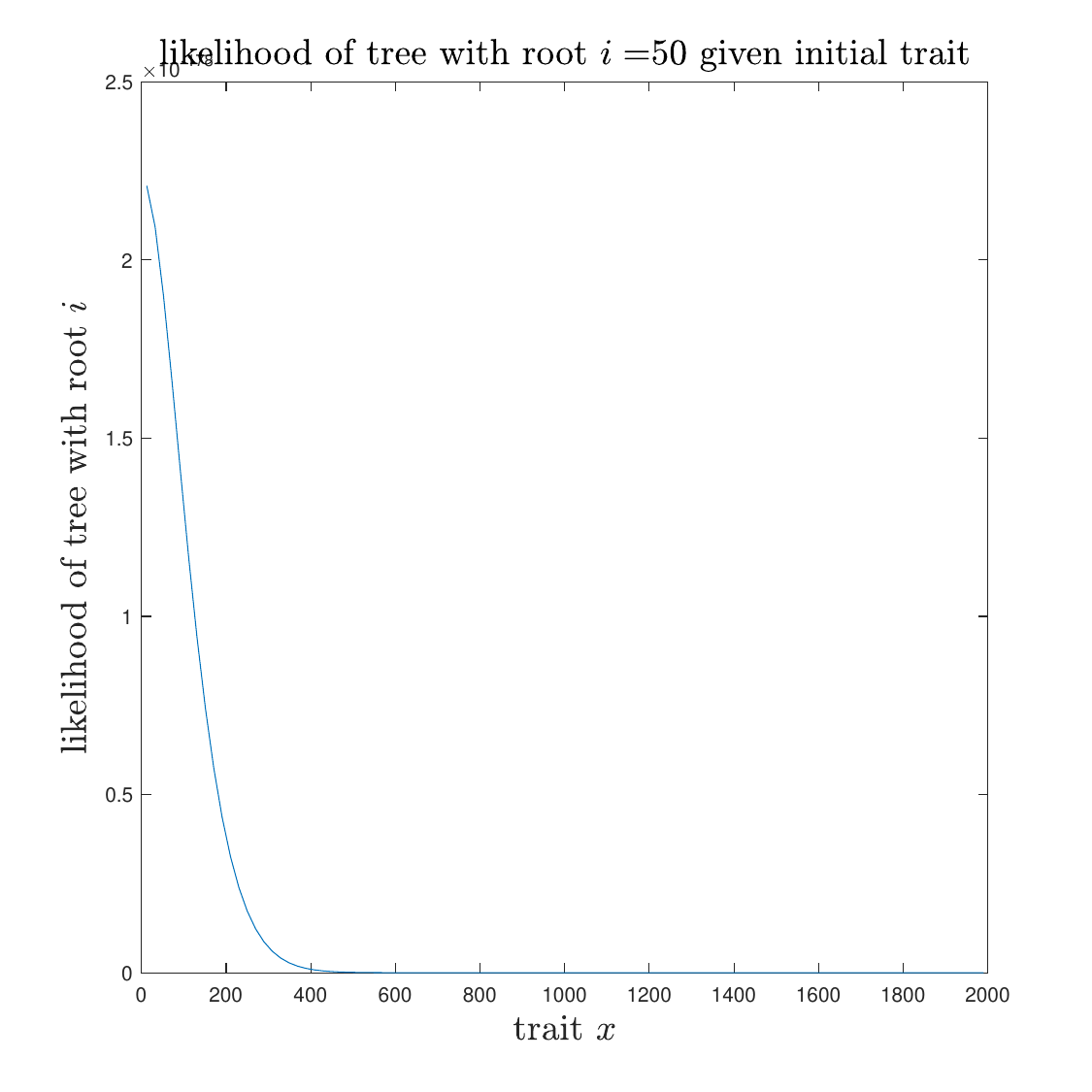}
\\
\includegraphics[scale=0.4]{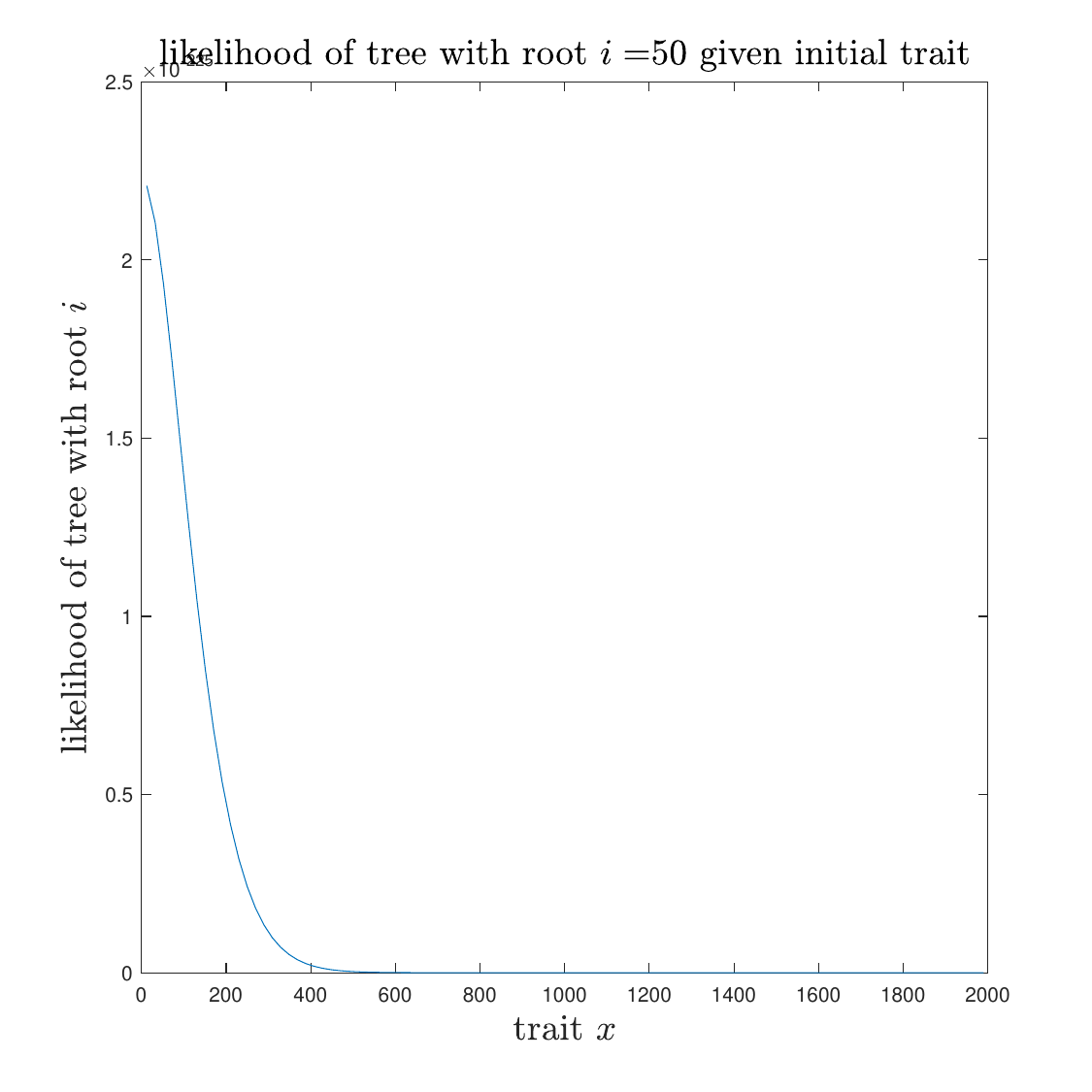}
\
\includegraphics[scale=0.4]{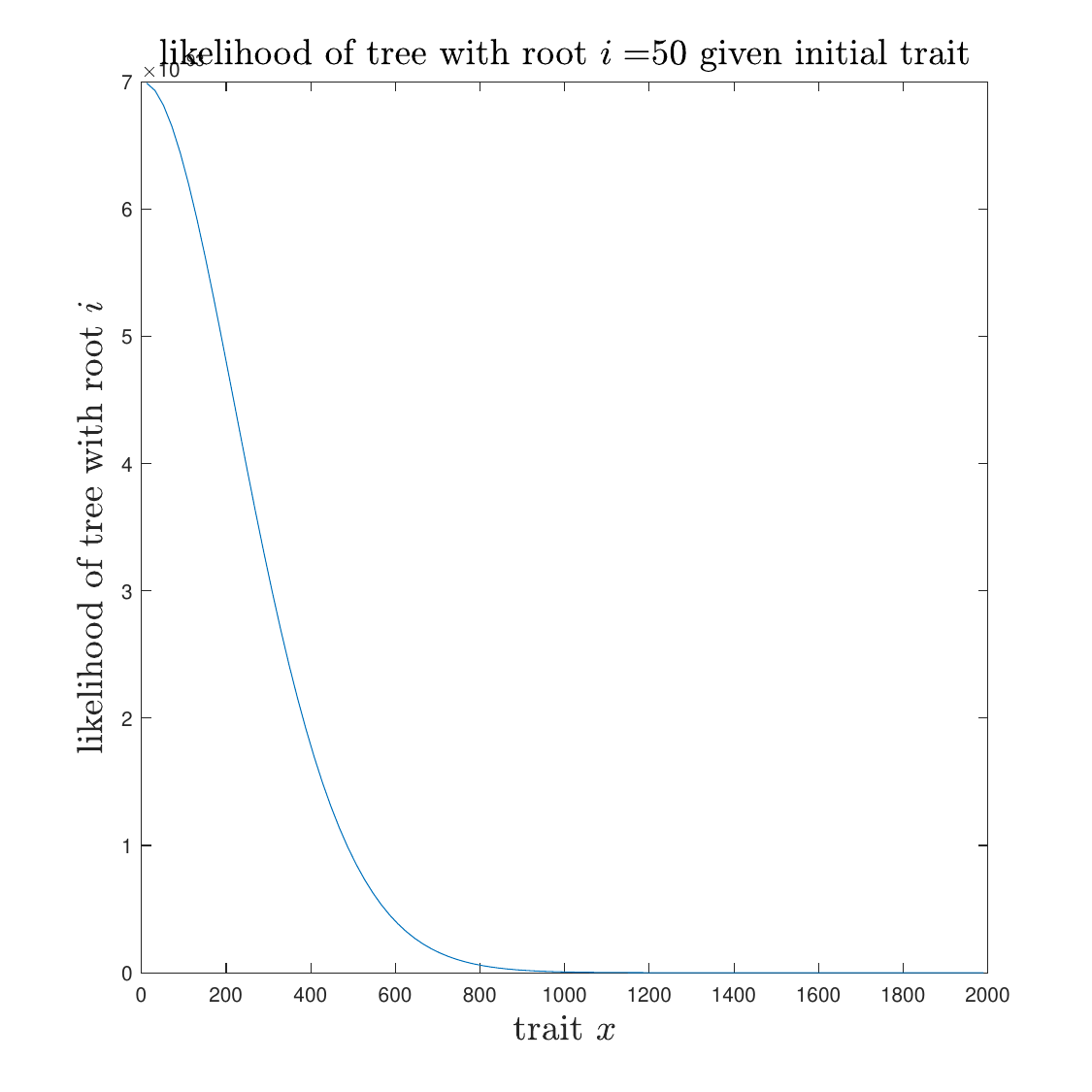}
\\
\includegraphics[scale=0.4]{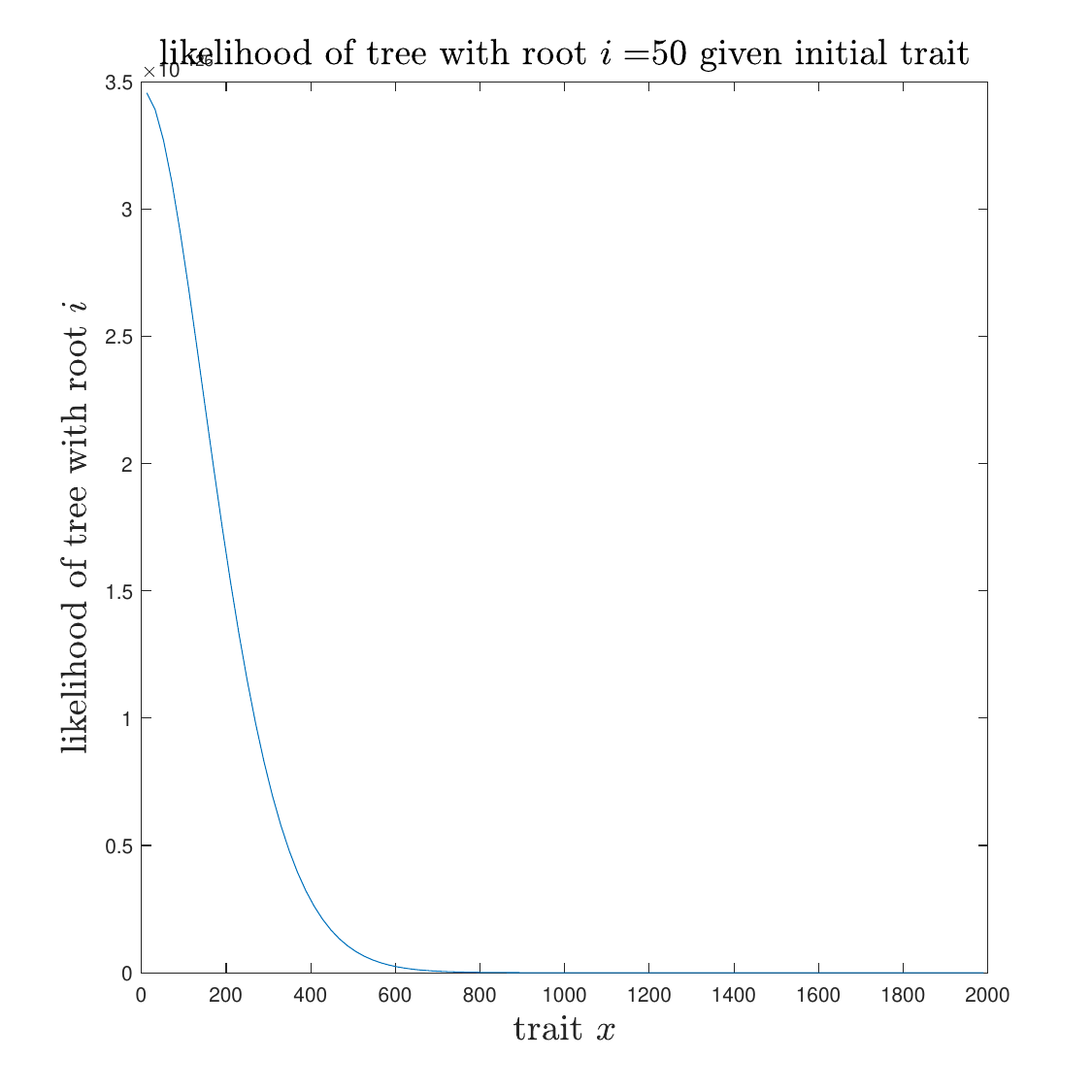}
\caption{From top left to the bottom right: The likelihood of observing the phylogenetic tree that started with parent $i$ given trait~$x$ observed at the start of the tree, in the \protect\hyperlink{QBD3}{QBD3} model in Section~\ref{sec:QBDmodels} and Empirical Dataset~1 (Figure~\ref{DataExample2}), for the ${\bf r}$ vectors $\#1-\#5$ in Table~\ref{tab:exampleDATA2}.}
\label{StatTraitsModel5phasesOverallParrentTraitsOnlyEmpirical}
\end{figure}

\newpage
\section{Output: Empirical Dataset~2 (Figure~\ref{DataExample3})}
\label{outputEmpiricalDataExample3}

Here, we consider the \protect\hyperlink{QBD3}{QBD3} model in Section~\ref{sec:QBDmodels} with five phases, and evaluate the likelihood of the Empirical Dataset~2 for a range of the QBD parameters.

We can see in Figure 25 that the 4 parameter sets with positive drift (${\bf r}$ vectors $\#2-\#5$ in Table~\ref{tab:exampleDATA2}) give broadly similar sets of likelihood curves. Curves vary in their width depending on the height of the branch leading to tip $i$. It also seems that parameter sets with drift closer to zero produce broader curves whereas the two parameter sets with the largest (positive) drift have narrower curves centered around lower initial trait values.

In Figures 26, 27 and 28 we can again see that the behaviour seems to be dominated by the drift with the panels corresponding to ${\bf r}$ vectors $\#2-\#5$ all giving higher likelihood to root states where the level is near the lower boundary whereas panel one prefers root states with levels near the upper boundary.

\begin{figure}
\centering
\includegraphics[scale=0.4]{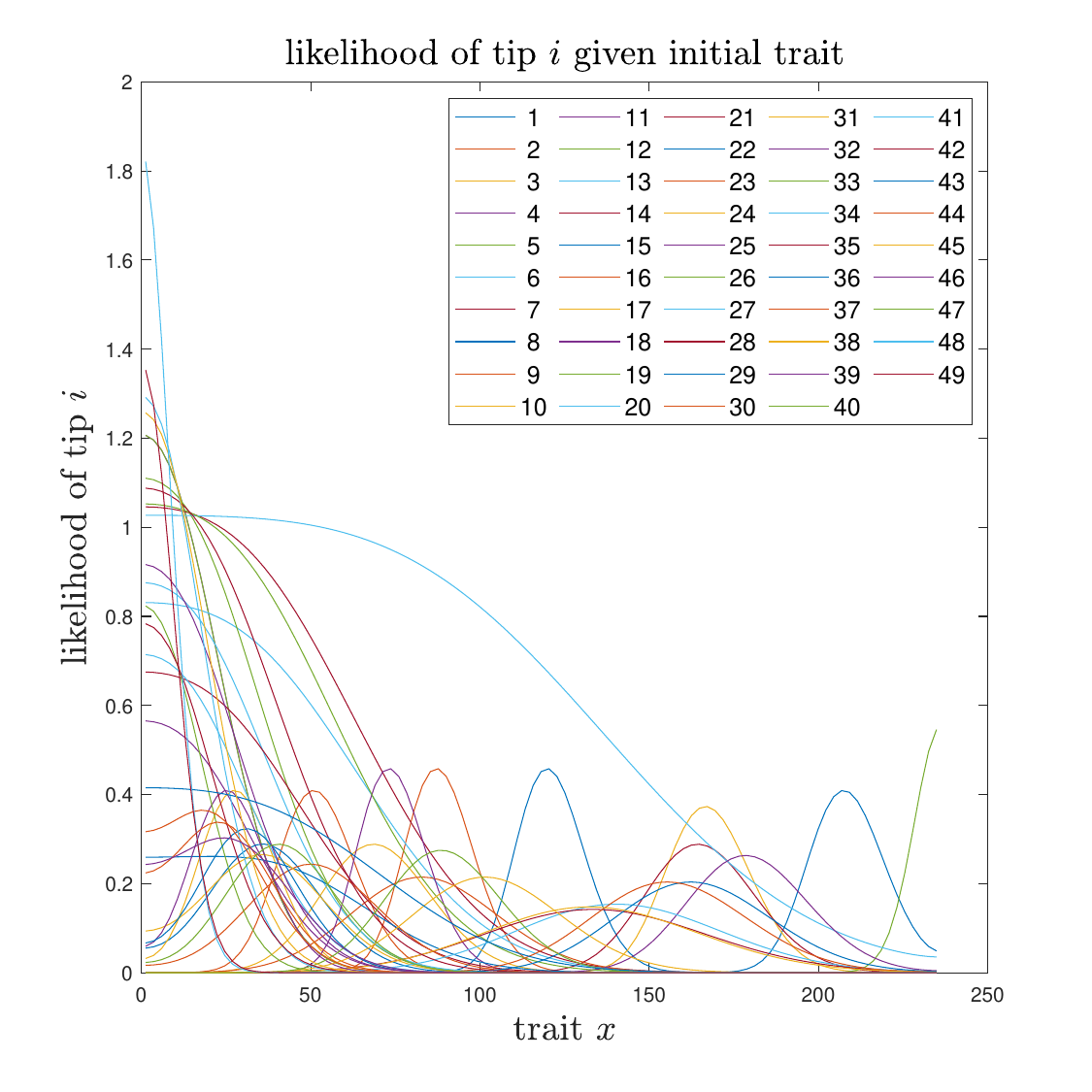}
\
\includegraphics[scale=0.4]{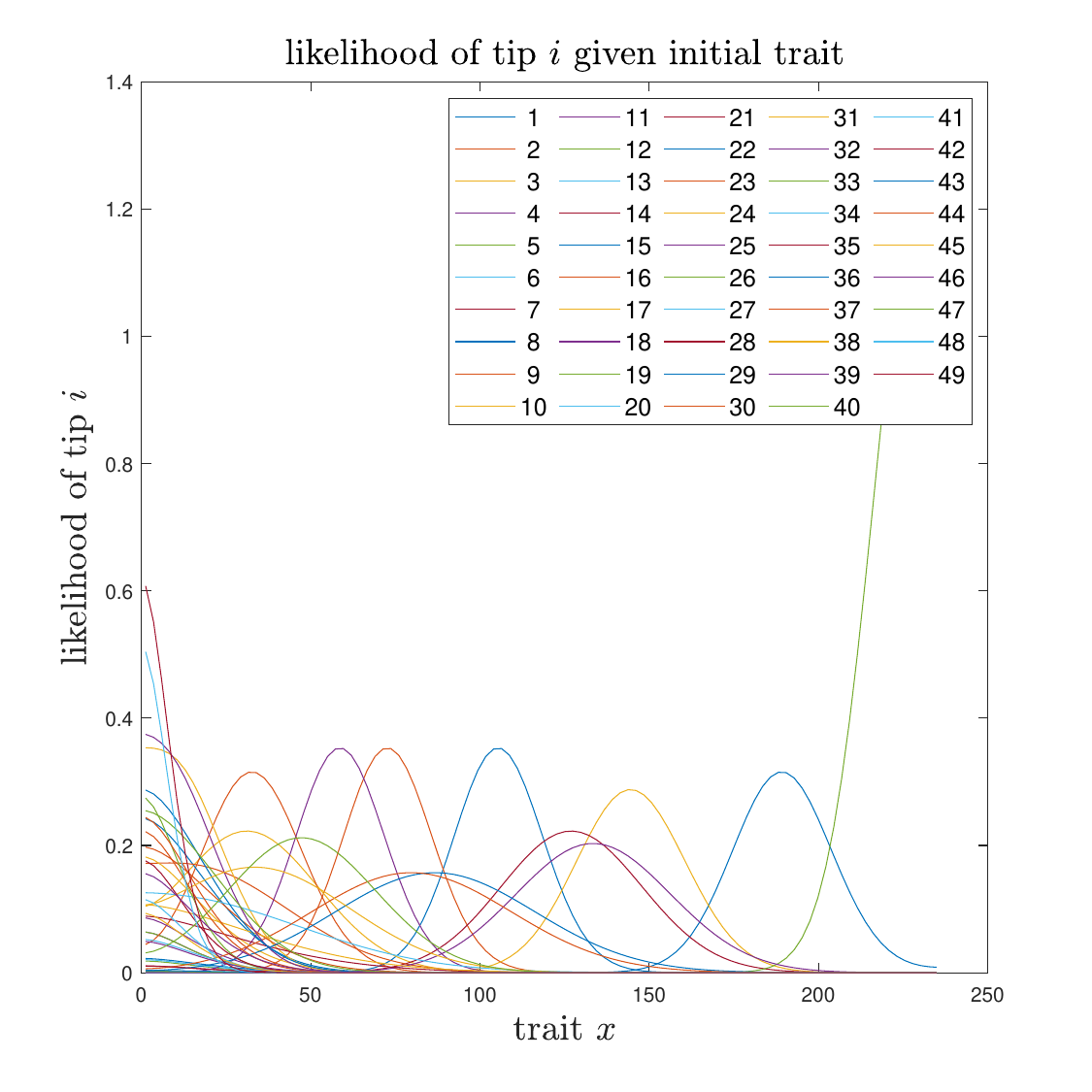}
\\
\includegraphics[scale=0.4]{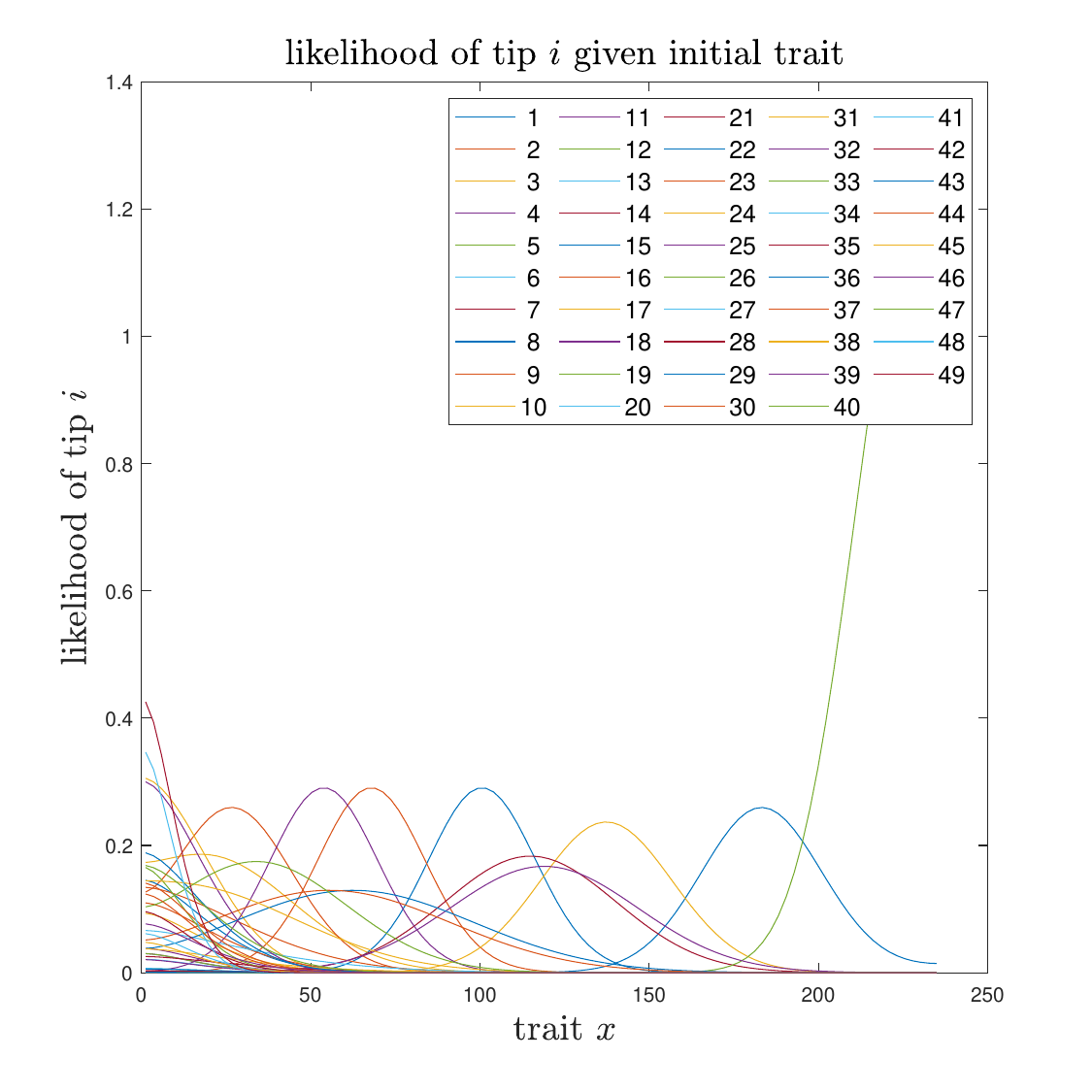}
\
\includegraphics[scale=0.4]{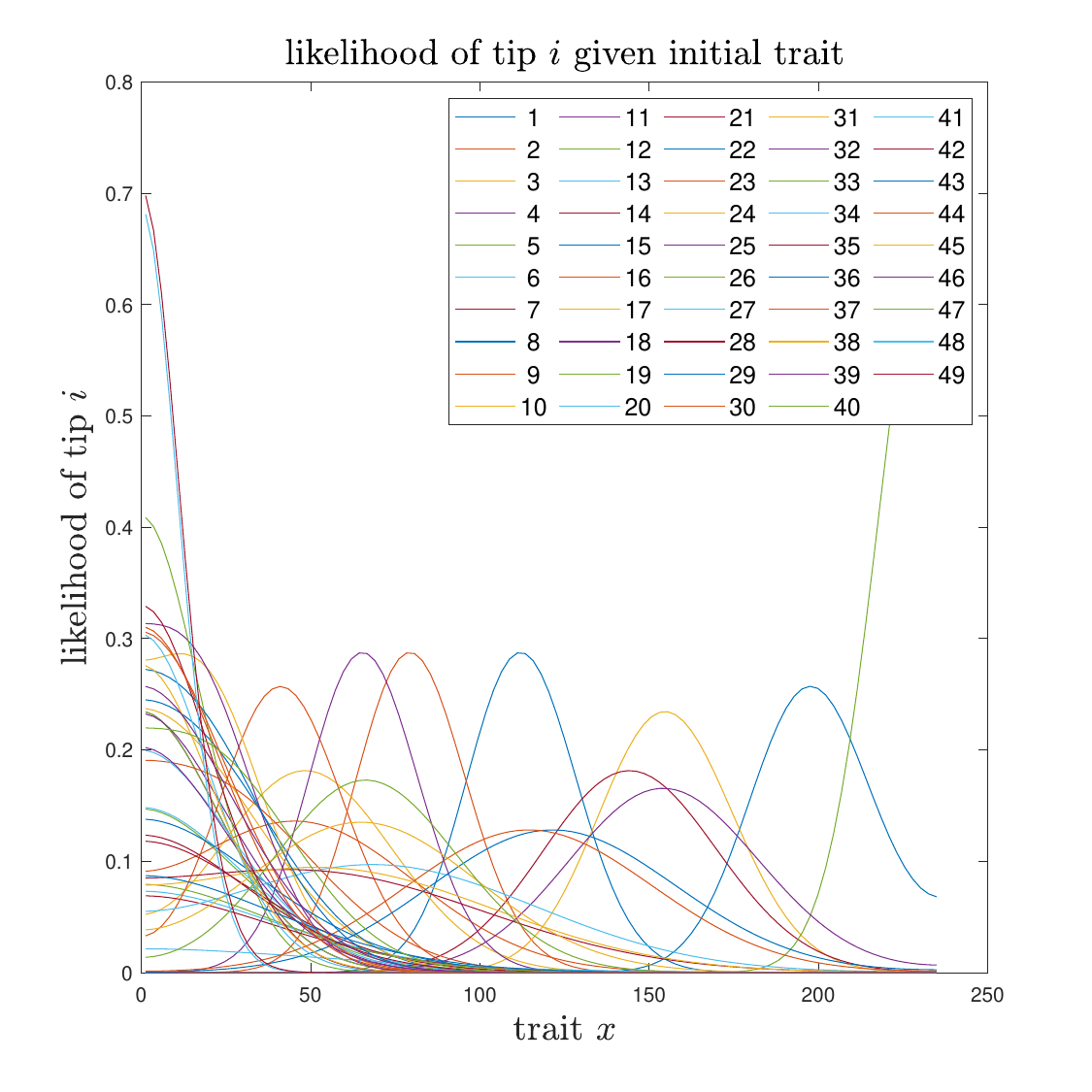}
\\
\includegraphics[scale=0.4]{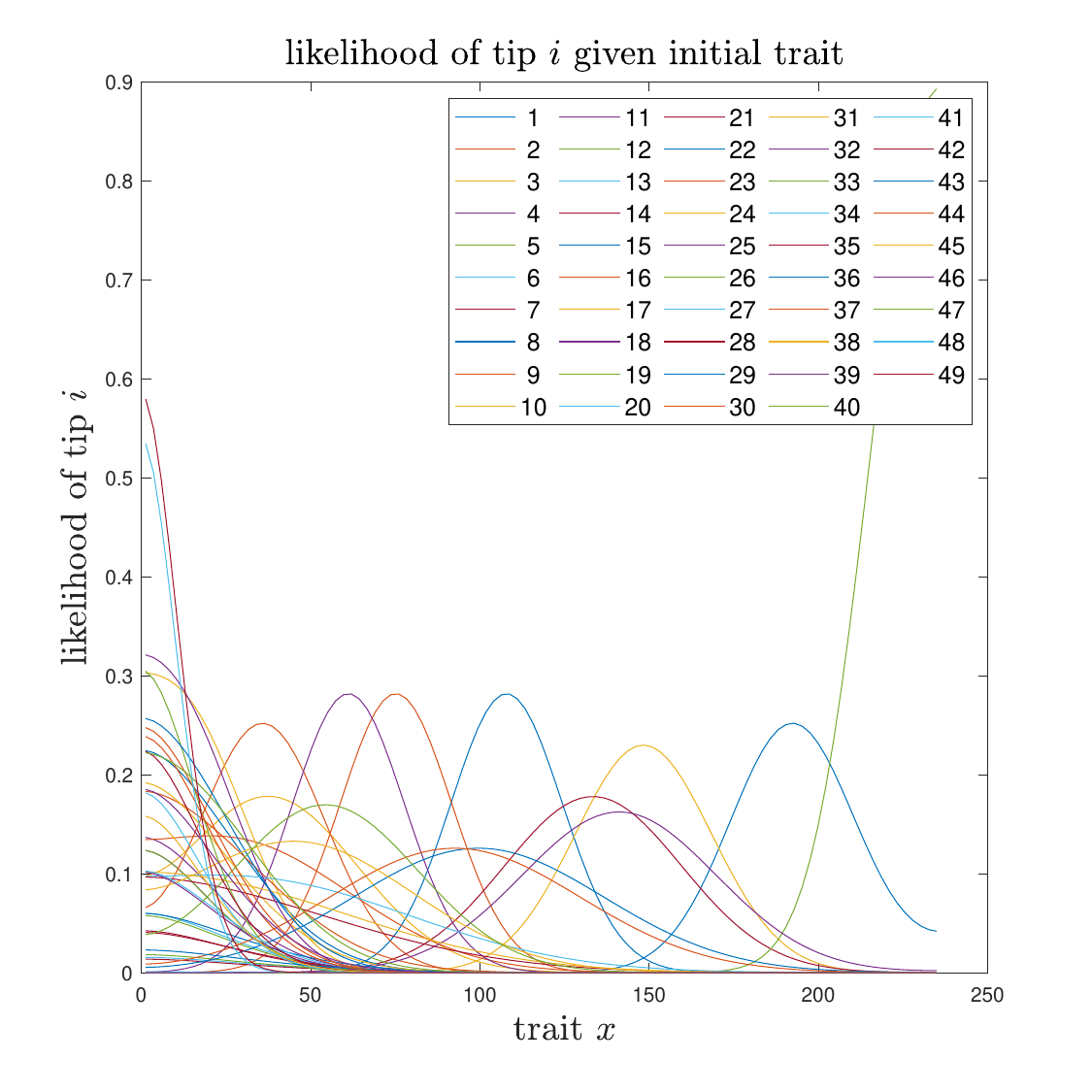}
\caption{From top left to the bottom right: The likelihood of observing tip $i$ given trait observed at the start of the branch corresponding to tip $i$, in the \protect\hyperlink{QBD3}{QBD3} model in Section~\ref{sec:QBDmodels} and Empirical Dataset~2 (Figure~\ref{DataExample3}), for the ${\bf r}$ vectors $\#1-\#5$ in Table~\ref{tab:exampleDATA2B}.}
\label{StatTraitsModel5phasesOverallcherriesEmpiricalData3}
\end{figure}

\begin{figure}
\centering
\includegraphics[scale=0.4]{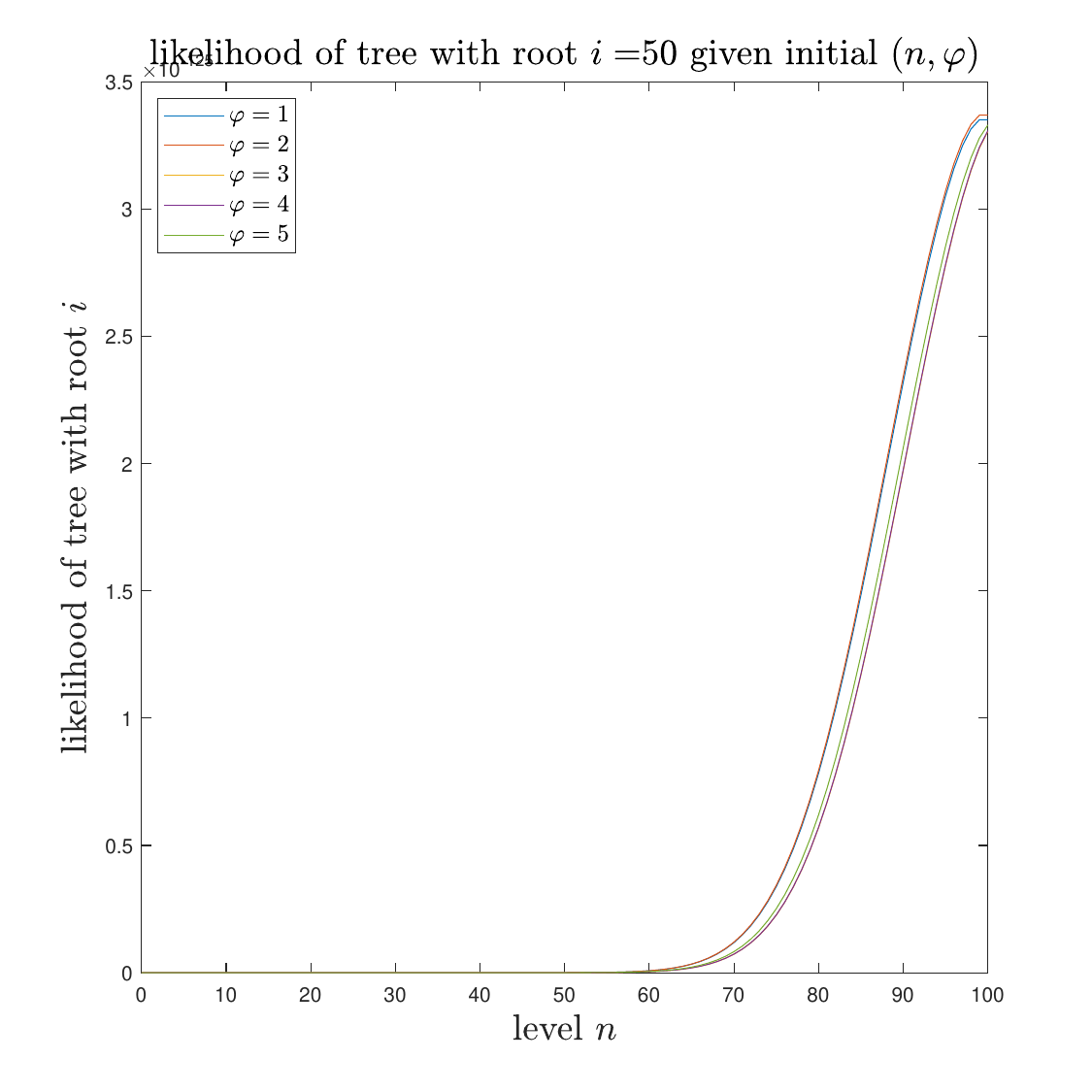}
\
\includegraphics[scale=0.4]{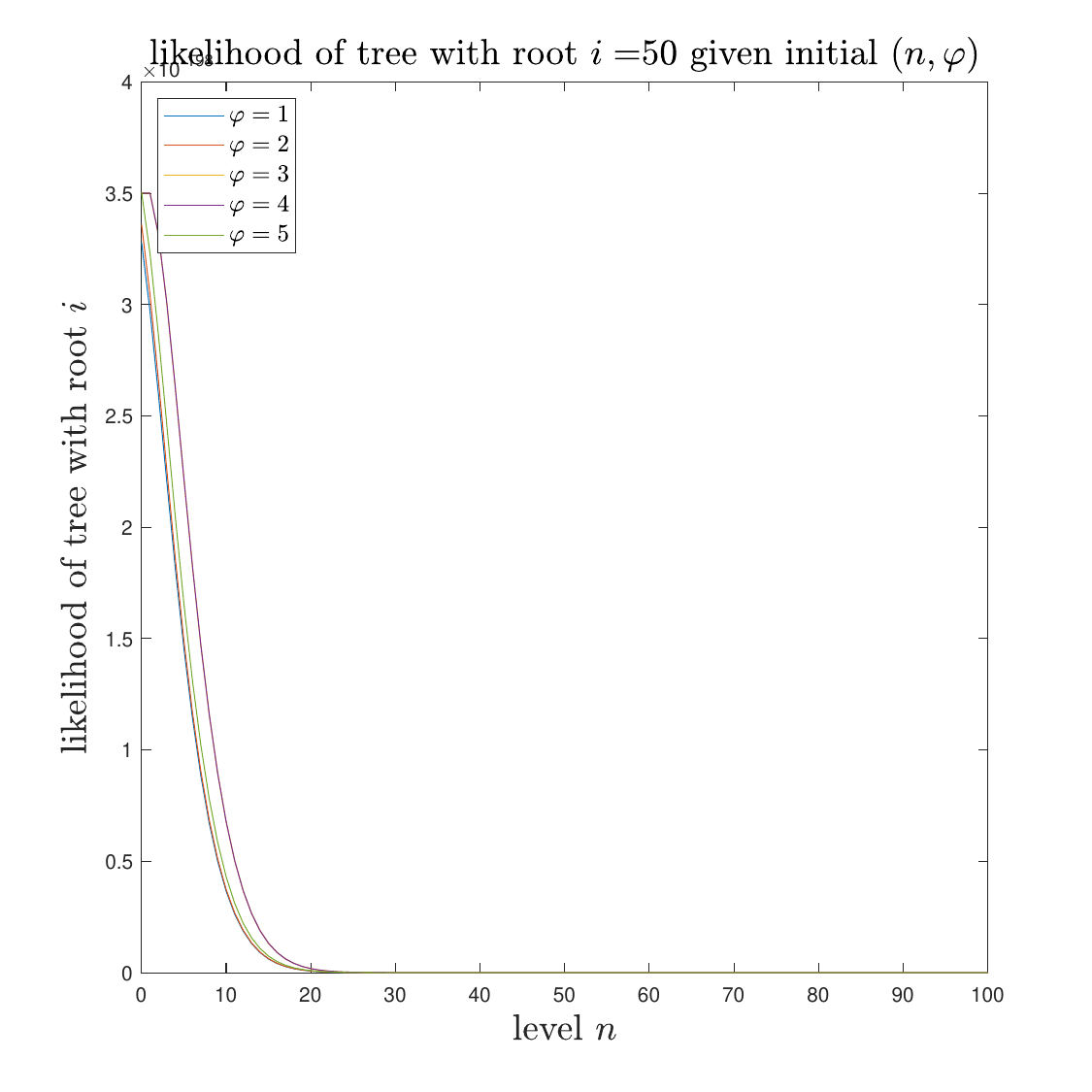}
\\
\includegraphics[scale=0.4]{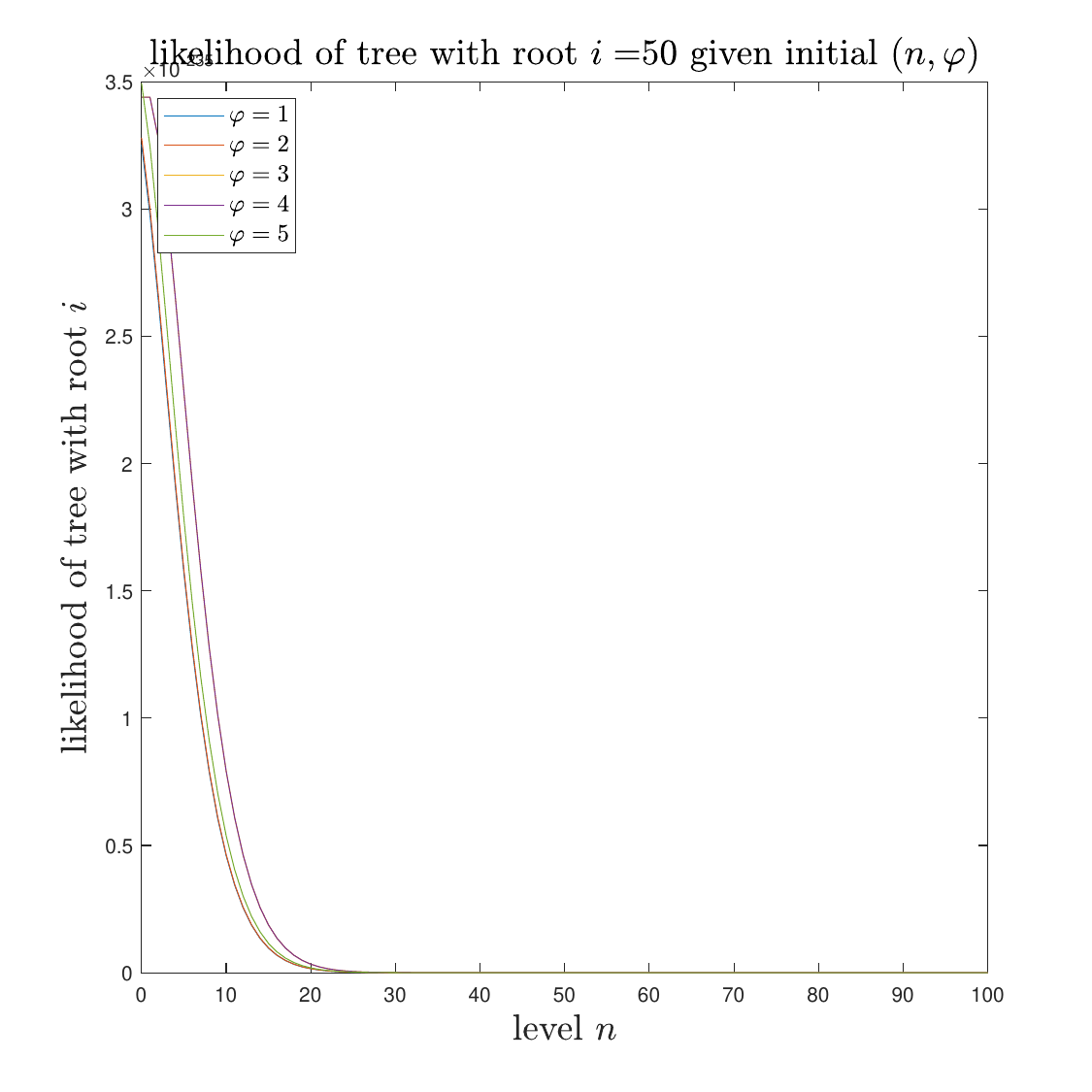}
\
\includegraphics[scale=0.4]{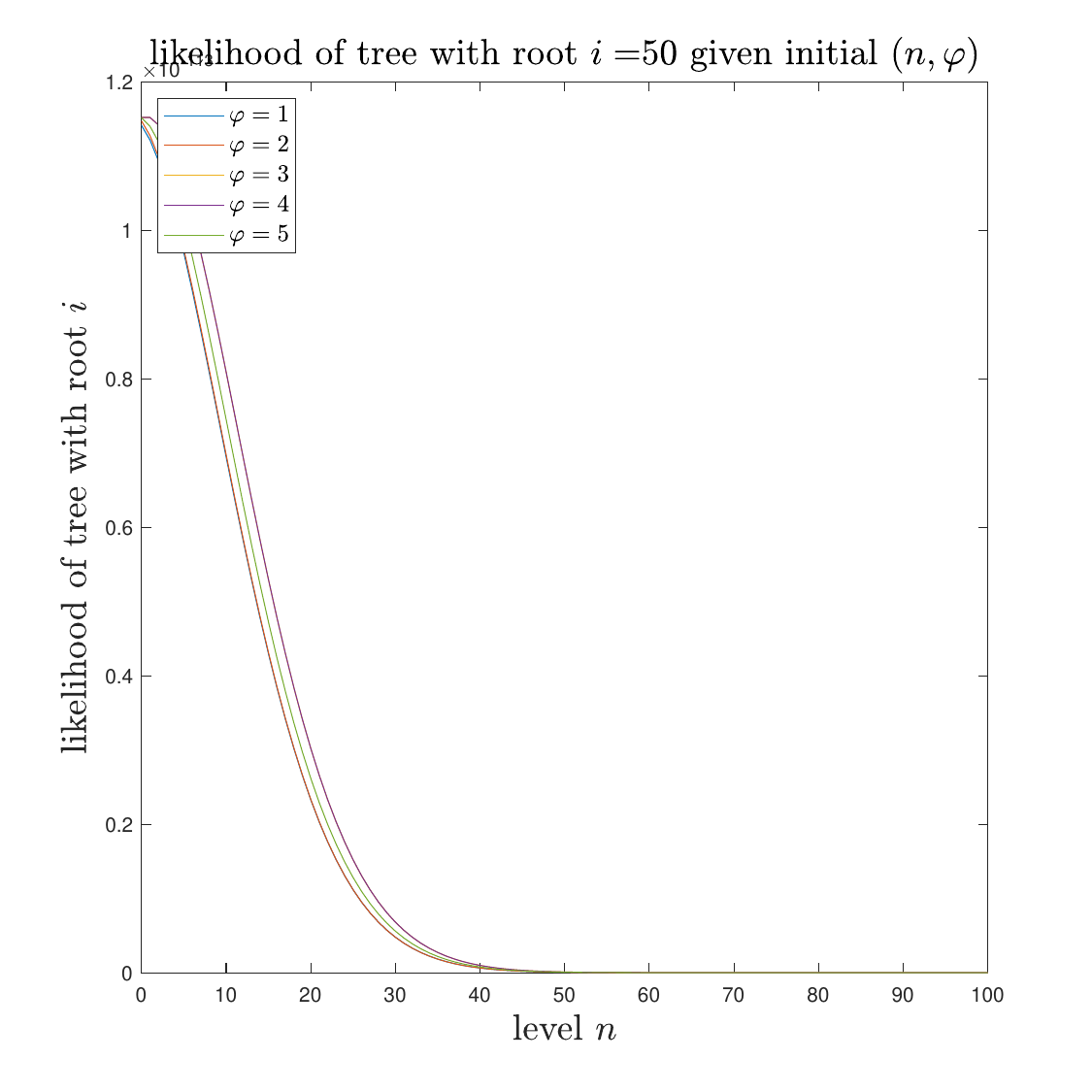}
\\
\includegraphics[scale=0.4]{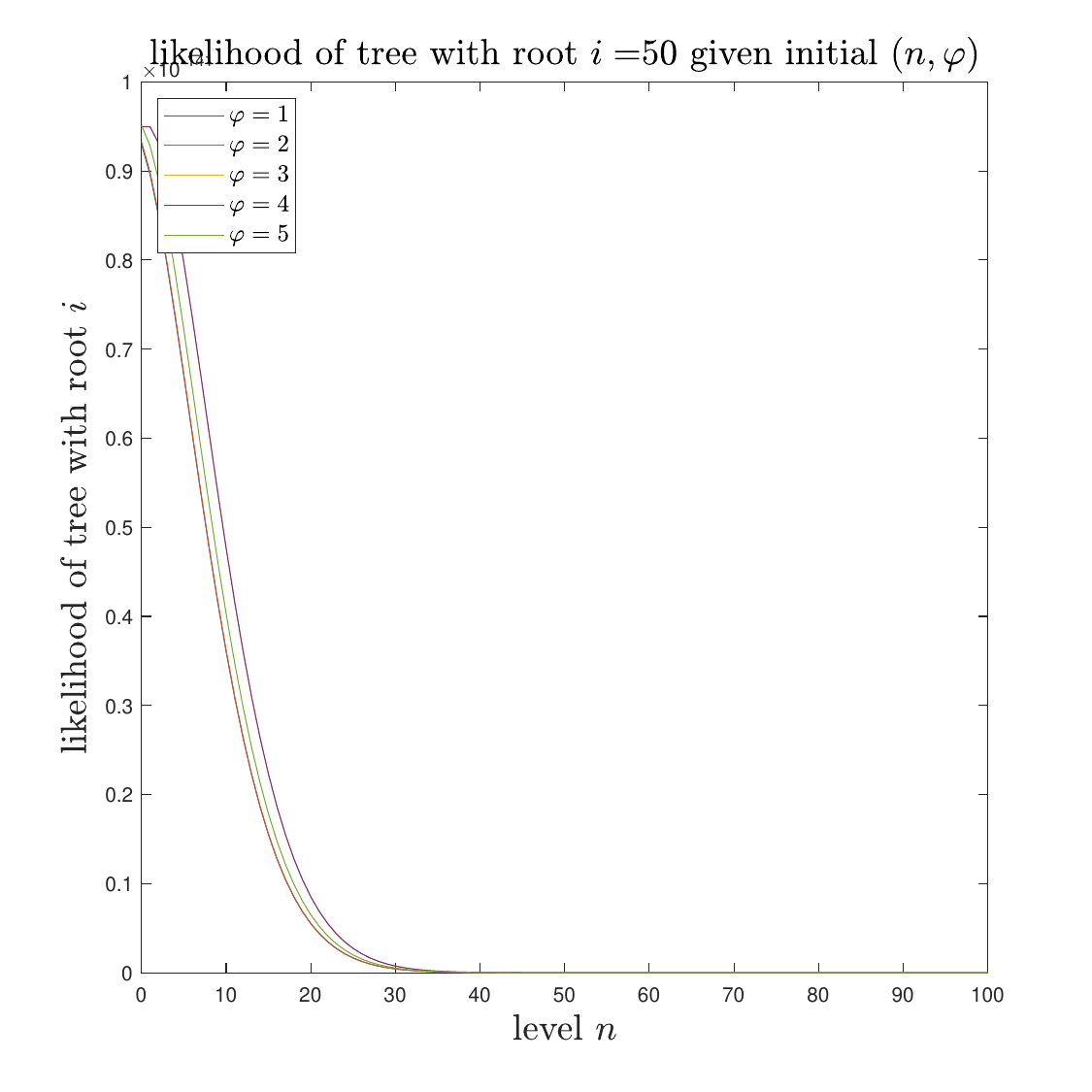}
\caption{From top left to the bottom right: The likelihood of observing the phylogenetic tree that started with parent $i$ given level $n$ and phase~$\varphi$ observed at the start of the tree, in the \protect\hyperlink{QBD3}{QBD3} model in Section~\ref{sec:QBDmodels} and Empirical Dataset~2 (Figure~\ref{DataExample3}), for the ${\bf r}$ vectors $\#1-\#5$ in Table~\ref{tab:exampleDATA2B}.}
\label{StatTraitsModel5phasesOverallParrentEmpiricalData3}
\end{figure}

\begin{figure}
\centering
\includegraphics[scale=0.4]{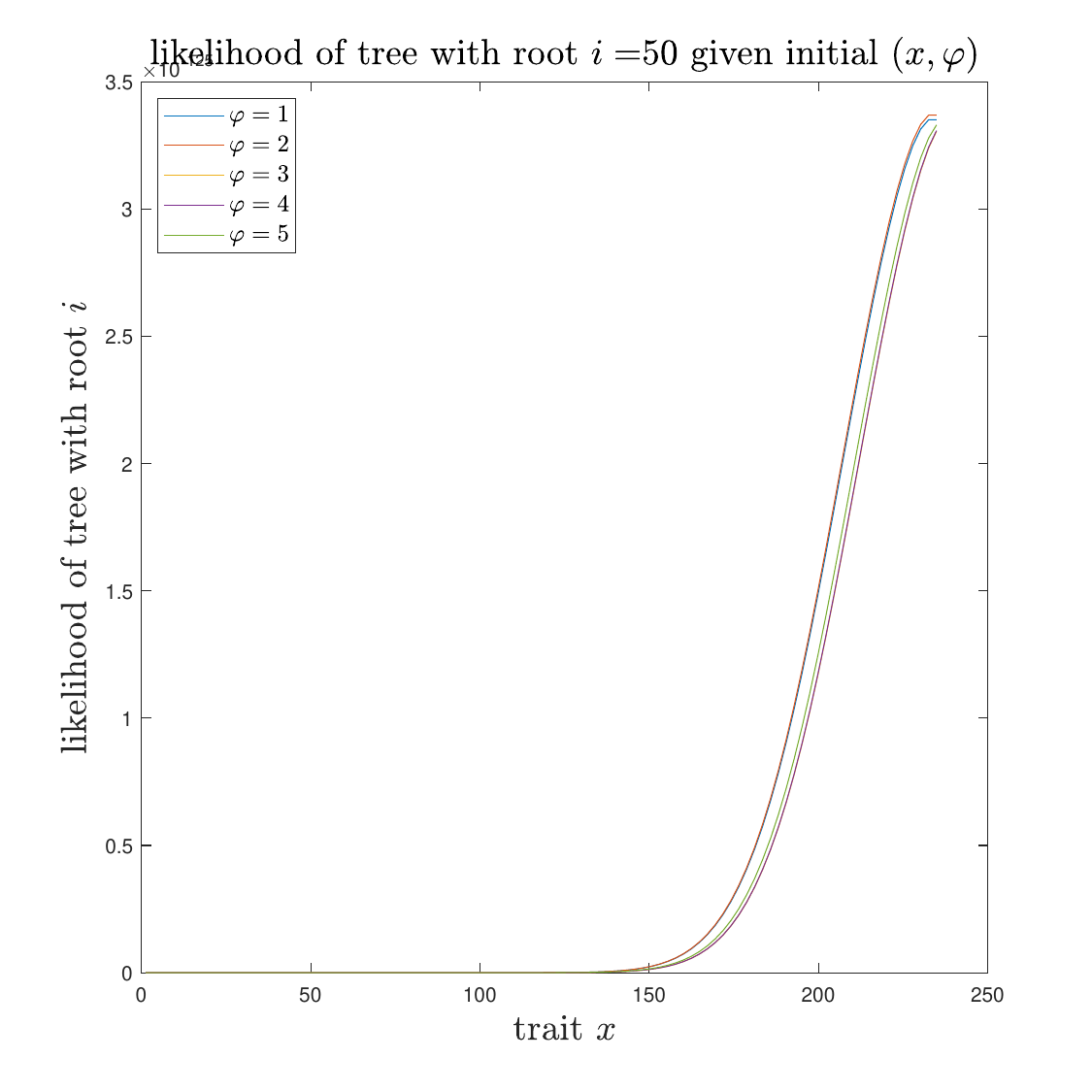}
\
\includegraphics[scale=0.4]{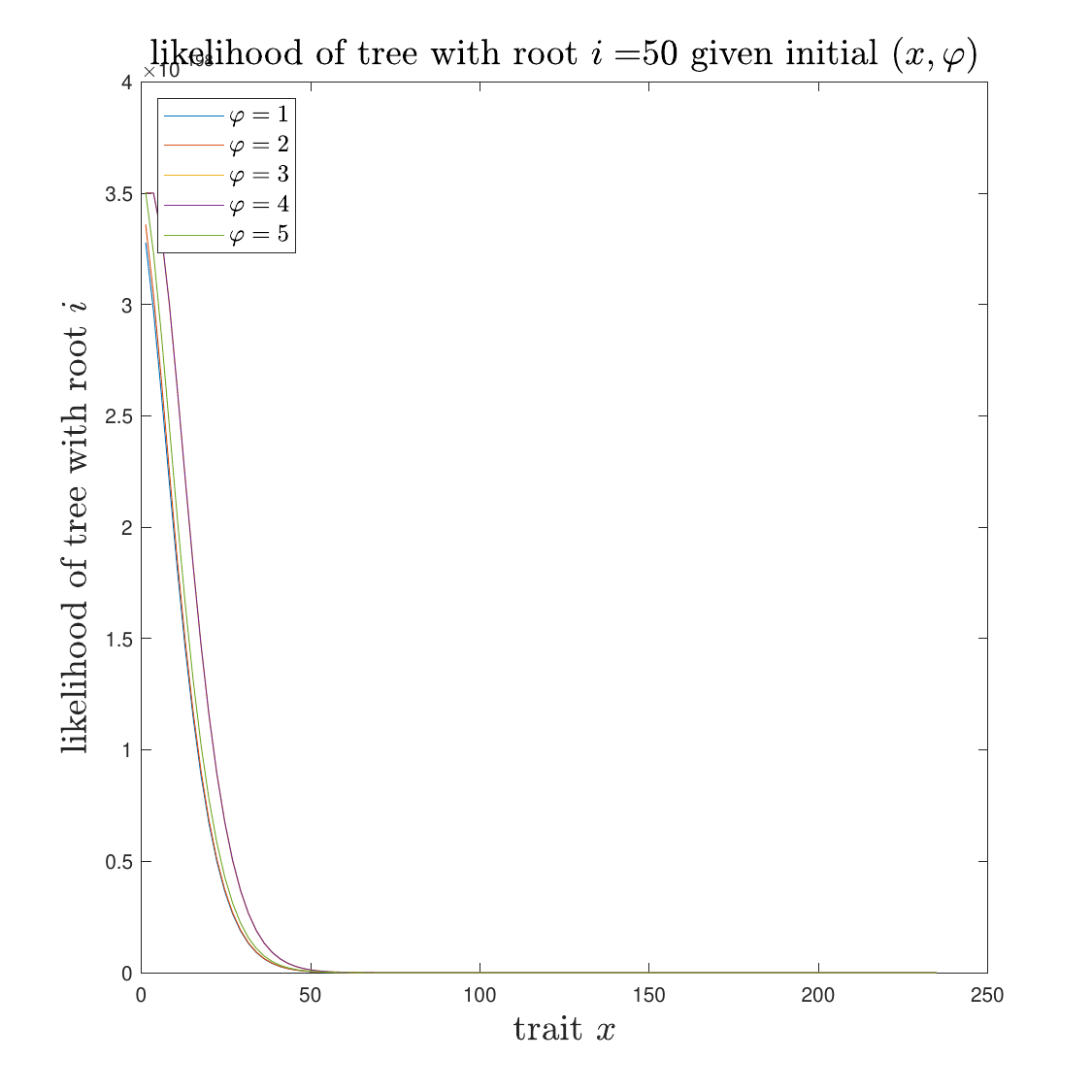}
\\
\includegraphics[scale=0.4]{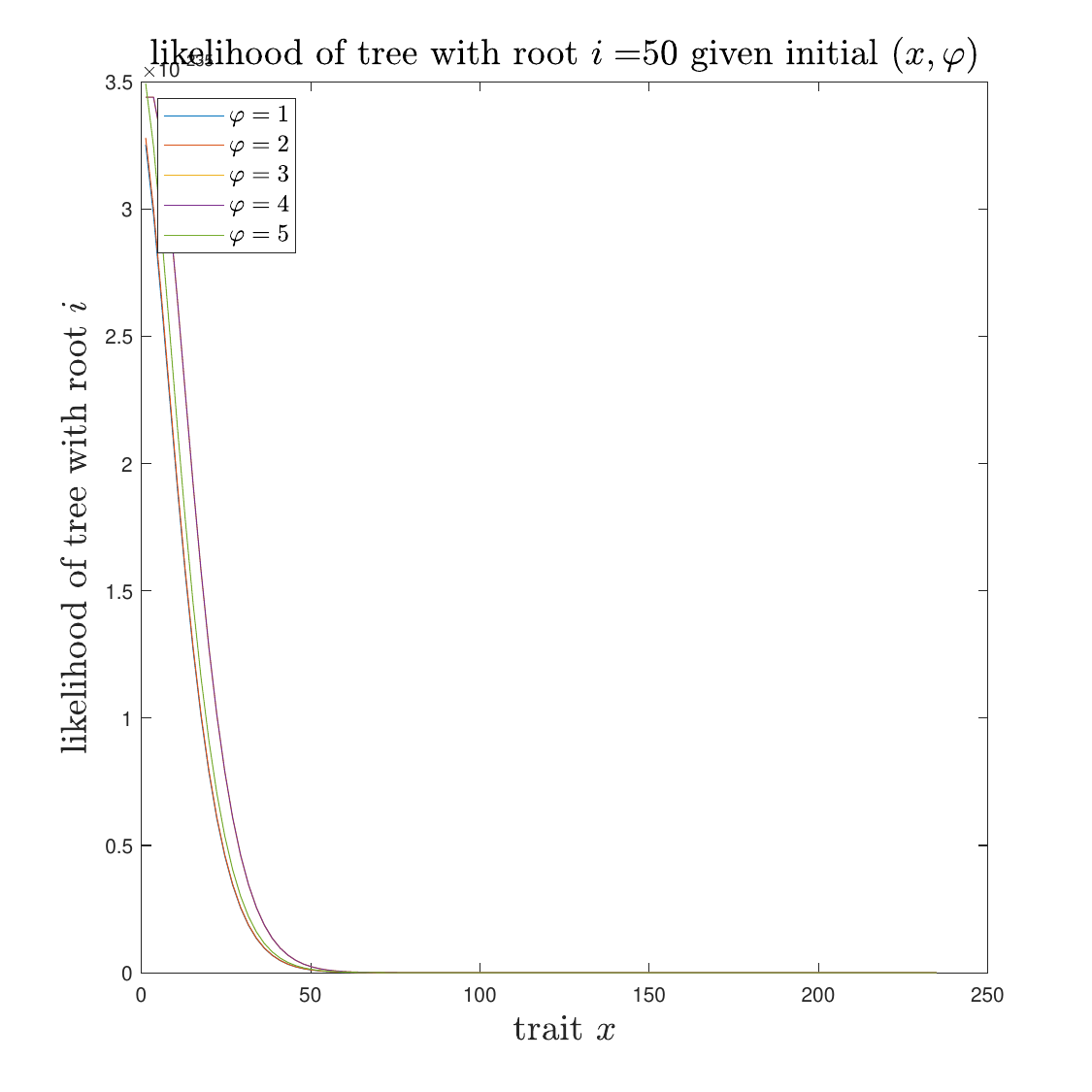}
\
\includegraphics[scale=0.4]{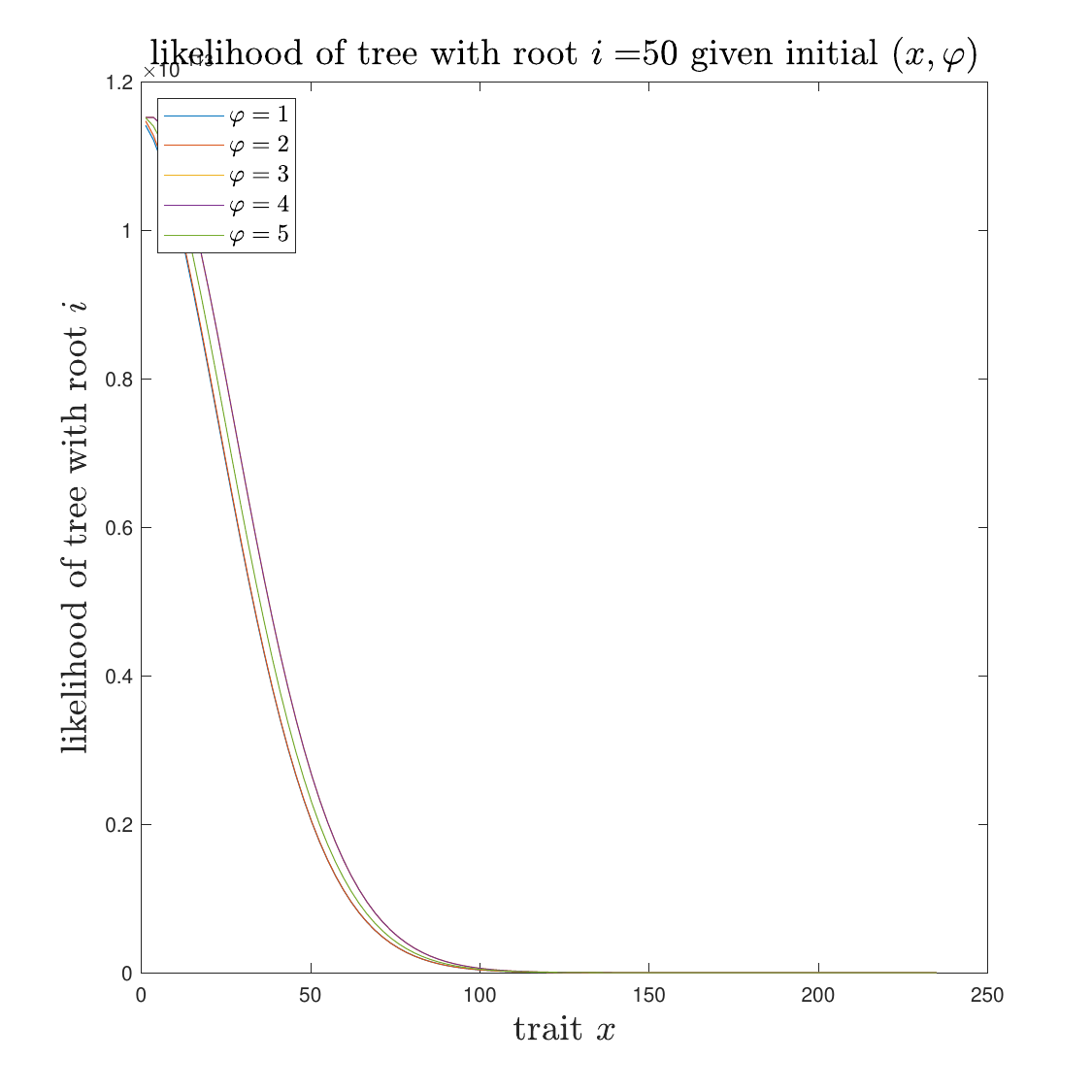}
\\
\includegraphics[scale=0.4]{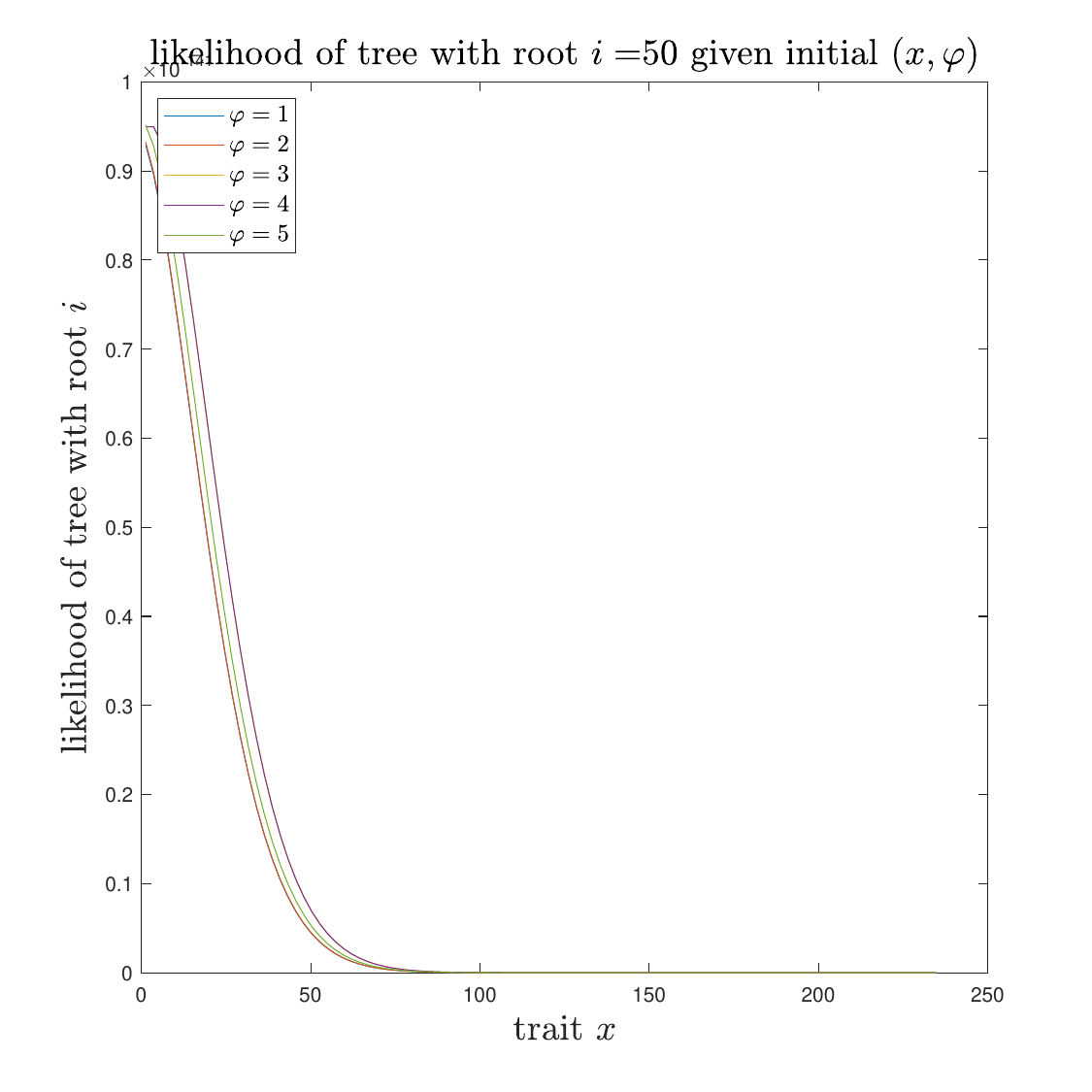}
\caption{From top left to the bottom right: The likelihood of observing the phylogenetic tree that started with parent $i$ given trait~$x$ and phase~$\varphi$ observed at the start of the tree,in the \protect\hyperlink{QBD3}{QBD3} model in Section~\ref{sec:QBDmodels} and Empirical Dataset~2 (Figure~\ref{DataExample3}), for the ${\bf r}$ vectors $\#1-\#5$ in Table~\ref{tab:exampleDATA2B}.}
\label{StatTraitsModel5phasesOverallParrentEmpiricalData3}
\end{figure}

\begin{figure}
\centering
\includegraphics[scale=0.4]{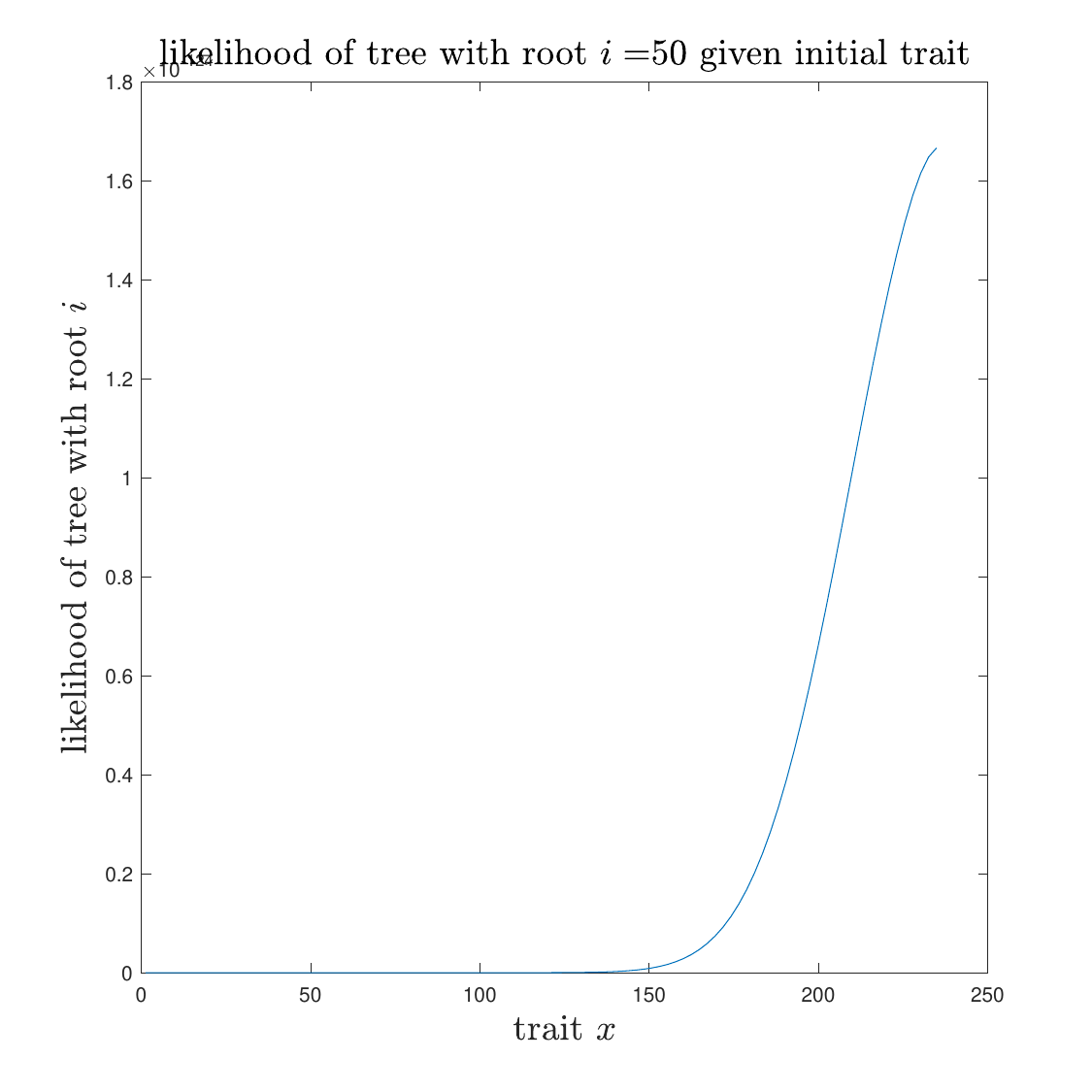}
\
\includegraphics[scale=0.4]{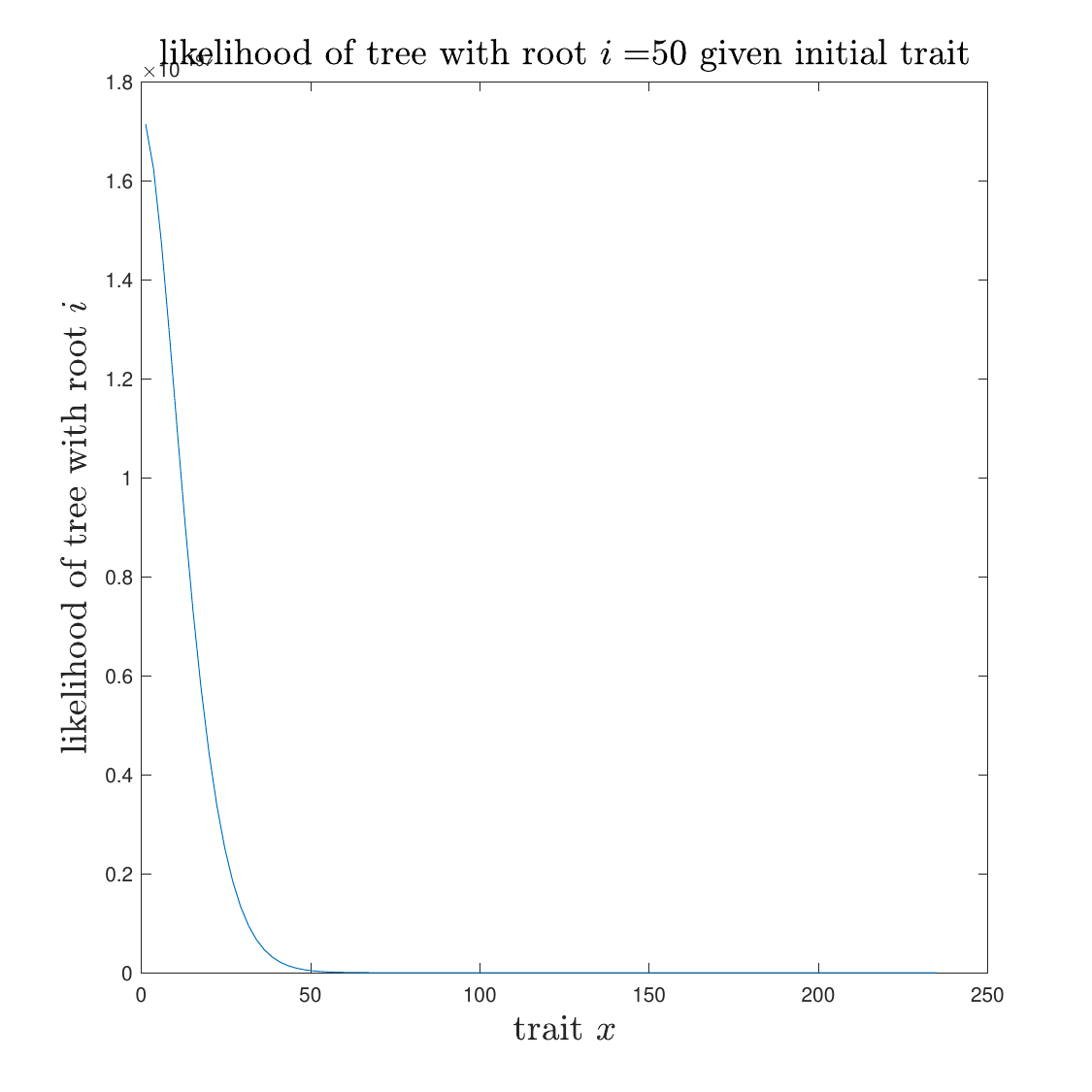}
\\
\includegraphics[scale=0.4]{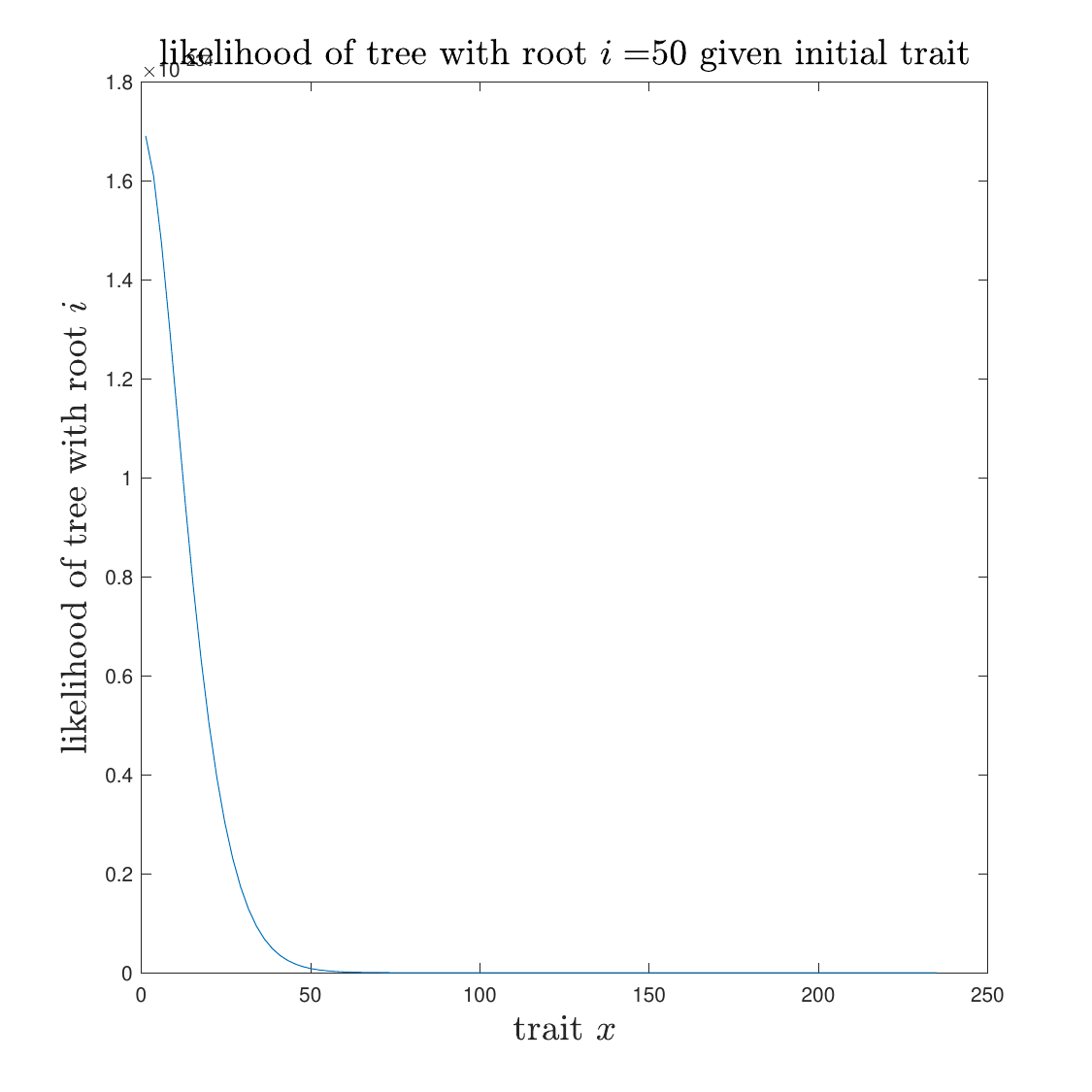}
\
\includegraphics[scale=0.4]{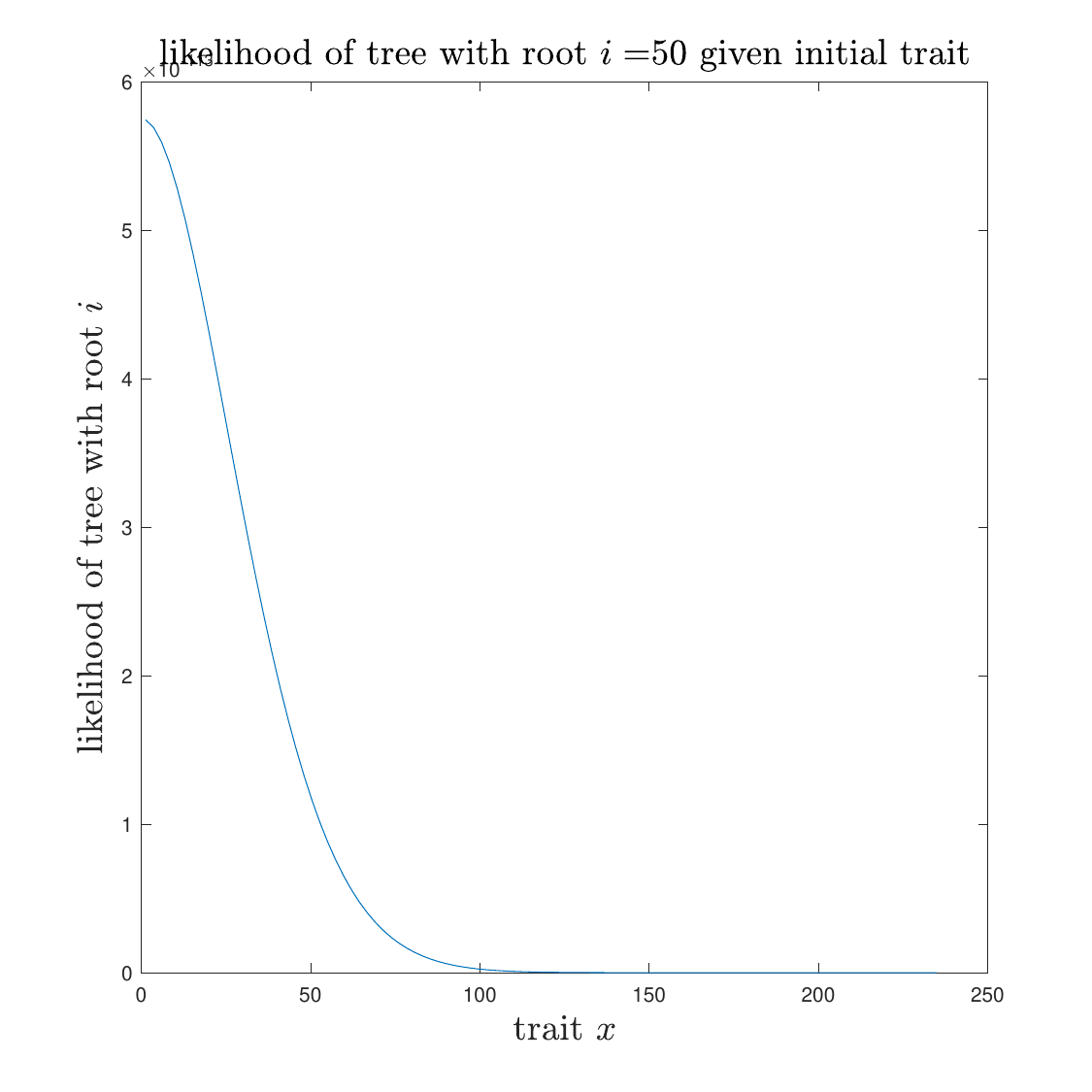}
\\
\includegraphics[scale=0.4]{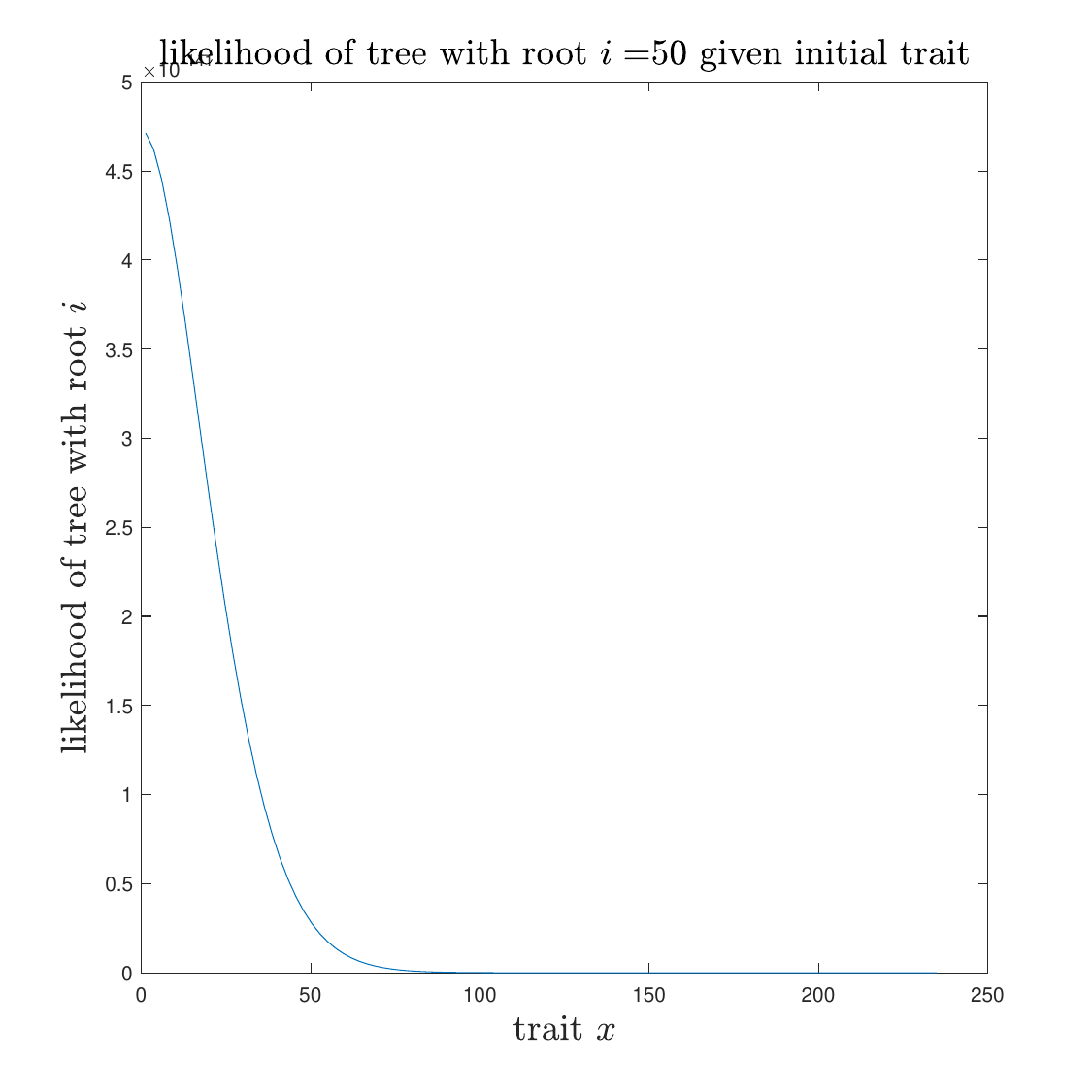}
\caption{From top left to the bottom right: The likelihood of observing the phylogenetic tree that started with parent $i$ given trait~$x$ observed at the start of the tree, in the \protect\hyperlink{QBD3}{QBD3} model in Section~\ref{sec:QBDmodels} and Empirical Dataset~2 (Figure~\ref{DataExample3}), for the ${\bf r}$ vectors $\#1-\#5$ in Table~\ref{tab:exampleDATA2B}.}
\label{StatTraitsModel5phasesOverallParrentTraitsOnlyEmpiricalData3}
\end{figure}

\end{document}